\documentclass[12pt,twoside]{report}
\usepackage{fullpage}
\usepackage{amsmath}
\usepackage{amssymb}
\usepackage{epsfig}
\usepackage{subfigure}
\usepackage{feynmf}

\begin{document}

\thispagestyle{empty}
\centerline{\sf\large\normalsize UNIVERSITY OF LJUBLJANA}

\centerline{\sf\large FACULTY OF MATHEMATICS AND PHYSICS}

\centerline{\sf\large PHYSICS DEPARTMENT}
\vspace{6cm}
\centerline{\it\Large Sa\v sa Prelov\v sek Komelj}
\vspace{1cm}
\centerline{\bf \Large Weak Decays of Heavy Mesons}
\vspace{0.4cm}
\centerline{\sf\normalsize DOCTORAL THESIS}
\vspace{65pt}
\centerline{{\tt\large SUPERVISOR:}{\it\large \, prof. dr. Svjetlana Fajfer}}
\vspace{4cm}
\centerline{\it\normalsize Ljubljana, 2000}

\newpage
\thispagestyle{empty}

\centerline{\bf Abstract}

\vspace{0.5cm}

 The weak decays of heavy mesons - bound states of a quark and an anti-quark, 
at least one of which carries heavy flavour $c$ or $b$ - enable us to probe the validity of the 
standard model of elementary particle interactions and   determine 
 several parameters of this model. 

I explore the possibility of using heavy meson decays as probes for  flavour 
changing neutral transitions (FCNC) $c\to u\gamma$ and $c\to ul^+l^-$. In the standard model, these  are the most frequent flavour changing neutral transitions among the quarks with charge $2/3e_0$  and  have enhanced sensitivity to conjectured physics beyond the standard model.  In  hadron decays, a flavour changing neutral transition  can be severely overshadowed by  long distance contributions.  I calculate the probabilities for the relevant  heavy meson decays  within the standard model and explore their sensitivity to different scenarios of physics  beyond the standard model.  The 
$B_c\to B_u^*\gamma$ decay is proposed as the most suitable case study of the 
$c\to u\gamma$ transition. Using the Isgur-Scora-Grinstein-Wise model, I predict the branching ratio for this decay to be of the order of $10^{-8}$. Its detection at a higher rate would signal new physics.  Weak decays of charm mesons to a light meson and a photon or a charged lepton pair are also studied for this purpose. Their probabilities are found to be dominated by the long 
distance contributions and   raise our hopes that they may be detected soon. A window of opportunity to probe the $c\to ul^+l^-$ transition  is found  in the decay $D\to \pi l^+l^-$, at the kinematical region of high di-lepton mass $\sqrt{(p_{l^+}\!+\!p_{l^-})^2}$.  In order to study the charm meson decays, I adapt a hybrid model which combines heavy quark effective theory and  chiral perturbation theory.  By predicting the nonleptonic decay rates, I demonstrate for the applicability of the model. I also propose a mechanism to incorporate the long distance contributions in a manifestly gauge invariant way. 

\newpage
\thispagestyle{empty}

\centerline{\bf Acknowledgements}

\vspace{0.5cm}

{\it   Special  thanks go to my advisor prof. dr. Svjetlana Fajfer. Her assistance, guidance and encouragement have been of crucial importance for this work. I would like to thank prof. dr. Paul Singer for all the enthusiasm  he has shared with me and for the successful collaboration.  I am grateful for the hospitality at the Israel Institute of Technology, where part of this work has been done. 
I have  benefited greatly from the numerous conversations with  prof. dr. Matja\v z Polj\v sak, Jure Zupan, prof. dr. Mitja Rosina and dr. Borut Bajc. 
I should especially thank to my closed ones for their love and support. 

The work has been performed at the ``Department of Theoretical Physics'' at ``Jozef \v Stefan Institute'', Ljubljana, Slovenia. It was financially supported by the Ministry of Science and technology of the Republic of Slovenia. }

\tableofcontents

\chapter{Introduction}

The present understanding of elementary particle interactions  is based on the 
quantum field theory. The form of the interactions among the quarks and leptons 
on one side, and the bosonic carriers of the interaction on the other,  
arises by imposing the invariance under the space-time
dependent symmetry group. The present knowledge contained in the standard model 
of electro-weak and strong interactions is based on the $U(1)_Y$ transformations 
according to the particle's hypercharge, $SU(2)_L$ transformations of weak 
isospin left-handed doublets and $SU(3)_c$ transformations in the color space of 
quarks \cite{SM}. The masses of the leptons, quarks and weak gauge bosons arise 
by  spontaneous symmetry breaking which keeps the equations of motion, but not 
the vacuum,  invariant under the symmetry group.  The Higgs mechanism of 
spontaneous symmetry breaking introduces an additional scalar weak doublet with 
a nonzero vacuum expectation value, invariant only under the subgroup $U(1)_{EM}$, 
but not on the whole group $SU(2)_L\times U(1)_Y$. The physical fluctuations around 
this vacuum are represented by the Higgs boson which has not been detected yet. The standard model constructed in this way is renormalizable, meaning that 
all the infinities arising from the theory can be removed in a physically 
sensible way by redefining the free parameters of the theory. 

The quark fields in the $SU(2)_L$ doublets are not  the mass 
eigenstates in general. The quarks of a given charge are rotated from the weak to the mass 
eigenstate basis by means of the unitary matrix. The 
coupling between a quark with  charge $-1/3~e_0$, a quark with charge $2/3~e_0$  and a $W$ 
boson is given by the unitary Cabibbo Kobayashi Maskawa (CKM) mixing matrix. 
This matrix can be parameterized in terms of three real and one imaginary 
parameter  and the nonzero value of the imaginary parameter  offers the 
possibility to account for the charge-parity (CP) violation in the standard 
model \cite{CKM}. The electromagnetic  and the neutral weak currents are 
rendered flavour diagonal and there are no flavour changing neutral currents at 
the three level.

The standard model interactions of elementary particles have been
extensively tested in accelerator facilities and agree well with the
data measured up to the energies available at present. Three generations of 
quarks and leptons are experimentally established, together with the 
corresponding gauge bosons. 
The only elementary building block lacking experimental detection is the Higgs 
boson.  Among the free parameters of the model\footnote{The free parameters of the standard model are masses of the elementary 
particles, CKM mixing angles and coupling constants for strong, electromagnetic 
and weak interactions},  the value of the imaginary phase in the CKM matrix 
is the most uncertain, while the Higgs mass is still unknown. At present, the 
value of the imaginary parameter in the CKM matrix is experimentally 
constrained by the CP conserving processes and the CP violating kaon decays. If 
the forthcoming measurements of the CP violation in $B$ meson decays will 
confirm this value, the CP violation is indeed due 
to the imaginary phase in the CKM matrix \cite{CKM}. 

In spite of the many experimental successes of the standard model in describing 
the
elementary particle interactions, it is widely
believed that it is not the ultimate fundamental theory. One of the  main
reasons for this is that it accounts only for electromagnetic, weak and
strong interactions. The  fourth fundamental interaction - gravity
- has not been quantized in the same way as the other three and has
not been included in the standard model. Even without the failure to
account for gravity, which is extremely weak among the elementary
particles anyway,  the model  fails to meet several aesthetics wishes. It does
not unify the other three fundamental interactions. The Higgs
mechanism of electroweak symmetry breaking  has
no dynamical explanation, it is imposed  and  renders all the masses
of the elementary particles as free parameters. 

Additional unsatisfactory
property from the aesthetic point of view is connected with the fact that some values of 
the free parameters
are much smaller than expected by considering
the symmetries of the model.  The first example of this kind is the Higgs boson 
mass.  While the masses of
fermions and gauge bosons are not invariant under the local
$SU(2)_L\times U(1)_Y$ transformation and are naturally of the order of the
$SU(2)_L\times U(1)_Y$ breaking scale $v=250$ GeV, the Higgs mass term
respects the whole symmetry group  $SU(3)_c\times
SU(2)_L\times U(1)_Y$ and would naturally be much larger than
$v$. Precision measurements of the electroweak parameters constrain the
Higgs mass through its effects in radiative corrections to be
$m_H<450$ GeV at the $95~\%$ confidence level \cite{PDG} and the standard
model does not offer any explanation as to why the Higgs mass is so
small. The second example is given by the neutrino masses. In the so
called minimal standard model, neutrinos are imposed to have  zero 
masses, although this is not required by any of the gauge
symmetries. Without the gauge 
symmetry preserving the  masses of the neutrinos, it has been
suspected that the neutrinos have tiny masses. This has indeed been
confirmed recently in the very convincing data on the oscillations of
the atmospheric neutrinos \cite{SK}. The massive neutrinos can be
accounted for by a slight extension of the minimal model, but the
neutrino mass parameters have to be assigned unnaturally small values 
in  this case. Another parameter, which has to be assigned an 
exceedingly small value required by the data on the neutron dipole 
moment, is connected to the CP violation
in the strong interactions \cite{cheng.li}. This kind of aesthetic
reasoning  does not 
exclude the standard model by itself,   but it may point to possible new 
symmetries
behind it. 

Potentially more serious threats come from the inability of
the model to
satisfy all the cosmological bounds coming from the unique high
energy experiment in the early universe. For example, it does
not explain  inflation - a period of exponential expansion in the early stage of 
the universe which solves many of the cosmological problems.  

These and many other reasons call for a physics beyond the standard model. 
If a scenario of new physics explains the mechanism of electroweak symmetry
breaking, it has to reside at an energy scale not far beyond
$1$ TeV. The scenarios of new physics in general predict a set of new
particles in addition to those present in the standard model. The
experimental challenge of finding the new physics follows two main
directions. In direct searches the idea is to produce the new
particles and detect them directly. This may take some time if the
states are set at several hundred GeV. A complementary idea is to  measure
 the effects of  the new particles in the processes where they enter as the
intermediate virtual states. These effects are 
expected to be relatively small in the processes that can occur in the standard 
model at the tree level. The effects, arising from the presence of  intermediate new states,  are expected to be  relatively more
significant in the rare processes that occur only at the loop level in the
standard model. Up to now, no significant signal for the physics beyond the
standard model has been seen at the available experimental facilities 
\footnote{The recent announcement of the
nonzero neutrino masses \cite{SK} may be regarded as the signal of the new
physics since the neutrinos are massless in the minimal standard
model. Although unnatural, it is extremely easy to account for the
tiny neutrino masses by the slight extension of the minimal model.}. 

In the present work I explore the possibility of probing the standard
model and the physics beyond it, in processes induced by the flavour changing 
neutral currents (FCNC). These are currents that change the flavor, but not the 
charge of the quark, and occur only through electro-weak loops in the standard 
model. Loop
processes are  sub-leading in the perturbative expansion, they are
 rare  and therefore relatively more
sensitive for the possible new physics. In the case of 
transitions among the down-like $2/3~e_0$ charged quarks $d$, $s$ and
$b$, the up-like $-1/3~e_0$ charged quarks $u$, $c$ and $t$  run in
the loop, and vice versa. If the masses of the intermediate quarks
were equal, their contributions would cancel due to the unitarity of
the CKM matrix. This is known as the Glashow, Iliopoulos and Maiani
mechanism \cite{GIM} and indicates that FCNC processes are  suppressed
even at the loop level, if the masses of the intermediate quarks are not very 
different. As a result, the
intermediate quarks of higher masses give higher rates to FCNC processes as long 
as their mixing
with the external quarks is not highly suppressed. Up to now, only the
transitions $s\to d$ and $b\to s$ have been experimentally
established. The transition between two quarks of the
same charge and different flavour has to be accompanied by the emission of 
particles with
zero net charge: photon, gluon or lepton-antilepton pair. 

Due to the confinement of the strong interaction in quantum chromodynamics (QCD) 
quarks can not be
observed as free particles. They are confined to hadrons and the
FCNC quark decays like $b\to s\gamma$  are probed in the corresponding hadron 
decays.  A 
hadron decay of interest is induced by the FCNC quark decay, but it may  also be 
induced by a different mechanism. Two mechanisms leading to 
the same initial and final states can not be distinguished by the basic
principles of quantum mechanics. The mechanism, which can potentially overshadow 
the
quark decay of interest, is called the long distance mechanism and involves the 
intermediate
hadrons which propagate over relatively large distances. The 
intermediate hadrons propagate
almost on shell, the strong interaction is in the nonperturbative
regime and the reliable theoretical treatment of the long distance
contribution from first principles is very difficult at present,
if not impossible. The part  induced by the FCNC quark decay is called the short 
distance contribution and  is theoretically under better control. It 
involves the quark decay via the
electroweak loop at short distances, the states in the loop are highly
virtual and they allow for the perturbative treatment. The only nonperturbative
physics entering in the short distance contribution is due to the  hadronization 
of the initial and final
quarks into the initial and final hadrons, respectively. The critical
aspect of probing the FCNC interactions in the hadron decays is
obviously related with our ability to disentangle the FCNC decay of
interest from the long distance dynamics.  

Until recently, only
the $s\to d$ flavour changing neutral transition has been experimentally 
established. 
Given the fact that the branching ratio for the $K_L^0\to \mu^+\mu^-$
decay was measured to be only of the order  of $10^{-9}$,  Glashow, Iliopoulos 
and Maiani predicted the existence of the charm 
quark  \cite{GIM}. At that time, the charm quark was the missing block of
the two $SU(2)_L$ doublets and with all the quarks  paired in the weak  doublets, the $s\bar d\to \mu^+\mu^-$  amplitude automatically 
vanishes at  the tree
level.  The possibility of probing the short distance process $s\to d
l\bar l$  ($l$ denotes a lepton) in $K\to \pi l\bar l$
decays  has also been under 
intense experimental and theoretical investigation. The $s\to dl\bar l$ occurs 
via the $W$ box and $\gamma,~Z$ penguin diagrams with the largest contribution 
arising from the intermediate top quark. The amplitude for the process is 
proportional to $V^*_{ts}V_{td}m_t^2$ and is small due to the relevant CKM matrix elements.   
The
$K^+\to \pi^+l^+l^-$ decay is dominated by the long distance contribution and 
can not serve as  a probe for the $s\to dl^+l^-$ decay. The long distance 
contribution in this channel arises  via $W$ exchange $u\bar s\to u\bar d$ which 
induces the  $K^+\to\pi^+$
transition, followed by the  photon  emission from
$K^+$ or $\pi^+$ and photon conversion to 
$l^+l^-$. This mechanism gives the branching ratio of the order of $10^{-7}$ \cite{rare.kaon,lichard.brem,burdman1}
and has been experimentally confirmed by the detection of this channel 
\cite{PDG}. This disturbing long distance contribution is 
absent in the CP
violating $K_L\to\pi^0 l^+l^-$ decay which can not proceed through CP
conserving one-photon exchange \cite{burdman1}. Due to the fact that $K_L$ is not a 
pure $CP$ odd state, there actually remains a small one-photon
exchange contribution and this decay remains in the list of long
distance polluted modes. The prediction for the branching ratio at the 
 level $10^{-11}$ is two orders of magnitude smaller than the present 
experimental limit \cite{isidori}. The processes with the neutrino final states 
$K^+\to \pi^+\nu\bar \nu$ and $K_L\to \pi^0\nu\bar\nu$
are almost completely determined by the transition $s\to d\nu\bar\nu$ at short 
distances, especially the  CP violating $K_L\to \pi^0\nu\bar\nu$ decay. The 
golden plated decay $K_L\to \pi^0\nu\bar \nu$ is predicted at the branching 
ratio $(3.1\pm 1.3)\times 10^{-11}$ \cite{isidori}, with the uncertainty coming 
from the present uncertainties of the CP violating parameters in the CKM matrix, 
while the theoretical uncertainty is of the order of $1\%$. The challenging 
experimental investigation  puts the upper limit $10^{-6}$ on its branching 
fraction at present \cite{KTeV}, but expects to be sensitive to the branching 
fractions $10^{-11}$ in the future. Theoretically a bit more uncertain decay $K^+\to 
\pi^+\nu\bar\nu$ is predicted at a branching ratio $(0.82\pm 0.32)\times 
10^{-10}$ \cite{isidori} which is to be compared with $4.2{+9.7\atop -3.5}\times 
10^{-10}$ based on the recent observation of one event in this channel 
\cite{E787}. 

Recently, CLEO and ALEPH
observed the $b\to s\gamma$ transition \cite{b.s.gamma}. The standard model 
amplitude for the $b\to s\gamma$ decay is proportional to $V^*_{tb}V_{ts}m_t^2$, it is enhanced due to 
the large top mass and not so $CKM$ suppressed as the $s\to d$ transition. The 
long distance background to this  decay arises via $W$ exchange in $b\to s c\bar 
c$ channel,  the $c\bar c$ hadronizes to a virtual $J/\psi$ and finally converts to a 
real photon. The magnitude of this long distance contribution can not be calculated  from  first 
principles at present, it varies from model to  model \cite{Soares96,GP,DHT} 
and can be as large as $20~\%$ compared to the  short distance part of the $b\to s\gamma$ rate.  
By applying the lower cut on the photon energy, the CLEO analysis 
\cite{b.s.gamma.new} picks out the photons coming only from $b\to s$,  and not 
from $b\to c$ decays, giving the inclusive rate $Br(B\to X_s\gamma)=(3.15\pm 
0.35\pm 0.32\pm 0.26)\times 10^{-4}$ with uncertainties arising from statistics, 
systematics and model dependence, respectively. The possibility to measure the decay
$b\to s\gamma$ inclusively is rather unique among the FCNC decays. It is 
especially welcome since the theoretical prediction for inclusive decay is 
largely free of the hadronization uncertainties giving $Br(B\to 
X_s\gamma)=(3.28\pm 0.22\pm 0.25)\times 10^{-4}$ \cite{blue} with errors arising 
form the uncertainty in the renormalization scale and the standard model 
parameters, respectively. The exclusive mode requires the knowledge of the form 
factors that describe the hadronization of the initial $b$ and final $s$ quarks. 
The predictions for the $B\to K^*\gamma$ rate vary between $6\%$ to $40\%$ of the 
inclusive mode, compared to the measured fractional rate of $(18\pm 6)\%$ 
\cite{blue}. The comparison of the standard model predictions and measured rates 
in these channels gives the stringent bounds on the new models and their parameter 
space. 

The remaining $b\to d$ transition in the down-like sector  is proportional to 
$V^*_{tb}V_{td}m_t^2$, it is CKM suppressed and has not been detected yet. The 
inclusive measurement of $b\to d\gamma$ is very difficult due to the large $b\to 
s\gamma$ background and the channels $B\to \rho\gamma$ and $B\to \omega \gamma$ 
are investigated instead. 

The FCNC transitions among the up-like quarks are
especially rare in the standard model due to the small masses of the
intermediate down-like quarks and  only the upper experimental limits
for this processes are available at present \cite{PDG}.  The  peculiar
feature of the $t$ quark is that it decays via $t\to bW^+$ before it has
time to form a bound state. The $t\to c$ is more interesting than the
$t\to u$ transition due to the more favorable CKM factors and has been
studied in the channels $t\to c\gamma$, $t\to cZ$, $t\to cg$, $t\to cW^+W^-$ and
$e^+e^-\to\bar c t, c\bar t$ \cite{tc}. In the standard model, 
all the FCNC top quark decays are extremely rare with  the branching ratios 
smaller than $10^{-12}$ and any observation of these decays at current or 
planned accelerators would signal physics beyond the standard model. 
The $t\to c$ rates are especially sensitive to the models  in which  the
tree level  coupling $t\bar cH_0$  is proportional to  the quark masses
$m_t$ and $m_c$. In  such models \cite{CS} the
$t\to c$ rates are  severely enhanced over the standard model
predictions and would be observable in the near future. 

\vspace{0.3cm}

In the present work I study the transitions among the $c$ and $u$ quarks $c\to u\gamma$ and 
$c\to ul^+l^-$ which are the most
frequent flavour changing neutral  transitions among the up-like
quarks in the standard model. 

\begin{figure}[h]

\centering
\mbox{
\subfigure[]
{
\begin{fmffile}{f1o1nnn}
  \fmfframe(0,3)(0,3){
  \begin{fmfgraph*}(23,20)
  \fmfpen{thin}
  \fmfleftn{l}{2}\fmfrightn{r}{2}\fmftop{t1}
  \fmf{fermion}{l1,v1}
  \fmf{fermion,tension=0.4,left=0.3,label=$d,,s,,b
         $,la.d=20,la.s=left}{v1,v2}
  \fmf{fermion,tension=0.4,left=0.3}{v2,v3}
  \fmf{fermion}{v3,r1}
  \fmf{boson}{v2,t1}
  \fmf{zigzag,label=$W$,tension=0.4}{v1,v3}
  \fmflabel{$c$}{l1}\fmflabel{$u$}{r1}\fmflabel{$\gamma$}{t1}
  \end{fmfgraph*} }
\end{fmffile}
}
\quad
\subfigure[]
{
\begin{fmffile}{f1o2nn}
  \fmfframe(0,3)(0,3){
  \begin{fmfgraph*}(23,20)
  \fmfpen{thin}
  \fmfleft{l1,l2}\fmfright{r1,r2}\fmftop{t1}
  \fmf{fermion}{l1,v1}
  \fmf{zigzag,label=$W$,la.s=left,tension=0.3,left=0.3}{v1,v2}
  \fmf{zigzag,tension=0.3,left=0.3}{v2,v3}
  \fmf{fermion,tension=0.5,label=$d,,s,,b$,la.d=15,la.s=right}{v1,v3}
  \fmf{fermion}{v3,r1}
  \fmf{boson}{v2,t1}
  \fmflabel{$c$}{l1}\fmflabel{$u$}{r1}\fmflabel{$\gamma$}{t1}
  \end{fmfgraph*} }
\end{fmffile}
}
\quad
\subfigure[]
{
\begin{fmffile}{f1o3nnn}
  \fmfframe(0,3)(0,3){
  \begin{fmfgraph*}(32,20)
  \fmfpen{thin}
  \fmfleft{l1}\fmfright{r1}\fmftopn{t}{4}
  \fmf{fermion,tension=1}{l1,v1}
  \fmf{fermion,tension=1}{v1,v2}
  \fmf{fermion,tension=0.5,label=$d,,s,,b$,la.s=right}{v2,v3}
  \fmf{fermion}{v3,r1}
  \fmffreeze
  \fmf{boson,tension=2}{v1,t2}
  \fmf{zigzag,label=$W$,tension=0.2,left=1}{v2,v3}
  \fmflabel{$c$}{l1}\fmflabel{$u$}{r1}\fmflabel{$\gamma$}{t2}
  \end{fmfgraph*} }
\end{fmffile}
}
\quad
\subfigure[]
{
\begin{fmffile}{f1o4nnn}
  \fmfframe(0,3)(0,3){
  \begin{fmfgraph*}(32,20)
  \fmfpen{thin}
  \fmfleft{l1}\fmfright{r1}\fmftopn{t}{4}
  \fmf{fermion,tension=1}{l1,v1}
  \fmf{fermion,tension=0.5,label=$d,,s,,b$,la.s=right}{v1,v2}
  \fmf{fermion}{v2,v3,r1}
  \fmffreeze
  \fmf{boson,tension=2}{v3,t3}
  \fmf{zigzag,label=$W$,tension=0.2,left=1}{v1,v2}
  \fmflabel{$c$}{l1}\fmflabel{$u$}{r1}\fmflabel{$\gamma$}{t3}
  \end{fmfgraph*} }
\end{fmffile}
}
     }
\caption{The diagrams for the $c\to u\gamma$ decay at the lowest order in the electro-weak theory. Unitary gauge is used.} 
\label{fig1}  
\end{figure}
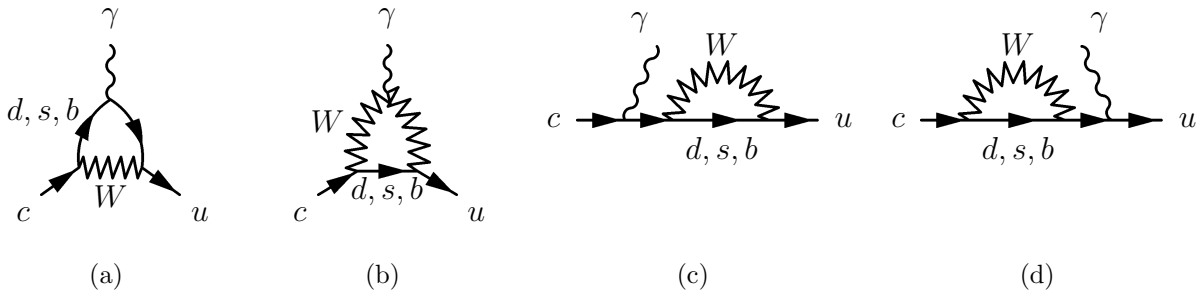

The $c\to u\gamma$ decay has the unique property  that the one loop
electroweak amplitude experiences a huge  enhancement after the effects of strong
interactions are incorporated \cite{BGHP,GHMW}. The one loop electroweak
amplitude arises due to the diagrams in Fig. \ref{fig1}, it is proportional to
$\sum_{q=d,s,b}V^*_{cq}V_{uq}m_q^2/m_W^2$ and gives the branching
ratio of the order of $10^{-17}$.  The branching ratio is enhanced by three 
orders of magnitude when the strong interactions are incorporated in the leading 
logarithmic approximation, it is enhanced by another six orders of magnitude 
when the strong corrections are incorporated at two-loops, while further 
enhancement is not expected \cite{GHMW}. The resulting $c\to u\gamma$ branching 
ratio  $10^{-8}$ in the standard model is still relatively small and  is on 
the verge of the experimental sensitivity of the planned accelerators. It can be 
used as an efficient  probe for models which could enhance the rate for this 
transition in comparison with the standard model rate. For this purpose the sensitivity of the $c\to u\gamma$  rate to several scenarios of physics beyond the standard model are reviewed in this work: the models with the extended Higgs sector, the minimal and non-minimal supersymmetric standard model, standard model with an extension of the fourth generation and left-right symmetric models. 
In order to observe the effects of new physics in the corresponding hadron decays, one has to
be able to disentangle the short distance $c\to u\gamma$ contribution
from the long distance background and this presents the main generic
problem in rare charm hadron decays. 
In the present work, I calculate the standard model predictions for the short
and long distance contributions in the relevant meson decays. I discuss briefly also the relevant baryon decays. In addition, I explore the sensitivity of  the hadron decay rates to different scenarios of physics  beyond the standard model.

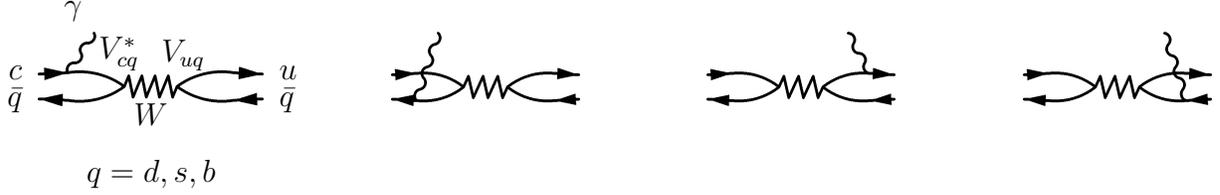
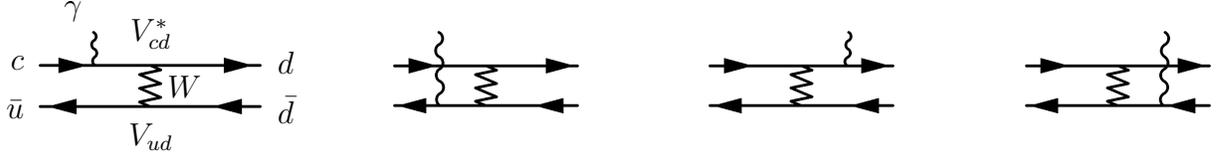
\begin{figure}[h]

\centering
\mbox{
\subfigure[The long distance weak annihilation contribution to the meson 
decays with the flavour structure $c\bar q\to u\bar q\gamma$ and $q=d,s,b$.]
{
\begin{fmffile}{f2a1}
  \fmfframe(3,3)(3,3){
  \begin{fmfgraph*}(30,15)
  \fmfpen{thin}
  \fmfleftn{l}{6}\fmfrightn{r}{6}\fmftopn{t}{5}\fmfbottomn{b}{5}
  \fmf{phantom,tension=0.5}{l5,ml}\fmf{plain,left=0.2,tension=0.7}{ml,pl}
  \fmf{plain,left=0.2,tension=0.7}{pl,al}\fmf{phantom,tension=0.5}{al,l2}
  \fmf{fermion}{l4,ml}\fmf{fermion}{al,l3}
  \fmf{zigzag,tension=1.5,label=$W$,la.s=right}{pl,pr}
  \fmf{phantom,tension=0.5}{r5,mr}\fmf{plain,right=0.2,tension=0.7}{mr,pr}
  \fmf{plain,right=0.2,tension=0.7}{pr,ar}\fmf{phantom,tension=0.5}{ar,r2}
  \fmf{fermion}{mr,r4}\fmf{fermion}{r3,ar}
  \fmffreeze
  \fmf{boson}{ml,t2}\fmflabel{$\gamma$}{t2}
  \fmflabel{$c$}{l4}\fmflabel{$\bar q$}{l3}
  \fmflabel{$u$}{r4}\fmflabel{$\bar q$}{r3}
  \fmfv{label=$V_{cq}^*$,la.d=3thick,la.a=110}{pl}
  \fmfv{label=$V_{uq}$,la.d=3thick,la.a=70}{pr}
   \fmfv{label=$q=d,,s,,b$}{b3}
  \end{fmfgraph*} }
\end{fmffile}
\quad
\begin{fmffile}{f2a2}
  \fmfframe(3,3)(3,3){
  \begin{fmfgraph*}(25,15)
  \fmfpen{thin}
  \fmfleftn{l}{6}\fmfrightn{r}{6}\fmftopn{t}{5}
  \fmf{phantom,tension=0.5}{l5,ml}\fmf{plain,left=0.2,tension=0.7}{ml,pl}
  \fmf{plain,left=0.2,tension=0.7}{pl,al}\fmf{phantom,tension=0.5}{al,l2}
  \fmf{fermion}{l4,ml}\fmf{fermion}{al,l3}
  \fmf{zigzag,tension=1.5}{pl,pr}
  \fmf{phantom,tension=0.5}{r5,mr}\fmf{plain,right=0.2,tension=0.7}{mr,pr}
  \fmf{plain,right=0.2,tension=0.7}{pr,ar}\fmf{phantom,tension=0.5}{ar,r2}
  \fmf{fermion}{mr,r4}\fmf{fermion}{r3,ar}
  \fmffreeze
  \fmf{boson}{al,t2}
  \end{fmfgraph*} }
\end{fmffile}
\quad
\begin{fmffile}{f2a3}
  \fmfframe(3,3)(3,3){
  \begin{fmfgraph*}(25,15)
  \fmfpen{thin}
  \fmfleftn{l}{6}\fmfrightn{r}{6}\fmftopn{t}{5}
  \fmf{phantom,tension=0.5}{l5,ml}\fmf{plain,left=0.2,tension=0.7}{ml,pl}
  \fmf{plain,left=0.2,tension=0.7}{pl,al}\fmf{phantom,tension=0.5}{al,l2}
  \fmf{fermion}{l4,ml}\fmf{fermion}{al,l3}
  \fmf{zigzag,tension=1.5}{pl,pr}
  \fmf{phantom,tension=0.5}{r5,mr}\fmf{plain,right=0.2,tension=0.7}{mr,pr}
  \fmf{plain,right=0.2,tension=0.7}{pr,ar}\fmf{phantom,tension=0.5}{ar,r2}
  \fmf{fermion}{mr,r4}\fmf{fermion}{r3,ar}
  \fmffreeze
  \fmf{boson}{mr,t4}
  \end{fmfgraph*} }
\end{fmffile}
\quad
\begin{fmffile}{f2a4}
  \fmfframe(3,3)(3,3){
  \begin{fmfgraph*}(25,15)
  \fmfpen{thin}
  \fmfleftn{l}{6}\fmfrightn{r}{6}\fmftopn{t}{5}
  \fmf{phantom,tension=0.5}{l5,ml}\fmf{plain,left=0.2,tension=0.7}{ml,pl}
  \fmf{plain,left=0.2,tension=0.7}{pl,al}\fmf{phantom,tension=0.5}{al,l2}
  \fmf{fermion}{l4,ml}\fmf{fermion}{al,l3}
  \fmf{zigzag,tension=1.5}{pl,pr}
  \fmf{phantom,tension=0.5}{r5,mr}\fmf{plain,right=0.2,tension=0.7}{mr,pr}
  \fmf{plain,right=0.2,tension=0.7}{pr,ar}\fmf{phantom,tension=0.5}{ar,r2}
  \fmf{fermion}{mr,r4}\fmf{fermion}{r3,ar}
  \fmffreeze
  \fmf{boson}{ar,t4}
  \end{fmfgraph*} }
\end{fmffile}
}
    }
\mbox{
\subfigure[The long distance weak annihilation contribution to the meson decays with the flavour structure $c\bar u\to d\bar d\gamma$. This mechanism presents background for $c\bar u\to u\bar u\gamma$ decay when the final state is mixture of $u\bar u$ and $d\bar d$ states.]
{
\begin{fmffile}{f2b1n}
  \fmfframe(3,3)(3,3){
  \begin{fmfgraph*}(30,15)
  \fmfpen{thin}
  \fmfleftn{l}{4}\fmfrightn{r}{4}\fmftopn{t}{5}
  \fmf{fermion}{l3,ml}\fmf{plain}{ml,pt,mr}\fmf{fermion}{mr,r3}
  \fmf{fermion}{al,l2}\fmf{plain}{al,pb,ar}\fmf{fermion}{r2,ar}
  \fmffreeze
   \fmf{zigzag,label=$W$}{pb,pt}
   \fmf{boson,tension=0.6}{ml,t2}\fmflabel{$\gamma$}{t2}
  \fmflabel{$c$}{l3}\fmflabel{$\bar u$}{l2}
  \fmflabel{$d$}{r3}\fmflabel{$\bar d$}{r2}
  \fmfv{label=$V_{cd}^*$,la.d=3thick,la.a=90}{pt}
  \fmfv{label=$V_{ud}$,la.d=3thick,la.a=-90}{pb}
  \end{fmfgraph*} }
\end{fmffile}
\quad
\begin{fmffile}{f2b2}
  \fmfframe(3,3)(3,3){
  \begin{fmfgraph*}(25,15)
  \fmfpen{thin}
  \fmfleftn{l}{4}\fmfrightn{r}{4}\fmftopn{t}{5}
  \fmf{fermion}{l3,ml}\fmf{plain}{ml,pt,mr}\fmf{fermion}{mr,r3}
  \fmf{fermion}{al,l2}\fmf{plain}{al,pb,ar}\fmf{fermion}{r2,ar}
  \fmffreeze
   \fmf{zigzag}{pb,pt}
   \fmf{boson,tension=0.6}{al,t2}
  \end{fmfgraph*} }
\end{fmffile}
\quad
\begin{fmffile}{f2b3}
  \fmfframe(3,3)(3,3){
  \begin{fmfgraph*}(25,15)
  \fmfpen{thin}
  \fmfleftn{l}{4}\fmfrightn{r}{4}\fmftopn{t}{5}
  \fmf{fermion}{l3,ml}\fmf{plain}{ml,pt,mr}\fmf{fermion}{mr,r3}
  \fmf{fermion}{al,l2}\fmf{plain}{al,pb,ar}\fmf{fermion}{r2,ar}
  \fmffreeze
   \fmf{zigzag}{pb,pt}
   \fmf{boson,tension=0.6}{mr,t4}
  \end{fmfgraph*} }
\end{fmffile}
\quad
\begin{fmffile}{f2b4}
  \fmfframe(3,3)(3,3){
  \begin{fmfgraph*}(25,15)
  \fmfpen{thin}
  \fmfleftn{l}{4}\fmfrightn{r}{4}\fmftopn{t}{5}
  \fmf{fermion}{l3,ml}\fmf{plain}{ml,pt,mr}\fmf{fermion}{mr,r3}
  \fmf{fermion}{al,l2}\fmf{plain}{al,pb,ar}\fmf{fermion}{r2,ar}
  \fmffreeze
   \fmf{zigzag}{pb,pt}
   \fmf{boson,tension=0.6}{ar,t4}
  \end{fmfgraph*} }
\end{fmffile}
}
     }
\caption{The long distance weak annihilation contribution to the weak 
radiative decays of mesons. }   
\label{fig2}
\end{figure}

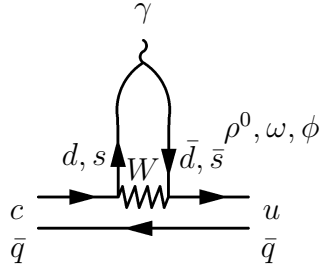
\begin{figure}[h]

\centering
\mbox{
\begin{fmffile}{f3o}
  \fmfframe(3,3)(3,3){
  \begin{fmfgraph*}(28,25)
  \fmfpen{thin}
  \fmfstraight
  \fmfleftn{l}{7}\fmfrightn{r}{7}\fmftopn{t}{9}
  \fmf{fermion}{l2,v1}
  \fmf{zigzag,label=$W$,tension=1.6,la.s=left}{v1,v3}
  \fmf{fermion}{v3,r2}
  \fmf{fermion}{r1,l1}
  \fmffreeze
  \fmf{fermion,tension=0.5,label=$d,,s$,la.d=35,la.s=left}{v1,p1}\fmf{phantom,tension=0.5}{p1,t4}
  \fmf{fermion,tension=0.5,label=$\bar d,,\bar s$,la.s=left,la.d=35}{p3,v3}\fmf{phantom,tension=0.5}{p3,t6}
  \fmffreeze
  \fmf{plain,left=0.3,tension=0.1}{p1,v2}
  \fmf{plain,left=0.3,tension=0.1}{v2,p3}
   \fmf{boson,tension=0.5}{v2,t5}
  \fmflabel{$c$}{l2}\fmflabel{$u$}{r2}\fmflabel{$\gamma$}{t5}
  \fmflabel{$\bar q$}{l1}\fmflabel{$\bar q$}{r1}
  \fmfv{label=$\rho^0,,\omega,,\phi$,la.d=10thick,la.a=-10}{p3}
  \end{fmfgraph*} }
\end{fmffile}
     }
\caption{The long distance penguin contribution to the meson decays with the flavour structure $c\bar q\to u\bar q\gamma$. The intermediate quark pairs $d\bar d$ and $s\bar s$ hadronize to the neutral vector mesons $\rho^0$, $\omega$, $\phi$ and finally convert to a photon.} 
\label{fig3}
\end{figure}

 The meson decays of interest have the flavour content $c\bar
q\to u\bar q\gamma$ where  $c\bar q$ is a pseudoscalar and $u\bar q$
is a vector meson (the pseudoscalar final state is forbidden by the
angular momentum conservation and transverse polarization of the
photon) and $q$ is of any flavor $u,~d,~s,~c$ or $b$. The short
distance contribution  is due to $c\to u\gamma$ decay and
the $\bar q$ is merely a spectator. In addition there are two types
of  long distance contributions. The most serious background presents the long
distance weak annihilation contribution illustrated in Fig. \ref{fig2}. Here the transition
between the initial and final meson is induced by the $W$ exchange and
the photon is emitted from the initial or the final state meson:
``$s$'' channel $W$ exchange $c\bar q\to u\bar q$ for $q=d,~s$ or $b$
is  illustrated in Fig. \ref{fig2}a and is proportional to $V_{cq}^*V_{uq}$;
``$t$'' channel $W$ exchange $c\bar u\to d\bar d$ is illustrated in
Fig. \ref{fig2}b and presents the background for $c\bar u\to u\bar u\gamma$
 decay when the final meson is mixture of $u\bar u$ and $d\bar d$ states. The
second is the long distance penguin contribution
sketched in Fig. \ref{fig3}. Here the $W$ exchange induces the $c\to ud\bar
d,us\bar s$ transition, $d\bar d$ and $s\bar s$ hadronize to
intermediate  $\rho^0$, $\omega$, $\phi$ mesons and finally convert to a real 
photon.

Experimentally, the most promising channels $c\bar q\to u\bar q\gamma$
are those where $q$ is the light quark $u$, $d$ or $s$, namely
$D^0\to \rho^0\gamma$, $D^+\to \rho^+\gamma$, $D^0\to \omega\gamma$
and $D^+_s\to K^{*+}\gamma$ decays. These decays have been studied
phenomenologically \cite{BGHP} and the first two also by using the QCD
sum rules \cite{KSW}. In the present work, a consistent theoretical
framework based on heavy quark and chiral symmetry is developed to
study these decays \cite{FPS1,FS,genova1}. They are shown to be dominated by
the long distance weak annihilation contributions  giving the branching ratios of
the order of $10^{-5}$ \cite{FPS1,FS} and even the most extreme enhancements 
arising from the 
possible new physics would hardly be visible in these decays. The
experimental upper bounds are at the $10^{-4}$ level at present
\cite{CLEO1}, so these decays may be detected soon. 

As the
most suitable probe to study the $c\to u\gamma$ transition I propose the beauty conserving decay $B_c\to B_u^*\gamma$  \cite{FPS3,moriond,southampton,cracow}. In this decay,  the potentially dangerous long
distance weak annihilation  contribution is suppressed  due to the
small CKM factor $V_{cb}^*V_{ub}$. Contrary to long distance dominated
charmed meson decays, the short and long distance contributions in
$B_c\to B_u^*\gamma$ decay are found to be of comparable size,  giving
the branching ratio of the order of $10^{-8}$ in the standard model
\cite{FPS3,moriond,southampton,cracow}.  The rate for this channel is sensitive to any
possible enhancements of $c\to u\gamma$ coming from the physics beyond the
standard model.  The $B_c\to B_u^*\gamma$ decay opens a new window for
future experiments and its detection   at a branching ratio well
above $10^{-8}$ would signal  new physics. Such effects could be
detected at LHC where $2.1\times 10^{8}$ mesons $B_c$ with
$p_T>20$ GeV$/c$ will be produced at integrated luminosity
$100~{\rm fb}^{-1}$. We have used the   
Isgur-Scora-Grinstein-Wise constituent quark model \cite{ISGW1}   to account for the nonperturbative strong dynamics within the mesons \cite{FPS3}. Later, the short 
distance contribution has been re-examined using the QCD sum rules method 
\cite{AS}. 

The baryon decays $cq_1q_2\to uq_1q_2\gamma$ are also examined and are found to 
be less suitable to probe the $c\to u\gamma$ transition.

\vspace{0.2cm} 

The kinematics of the three body decay $c\to ul^+l^-$ with $l=e,\mu$ offers more 
information  than the
kinematics of the two body decay $c\to u\gamma$. In addition to
the total rate, one can   measure  also the dependence of the
differential rate on the
invariant 
di-lepton mass $m_{ll}=\sqrt{(p_{l^+}\!+\!p_{l^-})^2}$ and the angle between the $u$ quark  and one of the 
leptons.   
 The leading term
in the one loop
electroweak amplitude has logarithmic dependence on the intermediate
quark masses 
$\sum_{q=d,s,b}V^*_{cq}V_{uq}\ln (m_q^2/m_W^2)$, it is not strongly
GIM suppressed and gives a branching ratio of the order of $10^{-9}$
\cite{FPS2}. The QCD corrections have not been calculated yet, but are
not expected to be sizable. The relevant meson decays have the
flavour content $c\bar q\to u\bar ql^+l^-$ where  $c\bar q$ forms a
pseudoscalar and $u\bar q$ forms a pseudoscalar or vector
meson. Experimentally the most promising are  the charmed meson
decays with $q=u,~d$ or $s$ and  here  the first theoretical study of the short and the long distance contributions to  these decays is presented\footnote{Only the long distance contribution to the channel $D^+\to \pi^+l^+l^-$ has been studied up to now \cite{singer}.}. For this purpose a framework, which combines the
heavy quark and chiral symmetry, is developed and applied to all
charmed meson decays to a light pseudoscalar or vector meson and a
lepton-antilepton pair \cite{FPS2,genova2}. All decays are found to be
dominated by the long distance contributions arising due to the
mechanism illustrated in Figs. \ref{fig2} and \ref{fig3} where  the real photon is
replaced by the virtual photon and converts to a lepton pair. The predicted rates indicate that some channels may be  detected soon. The
only charmed meson decay channel, in which the short distance transition $c\to
ul^+l^-$  is not overshadowed by the long distance dynamics, is
found to be  $D\to \pi l^+l^-$ at high $m_{ll}$. This is due to the fact that the long 
distance
part is increased at the resonances
$m_{ll}=m_{\rho},~m_{\phi},~m_{\phi}$ and dies out  at higher
$m_{ll}$. In $D\to \pi l^+l^-$ decays the kinematical upper bound
$m^{max}_{ll}=m_D-m_{\pi}$ is as high as possible and the region above
resonances, nonexistent in other decays, is dominated by the short
distance $c\to ul^+l^-$ process \cite{moriond}. Another place to
observe $c\to ul^+l^-$ is the  $B_c\to B_ul^+l^-$ decay in the region of
$m_{ll}$ below the resonances, since the long distance weak annihilation 
contribution is CKM suppressed by the factor $V_{cb}^*V_{ub}$ in this
channel.

Although the long distance dominated charmed meson decays may not be
used as probes for the flavour changing neutral processes, it is important to 
understanding their dynamics. Similar long distance contributions  present the
background to extract short distance transitions $b\to s$ and $s\to d$
in $B$ and $K$ meson decays. The theoretical and experimental
investigation of the long distance dominated charmed meson decays
serves as the controlled laboratory for nonperturbative strong dynamics which 
would help in understanding the similar  $B$ and $K$ meson decays. 

\vspace{0.3cm}

Motivated by the preceding discussion, the main subject of the
dissertation is the study of specific weak decays of heavy mesons -
bound states of quark and antiquark containing at least one heavy
quark $c$ or $b$. 
The main problem here is to account for the strong interaction of
quarks. The strong interactions are described by quantum chromodynamics 
and are understood in principle.  In practice, the
strong coupling is not small at low energies, perturbative expansion
is meaningless and the nonperturbative regime of strong interaction
presents a problem which has not been solved in its entirety, nor is it ever 
likely to be. Rather, what is available is a variety of
theoretical approaches and techniques, appropriate to a variety of
specific problems with varying levels of reliability. There are a few
situations in which one can do  rigorous and
predictive analyses, and many in which what can be said is more
imprecise and model dependent. Often one can not measure what one can
compute reliably, nor compute reliably what one can measure. While
approaches which are based directly on QCD are clearly preferred, more
model dependent methods are often all that is available and thus have an 
important role to play as well.

The theoretical methods to study hadron spectroscopy and their decays
fall roughly into four categories. The numerical calculation of path
integrals on discretized space-time is used in lattice QCD
\cite{lattice}. This method is based directly on QCD and will improve
with the availability of ever more powerful computers. At present it
is most successfully applied to the spectroscopy of hadrons. The
calculation of the quantities, in which several hadrons have to be fitted
on a lattice at the same time, is still numerically too demanding since
the coarse lattice can not resolve dynamics at short distances. Apart
from the numerical, this approach also has some fundamental
problems. Since the calculations are performed in the Euclidean space time, the 
lattice
QCD can not study complex quantities like strong scattering
phases. The QCD sum rules approach \cite{sum.rules} uses the analytical 
continuation to the perturbative QCD kinematical region,  where quantities are 
evaluated perturbatively. The third class of models is represented by different 
kinds of constituent quark models which are only inspired, but not derived, from 
QCD. 

To study the charmed meson decays I will exploit the fourth possibility represented by effective field
theories which follow rigorously from  QCD and are  popular in the
study of heavy meson decays. In addition to true symmetries of
quantum chromodynamics, in this approach also the approximate and spontaneously 
broken
symmetries are used. The Lagrangian is expressed in terms of
hadronic fields instead of in terms of the quark fields. The most general 
Lagrangian  invariant under given symmetries has, in general, infinitely many 
terms, each weighted by an unknown free parameter. Effective field theory has 
predictive power only if different terms present different orders in some small 
expansion parameter which is not a strong coupling constant, in this case. While 
this technique is rigorous in principle, it can  be used only in the kinematical 
regions where the additional symmetries are really valid. The predicted 
quantities depend on several free parameters  and successful application  should 
reproduce more data than there are free parameters. 

I will study the  relevant charmed meson weak decays  by using the
effective field theory which makes use of chiral and heavy quark
symmetries.  Chiral symmetry is the symmetry of QCD in the limit of
massless quarks and holds to a good approximation for light $u$, $d$
and $s$ quarks \cite{GL}. It corresponds to global rotations in the
three-dimensional flavour space of left and right handed quarks. It is
spontaneously broken and the corresponding Goldstone bosons are
represented by the octet of light pseudoscalar mesons. The chiral
perturbation theory  is an effective field theory with the systematic
perturbative expansion in orders of  energy of light pseudoscalar mesons
\cite{GL}. It converges when these energies  are smaller than the
energy of the chiral breaking scale which is of the order of $1$ GeV. Heavy
quark symmetry is the symmetry of QCD in the limit of infinitely heavy
quarks and can be applied to heavy quarks  $c$ and $b$ \cite{HQET1,neubert}. In this limit the dynamics depends only on the
velocity of the heavy quark and is independent of its flavour and
direction of the spin, so the heavy quark symmetry corresponds
to the rotations in the heavy quark spin and flavor space. Heavy quark effective 
field theory presents the systematic expansion in powers of inverse mass of the 
heavy quark \cite{HQET1,neubert}. 

Both symmetries can be combined to study the processes involving heavy
and light mesons in the kinematical region where the latter have small
energy \cite{heavy.chiral}. The most general Lagrangian invariant
under both symmetries describes the strong interactions among the
 heavy  and light mesons. It is expressed as  the perturbative
expansion in the energy of the light mesons and inverse power of heavy
quark mass. The number of free parameters of this effective field
theory increases  at higher orders in the perturbative
expansion. Eventually one has to compromise between a loss of the 
predictive power due to large number of free parameters and large
uncertainties due to the low order of the perturbative expansion. The
electromagnetic interactions are introduced by gauging the
symmetries of the Lagrangian. Since the interchange of the heavy and
light quarks is not the symmetry of the Lagrangian, the weak
interaction among the heavy and light quarks can not be obtained by
gauging. In the present approach, the weak current involving the heavy
and light quark is written as the most general expression that is
left-handed anti-tiplet under chiral transformation and is linear in
heavy meson fields.  Before
applying this approach to decays of interest, the model is
applied to charmed meson semileptonic decays \cite{BFO1} where its
free parameters are fitted. The applicability of the model is approved for in the
reasonable predictions for the charmed meson nonleptonic  rates
\cite{BFOP} where the factorization approximation is systematically
used. The approach is then adapted to study the decays $D\to V\gamma$,
$D\to Vl^+l^-$ and $D\to Pl^+l^-$ where $V$ and $P$ denote the light
vector and pseudoscalar mesons, respectively. Due to presence of the real or
virtual photon in the final state,  the amplitudes have to be invariant under the electro-magnetic gauge transformation. I propose a general mechanism to incorporate the long distance contributions in a manifestly gauge invariant way. 

\vspace{0.2cm}

The current chapter briefly illuminates the present status of the field and
introduces the motivation for  the problems that I  study. 
The standard model
predictions for  $c\to u\gamma$ and $c\to ul^+l^-$ decays at short
distances are presented in Chapter 2. The sensitivity of the $c\to u\gamma$ and $c\to ul^+l^-$ rates  on different scenarios of physics beyond the standard model are discussed as well. These include the models with the extended Higgs sector, the minimal and non-minimal supersymmetric standard model, standard model with an extension of the fourth generation and left-right symmetric models.   The general framework for long distance contributions in 
hadron decays of interest is presented in Chapter 3. The specific
hadronic decays, which are interesting as probes for flavour changing
neutral $c\to u\gamma$ and $c\to ul^+l^-$ transitions, are discussed in Chapter 
4 and 5. The $B_c\to B_u^*\gamma$ decay 
is proposed in Chapter 4 as the most suitable channel to probe the $c\to u\gamma$ transition. In Chapter 5, the charmed meson decays are studied using the effective 
field theory which combines heavy quark effective theory and chiral perturbation 
theory. After the presentation of the model and its applications   to the 
semileptonic and nonleptonic charm meson decays, the decays to a light meson and a photon or 
lepton-antilepton pair are studied.  The 
conclusions are gathered in Chapter 6.

\chapter{Flavour changing neutral transitions among $\boldsymbol{c}$ and $\boldsymbol{u}$ quarks 
\\ at short distances}

The flavour changing neutral transitions among $c$ and $u$ quarks are studied in this chapter. These processes have enhanced sensitivities to different scenarios of physics beyond the standard model and present an interesting probe for the standard model of the elementary particle interactions. The standard model predictions will be given in the first section. In the second section different scenarios of physics beyond the standard model will be studied. These include the models with the extended Higgs sector, the minimal and non-minimal supersymmetric standard model, standard model with an extension of the fourth generation and left-right symmetric models.  

The examples of the  most interesting flavour changing neutral transitions among $c$ and $u$ quarks are $c\to u\gamma$, $c\to ul^+l^-$, $c\to u ~gluon$, $c\to u\nu\bar \nu$, $c\bar u\to l^+l^-$, $c\bar u\to \nu\bar \nu$ and $c\bar u\leftrightarrow \bar cu$. Here  $l$ denotes a charged lepton $e$ or $\mu$, while $\tau$ is too heavy to be produced in the charm quark decay.
I will concentrate on the channels with a photon or a charged lepton pair in the final state due to the experimental difficulties connected with the observation of the neutrino and gluon. Neutrinos freely pass the detectors, while gluons hadronize due to the confinement and the corresponding decays would hardly give a distinctive experimental signature. The standard model prediction for $c\to u\gamma$ decay will be given in Section 1.1, while the predictions for $c\to ul^+l^-$ and $c\bar u\to l^+l^-$ decays will be given in Section 1.2. The sensitivity of these channels to possible scenarios of new physics will be explored in Section 2.

I will not study the  $c\bar u\to \bar c u$  transition responsible for $D^0-\bar D^0$ mixing, since this has been extensively studied elsewhere, among others in \cite{dmix}. The experimental upper bound on  $D^0-\bar D^0$ mixing will be used to constrain the parameter space for the scenarios  beyond the standard model  and  indirectly enter the predictions for the decays of interest here.

The weak interactions of quarks  are experimentally explored in the  corresponding decays of the hadronic states and the relevant hadron decays to probe the transitions among the $c$ and $u$ quarks will be studied in Chapters 3, 4 and 5.
 The predictions for the hadronic channels will unavoidably encounter the uncertainties connected with the nonperturbative regime of the strong interactions.
The main subject of this chapter are processes at the quark level. These can be calculated using the perturbative expansion in electro-weak and strong coupling constants in a well defined way.

\section{The standard model predictions}

\subsection{The $\boldsymbol{c\to u\gamma}$ decay} 

The general form of the amplitude for the decay $q_1\to q_2\gamma$, arising from  the theory with the left-handed charged current interactions and  invariant under the electromagnetic gauge transformation,  is derived in Appendix B 
$${\cal A}[q_1(p)\to q_2\gamma(q,\epsilon)]\propto \bar u_1(p-q)\sigma_{\mu\nu}[m_1(1+\gamma_5)+m_2(1-\gamma_5)]u_1(p)~\epsilon^{\mu}q^{\nu}~.$$
The mass of the quark $u$ is safely neglected compared to the mass of the quark $c$  and the amplitude is conveniently written as
\begin{equation}
\label{2.26}
{\cal A}(c\to u\gamma)=-{G_F\over \sqrt{2}}{e_0\over 4\pi^2}~ c_7^{eff}~m_c ~\epsilon^{\mu}\bar u_{u}(p-q)iq^{\nu}\sigma_{\mu\nu}(1+\gamma_5)u_c(p)~.
\end{equation}
In the above expression and throughout the whole work I systematically use the units in which $c=\hbar=1$ and $[m]=[p]=[E]=1/[l]$.  
The $c\to u\gamma$ rate is given in terms of a single coefficient $c_7^{eff}$, which is calculated in the subsequent subsections
\footnote{ The way $c_7^{eff}$ is defined in (\ref{2.26}), it also includes
the CKM factors. In general, all the coefficients $c_i$ defined in this
chapter include the CKM factors. In the next chapter I will use the 
coefficients $C_1$ and $C_2$ which will not include the CKM factors. }. The name indicates that $c_7^{eff}$ matches the Willson coefficient $c_7(\mu)$ in the leading logarithmic approximation. 
In the first step, the effects of strong interactions are neglected and the amplitude is evaluated in the leading order of the standard model electroweak Lagrangian. The  calculation at the one loop-electroweak order  gives $|c_7^{eff}|=(2.4\pm 2)\cdot 10^{-7}$. The strong interactions drastically increase the $c\to u\gamma$ rate. They are first taken into account in the leading logarithmic approximation, where amplitude is evaluated to all orders in $\alpha_s\log m_c/m_W$ giving  $|c_7^{eff}|=(8\pm 3)\cdot 10^{-6}$ \cite{BGHP,GHMW}. In the next step, all other contributions of the order of $\alpha_s$ are added to the amplitude and the value of $c_7^{eff}$ rises to  $|c_7^{eff}|=(4.7\pm 1.0)\cdot 10^{-3}$ \cite{GHMW}. Further increase is not expected.  The values of $c_7^{eff}$ in subsequent approximations and the corresponding branching ratios for $c\to u\gamma$ are gathered in Table \ref{tab1}. There is obviously huge enhancement due to the strong interactions and it is interesting to see how this comes about in three subsections to follow. 

\begin{table}[h]
\begin{center}
\begin{tabular}{|c|c|c|}
\hline
\hline
 $c\to u\gamma$ &  &$Br(c\to u\gamma)$\\
\hline
one loop electroweak diagrams& $c_7^{eff}=(-2.4\pm 2)\cdot 10^{-7}$&$(3.5{+7.5\atop -3.4})\cdot 10^{-17}$\\
leading logarithmic approximation&$|c_7^{eff}|=(8\pm 3)\cdot 10^{-6}$&$(3.9\pm 2)\cdot 10^{-14}$\\
two-loop diagrams&$c_7^{eff}=-(1.5+4.4i)[1\pm 0.2] 10^{-3}$&$(1.3\pm 0.6)\cdot 10^{-8}$\\
\hline
\hline
$c\to ul^+l^-$& &$Br(c\to ul^+l^-)$\\
\hline
one loop electroweak diagrams&$c_9^{eff}=0.24{+0.1\atop -0.06}$&$(1.7{+0.1\atop -0.7})\cdot 10^{-9}$\\
\hline
\hline
\end{tabular}
\caption{The standard model predictions for $c\to u\gamma$ and $c\to ul^+l^-$ decays. Coefficients $c_7^{eff}$ and $c_9^{eff}$ are defined via (\ref{2.26}, \ref{o7}) and (\ref{o9}), respectively. The coefficient $c_7^{eff}$  and the corresponding $c\to u\gamma$ branching ratios are given in three subsequent approximations related to the strong interactions. }
\label{tab1}
\end{center}
\end{table} 

\subsubsection{One loop electroweak amplitude}

The lowest order diagrams for the $c\to u\gamma$ decay in the unitary gauge are displayed in Fig. \ref{fig1} and the strong interaction are neglected. The amplitude  has the form (\ref{2.26}), so one has to work at least to the second order in the external momenta and one can ignore all the terms that can not be reduced to the Dirac form given in (\ref{2.26}). Using the Gordon decomposition, the amplitude (\ref{2.26}) can be decomposed to
$${\cal A}(c\to u\gamma)=-\frac{G_F}{\sqrt{2}}{e_0\over 4\pi^2}~ c_7^{eff}~m_c\bar u_{u}(p-q)(1+\gamma_5)[2p\epsilon-m_c \!\!\not{\! \epsilon}]u_c(p)~$$
and only the terms of the form $p\epsilon$ are evaluated. There is no need to calculate the divergent diagrams in Figs. \ref{fig1}c and \ref{fig1}d since they are all proportional to $\bar u_u\!\!\not{\! \epsilon}(1+\gamma_5)u_c$ and will be canceled by the terms of similar form coming from the diagrams in Figs. \ref{fig1}a and \ref{fig1}b. The exchange of a particular intermediate quark in Figs. \ref{fig1}a and \ref{fig1}b renders a finite and an infinite part of the form $p\epsilon$ in the amplitude. The infinite terms are independent on the mass of the intermediate quark $m_q$ and their sum vanishes due to the unitarity of the CKM matrix $\sum_{q=d,s,b}V_{cq}^*V_{uq}=0$.  The amplitude is then given by the finite terms. The calculation \cite{IL} at the second order in the external momenta and to all orders in the internal quark masses gives
\begin{eqnarray}
\label{gamma.one.c7}
c_7^{eff}&=&\sum_{q=d,s,b}V_{cq}^*V_{uq}F_2(x_q)/2\qquad {\rm with} \qquad x_q=m_q^2/m_W^2 \qquad{\rm and}\nonumber\\
F_2(x)&=&Q\biggl[{x^3-5x^2-2x\over 4(x-1)^3}+{3x^2\ln x\over 2(x-1)^4}\biggr]-\biggl[{2x^3+5x^2-x\over 4(x-1)^3}-{3x^3\ln x\over 2(x-1)^4}\biggl]~.
\end{eqnarray}
Here $Q=-1/3$ is the charge of the intermediate quark (the  minus sign in front of the second parenthesis was noticed recently in \cite{KP}; it was mistakenly taken as plus sign in  \cite{BGHP} and \cite{GHMW}). In the case of the intermediate $d$, $s$ and $b$ quarks, $x_q$ is small and  $F_2(x_q)\!\!\simeq\!\! -5x_q/12$ gives
\begin{equation}
\label{2.35}
c_7^{eff}\simeq-{5\over 24}\sum_{q=d,s,b}V_{cq}^*V_{uq}{m_q^2\over m_W^2}~.
\end{equation}
The amplitude is strongly GIM suppressed at this order: the contribution of $d$ and $s$ quarks are small due to the small masses $m_d$ and $m_s$; the mass $m_b$ is relatively larger but the contribution of $b$ quark is suppressed  by the small $V_{cb}^*V_{ub}$ factor. The contributions of different intermediate quarks $q$ to  $F_2(x_q)$ and $c_7^{eff}=\sum V_{cq}^*V_{uq}F_2(x_q)/2$ (\ref{gamma.one.c7}) are presented in Table \ref{tab2} for $|V_{cb}^*V_{ub}|\simeq (1.3\pm 0.4)\cdot 10^{-4}$, $V_{cs}^*V_{us}\simeq -V_{cd}^*V_{ud}\simeq 0.22$, $m_d=11$ MeV, $m_s=140\pm 30$ MeV, $m_b=5$ GeV \cite{PDG} and give
\begin{equation}
\label{2.18}
c_7^{eff}=(-2.4\pm 2)\cdot 10^{-7}~,
\end{equation}
where the uncertainty is due mainly  to the unknown relative phase of $V_{cb}^*V_{ub}$ and  $V_{cs}^*V_{us}$.
As a result, the $c\to u\gamma$ branching ratio is unobservably small at this order
\begin{eqnarray}
\label{2.29}
Br(c\to u\gamma)={\Gamma(c\to u\gamma)\over \Gamma(D^0)}&=&6~ \biggl\vert {e_0c_7^{eff}\over 2\pi}\biggr\vert^2~{\Gamma(c\to d e^+\nu_e)\over |V_{cd}|^2\Gamma(D^0)}\nonumber\\
&=&6~ \biggl\vert {e_0c_7^{eff}\over 2\pi}\biggr\vert^2{G_F^2m_c^5\over 192\pi^3 \Gamma(D^0)}\sim (3.5\textstyle{{+7.5\atop -3.4}})\cdot 10^{-17} ~,
\end{eqnarray}
where $m_c=1.25$ GeV is taken.

It is instructive to compare the decay $c\to u\gamma$ with the decay $b\to s\gamma$ this order. As can be seen from Table \ref{tab2}, the top quark  completely dominates among intermediate quarks $u$, $c$, $t$  ($m_u=3.25$ MeV, $m_c=1.25$ GeV and $m_t=174$ GeV are taken) and the corresponding coefficient $c_7^{eff}=7.6\cdot 10^{-3}$ is not so GIM suppressed.

\begin{table}[h]
\begin{center}
\begin{tabular}{|c|c|c||c|c|c|}
\hline
\multicolumn{3}{|c||}{$c\to u\gamma$} & \multicolumn{3}{|c|}{$b\to s\gamma$}\\ \hline 
$q$ & $F_2(x_q)$ & $V_{cq}^*V_{uq}F_2(x_q)/2$&$q$ & $F_2(x_q)$ & $V_{qb}V_{qs}^*F_2(x_q)/2$\\
\hline
$d$ & $-7.9\cdot 10^{-9}$ & $8.4\cdot 10^{-10}$&$u$ & $2.3\cdot 10^{-9}$ & $-8.3\cdot 10^{-13}$\\
$s$ & $-1.3\cdot 10^{-6}$ & $-1.4\cdot 10^{-7}$&$c$ & $1.4\cdot 10^{-4}$ & $-2.7\cdot 10^{-6}$\\
$b$ & $-1.6\cdot 10^{-3}$ & $-1.0\cdot 10^{-7}$&$t$ & $0.39$ & $7.6\cdot 10^{-3}$\\
\hline
\end{tabular}
\caption{Comparison of $c\to u\gamma$ and $b\to s\gamma$ decays at the one-loop electroweak order: the contributions arising from different intermediate quarks to $F_2$ and $ c_7^{eff}$ (\ref{gamma.one.c7}) are shown.}
\label{tab2}
\end{center}
\end{table}

\subsubsection{The effects of strong interaction in the leading logarithmic approximation}

The perturbative  QCD corrections are included by adding the gluons in all possible ways to the diagrams in Fig. \ref{fig1}. In this section I sketch the first step in this direction - the effects of strong interaction are included to all orders in $\alpha_s \ln m_c/m_W$ in the leading logarithmic approximation. 

At this point, it is convenient to take into account that the external particles in $c\to u\gamma$ decay have momentum of the order  of $m_c$ and we can  get rid of the degrees of freedom with much higher masses. At low energies, the $c\to u\gamma$ decay (\ref{2.26}) is effectively induced by the  local Lagrangian
\begin{equation}
\label{o7}
{\cal L}^{c\to u\gamma}(x)=-{4G_F\over \sqrt{2}}c_7^{eff}O_7(x)\quad {\rm with}\quad
O_7(x)={e_0\over 32\pi^2}m_c\bar u(x)\sigma_{\mu\nu}(1+\gamma_5)c(x)F^{\mu\nu}(x)
\end{equation}
and $c_7^{eff}$ is given in (\ref{gamma.one.c7}) at the one-loop electroweak order.

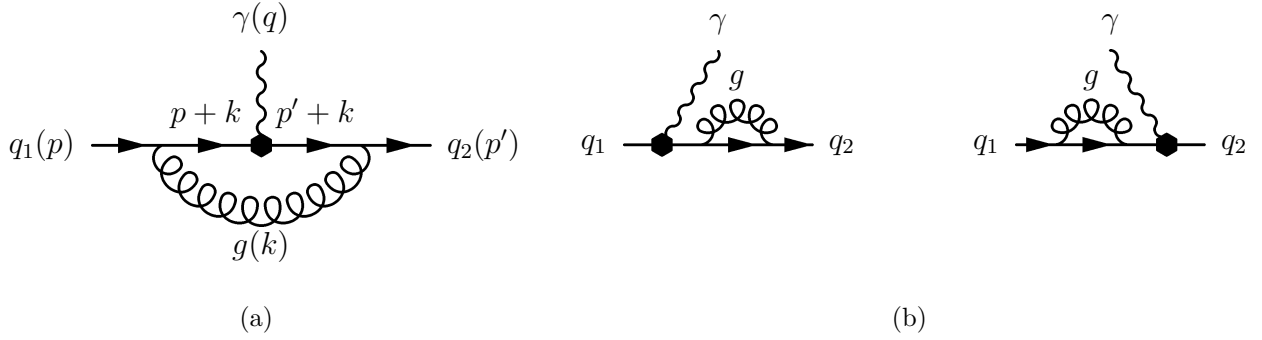
\begin{figure}[h]

\centering
\mbox{
\subfigure[]
{
\begin{fmffile}{f4a}
  \fmfframe(6,3)(6,3){
  \begin{fmfgraph*}(45,25)
  \fmfpen{thin}
  \fmfleft{l1}\fmfright{r1}\fmftop{t1}
  \fmf{fermion,tension=1}{l1,v1}
  \fmf{fermion,tension=0.7,label=$p+k$,la.s=left}{v1,v2}
  \fmf{fermion,tension=0.7,label=$p^\prime+k$,la.s=left}{v2,v3}
  \fmf{fermion}{v3,r1}
  \fmffreeze
  \fmf{boson}{v2,t1}
  \fmf{gluon,label=$g(k)$,la.d=80,tension=1.7,right=0.6}{v1,v3}
  \fmfv{de.sh=hexagon,de.filled=full,decor.size=4thick}{v2}
  \fmflabel{$q_1(p)$}{l1}\fmflabel{$q_2(p^\prime)$}{r1}\fmflabel{$\gamma(q)$}{t1}
  \end{fmfgraph*} }
\end{fmffile}
}
\quad
\subfigure[]
{
\begin{fmffile}{f4b1}
  \fmfframe(6,3)(6,3){
  \begin{fmfgraph*}(25,25)
  \fmfpen{thin}
  \fmfleft{l1}\fmfright{r1}\fmftop{t1}
  \fmf{plain,tension=1}{l1,v1}
  \fmf{plain,tension=1}{v1,v2}
  \fmf{fermion,tension=0.5}{v2,v3}
  \fmf{fermion}{v3,r1}
  \fmffreeze
  \fmf{boson,tension=2}{v1,t1}
  \fmf{gluon,label=$g$,tension=0.2,left=1}{v2,v3}
  \fmfv{de.sh=hexagon,de.filled=full,decor.size=4thick}{v1}
   \fmflabel{$q_1$}{l1}\fmflabel{$q_2$}{r1}\fmflabel{$\gamma$}{t1}
  \end{fmfgraph*} }
\end{fmffile}
\qquad
\begin{fmffile}{f4b2}
  \fmfframe(6,3)(6,3){
  \begin{fmfgraph*}(25,25)
  \fmfpen{thin}
  \fmfleft{l1}\fmfright{r1}\fmftop{t1}
  \fmf{fermion,tension=1}{l1,v1}
  \fmf{fermion,tension=0.5}{v1,v2}
  \fmf{plain}{v2,v3,r1}
  \fmffreeze
  \fmf{boson,tension=2}{v3,t1}
  \fmf{gluon,label=$g$,tension=0.2,left=1}{v1,v2}
  \fmfv{de.sh=hexagon,de.filled=full,decor.size=4thick}{v3}
   \fmflabel{$q_1$}{l1}\fmflabel{$q_2$}{r1}\fmflabel{$\gamma$}{t1}
  \end{fmfgraph*} }
\end{fmffile}
}
     }
\caption{Strong corrections to the effective operator $O_7$ (\ref{o7}) 
at the order of $\alpha_s$. The action of the operator $O_7$ (\ref{o7}) is 
dented by the hexagon. Only the diagrams which contribute to the anomalous 
dimension $\gamma_{77}$ are presented here.} 
\label{fig4} 
\end{figure}

One type of QCD corrections to the effective Lagrangian ${\cal L}^{c\to u\gamma}$ (\ref{o7}) at the order  $\alpha_s$ are incorporated by adding the gluon exchanges as shown in Fig. \ref{fig4}. For simplification, other QCD corrections at this order will be discussed later. 
The  transition  $O_7$ accompanied by the one gluon exchange in Fig. \ref{fig4} induces the effective operator $O_7^{loop}$, which is calculated in  Appendix A.2 and is infinite.   The low energy operator $O_7$ is
 renormalized at some renormalization scale $\mu$ by adding the
 counter-terms, which cancel the divergences: $O_7\to O_7(\mu)=O_7+O_7^{CT}(\mu)$.   After the cancellation of the divergences, the  amplitude for the diagrams in Fig. \ref{fig4}, calculated from $O_7(\mu)$, turns out to depend on the renormalization scale $\mu$ via $\alpha_s\ln (E/\mu)$ and $E$ is the typical energy of the external particles.
 The low energy effective operator $O_7(\mu)$ is matched to the
 operator given by the full theory  at the energy scale $\mu=m_W$, while the  energies of the external particles are of the order of $m_c$.  The factor  $\alpha_s\ln m_c/m_W$  in not small and arises since we have calculated the amplitude from the effective operator $O_7(m_W)$. Needless to say that $O_7(m_c)$  would be more suitable for the calculation, since the one-loop strong corrections arising from the gluons with the virtualities from $m_W$ down to $m_c$ are proportional to   $\alpha_s\ln m_c/m_c$  and therefore vanish. As the bare Lagrangian ${\cal L}=-4G_F/\sqrt{2}~c_7(\mu)O_7(\mu)$ does not depend on the renormalization scale, the running of the coefficient  $c_7(\mu)$ is introduced in such a way as to cancel the $\mu$ dependence of $O_7(\mu)$. In Appendix A.1 this
 condition is used to determine the running of $c_7(\mu)$ from the
 diagrams in Fig. \ref{fig4}, giving
\begin{equation}
\label{evolution.77}
[\mu{d\over d\mu}-\gamma_{77}(\mu)]c_7(\mu)=0\quad {\rm with}\quad
\gamma_{77}(\mu)={16\over 3}~{g_s(\mu)^2\over 8\pi^2} 
\end{equation}
originally calculated  in \cite{SVZ}. 
The effective operator $O_7$ is matched at $\mu\!=\!m_W$ with the operator given by the full theory at the one loop electroweak order, so $c_7(\mu\!=\!m_W)$ is given by (\ref{gamma.one.c7}). 
The evolution of $c_7$ down to $\mu\!=\!m_c$ is obtained by solving  Eq. (\ref{evolution.77}). Defining the running of the strong coupling $g_s$ via the coefficient $\beta=\mu ~dg_s/d\mu$, the solution to (\ref{evolution.77}) is given by \begin{equation}
\label{2.27}
c_7(\mu)=\exp \biggl[\int_{g_s(m_W)}^{g_s(\mu)}dg_s{\gamma_{77}(g_s)\over \beta(g_s)}\biggr]~c_7(m_W)~
\end{equation}
with $\beta_0=11-2n_f/3$. The number of active quark flavours  $n_f$ changes from five to four at $\mu=m_b$. Defining $\gamma_{77}=(g_s^2/8\pi^2)b_{77}$ with $b_{77}=16/3$ this integrates to  \begin{equation}
\label{2.11}
c_7(\mu)\!=\!\biggl[{\alpha_s(m_W)\over \alpha_s(\mu)}\biggr]^{{b_{77}\over \beta_0}}\!c_7(m_W)\qquad {\rm or} \qquad c_7(m_c)\!=\!\biggl[{\alpha_s(m_W)\over \alpha_s(m_b)}\biggr]^{{16\over 23}}\biggl[{\alpha_s(m_b)\over \alpha_s(m_c)}\biggr]^{{16\over 25}}\!c_7(m_W)
\end{equation}
and $c_7(m_W)$ is given by (\ref{gamma.one.c7}, \ref{2.18}). 
If diagrams in Fig. \ref{fig4} presented the only QCD correction to $O_7$ at this order, then the Lagrangian (\ref{o7}) would  present the effective Lagrangian for $c\to u\gamma$ decay within the described approximation.  The suitable renormalization scale is $\mu=m_c$ and the $c\to u\gamma$ amplitude (\ref{2.26}) would be given by the coefficient $c_7^{eff}=c_7(m_c)$ (\ref{2.11}). 
If we renormalized the Lagrangian  at $\mu=m_W$ instead, we would have to sum the contributions to all orders in $\alpha_s\ln(m_c/m_W)$ in order to get the same result. For this reason this approximation is called the leading logarithmic approximation. The idea to incorporate the effects of highly virtual particles  in low energy phenomena into coefficients rather than in the operators  was introduced by K.G. Willson \cite{willson} and the corresponding coefficients are called the Willson coefficients.

\begin{figure}[h]

\centering
\mbox{
\subfigure[]
{
\begin{fmffile}{f5a1n}
  \fmfframe(6,3)(6,3){
  \begin{fmfgraph*}(25,20)
  \fmfpen{thin}
  \fmfleft{l1,l2}\fmfright{r1,r2}
  \fmf{fermion}{l1,v1,r1}
  \fmf{fermion}{l2,v2,r2}
  \fmf{zigzag,label=$W$,tension=0.7}{v1,v2}
  \fmflabel{$c$}{l2}\fmflabel{$u$}{r1}
   \fmflabel{$d,s,b$}{l1}\fmflabel{$d,s,b$}{r2}
  \end{fmfgraph*} }
\end{fmffile}
\qquad
\begin{fmffile}{f5a2n}
  \fmfframe(6,3)(6,3){
  \begin{fmfgraph*}(25,20)
  \fmfpen{thin}
  \fmfleft{l1,l2}\fmfright{r1,r2}
  \fmf{fermion}{l1,v1,m1,r1}
  \fmf{fermion}{l2,v2,m2,r2}
  \fmf{zigzag,label=$W$,tension=0.7,la.s=left}{v1,v2}
  \fmffreeze
  \fmf{gluon,label=$g$,la.s=right,la.d=80}{m1,m2}
  \fmflabel{$c$}{l2}\fmflabel{$u$}{r1}
   \fmflabel{$d,s,b$}{l1}\fmflabel{$d,s,b$}{r2}
  \end{fmfgraph*} }
\end{fmffile}
}
\quad
\subfigure[]
{
\begin{fmffile}{f5bn}
  \fmfframe(3,3)(3,3){
  \begin{fmfgraph*}(25,20)
  \fmfpen{thin}
  \fmfleft{l1,l2}\fmfright{r1,r2}\fmftop{t1,t2}
  \fmf{fermion}{l1,v1}
  \fmf{plain,tension=0.03}{v1,v2}
  \fmf{plain,tension=0.03,label=$d,,s,,b$,la.d=2,la.s=left}{v2,v3}
  \fmf{fermion}{v3,r1}
  \fmf{gluon,tension=0.1}{v2,a}
  \fmf{zigzag,label=$W$,tension=0.4}{v1,v3}
  \fmf{fermion}{t1,a,t2}
  \fmflabel{$c$}{l1}\fmflabel{$u$}{r1}\fmflabel{$q$}{t1}\fmflabel{$q$}{t2}
  \end{fmfgraph*} }
\end{fmffile}
}
    }
\mbox{
\subfigure[]
{
\begin{fmffile}{f5cn}
  \fmfframe(3,3)(3,3){
  \begin{fmfgraph*}(25,15)
  \fmfpen{thin}
  \fmfleft{l1,l2}\fmfright{r1,r2}\fmftop{t1}
  \fmf{fermion}{l1,v1}
  \fmf{plain,tension=0.2}{v1,v2}
  \fmf{plain,tension=0.2,label=$d,,s,,b$,la.d=2,la.s=left}{v2,v3}
  \fmf{fermion}{v3,r1}
  \fmf{boson}{v2,t1}
  \fmf{zigzag,label=$W$,tension=0.4}{v1,v3}\fmflabel{$\gamma$}{t1}
  \fmflabel{$c$}{l1}\fmflabel{$u$}{r1}
  \end{fmfgraph*} }
\end{fmffile}
}
\quad
\subfigure[]
{
\begin{fmffile}{f5dnn}
  \fmfframe(3,3)(3,3){
  \begin{fmfgraph*}(25,15)
  \fmfpen{thin}
  \fmfleft{l1,l2}\fmfright{r1,r2}\fmftop{t1}
  \fmf{fermion}{l1,v1}
  \fmf{plain,tension=0.2}{v1,v2}
  \fmf{plain,tension=0.2,label=$d,,s,,b$,la.d=2,la.s=left}{v2,v3}
  \fmf{fermion}{v3,r1}
  \fmf{gluon,tension=0.7}{v2,t1}
  \fmf{zigzag,label=$W$,tension=0.4}{v1,v3}\fmflabel{$g$}{t1}
  \fmflabel{$c$}{l1}\fmflabel{$u$}{r1}
  \end{fmfgraph*} }
\end{fmffile}
}
     }
\caption{The diagrams that give rise to the effective operators $O_{1,..8}$ (\ref{operatorji}). Only some typical respresentatives of the classes are shown.} \label{fig5}  
\end{figure}
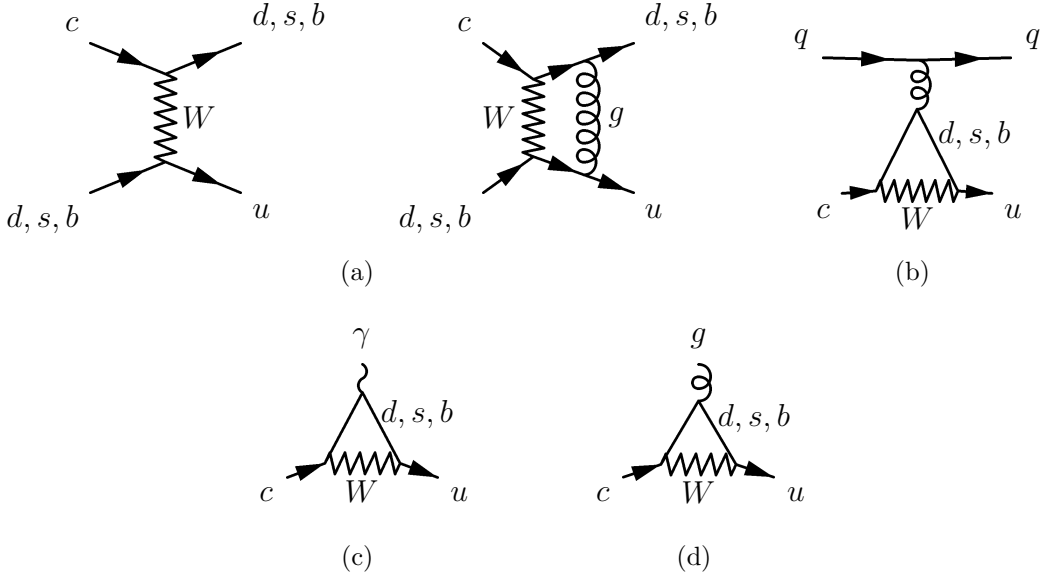

This was, however, overly simplified in order to illustrate the idea behind the leading logarithmic approximation. In addition to $O_7$, there are several effective local operators which mix among themselves when they are evolved from $\mu=m_W$ down to some lower scale \cite{GHMW,buras}. The current-current operators $O_{1,2}$ arise from the diagrams Fig. \ref{fig5}a, the QCD-penguin operators $O_{3,4,5,6}$ arise form the diagram Fig. \ref{fig5}b and the magnetic penguin operators $O_7$ and $O_8$ arise form the diagrams Fig. \ref{fig5}c and \ref{fig5}d, respectively (only some typical representatives of the classes are presented in Fig. \ref{fig5}) \footnote{In comparison to \cite{GHMW} and \cite{buras} I change the definitions of $O_1(\mu)$ and $O_2(\mu)$ in the way it is frequently used for the study of the nonleptonic decays.} 
\begin{eqnarray}
\label{operatorji}
O_{1}^q&=&4(\bar u_{\alpha}\gamma_{\mu}P_Lq_{\alpha})~(\bar q_{\beta}\gamma^{\mu}P_Lc_{\beta})~,\qquad \qquad q=d,~s,~b\nonumber\\
O_{2}^q&=&4(\bar u_{\alpha}\gamma_{\mu}P_Lc_{\alpha})~(\bar q_{\beta}\gamma^{\mu}P_Lq_{\beta})~,\qquad \qquad q=d,~s,~b\nonumber\\
O_3&=&(\bar u_{\alpha}\gamma_{\mu}P_Lc_{\alpha})~~\textstyle{\sum_q}~(\bar q_{\beta}
\gamma^{\mu}P_Lq_{\beta})~,\nonumber\\
O_4&=&(\bar u_{\alpha}\gamma_{\mu}P_Lc_{\beta})~~\textstyle{\sum_q}~(\bar q_{\beta}
\gamma^{\mu}P_Lq_{\alpha})~,\nonumber\\
O_5&=&(\bar u_{\alpha}\gamma_{\mu}P_Lc_{\alpha})~~\textstyle{\sum_q}~(\bar q_{\beta}
\gamma^{\mu}P_Rq_{\beta})~,\nonumber\\
O_6&=&(\bar u_{\alpha}\gamma_{\mu}P_Lc_{\beta})~~\textstyle{\sum_q}~(\bar q_{\beta}
\gamma^{\mu}P_Rq_{\alpha})~,\nonumber\\
O_7&=&e_0\tfrac{1}{16\pi^2}m_c(\bar u_{\alpha}\sigma_{\mu\nu}P_Rc_{\alpha})F^{\mu\nu}~,\nonumber\\
O_8&=&g\tfrac{1}{ 16\pi^2}m_c(\bar u_{\alpha}\sigma_{\mu\nu}T^a_{\alpha\beta}P_Rc_{\beta})G^{a\nu\mu}
\end{eqnarray}
with $P_{R,L}=(1\pm\gamma_5)/2$ and color indices $\alpha$, $\beta$.  The effective bare Lagrangian for the $c\to u$ transition is represented by the sum of all these effective operators, multiplied by the corresponding Willson coefficients 
\begin{equation}
\label{2.13}
{\cal L}=-{4G_F\over \sqrt{2}} \biggl(\sum_{q=d,s,(b)} \!\!{1\over 4}[c_1^q(\mu)O_1^q(\mu)+c_2^q(\mu)O_2^q(\mu)]+\sum_{i=3,..,8}c_i(\mu)O_i(\mu)\biggr)~.
\end{equation}
The $b$ quark in the first sum is active only in the region $\mu>m_b$.

\begin{figure}[h]

\centering
\mbox{
\begin{fmffile}{f6o1mn}
  \fmfframe(3,6)(3,6){
  \begin{fmfgraph*}(30,30)
  \fmfpen{thin}
  \fmfleft{l3,l2,l1,l3}\fmfright{r4,r2,r1,r3}\fmftop{t1}\fmfbottom{b1}
  \fmfv{de.sh=hexagon,de.filled=full,decor.size=4thick,label=$
   O_1^{q=d,,s,,(b)}$,la.d=3thick,la.a=90}{v}
  \fmf{plain,tension=1}{l1,al}\fmf{plain,tension=0.3}{al,v}
  \fmf{plain,tension=0.3}{v,ar}
  \fmf{plain,tension=1}{ar,r1}
  \fmffreeze
  \fmf{phantom,tension=2}{ml,l2}\fmf{phantom,tension=2}{mr,r2}
  \fmf{plain,left=0.4}{b,ml}
  \fmf{plain,right=0.4,label=$d,,s,,(b)$,la.si=right}{b,mr}
  \fmf{plain,left=0.5}{ml,v}\fmf{plain,right=0.5}{mr,v}
  \fmf{boson,tension=2}{b,b1}
  \fmffreeze
  \fmf{gluon,label=$g$,la.si=left}{ml,al}
  \fmflabel{$c$}{l1}\fmflabel{$u$}{r1}\fmflabel{$\gamma$}{b1}
  \end{fmfgraph*} }
\end{fmffile}
\quad
\begin{fmffile}{f6o2mn}
  \fmfframe(3,6)(3,6){
  \begin{fmfgraph*}(30,30)
  \fmfpen{thin}
  \fmfleft{l3,l2,l1,l3}\fmfright{r4,r2,r1,r3}\fmftop{t1}\fmfbottom{b1}
  \fmfv{de.sh=hexagon,de.filled=full,decor.size=4thick}{v}
  \fmf{plain,tension=1}{l1,al}\fmf{plain,tension=0.3}{al,v}
  \fmf{plain,tension=0.3}{v,ar}\fmf{plain,tension=1}{ar,r1}
  \fmffreeze
  \fmf{phantom,tension=2}{ml,l2}\fmf{phantom,tension=2}{mr,r2}
  \fmf{plain,left=0.4}{b,ml}\fmf{plain,right=0.4}{b,mr}
  \fmf{plain,left=0.5}{ml,v}\fmf{plain,right=0.5}{mr,v}
  \fmf{boson,tension=2}{b,b1}
  \fmffreeze
  \fmf{gluon}{mr,ar}
  \fmflabel{$c$}{l1}\fmflabel{$u$}{r1}\fmflabel{$\gamma$}{b1}
  \end{fmfgraph*} }
\end{fmffile}
\quad
\begin{fmffile}{f6o3mn}
  \fmfframe(3,6)(3,6){
  \begin{fmfgraph*}(30,30)
  \fmfpen{thin}
  \fmfleft{l3,l2,l1,l3}\fmfright{r4,r2,r1,r3}\fmftop{t1}\fmfbottom{b1}
  \fmfv{de.sh=hexagon,de.filled=full,decor.size=4thick}{v}
  \fmf{plain,tension=1}{l1,al}\fmf{plain,tension=0.3}{al,v}
  \fmf{plain,tension=0.3}{v,ar}\fmf{plain,tension=1}{ar,r1}
  \fmffreeze
  \fmf{phantom,tension=2}{ml,l2}\fmf{phantom,tension=2}{mr,r2}
  \fmf{plain,left=0.4}{b,ml}\fmf{plain,right=0.4}{b,mr}
  \fmf{plain,left=0.5}{ml,v}\fmf{plain,right=0.5}{mr,v}
  \fmf{boson,tension=2}{b,b1}
  \fmffreeze
  \fmf{gluon}{ml,ar}
  \fmflabel{$c$}{l1}\fmflabel{$u$}{r1}\fmflabel{$\gamma$}{b1}
  \end{fmfgraph*} }
\end{fmffile}
     }
\mbox{
\begin{fmffile}{f6o4n}
  \fmfframe(3,6)(3,6){
  \begin{fmfgraph*}(30,30)
  \fmfpen{thin}
  \fmfleft{l3,l2,l1,l3}\fmfright{r4,r2,r1,r3}\fmftop{t1}\fmfbottom{b1}
  \fmfv{de.sh=hexagon,de.filled=full,decor.size=4thick}{v}
  \fmf{plain,tension=1}{l1,al}\fmf{plain,tension=0.3}{al,v}
  \fmf{plain,tension=0.3}{v,ar}\fmf{plain,tension=1}{ar,r1}
  \fmffreeze
  \fmf{phantom,tension=2}{ml,l2}\fmf{phantom,tension=2}{mr,r2}
  \fmf{plain,left=0.4}{b,ml}\fmf{plain,right=0.4}{b,mr}
  \fmf{plain,left=0.5}{ml,v}\fmf{plain,right=0.5}{mr,v}
  \fmf{boson,tension=2}{b,b1}
  \fmffreeze
  \fmf{gluon}{mr,al}
  \fmflabel{$c$}{l1}\fmflabel{$u$}{r1}\fmflabel{$\gamma$}{b1}
  \end{fmfgraph*} }
\end{fmffile}
\quad
\begin{fmffile}{f6o5n}
  \fmfframe(3,6)(3,6){
  \begin{fmfgraph*}(30,30)
  \fmfpen{thin}
  \fmfleft{l3,l2,l1,l3}\fmfright{r4,r2,r1,r3}\fmftop{t1}\fmfbottom{b1}
  \fmfv{de.sh=hexagon,de.filled=full,decor.size=4thick}{v}
  \fmf{plain,tension=1}{l1,al}\fmf{plain,tension=0.5}{al,f,v}
  \fmf{plain,tension=0.3}{v,ar}\fmf{plain,tension=1}{ar,r1}
  \fmffreeze
  \fmf{phantom,tension=2}{ml,l2}\fmf{phantom,tension=2}{mr,r2}
  \fmf{plain,left=0.4}{b,ml}\fmf{plain,right=0.4}{b,mr}
  \fmf{plain,left=0.5}{ml,v}\fmf{plain,right=0.5}{mr,v}
  \fmf{phantom,tension=2}{b,b1}
  \fmffreeze
  \fmf{gluon}{ml,al}\fmf{boson}{f,t1}
  \fmflabel{$c$}{l1}\fmflabel{$u$}{r1}\fmflabel{$\gamma$}{t1}
  \end{fmfgraph*} }
\end{fmffile}
\quad
\begin{fmffile}{f6o6n}
  \fmfframe(3,6)(3,6){
  \begin{fmfgraph*}(30,30)
  \fmfpen{thin}
  \fmfleft{l3,l2,l1,l3}\fmfright{r4,r2,r1,r3}\fmftop{t1}\fmfbottom{b1}
  \fmfv{de.sh=hexagon,de.filled=full,decor.size=4thick}{v}
  \fmf{plain,tension=1}{l1,al}\fmf{plain,tension=0.3}{al,v}
  \fmf{plain,tension=0.5}{v,f,ar}\fmf{plain,tension=1}{ar,r1}
  \fmffreeze
  \fmf{phantom,tension=2}{ml,l2}\fmf{phantom,tension=2}{mr,r2}
  \fmf{plain,left=0.4}{b,ml}\fmf{plain,right=0.4}{b,mr}
  \fmf{plain,left=0.5}{ml,v}\fmf{plain,right=0.5}{mr,v}
  \fmf{phantom,tension=2}{b,b1}
  \fmffreeze
  \fmf{gluon}{mr,ar}\fmf{boson}{f,t1}
  \fmflabel{$c$}{l1}\fmflabel{$u$}{r1}\fmflabel{$\gamma$}{t1}
  \end{fmfgraph*} }
\end{fmffile}
    }
\caption{The dominant contribution to the decay $c\to u\gamma$ at the order 
$\alpha_s$ comes from the diagrams above when the functional dependence
on the intermediate quark masses  $d$, $s$ and $b$ is taken into account. The
hexagon denotes the action of the operator  $O_1^{q=d,s,(b)}$
(\ref{operatorji}) evaluated at the renormalization scale $m_c$.}  
\label{fig6} 
\end{figure}
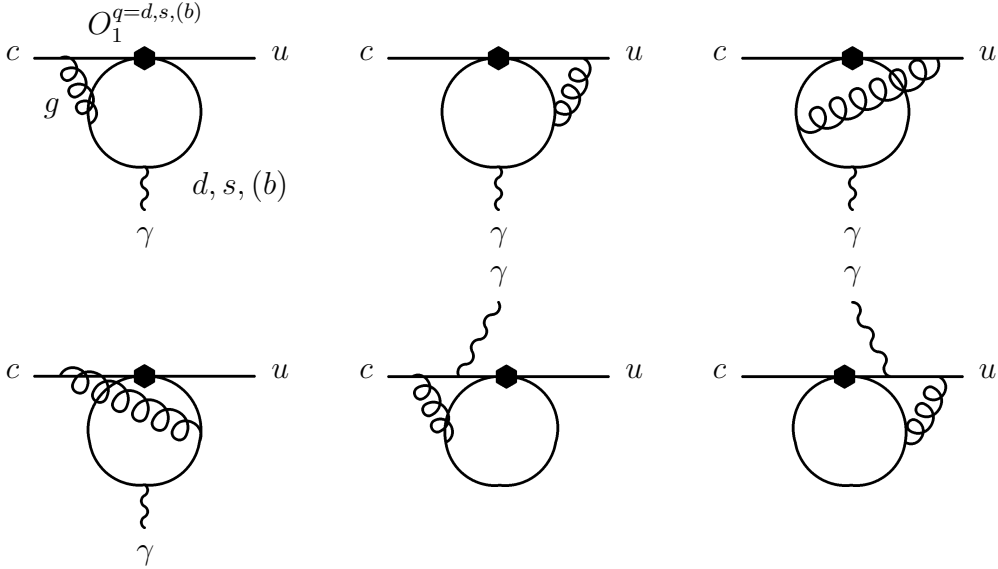

In  order to calculate the $c\to u\gamma$ amplitude (\ref{2.26}, \ref{o7}) in the leading logarithmic approximation we have to evaluate $c_7^{eff}=c_7(m_c)$.  The evolution of the coefficients $c_i(\mu)$ is obtained by imposing the $\mu$-dependences of $c_i(\mu)$ and $O_i(\mu)$ to cancel. The running can be expressed in terms of the  anomalous dimension matrix $\gamma$, which is determined by the explicit renormalization of the operators $O_i(\mu)$
\begin{equation}
\label{2.13a}
[\delta_{ji}~\mu{d\over d\mu}-\gamma_{ij}(\mu)]~c_i(\mu)=0~.
\end{equation}
The general idea of mixing is illustrated on the example of mixing between the operators  $O_1(\mu)$ and $O_2(\mu)$ in Appendix A, where matrix elements $\gamma_{ij}$ for $i,j\!=\!1,2$  are calculated. The operators $O_{1-6,8}$, which are of zero-th order in electromagnetic coupling $e$, can mix with $O_{7}$, which is of the first order in $e$, only if the mixing diagrams involve an additional electromagnetic vertex. The operator $O_7$, for example, mixes strongly with operator $O_1$ through the diagrams in Fig. \ref{fig6} and the detailed calculation of $\gamma_{17}$ is presented in \cite{grinstein}. Other entries in the matrix $\gamma=(g_s^2/8\pi^2)b$ have been calculated in series of papers (see the references in Chapter 2 of \cite{buras}) and  depend on the number of the active quark flavors $n_f$ and the charge of the quarks, but not on the quark masses \cite{GHMW}\footnote{Note that I have changed the definition of $O_1\leftrightarrow O_2$ compared to \cite{GHMW}, so the matrix looks different.}
\begin{equation}
\label{2.15}
b={1\over 2}\begin{pmatrix}-2&6&-{2\over 9}&{2\over 3}&-{2\over 9}&{2\over 3}&8Q_1+{16\over 27}Q_2&{70\over 27}\\
6&-2&0&0&0&0&0&3\\
0&0&-{22\over 9}&{22\over 3}&-{4\over 9}&{4\over 3}&{464\over 27}Q_2&{140\over27}+3n_f\\
0&0&6-{2\over 9}n_f&-2+{2\over 3}n_f&-{2\over 9}n_f&{2\over 3}n_f&8\bar Q+{16\over 27}n_fQ_2&6+{70\over 27}n_f\\
0&0&0&0&2&-6&-{32\over 3}Q_2&-{14\over 3}-3n_f\\
0&0&-{2\over 9}n_f&{2\over 3}n_f&-{2\over 9}n_f&-16+{2\over 3}n_f&-8\bar Q+{16\over 27}n_fQ_2&-4-{119\over 27}n_f\\
0&0&0&0&0&0&{32\over 3}&0\\
0&0&0&0&0&0&{32\over 3}Q_2&{28\over 3}\\\end{pmatrix}~.
\end{equation} 
with $Q_1=-1/3$, $Q_2=2/3$ and $\bar Q=2/3$ for $c\to u\gamma$  decay\footnote{In the case of  $b\to s\gamma$ decay  $Q_1=2/3$, $Q_2=-1/3$ and $\bar Q=1/3$ have to be taken.}.

The evolution of $c_i$ from  $\mu_0$ to $\mu$ is given by the solution to Eq. (\ref{2.13a})
\begin{equation}
\label{2.14}
c_i(\mu)=(M~D~M^{-1})_{ij}c_j(\mu_0)~,
\end{equation}
where $M$ is the orthogonal matrix $b^T\!\!=\!\!Mb^{diag}M^{-1}$ that diagonalizes $b^T$ and $D$ is the diagonal matrix given by the eigenvalues of $b^{diag}$ 
$$D=\biggl(\biggl[{\alpha_s(\mu_0)\over \alpha_s(\mu)}\biggr]^{b^{diag}_{1}/ \beta_0},...~,\biggl[{\alpha_s(\mu_0)\over \alpha_s(\mu)}\biggr]^{b^{diag}_{8}/\beta_0}\biggr
)~.$$

The operators $O_i(\mu)$ are renormalized at $\mu\!=\!m_W$ and the corresponding coefficients $c_i(m_W)$ are given by the electroweak theory. Coefficient $c_1^q(m_W)=V_{cq}^*V_{uq}$ is given by the current-current interaction, $c_7(m_W)$ is proportional to $\sum V_{cq}^*V_{uq}m_q^2/m_W^2$ (\ref{gamma.one.c7})
 and $c_8(m_W)$ is also found to be proportional to $ m_q^2/m_W^2$ \cite{buras}. All the remaining coefficients are equal to zero in the absence of the strong interactions, $c_{2-6}(m_W)\!=\!0$. 

The evolution of $c_i(\mu)$ from $\mu\!=\!m_W$ to $m_c$ is performed in two steps as the number of active quark flavors changes from $n_f\!=\!5$ to $4$ at  $\mu\!=\!m_b$.
 The evolution (\ref{2.14}) from $\mu\!=\!m_W$ to $m_b$  gives $c_i(m_b)$ that are proportional to $c_j(m_W)$. At $\mu\!=\!m_W$ only $c_1^q(m_W)=V_{cq}^*V_{uq}$ is seizable, while the others are equal to zero or proportional to $m_q^2/m_W^2$ and therefore negligible. The evolution from $\mu\!=\!m_W$ to $m_b$ with $n_f\!=\!5$ active flavors gives \cite{GHMW}
\begin{equation}
\label{2.29.1}
c^q_{1,2}(m_b)={1\over 2}V_{cq}^*V_{uq}\biggl(\biggl[{\alpha_s(m_W)\over \alpha_s(m_b)}\biggr]^{{6\over 23}}\pm \biggl[{\alpha_s(m_W)\over \alpha_s(m_b)}\biggr]^{-{12\over 23}}\biggr)\ ,\qquad
c_{3-8}(m_b)\propto \sum_{q=d,s,b}V_{cq}^*V_{uq}=0~.
\end{equation}
The coefficient $c_7^{eff}=c_7(m_c)$, which  is responsible for the $c\to u\gamma$ decay in the leading logarithmic approximation, is given by evolving (\ref{2.29.1})   down to $\mu=m_c$ with $n_f=4$ active flavors \cite{GHMW}
$$c_7^{eff}=c_7(m_c)=\sum_{q=d,s}~\sum_{i=1}^8~ \biggr[{\alpha_s(m_b)\over \alpha_s(m_c)}\biggr]^{d_i}[a_ic_1^q(m_b)+b_ic_2^q(m_b)]~$$
 and using $V_{cd}^*V_{ud}+V_{cs}^*V_{us}=-V_{cb}^*V_{ub}$
\begin{align}
\label{2.28}
c_7^{eff}=c_7(m_c)=-{1\over 2}V_{cb}^*V_{ub}~\sum_{i=1}^8 &\biggl[{\alpha_s(m_b)\over \alpha_s(m_c)}\biggr]^{d_i}\Biggl\{a_i\biggl(\biggl[{\alpha_s(m_W)\over \alpha_s(m_b)}\biggr]^{{6\over 23}}- \biggl[{\alpha_s(m_W)\over \alpha_s(m_b)}\biggr]^{{-12\over 23}}\biggr)\nonumber\\
&+b_i\biggl(\biggl[{\alpha_s(m_W)\over \alpha_s(m_b)}\biggr]^{{6\over 23}}+ \biggl[{\alpha_s(m_W)\over \alpha_s(m_b)}\biggr]^{{-12\over 23}}\biggr)\Biggr\}~.
\end{align}
The coefficients  $a_i$, $b_i$ and $d_i$ are rational numbers and are listed in \cite{GHMW}. 
The proportionality to $V_{cb}^*V_{ub}$ (\ref{2.28}) arises due to the independence of the anomalous dimension matrix (\ref{2.15}) on the quark masses $m_d$, $m_s$. In the evolution from $\mu=m_b$ down to $m_c$, the contributions from intermediate $d$ and $s$ quarks differ only in the CKM factors $V_{cd}^*V_{ud}$ and $V_{cs}^*V_{us}$ (\ref{2.29.1}), respectively. Their sum  is proportional to $V_{cd}^*V_{ud}+V_{cs}^*V_{us}=-V_{cb}^*V_{ub}$ and  therefore small.  Taking $|V_{cb}^*V_{ub}|\simeq (1.3\pm 0.4)\cdot 10^{-4}$, $m_b=5$ GeV, $m_c=1.5$ GeV and $\alpha_s(m_Z)=0.12$ the authors of  \cite{GHMW} get
\begin{equation}
\label{2.17}
|c_7^{eff}|=|c_7(m_c)|=0.060~|V_{cb}^*V_{ub}|\simeq (8\pm 3)\cdot 10^{-6}~.
\end{equation}

\subsubsection{The strong corrections at two-loop level}

Although the leading logarithmic result (\ref{2.17}) is much larger than the purely electroweak contribution (\ref{2.18}), there remains a strong cancellation between $s$ and $d$ loops whose CKM factors are similar in magnitude but have opposite signs. This renders suppression by $V_{cb}^*V_{ub}$ in the leading logarithmic result (\ref{2.17}). The cancellation is circumvented when the functional dependence of the amplitude on the $s$ and $d$ quark masses becomes substantial. This happens when the amplitude is calculated at the order of $\alpha_s$  together with its dependence on the internal quark masses. In Ref. \cite{GHMW} this has been done for  the dominant contributions to $c\to u\gamma$ decay at the order of $\alpha_s$.  These are induced by the operators $O_{1,2}^q(m_c)$ (\ref{operatorji}) - the only operators whose corresponding coefficients $c_{1,2}(m_c)$ are not suppressed by $V_{cb}^*V_{ub}$ (like it is the case for $c_7(m_c)$ (\ref{2.17})). The evolution of $c_{1,2}(m_b)$ (\ref{2.29.1}) down to $\mu=m_c$ is given by (\ref{2.14}) with $n_f=4$ 
\begin{equation}
\label{2.19}
c_{1,2}^q(m_c)={1\over 2}V_{cq}^*V_{uq}\biggl(\biggl[{\alpha_s(m_W)\over \alpha_s(m_b)}\biggr]^{{6\over 23}}\biggl[{\alpha_s(m_W)\over \alpha_s(m_b)}\biggr]^{{6\over 25}}\pm \biggl[{\alpha_s(m_W)\over \alpha_s(m_b)}\biggr]^{-{12\over 23}}\biggl[{\alpha_s(m_W)\over \alpha_s(m_b)}\biggr]^{-{12\over 25}}\biggr)~
\end{equation}
and the terms of the order of $V_{cb}^*V_{ub}$ are neglected. 
 The $c\to u\gamma$ diagrams induced by $O_1^q(m_c)$ are shown in Fig. \ref{fig6}, while the similar diagrams with $O_2^q(m_c)$ insertions vanish due to their color structure. The coefficient $c_7^{eff}$, that represents the magnitude of $c\to u\gamma$ transition at this order, is obtained by the explicit evaluation of the two-loop diagrams in Fig. \ref{fig6}  giving \cite{GHMW}
\begin{align}
\label{2.20}
c_7^{eff}&={\alpha_s(m_c)\over 4\pi}\biggl[c_2^s(m_c)f({m_s^2\over m_c^2})+c_2^d(m_c)f({m_d^2\over m_c^2})\biggr]+{\cal O}(V_{ub}^*V_{cb})\\   
&=V_{cs}^*V_{us}{\alpha_s(m_c)\over 8\pi}\biggl[f({m_s^2\over m_c^2})-f({m_d^2\over m_c^2})\biggr] \nonumber\\
&\qquad\times\biggl(\biggl[{\alpha_s(m_W)\over \alpha_s(m_b)}\biggr]^{{6\over 23}}\biggl[{\alpha_s(m_W)\over \alpha_s(m_b)}\biggr]^{{6\over 25}}+ \biggl[{\alpha_s(m_W)\over \alpha_s(m_b)}\biggr]^{-{12\over 23}}\biggl[{\alpha_s(m_W)\over \alpha_s(m_b)}\biggr]^{-{12\over 25}}\biggr)+{\cal O}(V_{ub}^*V_{cb})~.\nonumber
\end{align}
The function $f$ was extracted \cite{GHMW} from an analogous computation performed for the $b\to s\gamma$ decay \cite{GHW}
\begin{align*}
&f(z)=-{1\over 243}\lbrace576\pi^2z^{3/2}+[3672-288\pi^2-1296\zeta(3)+(1944-324\pi^2)L+108L^2+36L^3]z\\
&+[324-576\pi^2+(1728-216\pi^2)L+324L^2+36L^3]z^2+[1296-12\pi^2+1776L-2052L^2]z^3\rbrace\\
&-{4\pi i\over 81}\lbrace[144-6\pi^2+18L+18L^2]z+[-54-6\pi^2+108L+18L^2]z^2+[116-96L]z^3\rbrace\\
&+{\cal O}(z^4L^4)~
\end{align*}
with $L=\ln z$ and $\zeta(3)\!\simeq\! 1.20$. The imaginary part of the function $f$ arises from the part of the integration over the loop momenta when  $d$ or $s$ quarks are on-shell. The function is renormalization scheme independent. All scheme dependent terms undergo the GIM mechanism and only affect the ${\cal O}(V_{ub}^*V_{cb})$ part of (\ref{2.20}). Using $m_s=140\pm 30$ MeV, $m_d=1$1 MeV and $m_c=1.25$ GeV Eq. (\ref{2.20}) gives 
\begin{equation}
\label{2.22}
f({m_s^2\over m_c^2})-f({m_d^2\over m_c^2})= -(0.24+0.68i)[1\pm 0.2]
\end{equation}
and together with (\ref{2.20}) the authors of  \cite{GHMW} get 
\begin{equation}
\label{c7}
c_7^{eff}=-(1.5+4.4i)[1\pm 0.2] 10^{-3}~.
\end{equation} 
The corresponding branching ratio at this order is given by (\ref{2.29})
\begin{equation}
\label{2.60}
Br(c\to u\gamma)=(1.3\pm 0.6)\cdot 10^{-8}~.
\end{equation}
The amplitude at the order of $\alpha_s$ (\ref{c7}) is more than two orders of magnitude larger than the leading logarithmic result (\ref{2.17}) and four orders of magnitude larger than the electroweak result (\ref{2.18}). The form of Eq. (\ref{2.20}) ensures that no further enhancement of the $c\to u\gamma$ rate is expected at even higher orders of strong interaction \cite{GHMW}: the amplitude of $c\to u\gamma$ must be of order $eG_F$; any  $\Delta S=0$ charm decay  must be  Cabibbo suppressed by $V_{cs}^*V_{us}$; the suppression by $m_s^2/m_c^2$ is rather mild as can be seen from (\ref{2.22}). The possibilities of finding large contributions by considering higher orders in perturbation theory have been therefore exhausted. 

\vspace{0.1cm}

The standard model predictions for $c\to u\gamma$ in three subsequent approximations are gathered in Table \ref{tab1}. 

\subsection{The $\boldsymbol{c\to ul^+l^-}$ decay in the standard model} 

The diagrams for the $c\to ul^+l^-$  decay at the lowest order in the electro-weak theory are presented in Fig. \ref{fig8}. In order to calculate the $c\to ul^+l^-$ rate, the amplitudes for $c\to u\gamma^*$, $c\to uZ^*$  and $W$ box diagrams\footnote{Superscript $*$ denotes a virtual particle.}  have to be evaluated  and the general forms are derived in Appendix B
\begin{align}
\label{2.decomp}
{\cal  A}[c\to u\gamma^*(q,\epsilon)]&=\epsilon^{\mu}\bar u_u[Aim_cq^{\nu}\sigma_{\mu\nu}(1+\gamma_5)+B(q^2\gamma_{\mu}-q_{\mu}\!\!\not {\!q})(1-\gamma_5)]u_c~,\\
{\cal  A}[c\to uZ^*]&=C ~\bar u_u\gamma_{\mu}(1-\gamma_5)u_c~\epsilon^{\mu}~,\nonumber\\
{\cal A}^{box}&=D ~\bar u_u\gamma_{\mu}(1-\gamma_5)u_c~\bar u_l\gamma^{\mu}(1-\gamma_5)u_l~.\nonumber
\end{align}
 The Lorentz invariant coefficients $A$, $B$, $C$ and $D$ have been calculated to all orders in the internal quark masses for analogous $s\to d$ diagrams in \cite{IL}. In the case of $c\to u$ transition the coefficients are expressed in terms of the masses $m_d$, $m_s$ and $m_b$ and the terms proportional to $m_q^2/m_W^2$ can be safely neglected. The leading contribution comes from the term $\ln m_q^2/m_W^2$, which appears in the coefficient $B$ and arises due to the virtual photon emission from the intermediate quark in Fig. \ref{fig8}a. Note that the coefficient $B$ does not contribute in the case of the real photon and the $c\to u\gamma$ amplitude is proportional only to $m_q^2/m_W^2$ (which is much smaller than  $|\ln m_q^2/m_W^2|$ responsible for $c\to ul^+l^-$ decay). 
Neglecting the terms proportional to $m_q^2/m_W^2$, the effective local Lagrangian that induces the $c\to ul^+l^-$ transition is given by 
\begin{align}
\label{o9}
{\cal L}^{c\to ul^+l^-}&=-4~{G_F\over \sqrt{2}}~c_9^{eff}~O_9\qquad {\rm with}\nonumber\\
O_9&={e_0^2\over 32\pi^2}~\bar u_{\alpha}\gamma_{\mu}(1-\gamma_5)c_{\alpha}~\bar l\gamma^{\mu}l\qquad {\rm and}\qquad
c_9^{eff}={2\over 9}\sum_{q=d,s,b}V_{cq}^*V_{uq}\ln{m_q^2\over m_W^2}~.
\end{align}
In this approximation, the standard model prediction for the $c\to ul^+l^-$ rate  depends  on a single coefficient $c_9^{eff}$ (the name indicates that $c_9^{eff}$ matches the Willson coefficient $c_9(\mu)$ in the leading logarithmic approximation). 
Taking $m_d=11\pm 5$ MeV and $m_s=140\pm 30$ MeV \cite{PDG} this gives \cite{FPS2,genova2} 
\begin{equation}
\label{c9}
c_9^{eff}=0.24{+0.01\atop -0.06} 
\end{equation}
 and with $m_c=1.25$ GeV
\begin{eqnarray}
\label{2.33}
Br(c\to ul^+l^-)={\Gamma(c\to ul^+l^-)\over \Gamma(D^0)}&=&2~\biggl\vert{e_0^2c_9^{eff}\over 16\pi^2}\biggr\vert^2{\Gamma(c\to d e^+\nu_e)\over |V_{cd}|^2\Gamma(D^0)}\\
&=&2~\biggl\vert{e_0^2c_9^{eff}\over 16\pi^2}\biggr\vert^2{G_F^2m_c^5\over 192\pi^3 \Gamma(D^0)}\sim (1.7\textstyle{{+0.1\atop -0.7}})\cdot 10^{-9}~.\nonumber 
\end{eqnarray}

\begin{figure}[h]
\centering
\mbox{
\subfigure[]
{
\begin{fmffile}{f8a1nn}
  \fmfframe(0,3)(0,3){
  \begin{fmfgraph*}(25,25)
  \fmfpen{thin}
  \fmfleftn{l}{2}\fmfrightn{r}{2}\fmftopn{t}{4}
  \fmf{fermion}{l1,v1}
  \fmf{fermion,tension=0.4,left=0.3,label=$d,,s,,b
         $,la.d=20,la.s=left}{v1,v2}
  \fmf{fermion,tension=0.4,left=0.3}{v2,v3}
  \fmf{fermion}{v3,r1}
  \fmf{boson,label=$\gamma,,Z$}{v2,a}
  \fmf{zigzag,label=$W$,tension=0.4}{v1,v3}
  \fmflabel{$c$}{l1}\fmflabel{$u$}{r1}\fmf{fermion}{t2,a,t3}
  \fmflabel{$l^+$}{t2}\fmflabel{$l^-$}{t3}
  \end{fmfgraph*} }
\end{fmffile}
\quad
\begin{fmffile}{f8a2n}
  \fmfframe(0,3)(0,3){
  \begin{fmfgraph*}(25,25)
  \fmfpen{thin}
  \fmfleft{l1,l2}\fmfright{r1,r2}\fmftopn{t}{4}
  \fmf{fermion}{l1,v1}
  \fmf{zigzag,label=$W$,la.s=left,tension=0.3,left=0.3}{v1,v2}
  \fmf{zigzag,tension=0.3,left=0.3}{v2,v3}
  \fmf{fermion,tension=0.5,label=$d,,s,,b$,la.d=15,la.s=right}{v1,v3}
  \fmf{fermion}{v3,r1}
  \fmf{boson}{v2,a}
  \fmflabel{$c$}{l1}\fmflabel{$u$}{r1}\fmf{fermion}{t2,a,t3}
  \fmflabel{$l^+$}{t2}\fmflabel{$l^-$}{t3}
  \end{fmfgraph*} }
\end{fmffile}
\quad
\begin{fmffile}{f8a3n}
  \fmfframe(0,3)(0,3){
  \begin{fmfgraph*}(35,20)
  \fmfpen{thin}
  \fmfleft{l1}\fmfright{r1}\fmftopn{t}{4}
  \fmf{fermion,tension=1}{l1,v1}
  \fmf{fermion,tension=1}{v1,v2}
  \fmf{fermion,tension=0.5,label=$d,,s,,b$,la.s=right}{v2,v3}
  \fmf{fermion}{v3,r1}
  \fmffreeze
  \fmf{boson,tension=2}{v1,a}
  \fmf{zigzag,label=$W$,tension=0.2,left=1}{v2,v3}
  \fmflabel{$c$}{l1}\fmflabel{$u$}{r1}\fmf{fermion}{t1,a,t2}
  \fmflabel{$l^+$}{t1}\fmflabel{$l^-$}{t2}
  \end{fmfgraph*} }
\end{fmffile}
\quad
\begin{fmffile}{f8a4n}
  \fmfframe(0,3)(0,3){
  \begin{fmfgraph*}(35,20)
  \fmfpen{thin}
  \fmfleft{l1}\fmfright{r1}\fmftopn{t}{4}
  \fmf{fermion,tension=1}{l1,v1}
  \fmf{fermion,tension=0.5,label=$d,,s,,b$,la.s=right}{v1,v2}
  \fmf{fermion}{v2,v3,r1}
  \fmffreeze
  \fmf{boson,tension=2}{v3,a}
  \fmf{zigzag,label=$W$,tension=0.2,left=1}{v1,v2}
  \fmflabel{$c$}{l1}\fmflabel{$u$}{r1}\fmf{fermion}{t3,a,t4}
  \fmflabel{$l^+$}{t3}\fmflabel{$l^-$}{t4}
  \end{fmfgraph*} }
\end{fmffile}
}
     }
\mbox{
\subfigure[]
{
\begin{fmffile}{f8bn}
  \fmfframe(3,3)(3,3){
  \begin{fmfgraph*}(30,20)
  \fmfpen{thin}
  \fmfleft{l1,l2}\fmfright{r1,r2}\fmftop{t1,t2}
  \fmf{fermion,tension=1}{l1,v1}
  \fmf{fermion,tension=0.5,label=$d,,s,,b$,la.s=right}{v1,v2}
  \fmf{fermion}{v2,r1}
  \fmffreeze  
  \fmf{fermion,tension=1}{t1,m1}
  \fmf{fermion,tension=0.8,label=$\nu$,la.s=left}{m1,m2}
  \fmf{fermion}{m2,t2}
  \fmf{zigzag,label=$W$,la.s=left,tension=0.2}{v1,m1}
  \fmf{zigzag,label=$W$,la.s=right,tension=0.2}{v2,m2}
  \fmflabel{$c$}{l1}\fmflabel{$u$}{r1}
  \fmflabel{$l^+$}{t1}\fmflabel{$l^-$}{t2}
  \end{fmfgraph*} }
\end{fmffile}
}
     }
\caption{The diagrams for $c\to ul^+l^-$ decay at the lowest order in the electro-weak theory. Unitary gauge in used.} 
\label{fig8}   
\end{figure}

The effects of strong interactions in $c\to ul^+l^-$ decay have not been studied in detail yet. The form of effective Lagrangian (\ref{o9}) ensures that the effects of the strong interactions could not drastically enhance the $c\to ul^+l^-$ rate. The amplitude for $c\to ul^+l^-$ must be of the order of $e^2G_F$ and so by a factor $e$ suppressed compared to $c\to u\gamma$ amplitude. Any  $\Delta S=0$ charm decay  must be also Cabibbo suppressed by $V_{cs}^*V_{us}$. The factor $\ln m_q^2/m_W^2$ is of the order of unity and  is not expected to be drastically affected by the strong interactions. This is in contrast to $c\to u\gamma$ decay, where the QCD corrections drastically enhance  a tiny factor $m_q^2/m_W^2$ present in the one-loop electroweak amplitude. 

\subsubsection{The short distance $\boldsymbol{c\bar u\to l^+l^-}$ contribution to $\boldsymbol{D^0\to l^+l^-}$ decay}

The transition $c\bar u\to l^+l^-$ can be experimentally studied only in the decay $D^0\to l^+l^-$ of the pseudoscalar state $D^0$, which is stable against the strong and electromagnetic decays.  
Although the hadronic decays will be the main concern of the following chapters, I briefly discuss the  $D^0\to l^+l^-$ decay here since this decay will be considered in various scenarios of physics beyond the standard model in the next section. The short distance contribution is induced by the transition $c\bar u\to l^+l^-$ and all the nonperturbative strong effects are incorporated into a single parameter $f_D$, describing how $c$ and $\bar u$ quarks are bounded to the meson $D^0$ 
\begin{equation}
\label{2.45}
\langle 0|\bar u\gamma^{\mu}\gamma_5c|D(p)\rangle\equiv if_Dp^{\mu}~.
\end{equation}
The general form of the $c\bar u\to l^+l^-$ amplitude is given by the expressions in (\ref{2.decomp}) with $\bar u_u$ replaced by $\bar v_u$.  
Due to the kinematic structure in $D^0\to l^+l^-$  decay only  the $W$ box diagram in Fig. \ref{fig8}b  contributes and  the corresponding amplitude is proportional to a very small factor $\sum
V_{cq}^*V_{uq}m_q^2/m_W^2$ (in the $c\to ul^+l^-$ decay we have entirely neglected the box diagram contribution in comparison with the virtual photon emission form the intermediate quark giving $\sum V_{cq}^*V_{uq}\ln m_q^2/m_W^2$). Additional suppression of the  $D^0\to l^+l^-$ rate comes from the helicity suppression: the $W$ bosons in the box couple only to left-handed lepton and left-handed anti-lepton, while the conservation of angular momentum in $D^0\to l^-l^+$  requires one of them to  be left and the other to be right-handed. The resulting short distance branching ratio is suppressed by $(m_l/m_D)^2(m_q/m_W)^4$ and is of the order of $10^{-19}$. Although the long distance contributions enhance the branching ratio up to about $3\cdot 10^{-15}$ \cite{pakvasa} this decay is beyond the present experimental sensitivity at $10^{-6}$ \cite{PDG}.  For this reason I will not study this decay  in Chapters 3, 4 and 5 and I will concentrate only on the meson  decays induced by the flavour changing neutral transitions $c\to ul^-l^-$ and $c\to u\gamma$.   

\section{Beyond the standard model}

\subsection{Models with extended Higgs sector}

In multi-Higgs doublet models, the new sources of flavour changing neutral interactions arise via the interactions of quarks with the new physical Higgses. 
The typical effects that arise from adding one or more scalar doublets to the
standard model can be inferred by
studying the two Higgs doublet model. The Yukawa part 
of the standard model Lagrangian with one Higgs doublet   
\begin{equation}
\label{2.30}
{\cal L}_Y=\bar Q^{\prime}_L\lambda^u\tilde\Phi U_R^{\prime}+
\bar Q^{\prime}_L\lambda^d\Phi D_R^{\prime}+\bar \Psi^{\prime}_L\lambda^l\Phi l_R^{\prime}+h.c.
\end{equation}
is replaced by
\begin{eqnarray}
\label{2.32}
 {\cal L}_Y=\bar Q^{\prime}_L(\lambda^u_1\tilde\Phi_1+\lambda^u_2\tilde\Phi_2) U_R^{\prime}+
\bar Q^{\prime}_L(\lambda^d_1\Phi_1+\lambda^d_2\Phi_2) D_R^{\prime}+\bar
\Psi^{\prime}_L(\lambda^l_1\Phi_1+\lambda^l_2\Phi_2) l_R^{\prime}+h.c.
\end{eqnarray}
in the case of two Higgs doublets 
 $$\Phi_1={\phi_1^+\brack {v_1+H_1+iA_2\over \sqrt{2}}}\qquad{\rm and}\qquad \Phi_2={\phi_2^+\brack {v_2+H_2+iA_2\over \sqrt{2}}}~.$$
Here
$Q_L^{\prime}\!=\!(U_L^{\prime},D_L^{\prime})$,  $\psi_L^{\prime}\!=\!(\nu_l^{\prime},l_L^{\prime})$, $U^{\prime}$,
$D^{\prime}$, $\nu^{\prime}$ and $l^{\prime}$ are triplets in the good weak isospin
bases denoted by prime. The $\lambda^{u,d,l}$ and
$\lambda^{u,d,l}_{1,2}$ are general $3\times 3$ complex matrices. The $SU(2)_L\times
U(1)_Y$ symmetry is spontaneously broken by the nonzero vacuum expectation values of 
the lower components $\langle\Phi_1\rangle=(0,v_1/\sqrt{2})$
and $\langle\Phi_2\rangle=(0,v_2/\sqrt{2})$ and the 
$U(1)_{em}$ symmetry is preserved\footnote{The $v_1$ and $v_2$ are in general two complex numbers with different phases. In order to simplify the discussion the complex phases are taken to be equal, since there is no experimental data to constrain the phase difference at present.}.  After breaking the symmetry spontaneously five physical Higgs fields are present in the Lagrangian: positively and negatively charged scalars $H^+$ and $H^-$; two neutral scalars $H^0$ and $h^0$ with different masses and a neutral pseudoscalar $A^0$, as discussed in Appendix C in detail. The masses for the quarks are contained in  
\begin{align}
\label{2.31}
 {\cal L}_M=\bar U^{\prime}_L\Gamma^u U_R^{\prime}+\bar D^{\prime}_L\Gamma^d D_R^{\prime}+h.c.\qquad {\rm with} \qquad &\Gamma^u=(\lambda^u_1v_1+\lambda^u_2v_2)/\sqrt{2}\nonumber\\
&\Gamma^d=(\lambda^d_1v_1+\lambda^d_2v_2)/\sqrt{2}~.
\end{align}
 The mass matrices  in the weak isospin basis $\Gamma^{u,d}$ are diagonalized via a bi-unitary transformation, but  $\lambda^{u,d}_1$ and $\lambda_2^{u,d}$ do not get diagonalized at the same time. The Lagrangian (\ref{2.32}) therefore induces the tree level flavour changing neutral couplings (FCN) with the neutral physical Higgs bosons. This is in contrast to the standard model 
with one Higgs doublet, where the bi-unitary transformation diagonalizes the mass matrices $\lambda^{u,d}v/\sqrt{2}$ and the matrices $\lambda^{u,d}$ at the same time and there are no flavour changing couplings with neutral Higgs bosons in (\ref{2.30}).
The flavour changing neutral couplings in the two Higgs doublet model have to be suppressed in order to satisfy the experimental evidence for the smallness of FCNC's. One way to achieve this is by introducing the very large Higgs boson masses, which could lead to the violation of the unitarity and I will not consider the scenario where the mass of the Higgses are much greater than $1$ TeV.  Natural suppression of the flavour changing neutral processes and lepton-flavour violating processes can be achieved also by imposing an ``ad hoc''  discrete symmetry \cite{GW}, which forbids this processes at the tree level.  

\begin{itemize}
\item
In the so called {\bf Model I}, the invariance under $\Phi_1\to -\Phi_1$, $\Phi_2\to \Phi_2$, $D_R\to -D_R$, $U_R\to -U_R$, $l_R\to \pm l_R$ is imposed and $\lambda^u_2$, $\lambda_2^d$ vanish. The $Q=2/3$ and $Q=-1/3$ quarks couple to a single Higgs doublet and no FCN transition can occur at the tree level. 
\item In {\bf Model II} the invariance under  $\Phi_1\to -\Phi_1$, $\Phi_2\to \Phi_2$, $D_R\to -D_R$, $U_R\to U_R$, $l_R\to \pm l_R$ is imposed  and  $\lambda^u_1$, $\lambda_2^d$  vanish. The $Q=2/3$ and $Q=-1/3$ quarks couple to the different Higgs doublets and tree level FCNC transitions are absent\footnote{This scenario naturally arises in supersymmetric models as discussed in the next section.}. 
\item
In {\bf Model III} no ad hoc discrete symmetry is imposed and (\ref{2.32}) induces FCN couplings with the neutral Higgses at the tree level. In this case the magnitude of the FCNC couplings have to be arranged so that the prediction of this model do not contradict the measurements. The experimental data on the processes  $K^0\leftrightarrow \bar K^0$ and $K^0\to \mu^+\mu^-$  severely constrain the tree level Higgs coupling with the light $s$ and $d$ quarks. Other FCN couplings, especially those containing heavy quarks, are not so severely constrained by the experimental data. An example of Model III was proposed by 
 Cheng and Sher \cite{CS}, where for the magnitude of $q_i-q_i-H$ couplings is given by 
\begin{equation}
\label{2.34}
\xi_{i,j}\sim {\sqrt{m_im_j}\over v}\Delta_{ij}
\end{equation}
with the coefficients $\Delta_{ij}$ of the order of one. This model naturally suppresses the FCN transitions among the light quarks  (as required by the data in the down quark sector) and leaves the possibility for the  bigger FCN couplings among heavy quarks, especially among $t$ and $c$ quarks (the $t\leftrightarrow c$ transitions have been extensively studied in \cite{tc}). This Cheng-Sher ansatz (\ref{2.34}) naturally arises in the models with the  Fritzsch type of the Yukawa couplings \cite{CS}. 
\end{itemize}

\begin{figure}[h]
\centering
\mbox{
\subfigure[]
{
\begin{fmffile}{f9an}
\fmfframe(3,3)(3,3){
  \begin{fmfgraph*}(23,23)
  \fmfpen{thin}
  \fmfleft{l2,l1,l3,l4} \fmfrightn{r}{3}
  \fmf{fermion}{l1,v1,r1}
  \fmf{dashes,label=$h^0,,H^0,,A^0$,la.s=left,la.d=0.6,tension=0.6}{v1,v2} 
  \fmf{fermion}{r3,v2,r2}
  \fmflabel{$c$}{l1}\fmflabel{$u$}{r1}\fmflabel{$l^+$}{r3}\fmflabel{$l^-$}{r2}
  \end{fmfgraph*} }
\end{fmffile}
}
\quad
\subfigure[]
{
\begin{fmffile}{f9b}
  \fmfframe(3,3)(3,3){
  \begin{fmfgraph*}(23,23)
  \fmfpen{thin}
  \fmfleft{l1,l2}\fmfright{r1,r2}\fmftop{t1}
  \fmf{fermion}{l1,v1}
  \fmf{fermion,tension=0.4}{v1,v2}
  \fmf{fermion,tension=0.4,label=$d,,s,,b$,la.d=30,la.s=left}{v2,v3}
  \fmf{fermion}{v3,r1}
  \fmf{boson}{v2,t1}
  \fmf{dashes,label=$H^+$,tension=0.4}{v1,v3}
  \fmflabel{$c$}{l1}\fmflabel{$u$}{r1}\fmflabel{$\gamma$}{t1}
  \end{fmfgraph*} }
\end{fmffile}
}
\quad
\subfigure[]
{
\begin{fmffile}{f9c}
\fmfframe(3,3)(3,3){
  \begin{fmfgraph*}(23,23)
  \fmfpen{thin}
  \fmfleft{p1,p2,l1,q,l2,p3,p4} \fmfrightn{r}{2}
  \fmf{fermion}{l2,v1,l1}\fmf{dashes,label=$A^0$,la.s=left,la.d=20,tension=1}{v1,v2} \fmf{fermion}{r2,v2,r1}
  \fmflabel{$c$}{l2}\fmflabel{$\bar u$}{l1}\fmflabel{$l^+$}{r2}
  \fmflabel{$l^-$}{r1}  \fmflabel{$D^0$}{q}
  \end{fmfgraph*} }
\end{fmffile}
}
\quad 
\subfigure[]
{
\begin{fmffile}{f9d}
\fmfframe(5,5)(5,5){
  \begin{fmfgraph*}(23,23)
  \fmfpen{thin}
  \fmfleft{p1,p2,l1,q,l2,p3,p4} \fmfright{q1,q2,r1,qq,r2,q3,q4}
  \fmf{fermion}{l2,v1,l1}\fmf{dashes,label=$A^0$,la.s=left,la.d=20,tension=1.5}{v1,v2} \fmf{fermion}{r2,v2,r1}
  \fmflabel{$c$}{l2}\fmflabel{$\bar u$}{l1}\fmflabel{$\bar c$}{r2}\fmflabel{$u$}{r1}\fmflabel{$D^0$}{q}\fmflabel{$\bar D^0$}{qq}
  \end{fmfgraph*} }
\end{fmffile}
}
    }
\caption{Two Higgs doublet model: scenario III with flavour changing 
neutral couplings at tree level. Physical Higgs bosons in this model are 
 neutral scalars $H^0$ and $h^0$, neutral pseudoscalar $A^0$ and charged scalars 
 $H^{\pm}$. In Fig.  (b) the photon can be emitted from any charged line.}
\label{fig9}
\end{figure}
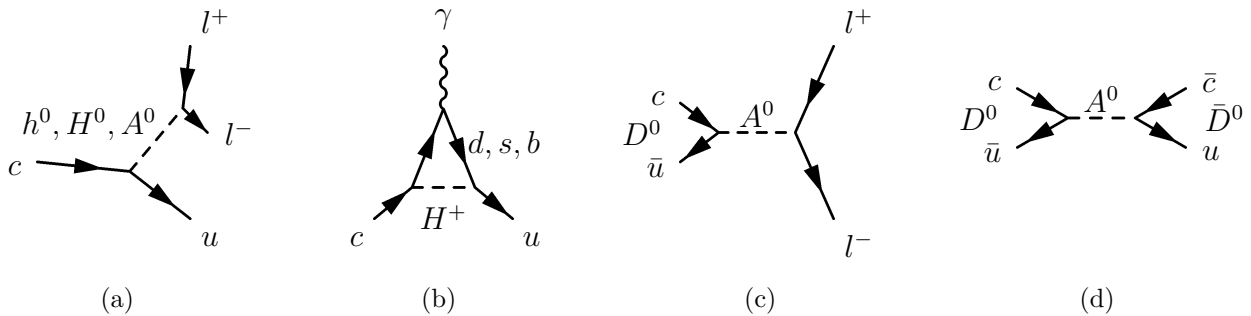

In any case, the tree level FCN couplings in the two Higgs doublet model are free parameters until they are constrained by the experimental data. The Models I and II are expected to induce only small modifications to $c\leftrightarrow u$ transition, as they arise only at the loop level. I consider only the Model III with the tree level $c-u-H$ couplings, where the leading new  contributions to the transitions $c\to ul^+l^-$, $c\to u\gamma$ and $D^0\to l^+l^-$ arise via the diagrams in Figs. \ref{fig9}a, \ref{fig9}b and \ref{fig9}c, respectively.

The process $c\to u\gamma$ was studied in \cite{CNS} and the leading contribution comes from the intermediate $b$ quark and charged Higgs. The upper limits on the $c-b-H^+$ and $b-u-H^-$ couplings were obtained from $b\to c\tau\nu_{\tau}$ and $b\to ul\nu_l$ decays, respectively, and the corresponding contribution is unobservably small
$$Br(c\to u\gamma)^H<1.8\cdot 10^{-11}~.$$   

One would expect that the effects of these couplings would be much more significant in the $c\to ul^+l^-$ decay, which occurs through the exchange of the scalar or pseudoscalar Higgs at the tree level. It turns out, however, that the smallness of the $~l-l-H$ coupling  $m_l/v$ renders even the $c\to u\mu^+\mu^-$ rate quite small. The upper limit on the coupling $\xi_{cu}$ among $c$, $u$ and the pseudoscalar Higgs $A^0$ is obtained when the experimental upper bound on $\Delta m_D<1.6\cdot 10^{-13}$ GeV  \cite{PDG} is saturated by the mechanism presented on Fig. \ref{fig9}d. The standard model prediction for $\Delta m_D$, discussed in Appendix D, is  namely  three order of magnitudes smaller than the present experimental upper bound and there is a large discovery window for new effects in this region. In the factorization approximation, the mechanism on Fig. \ref{fig9}d renders $\Delta m_D$ equal to  \cite{burdman}
$$\Delta m_D^{A^0}\simeq {5\over 12} {\xi_{cu}^2\over m_{A^0}^2}f_D^2m_D$$
and $\xi_{cu}\lesssim 7\cdot 10^{-4}$ is obtained for $m_{A^0}=300$ GeV and $f_D=0.2$ GeV. In terms of the relation (\ref{2.34}), this corresponds to the  upper bound  $\Delta_{cu}\lesssim 2$ of the order of unity for $m_c=1.25$ GeV and $m_u=3.25$ MeV, as expected. I use the upper bound on the $c-u-A^0$ coupling $\xi_{cu}$ to predict the $c\to u\mu^+\mu^-$ and $D^0\to\mu^+\mu^-$ rates corresponding to the mechanisms shown in Figs. \ref{fig9}a and \ref{fig9}c, respectively. The $D^0\to \mu^+\mu^-$ branching ratio is bounded at  
$$Br(D^0\to \mu^+\mu^-)^{A_0}={f_D^2m_D^3\over 8\pi\Gamma(D^0)}\biggl({\xi_{cu}m_{\mu}\over vm_{A^0}^2}\biggr)^2\lesssim 6\cdot 10^{-14}~,$$
which is about an order of magnitude larger than the standard model prediction but it is far bellow the present experimental sensitivity at the level  of $10^{-6}$ \cite{PDG}. 
The process $c\to u\mu^+\mu^-$ shown in Fig. \ref{fig9}a is mediated by a pseudoscalar and two scalar Higgses, which have different masses and couplings to $c$ and $u$ quarks discussed in Appendix C. For the reasons of simplicity the masses and couplings are taken to be equal and all three neutral Higgses  contribute to $c\to u\mu^+\mu^-$ an equal amount
$$Br(c\to u\mu^+\mu^-)^{H}={9m_c^5\over 768\pi^3\Gamma(D^0)}\biggl({\xi_{cu}m_{\mu}\over vm_{H}^2}\biggr)^2\lesssim 6.5 \cdot 10^{-15}~,$$
which is much smaller than the standard model prediction $1.7\cdot 10^{-9}$ (\ref{2.33}). I conclude by stressing that the $c-u-H$ coupling in model III is severely constrained by the experimental upper bound on $\Delta m_D$, which renders the new contributions to $c\to u\mu^+\mu^-$ and $D^0\to \mu^+\mu^-$ insignificant. 

The general model with tree level FCN couplings can violate also the lepton number at the tree level, as can be seen from (\ref{2.32}). My predictions above are based on the lepton number conserving $H-l-l$ couplings given by $m_l/v$. The FCN transitions accompanied with lepton number violation in Model III have been studied in \cite{CMM}. The authors of \cite{CMM} have obtained rather mild bounds on the $c-u-H$ and $l-l^{\prime}-H$ couplings
from the data on the charm meson leptonic and semileptonic decays, since they have not taken into account the upper bound on the $c-u-H$ coupling coming from $\Delta m_D$.

\subsection{Supersymmetric models}

Supersymmetry \footnote{The general discussion on supersymmetry is taken  
from the reviews on this subject listed in \cite{SUSY}.} is a symmetry between fermions and bosons. In a supersymmetric model each fermion of given quantum numbers with  respect to the gauge group has a bosonic superpartner with equal mass and quantum numbers; and vice versa. As we look at the list of the fermions and bosons in the standard model, we notice that none of them can be superpartner to the other. In the minimal supersymmetric standard model (MSSM) the minimal number of particles is added in order to supersymmetrize the standard model. None of the superpartners of the standard model particles have not been observed yet, so it is anticipated that they all have higher masses. As the mass of a particle and its superpartner are not equal, supersymmetry must be broken in any realistic supersymmetric model. One might wonder if there is a good reason why all of the superpartners of the standard model particles are heavy enough to have avoided discovery so far. The reason is that all the particles in MSSM, which have been observed so far, would be necessarily massless if the symmetry $SU(2)_L\times U(1)_Y$ was not broken. On the other hand, the mass terms for the Higgs and all the superpartners respect the full symmetry $SU(3)\times SU(2)_L\times U(1)_Y$ and have naturally higher masses. 

One of the motivations to study  supersymmetry is a hope that the hierarchy problem can be solved.  In the standard model the Higgs mass depends quadratically on the ultraviolet cut off for the radiative corrections and is expected to be large. Yet, it is experimentally constrained through its radiative effects to be less than $m_H<450$ GeV at the $95~\%$ confidence level \cite{PDG}. In the MSSM with exact supersymmetry, the loop corrections to the Higgs boson mass vanish, which can be considered as a natural explanation of the small Higgs mass in the broken MSSM. The second nice feature of MSSM is that it contains just the right proportions of the new particles, so the three gauge couplings corresponding to $SU(3)$, $SU(2)_L$ and $U(1)_Y$ groups run with energy a bit differently  and they seem to meet at the grand unification scale at about $10^{16}$ GeV. From the aesthetic theoretical point of view supersymmetry is attractive as it presents the only possible non-trivial unification\footnote{The supersymmetry group is not the direct product of the internal and space-time symmetry groups.} of the internal and space-time symmetries compatible with quantum field theory \cite{CM}.

\vspace{0.2cm}

Let me now turn to the flavour changing neutral (FCN) currents in the supersymmetric models, in particular to the FCN currents among the $c$ and $u$ quarks. We shall see, that  the FCN transitions in MSSM occur at the rates comparable to those in the standard model. In MSSM these rates turn out to be small due to the Super GIM mechanism which resembles the GIM mechanism in the standard model. The FCN transitions can however be severely enhanced over the standard model predictions in some non-minimal versions of the supersymmetric models \cite{BGM}. 

\vspace{0.2cm}

There are no tree level FCN couplings in MSSM. This is true in spite of the fact that MSSM necessarily contains two Higgs doublets and one might expect the tree level FCN couplings with Higgs, as discussed in Section 2.2.1. The absence of these couplings is connected with the supersymmetric nature of the theory, which requires the  super-potential to be an analytic function of the fields and can depend on the Higgs fields only through $\Phi$ and not through $\tilde \Phi$. The $\bar Q_L\tilde \Phi U_R$ term, for example, is forbidden and two different scalar doublets $\Phi_u$ and $\Phi_d$ are needed to give the masses to $Q=2/3$ and $Q=-1/3$ quarks \footnote{Two different scalar doublets of different hypercharge are needed also for the reasons of anomaly cancelations.}. Field $\Phi_u$ couples only to quarks with charge $2/3$, while  $\Phi_d$ couples only to quarks with charge $-1/3$
\begin{equation}
 {\cal L}_Y=\bar Q^{\prime}_L\lambda^u\Phi_uU_R^{\prime}+
\bar Q^{\prime}_L\lambda^d\Phi_dD_R^{\prime}+h.c.~.
\end{equation}
 The MSSM presents an example  of Model II discussed in Section 2.2.1 and has  no tree level FNC Higgs couplings.

Among the FCN interactions at the loop level, there are the standard model contributions, where the standard model particles run in the loop. In MSSM there are other particles that can run in the loop and couple to the external quarks thereby inducing new contributions to FCN processes. One possibility comes from the intermediate Higgs fields $\Phi_u$ and $\Phi_d$ that couple to the external quarks via the usual Yukawa type interaction and a typical representative of such process is shown in Fig. \ref{fig9}b. This kind of diagrams were discussed in Section 2.2.1 and were shown to have small effects on the FCN rates. In a supersymmetric theory the quarks do not couple only to the gauge bosons and Higgses, but also to their supersymmetric partners (the supersymmetric partner of a given standard model particle will be denoted by tilde). The corresponding vertices are the supersymmetrized version of the standard model vertices with the same strength and with two particles replaced by their superpartners, as shown schematically in Fig. \ref{fig10}. Among the new quark interactions in MSSM, the only strong interaction is given by the quark-squark-gluino vertex. This vertex induces the dominant contributions to the FCN processes in the supersymmetric models. The corresponding diagram for the $c\to u\gamma$ decay is shown in Fig. \ref{fig11}. In Fig. \ref{fig13}a, this diagram is redrawn in the good weak isospin quarks basis, which is  more instructive for further discussion. In the weak isospin basis, the gauge interactions are diagonal, while the mass terms are nondiagonal and couple left and right handed quarks. For the purpose of further discussion let me first derive the convenient  expressions for the quark mass matrices in the weak isospin basis. 
The mass  and the weak eigenstates are denoted by $U^T=(u,c,t)$, $D^T=(d,s,b)$ and $U^{\prime T}=(u^{\prime},c^{\prime},t^{\prime})$,  $D^{\prime T}=(d^{\prime},s^{\prime},b^{\prime})$, respectively, and the quark mass term is given by
\begin{equation*}
{\cal L}_M=\bar U^{\prime}_L\Gamma^u U_R^{\prime}+\bar D^{\prime}_L\Gamma^d D_R^{\prime}+h.c.~.
\end{equation*}
We can choose to work in the basis where $\Gamma^u=\Gamma^u_D$ is diagonal by performing the unitary transformations on $Q_L^\prime=(U_L^\prime,D_L^\prime)$ and $U_R^\prime$. It is not possible to do the same for $\Gamma^d$ since we have already chosen a particular form of $Q_L$ to diagonalize $\Gamma^u$. In this case  $U^{\prime}=U$, while $D^{\prime}$ is related to the mass eigenstate $D$ via the unitary transformation $D_{L,R}=A_{L,R}D^{\prime}_{L,R}$. We note also that $A_L^{\dagger}=V^{CKM}$ since $j^{\mu}_W=\bar U_L^{\prime}\gamma^{\mu}D_L^{\prime}=\bar U_L\gamma^{\mu}A_L^{d\dagger}D_L\equiv \bar U_L\gamma^{\mu}V^{CKM}D_L$. So the quark mass matrices in the weak isospin basis are given by
\begin{equation}
\label{2.36}
\Gamma^u=\Gamma^u_D~,\qquad\qquad \Gamma^d=V^{CKM}\Gamma_D^dA_R~.
\end{equation}

\begin{figure}[h]
\centering
\mbox{
\subfigure[]
{
\begin{fmffile}{f10o1nn}
\fmfframe(5,2)(5,2){
  \begin{fmfgraph*}(20,12)
  \fmfpen{thin}
  \fmfleftn{l}{1} \fmfrightn{r}{2}
  \fmf{fermion}{l1,v1}\fmf{fermion}{v1,r2} 
  \fmf{gluon}{v1,r1}
  \fmflabel{$q$}{l1}\fmflabel{$q$}{r2}\fmflabel{$G$}{r1}
  \end{fmfgraph*} }
\end{fmffile}
\quad
\begin{fmffile}{f10o3nn}
\fmfframe(5,2)(5,2){
  \begin{fmfgraph*}(20,12)
  \fmfpen{thin}
  \fmfleftn{l}{1} \fmfrightn{r}{2}
  \fmf{fermion}{l1,v1}\fmf{fermion}{v1,r2} 
  \fmf{dashes}{v1,r1}
  \fmflabel{$q$}{l1}\fmflabel{$q$}{r2}\fmflabel{$H$}{r1}
  \end{fmfgraph*} }
\end{fmffile}
}
\subfigure[]
{
\begin{fmffile}{f10o2nnn}
\fmfframe(5,2)(5,2){
  \begin{fmfgraph*}(20,12)
  \fmfpen{thin}
  \fmfleftn{l}{1} \fmfrightn{r}{2}
  \fmf{fermion}{l1,v1}\fmf{fermion}{v1,r1} 
  \fmf{dashes}{v1,r2}
  \fmflabel{$q$}{l1}\fmflabel{$\tilde q$}{r2}\fmflabel{$\tilde G$}{r1}
  \end{fmfgraph*} }
\end{fmffile}   
\quad
\begin{fmffile}{f10o4nn}
\fmfframe(5,2)(5,2){
  \begin{fmfgraph*}(20,12)
  \fmfpen{thin}
  \fmfleftn{l}{1} \fmfrightn{r}{2}
  \fmf{fermion}{l1,v1}\fmf{fermion}{v1,r1} 
  \fmf{dashes}{v1,r2}
  \fmflabel{$q$}{l1}\fmflabel{$\tilde q$}{r2}\fmflabel{$\tilde H$}{r1}
  \end{fmfgraph*} }
\end{fmffile}
}
    }
\caption{Quark interactions in minimal supersymmetric standard model. In Fig. (a)  the standard model interactions are shown, while in Fig. (b)  their supersymmetric counterparts are added. The $q$, $G$ and $H$ denote quark, gauge boson and Higgs, respectively. The $\tilde q$, $\tilde G$ and $\tilde H$ denote their superpartners. Solid, wavy and dashed lines denote spin $1/2$, $1$ and $0$ particles, respectively.}
\label{fig10}
\end{figure}
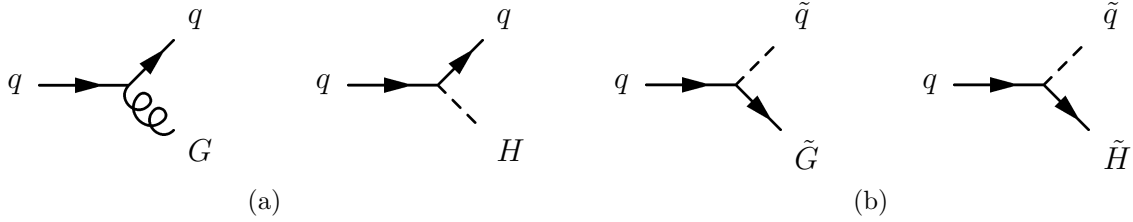

\begin{figure}[h]

\centering
\mbox{
\begin{fmffile}{f11n}
  \fmfframe(3,3)(3,3){
  \begin{fmfgraph*}(35,20)
  \fmfpen{thin}
  \fmfleft{l1}\fmfright{r1}\fmftop{p1,t1,p3,p2,p4}
  \fmf{fermion,tension=1}{l1,v1,v2}
  \fmf{dashes,tension=0.3,label=$\tilde u,,\tilde c,,\tilde t$,la.s=left}{v2,v3}
  \fmf{fermion}{v3,r1}
  \fmffreeze
  \fmf{boson,tension=2}{v1,t1}
  \fmf{fermion,label=$\tilde g$,tension=0.2,right=1}{v2,v3}
  \fmflabel{$c$}{l1}\fmflabel{$u$}{r1}\fmflabel{$\gamma$}{t1}
  \end{fmfgraph*} }
\end{fmffile}
     }
\caption{Dominant diagram for the $c\to u\gamma$ decay in the minimal 
supersymmetric standard model (in addition to the standard model diagrams). 
The photon can be emitted from any of the charged lines.}  
\label{fig11}
\end{figure}
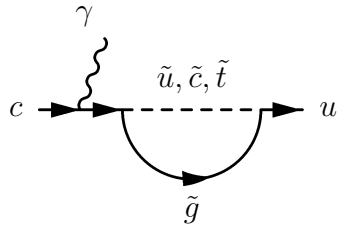

In order to get acquainted with the weak isospin quark basis, it is instructive to see how the standard model $c\to u\gamma$ amplitude  of the form $m_c\sum_{d,s,b} V_{ci}^*V_{ui}m_i^2$ (\ref{2.26}, \ref{2.35})  naturally arises in this basis. In the standard model the $c\to u\gamma$ transition couples the left-handed $u$ quark and the right handed $c$ quark, as shown in Appendix B for $m_u\to 0$. In the weak isospin bases the amplitude is expanded in orders of the mass vertices inserted in the diagram. An odd number of mass insertions is necessary to couple $u_L-c_R$ and the first order in this expansion vanishes. The lowest non-vanishing order is shown in Fig. \ref{fig16} with three mass insertions. By using  (\ref{2.36})
\begin{equation}
\label{2.38}
m_c\Gamma_{1i}^d\Gamma_{i2}^{d\dagger}=m_c(V^{CKM}\Gamma_D^{d}A_RA_R^{\dagger}\Gamma_D^{d\dagger}V^{CKM\dagger})_{12}=m_c\sum_{d,s,b} V_{ci}^*V_{ui}m_i^2
\end{equation}
and the same form of the $c\to u\gamma$ amplitude  as in the good mass basis (\ref{2.26}, \ref{2.35}) is obtained.

\begin{figure}[h]

\centering
\mbox{
\begin{fmffile}{f12}
  \fmfframe(3,3)(3,3){
  \begin{fmfgraph*}(65,20)
  \fmfpen{thin}
  \fmfleft{l1}\fmfright{r1}\fmftop{t1,p1,p2,p3,p4}
  \fmf{plain,tension=1}{l1,a}
  \fmf{plain}{a,v1}
  \fmf{plain,tension=1,label=$c_L$,la.s=left}{v1,v2}
  \fmf{plain,tension=1,label=$c_L^\prime$,la.s=left}{v2,v3}
  \fmf{plain,tension=0.5,label=$s_L^\prime$,la.s=left}{v3,v4}
  \fmf{plain,tension=0.4,label=$d_L^\prime,,s_L^\prime,,b_L^\prime$,
            la.s=left}{v4,v5}
  \fmf{plain,tension=0.5,label=$d_L^\prime$,la.s=left}{v5,v6}
  \fmf{plain,tension=1,label=$u_L^\prime$,la.s=left}{v6,v7}
  \fmf{plain,tension=1}{v7,r1}
  \fmffreeze
  \fmf{boson,tension=2}{a,t1}
  \fmf{zigzag,label=$W$,la.s=right,right=0.5}{v3,v6}
  \fmflabel{$c_R$}{l1}\fmflabel{$u_L$}{r1}\fmflabel{$\gamma$}{t1}
  \fmfv{de.sh=pentagram,de.si=3thick,de.fill=full}{v1,v4,v5}
  \fmfv{de.sh=circle,de.si=3thick,de.fill=full}{v2,v7}
  \end{fmfgraph*} }
\end{fmffile}
     }
\caption{The $c\to u\gamma$ process in the standard model at the lowest order of
the mass insertion expansion. Quarks with primes are in the weak isospin basis,
while those without primes are in the mass eigenstate basis. The crosses denote the
mass insertion, while the dots denote the unitary transformation among weak and
the mass basis. The photon can be emitted from any charged line. } 
\label{fig16} 
\end{figure}
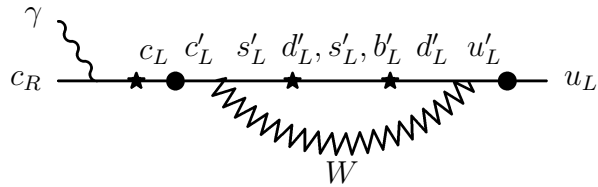

\begin{figure}[h]

\centering
\mbox{ 
\subfigure[The process $c\to u\gamma$. ]
{
\begin{fmffile}{f13a1}
  \fmfframe(3,3)(3,3){
  \begin{fmfgraph*}(40,25)
  \fmfpen{thin}
  \fmfleft{l1}\fmfright{r1}\fmftop{t1,p1,p2,p3,p4}
  \fmf{plain,tension=1}{l1,a}
  \fmf{plain}{a,v1}
  \fmf{plain,tension=0.5,label=$c_L$,la.s=right,la.d=20}{v1,v3}
  \fmf{dashes,tension=0.5,label=$\tilde c_L^\prime$,la.d=20,la.s=right}{v3,v4}
  \fmf{dashes,tension=0.5,label=$\tilde u_L^\prime$,la.d=20,la.s=right}{v4,v6}
  \fmf{plain,tension=1}{v6,r1}
  \fmffreeze
  \fmf{boson,tension=2}{a,t1}
  \fmf{plain,label=$\tilde g$,la.s=right,right=1}{v3,v6}
  \fmflabel{$c_R$}{l1}\fmflabel{$u_L$}{r1}\fmflabel{$\gamma$}{t1}
  \fmfv{de.sh=pentagram,de.si=3thick,de.fill=full}{v1,v4}
  \fmfv{la.d=3thick,la.a=90,label=$m_c$}{v1}
  \fmfv{la.d=3thick,la.a=90,label=$(\Gamma^{\tilde u_L})_{12}$}{v4}
  \fmflabel{{\bf I}}{p2}
  \end{fmfgraph*} }
\end{fmffile}
\quad
\begin{fmffile}{f13a2}
  \fmfframe(3,3)(3,3){
  \begin{fmfgraph*}(40,25)
  \fmfpen{thin}
  \fmfleft{l1}\fmfright{r1}\fmftop{t1,p1,p2,p3,p4}
  \fmf{plain,tension=1}{l1,a}
  \fmf{plain}{a,v3}
  \fmf{dashes,tension=0.5,label=$\tilde c_R^\prime$,la.d=20,la.s=right}{v3,v4}
  \fmf{dashes,tension=0.3,label=$\tilde c_L^\prime$,la.d=20,la.s=right}{v4,v5}
  \fmf{dashes,tension=0.5,label=$\tilde u_L^\prime$,la.d=20,la.s=right}{v5,v6}
  \fmf{plain,tension=1}{v6,r1}
  \fmffreeze
  \fmf{boson,tension=2}{a,t1}
  \fmf{plain,label=$\tilde g$,la.s=right,right=1}{v3,v6}
  \fmflabel{$c_R$}{l1}\fmflabel{$u_L$}{r1}\fmflabel{$\gamma$}{t1}
  \fmfv{de.sh=pentagram,de.si=3thick,de.fill=full}{v4,v5}
  \fmfv{la.d=3thick,la.a=70,label=$(\Gamma^{\tilde u_L\tilde u_R})_{22}$}{v4}
  \fmfv{la.d=8thick,la.a=70,label=$(\Gamma^{\tilde u_L})_{12}$}{v5}
  \fmflabel{{\bf II}}{p2}
  \end{fmfgraph*} }
\end{fmffile}
\quad
\begin{fmffile}{f13a3}
  \fmfframe(3,3)(3,3){
  \begin{fmfgraph*}(40,25)
  \fmfpen{thin}
  \fmfleft{l1}\fmfright{r1}\fmftop{t1,p1,p2,p3,p4}
  \fmf{plain,tension=1}{l1,a}
  \fmf{plain}{a,v3}
  \fmf{dashes,tension=0.5,label=$\tilde c_R^\prime$,la.d=20,la.s=right}{v3,v4}
  \fmf{dashes,tension=0.5,label=$\tilde u_L^\prime$,la.d=20,la.s=right}{v4,v6}
  \fmf{plain,tension=1}{v6,r1}
  \fmffreeze
  \fmf{boson,tension=2}{a,t1}
  \fmf{plain,label=$\tilde g$,la.s=right,right=1}{v3,v6}
  \fmflabel{$c_R$}{l1}\fmflabel{$u_L$}{r1}\fmflabel{$\gamma$}{t1}
  \fmfv{de.sh=pentagram,de.si=3thick,de.fill=full}{v4}
  \fmfv{la.d=3thick,la.a=90,label=$(\Gamma^{\tilde u_L\tilde u_R})_{12}$}{v4}
  \fmflabel{{\bf III}}{p2}
  \end{fmfgraph*} }
\end{fmffile}
}
     }
\mbox{
\subfigure[The process $D^0\to \bar D^0$. ]
{
\begin{fmffile}{f13b1}
  \fmfframe(6,3)(6,3){
  \begin{fmfgraph*}(45,25)
  \fmfpen{thin}
  \fmfleft{l1,l2}\fmfright{r1,r2}
  \fmf{fermion}{l2,v2}\fmf{plain,label=$\tilde g$,la.s=left,tension=0.5}{v2,a2}\fmf{plain}{a2,r2}
  \fmf{fermion}{v1,l1}\fmf{plain,label=$\tilde g$,la.s=right,tension=0.5}{v1,a1}\fmf{plain}{a1,r1}
   \fmffreeze  
  \fmf{dashes,tension=0.5,label=$\tilde u_L^\prime$,la.d=20,la.s=right}{v1,v3}
  \fmf{dashes,tension=0.5,label=$\tilde c_L^\prime$,la.d=20,la.s=right}{v3,v2}
  \fmf{dashes,tension=0.5,label=$\tilde c_L^\prime$,la.d=20,la.s=left}{a1,a3}
  \fmf{dashes,tension=0.5,label=$\tilde u_L^\prime$,la.d=20,la.s=left}{a3,a2}
  \fmflabel{$c_L$}{l2}\fmflabel{$u_L$}{r2}
  \fmflabel{$\bar u_L$}{l1}\fmflabel{$\bar c_L$}{r1}
  \fmfv{de.sh=pentagram,de.si=3thick,de.fill=full}{v3,a3}
  \fmfv{la.d=3thick,la.a=0,label=$(\Gamma^{\tilde u_L})_{12}$}{v3}
  \end{fmfgraph*} }
\end{fmffile}
\quad
\begin{fmffile}{f13b2}
  \fmfframe(6,3)(6,3){
  \begin{fmfgraph*}(50,25)
  \fmfpen{thin}
  \fmfleft{l1,l2}\fmfright{r1,r2}
  \fmf{fermion}{l2,v2}\fmf{plain,label=$\tilde g$,la.s=left,tension=0.5}{v2,a2}\fmf{plain}{a2,r2}
  \fmf{fermion}{v1,l1}\fmf{plain,label=$\tilde g$,la.s=right,tension=0.5}{v1,a1}\fmf{plain}{a1,r1}
   \fmffreeze  
  \fmf{dashes,tension=0.5,label=$\tilde u_R^\prime$,la.d=20,la.s=right}{v1,v3}
  \fmf{dashes,tension=0.5,label=$\tilde c_L^\prime$,la.d=20,la.s=right}{v3,v2}
  \fmf{dashes,tension=0.5,label=$\tilde c_R^\prime$,la.d=20,la.s=left}{a1,a3}
  \fmf{dashes,tension=0.5,label=$\tilde u_L^\prime$,la.d=20,la.s=left}{a3,a2}
  \fmflabel{$c_L$}{l2}\fmflabel{$u_L$}{r2}
  \fmflabel{$\bar u_R$}{l1}\fmflabel{$\bar c_R$}{r1}
  \fmfv{de.sh=pentagram,de.si=3thick,de.fill=full}{v3,a3}
  \fmfv{la.d=3thick,la.a=0,label=$(\Gamma^{\tilde u_L\tilde u_R})_{12}$}{v3}
  \end{fmfgraph*} }
\end{fmffile}
}
     } 
\caption{The   dominant contributions to the processes (a) $c\to u\gamma$ and (b) $D^0\to
\bar D^0$ in the supersymmetric 
models.  The crosses denote the
quark or squark mass insertions $(\Gamma^{\tilde u_L})_{12}$ and $(\Gamma^{\tilde u_L\tilde u_R})_{12}$ - the lowest order in the 
mass insertion expansion is given. 
States with primes are in the weak isospin basis,
while those without primes are in the mass eigenstate basis. I choose to work in
the basis where the  mass and weak eigenstates for the up-like quarks match 
(\ref{2.36}). The mass matrix for the up-like squarks has to be non-diagonal 
 (the mass and the weak eigenstates for the up-like squarks 
have to be different) in order to get the flavour changing neutral transition. }
\label{fig13}
\end{figure}
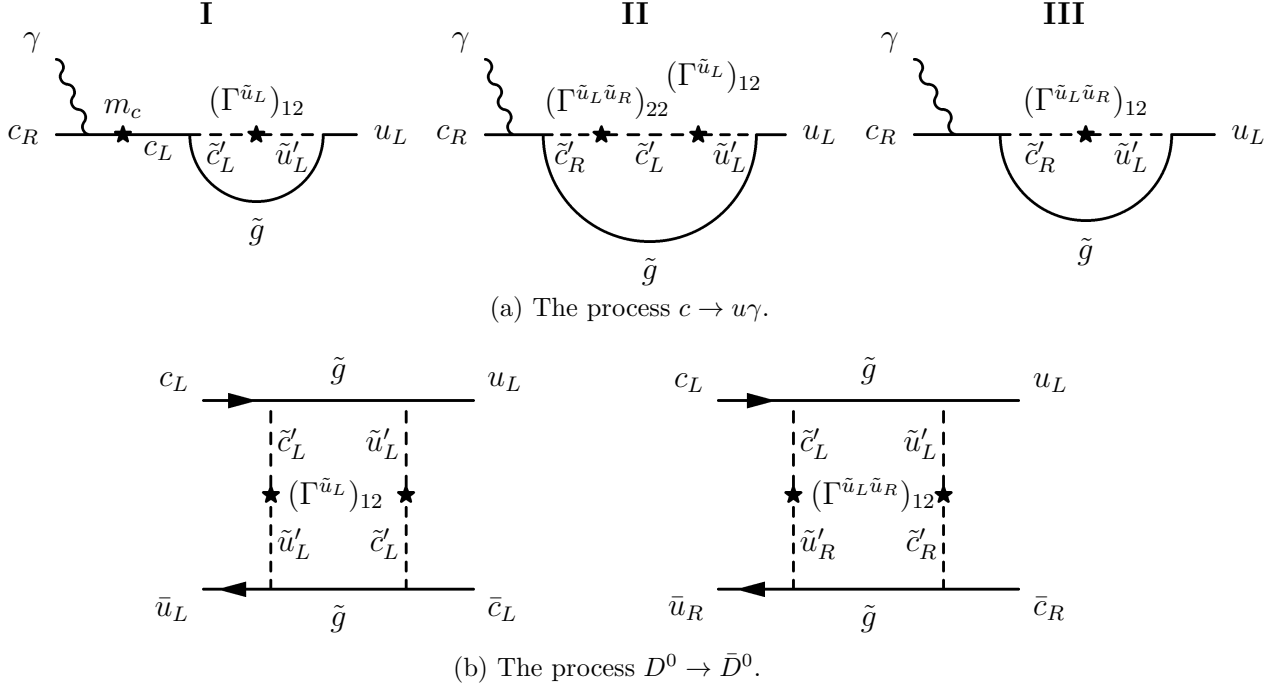

\vspace{0.2cm}

The dominant $c\to u\gamma$ diagrams in MSSM are shown in Fig. \ref{fig13}a. The crosses represent the squark mass matrix $\Gamma^{\tilde u_L}$ for the $\tilde U_L^{\prime}$ squarks or the matrix  $\Gamma^{\tilde u_L\tilde u_R}$ responsible for the $\tilde U_L^{\prime}-\tilde U_R^{\prime}$ squark mixing\footnote{The $\tilde q_L$ and $\tilde q_R$ are two different scalar particles.}. In order to get a flavour changing neutral transition among $c$ and $u$ quarks, these matrices  must have a nonzero matrix element $\{12\}$ in the weak eigenstate bases. The main subject to follow is therefore to determine the {\bf squark mass matrices} in different supersymmetric scenarios.
\begin{itemize}
\item
In the model with the {\bf exact supersymmetry}, the masses of squarks are equal to the masses of quarks. The mass term for the squarks in the weak eigenstate basis 
\begin{equation}
\label{2.37}
{\cal L}_M=\tilde U_L^{\prime\dagger}\Gamma^{u\dagger}\Gamma^u\tilde U_L^{\prime}+\tilde U_R^{\prime\dagger}\Gamma^{u\dagger}\Gamma^u\tilde U_R^{\prime}+\tilde D_L^{\prime\dagger}\Gamma^{d\dagger}\Gamma^d\tilde D_L^{\prime}+\tilde D_R^{\prime\dagger}\Gamma^{d\dagger}\Gamma^d\tilde D_R^{\prime}+...
\end{equation}
is diagonal due to (\ref{2.36}). In addition there is no $\tilde U_L^{\prime}-\tilde U_R^{\prime}$ mixing and no FCN contribution can arise until the supersymmetry is broken.
\item 
In any realistic model supersymmetry has to be broken. It has been found that it is almost impossible to break the supersymmetry spontaneously by giving the non-zero vacuum expectation value to one of the scalar fields present in the  MSSM \cite{SUSY}. Supersymmetry is broken spontaneously by a nonzero vacuum expectation value of some fields not present in MSSM, but in a  ``hidden'' sector at higher energies. In the ``low'' energy MSSM sector this mechanism is effectively described by the Lagrangian ${\cal L}_{soft}$, which is added to the supersymmetric part of the Lagrangian  and explicitly breaks the supersymmetry. In order to keep the theory renormalizable, ${\cal L}_{soft}$ can contain only the so-called soft breaking terms. Among the soft terms there are the squark mass terms and $\tilde q_L-\tilde q_R$ mixing terms with completely general couplings. Such {\bf general softly broken supersymmetric model}\footnote{A general softly broken minimal supersymmetric standard model has more than 100 free parameters.} could render huge contributions to FCN processes and  we can not say much about the predictions of the model  until we find some organizing principle for the values of the soft breaking couplings. This is achieved by invoking some dynamical model of supersymmetry breaking in the ``hidden'' sector, which gives the values of all the soft breaking terms in terms of few parameters of the dynamical model. 
\item
I discuss the squark mass matrices in the models where the supersymmetry is broken via the {\bf gauge mediation} . 
In addition to the MSSM particles the model contains also the scalar particles $S$ in the ``hidden'' sector. These scalars have the nonzero vacuum expectation value and do not couple to any of the MSSM particles directly. There are additional messenger fermions and scalars, that couple to the scalar particles $S$ and also to the gauge bosons and the gauginos of the MSSM. In this way the messenger particles communicate the supersymmetry breaking from the ``hidden'' sector through the flavour blind gauge interactions to MSSM.

\begin{figure}[h]

\centering
\mbox{
\subfigure[The origin of $(\Gamma^{\tilde u_L})_{12}$ (\ref{2.39}): the cross denotes the matrix element $\{12\}$ of the down-like squark mass matrix $(\Gamma^d\Gamma^{d\dagger})_{12}$  given by the supersymmetric part of the Lagrangian (\ref{2.37}).  ]
{
\begin{fmffile}{f14ann}
  \fmfframe(8,2)(8,2){
  \begin{fmfgraph*}(45,25)
  \fmfpen{thin}
  \fmfleft{l1,l2}\fmfright{r1,r2}\fmftopn{t}{4}
  \fmf{dashes,tension=1}{l1,v3}
  \fmf{dashes,tension=0.5,label=$\tilde s_L^\prime$,la.d=20,la.s=right}{v3,v4}
  \fmf{dashes,tension=0.5,label=$\tilde d_L^\prime$,la.d=20,la.s=right}{v4,v6}
  \fmf{dashes,tension=1}{v6,r1}
  \fmffreeze
  \fmf{zigzag,label=$W$,la.s=left,left=0.5}{v3,a1}
  \fmf{phantom,tension=8}{a1,t2}\fmf{phantom,tension=8}{a2,t3}
  \fmf{zigzag,label=$W$,la.s=left,left=0.5}{a2,v6}
  \fmf{dbl_dots,right}{a1,a2}\fmf{dbl_dots,left}{a1,a2}
  \fmflabel{$\tilde c_L^\prime$}{l1}\fmflabel{$\tilde u_L^\prime$}{r1}
  \fmfv{de.sh=pentagram,de.si=3thick,de.fill=full}{v4}
  \fmfv{la.d=3thick,la.a=90,label=$(\Gamma^d\Gamma^{d\dagger})_{12}$}{v4}
  \end{fmfgraph*} }
\end{fmffile}
}
\quad
\subfigure[The origin of $(\Gamma^{\tilde u_L\tilde u_R})_{22}$ (\ref{2.40}): the cross denotes the matrix element $\{22\}$ of the up-like quark mass matrix $(\Gamma^u)_{22}=m_c$. ]
{
\begin{fmffile}{f14bnn}
  \fmfframe(8,2)(8,2){
  \begin{fmfgraph*}(45,25)
  \fmfpen{thin}
  \fmfleft{l1,l2}\fmfright{r1,r2}\fmftopn{t}{4}
  \fmf{dashes,tension=1}{l1,v3}
  \fmf{plain,tension=0.5,label=$c_L^\prime$,la.d=20,la.s=right}{v3,v4}
  \fmf{plain,tension=0.5,label=$c_R^\prime$,la.d=20,la.s=right}{v4,v6}
  \fmf{dashes,tension=1}{v6,r1}
  \fmffreeze
  \fmf{fermion,label=$\tilde g$,la.s=left,left=0.5}{v3,a1}
  \fmf{phantom,tension=8}{a1,t2}\fmf{phantom,tension=8}{a2,t3}
  \fmf{plain,label=$\tilde g$,la.s=left,left=0.5}{a2,v6}
  \fmf{dbl_dots,right}{a1,a2}\fmf{dbl_dots,left}{a1,a2}
  \fmflabel{$\tilde c_L^\prime$}{l1}\fmflabel{$\tilde c_R^\prime$}{r1}
  \fmfv{de.sh=pentagram,de.si=3thick,de.fill=full}{v4}
  \fmfv{la.d=3thick,la.a=90,label=$(\Gamma^u)_{22}$}{v4}
  \end{fmfgraph*} }
\end{fmffile}
}
     }
\caption{The mechanism that gives rise to the squark mass matrices in the gauge-mediated supersymmetric models. The doted lines denote the messenger particles which  comunicate the  supersymmetry breaking from the hidden sector via the flavour blind  gauge interactions to the MSSM.}
\label{fig14}
\end{figure}
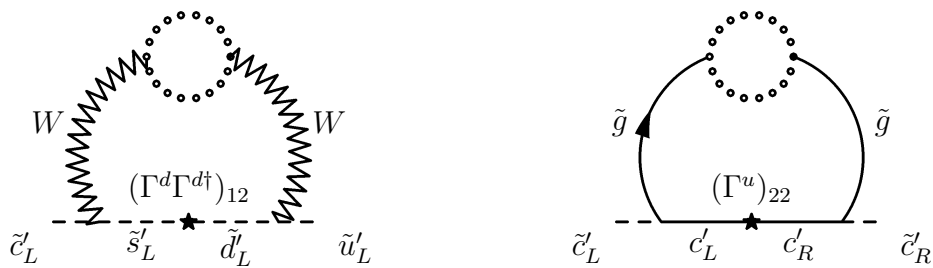

The nondiagonal squark mass terms $\Gamma^{\tilde u_L}$, in particular, arise via the loop diagram in Fig. \ref{fig14}a. The dotted lines denote the messenger particles, while others are the ordinary MSSM particles. The cross denotes the matrix element $\{12\}$ of the down-like squark mass matrix given by the supersymmetric part of the Lagrangian (\ref{2.37})  
\begin{equation}
\label{2.39}
(\Gamma^{\tilde u_L})_{12}\propto (\Gamma^d\Gamma^{d\dagger})_{12}=(V^{CKM}\Gamma_D^{d}A_RA_R^{\dagger}\Gamma_D^{d\dagger}V^{CKM\dagger})_{12}=\sum_{d,s,b} V_{ci}^*V_{ui}m_i^2
\end{equation}
and the proportionality coefficient turns out to be of the order of one \cite{duncan}. 

The diagonal squark $\tilde U_L^{\prime}-\tilde U_R^{\prime}$ mixing terms $\Gamma^{\tilde u_L\tilde u_R}$ arise among others via the loop diagram in Fig. \ref{fig14}b. The cross denotes the  matrix element $\{22\}$ of the up-like quark mass matrix and
\begin{equation}
\label{2.40}
(\Gamma^{\tilde u_L\tilde u_R})_{22}\propto (\Gamma^u)_{22}=m_c
\end{equation}
with the proportionality coefficient of the order of one \cite{duncan}. It turns out that one can not generate any off diagonal term  
\begin{equation}
\label{2.41}
(\Gamma^{\tilde u_L\tilde u_R})_{12}=0
\end{equation}
 and amplitude for the diagram $III$ in Fig. \ref{fig13}a vanishes.

 The mass matrix elements $(\Gamma^{\tilde u_L})_{12}$ (\ref{2.39}) and $(\Gamma^{\tilde u_L\tilde u_R})_{22}$ (\ref{2.40}) enter the $c\to u\gamma$ amplitude at the crosses in diagrams $I$ and $II$ of Fig. \ref{fig13}a giving 
\begin{equation}
A(c\to u\gamma)^{MSSM}\propto m_c\sum_{d,s,b} V_{ci}^*V_{ui}m_i^2~.
\end{equation}
The resulting $c\to u\gamma$ amplitude has the same form as  the standard model one-loop electroweak amplitude  given in (\ref{2.26}, \ref{2.35},  \ref{2.38}). In MSSM with this specific scenario of the supersymmetry breaking, the flavour changing neutral currents are subject to the similar GIM suppression  as in the standard model (Super GIM). The FCN  rates among the up-like quarks in MSSM are small as they depend on the down-like quark masses and are  comparable to those in the standard model.  Due to the  uncertainties of the theoretical predictions concerning the additional long distance effects and due to the experimental difficulties connected with measuring the small rates, the MSSM effects would hardly be observable in the $c
\to u$ transitions in the near future. 
\item
There is a lot to be said for the minimal realization of certain symmetry. It has the most stringent constraints and therefore the most predictive power. Nevertheless it would be to dogmatic to consider such a minimal ansatz only - in particular if that would prompt us to overlook potentially promising areas where new physics could emerge. The FCN transitions $c\to u$ have been found to be significantly affected in various {\bf non-minimal supersymmetric models} \cite{BGM}. The non-minimal models contain additional particles not present in MSSM (the MSSM contains a minimal set of particles that supersymmetrizes the standard model). 
 In order to get FCNC of a significant strength in the charm sector, the authors of  \cite{BGM} have constructed models\footnote{A model where the squark mass matrix $\Gamma^{\tilde u_L}$ is  not proportional to $|\Gamma_D^d|^2$ is obtained, for example,  by adding to MSSM with Higgs doublets $\Phi_u$ and $\Phi_d$ another pair of Higgs doublets $\Phi_u^\prime$ and $\Phi_d^\prime$. The $\Phi_u$ and $\Phi_u^\prime$ give masses to the up-like quarks and  $\Phi_d$ and $\Phi_d^\prime$ give masses to the down-like quarks \cite{BGM}. A realistic example of such kind is provided by the supersymmetric version of the model based on $SU(5)$ symmetry.} where the squark mass matrix $\Gamma^{\tilde u_L}$ is  not proportional to $|\Gamma_D^d|^2$ or not proportional to $V^{CKM}$ (in contrast to (\ref{2.39})) and/or where non-diagonal terms of $\Gamma^{\tilde u_L\tilde u_R}$ are  not equal to zero (in contrast to (\ref{2.41})). In the end the authors of  \cite{BGM} have treated the matrix elements $(\Gamma^{\tilde u_L})_{12}$  and $(\Gamma^{\tilde u_L\tilde u_R})_{12}$ as free parameters. These two parameters enter at the place of the crosses in the $c\to u\gamma$ diagrams in Fig. \ref{fig13}a as well as the $D^0-\bar D^0$ diagrams in Fig. \ref{fig13}b. By saturating the upper experimental upper bound on $\Delta m_D$ via the mechanism in Fig. \ref{fig13}b, they have obtained\footnote{The squark and gluino masses are taken to be of the order of $100$ GeV.} the upper bounds on $(\Gamma^{\tilde u_L})_{12}$  and $(\Gamma^{\tilde u_L\tilde u_R})_{12}$ and predicted the upper bounds on the $c\to u\gamma$ rate. With the present experimental upper bound $\Delta m_D<1.6\cdot 10^{-13}$ \cite{PDG} I get 
\begin{equation}
\label{2.43}
(\Gamma^{\tilde u_L})_{12}\le 72\ {\rm GeV}\qquad {\rm  and}\qquad (\Gamma^{\tilde u_L\tilde u_R})_{12}\le 260\ {\rm GeV}~.
\end{equation}
These squark mass matrix elements, arising from the non-minimal supersymmetric model, induce the $c\to u\gamma$ decay through  the mechanism in Fig. \ref{fig13}a 
\begin{equation}
\label{brsusy}
Br(c\to u\gamma)^{{non-minimal\atop SUSY}}\le 1.2 \cdot 10^{-5}
\end{equation}
with the dominant contribution coming from the diagram $III$. 
The bound on the branching ratio can be translated to the bound on the coefficient $c_7^{eff}$ responsible for the magnitude of  the effective Lagrangian (\ref{o7}) 
\begin{equation}
\label{c7susy}
|c_7^{eff}|^{{non-minimal\atop SUSY}}\le 0.14~.
\end{equation}
In the non-minimal supersymmetric model, the $c\to u\gamma$ branching ratio can be enhanced up to three orders of magnitude compared to the standard model prediction $(1.3\pm 0.6)\cdot 10^{-8}$ (\ref{2.60}). The main reason for this enhancement can be traced back the fact that the proportionality factor $m_q^2/m_W^2$ in the standard model and $m_q^2/m_H^2$ in two Higgs doublet model is replaced by the factor $m_{\tilde g}^2/m_{\tilde q}^2$ in the supersymmetric model  ($m_q$, $m_{\tilde q}$ and $m_{\tilde g}$ denote the masses of quark, squark and gluino in the loop).

The $c\to ul^+l^-$ and $D^0\to l^+l^-$ decays rates also depend on the matrix elements bounded by (\ref{2.43}) in the non-minimal supersymmetric model. The upper bounds for the $c\to ul^+l^-$ and $D^0\to l^+l^-$ decays rates in this model have been studied in \cite{BGM} and were found to be smaller than the corresponding standard model rates, so there is no hope for SUSY manifestation the leptonic channels. 
\end{itemize}

\subsection{Fourth-generation signatures}

The rate of the flavor changing transitions in the up-quark sector increases as the masses of the down-like quarks increase (\ref{gamma.one.c7}). In a three generation model $m_d$, $m_s$ and $m_b$ are small and the corresponding FCN transitions  among the up-like quarks are rare. These processes are naturally sensitive to a possible presence of the heavy quark $\hat{b}$  with charge $Q=-1/3$ belonging to the fourth generation\footnote{If a fourth generation exists, its neutrino must be heavy enough not to contradict the LEP result on the  invisible width $Z\to\bar\nu\nu$.}. In this respect, the FCN transitions among the down-like quarks are much less sensitive to the presence of $\hat{t}$, as its effects could not be easily distinguished from those of the heavy top quark. The effects of a heavy quark $\hat{b}$ in the observables $c\to u\gamma$, $c\to ul^+l^-$, $D^0\to \mu^+\mu^-$ and $\Delta m_D$  have been studied in  \cite{BHLP}; the effects in $\Delta m_D$ and $c\to u\gamma$ have been studied also in \cite{burdman,hewett}. The expressions for the amplitudes in the four  generation model can be generalized from the results in the three generation model presented in Section 2.1, but the approximation $m_{\hat{b}}^2/m_W^2\ll 1$ can not be used in this case. The rates are proportional to $|V_{c\hat{b}}^*V_{u\hat{b}}|^2$ and increase with increasing $m_{\hat b}$. I have updated the results of \cite{BHLP} by taking into account the new experimental limits $m_{\hat{b}}>128$ GeV  \cite{PDG} and 
\begin{equation}
\label{2.50}
|V_{c\hat{b}}|<0.5~,~~~ |V_{u\hat{b}}|<0.08\quad \Rightarrow \quad |V_{c\hat{b}}^*V_{u\hat{b}}|<0.04~,
\end{equation}
which arise by applying the unitary condition to the four-dimensional CKM matrix 
$$\sum_{q=d,s,b,{\hat{b}}}|V_{cq}|^2=\sum_{q=d,s,b,{\hat{b}}}|V_{uq}|^2=1~.$$
 For this purpose one has to use the directly measured values of $V_{cq}$ and $V_{uq}$,  presented in Eq. (11.16) of \cite{PDG}, and not the values obtained by applying the unitarity of the three-dimensional CKM matrix.  

The present experimental upper bound on $\Delta m_D<1.6\cdot 10^{-13}$ GeV \cite{PDG} gives stronger constraint on $|V_{c\hat{b}}^*V_{u\hat{b}}|$ than the unitarity condition. If the experimental upper bound is saturated by the $\hat{b}$ quark contribution, in which case the standard model contribution of the order of $10^{-17}-10^{-16}$ can be safely neglected, we have
\begin{equation}
\label{2.51}
|V_{c\hat{b}}^*V_{u\hat{b}}|<0.015\qquad {\rm for}\quad m_{\hat{b}}=250~{\rm GeV}~.
\end{equation}
 This bound renders the following $\hat{b}$ quark contributions to the branching ratios \cite{BHLP} 
\begin{eqnarray}
\label{c7four}
Br(D^0\to \mu^+\mu^-)^{\hat{b}}&\le& 1\cdot 10^{-10}~,\nonumber\\ 
  Br(c\to ul^+l^-)^{\hat{b}}&\le& 8\cdot 10^{-10}~,\nonumber\\
  Br(c\to u\gamma)^{\hat{b}}&\le& 5\cdot 10^{-7}\qquad (c_7^{eff}\le 2.8\cdot 10^{-2})~,
\end{eqnarray}
where the   $D^0\to \mu^+\mu^-$ and $c\to ul^+l^-$ decays are treated at the one-loop electroweak order and the  $c\to u\gamma$ decay  is treated at the leading logarithmic QCD approximation \cite{BHLP}. The  $\hat{b}$ quark has an unobservably small effect on the $c\to ul^+l^-$ rate. The effects of fourth generation could be observable  in  $D^0\to\mu^+\mu^-$ and $c\to u\gamma$ decays when the  experiments reach the corresponding sensitivity: the branching ratio of first can be enhanced as much as three orders of magnitudes, while the branching ratio of the second can be enhanced  by more than an order of magnitude. 

\subsection{Left-right symmetric models}

Finally, let me briefly comment on the left-right symmetric models, which have also been studied \cite{eeg} in the connection with the FCN transitions in the charm sector. The left-right symmetric models are based on the $SU(3)_c\times SU(2)_L\times SU(2)_R\times U(1)_Y$ group and contain the gauge boson  $W_R$, which couples to the right handed currents of $SU(2)_R$. 

In the {\it ordinary left-right symmetric model}, the $W_R$ couples the right-handed up-like quark with the right-handed down-like quark. The parameters of this scenario are  severely constrained by the present experimental data and its effects to the $c\to u$ transitions  are  small \cite{eeg}. 

More significant effects could arise from the so-called {\it flipped left-right symmetric model} \cite{eeg}, where the $W_R$ couples the right-handed up-like quark with an exotic quark. The exotic  quarks arise via the breaking of some unified group (for example $E_6$ \cite{eeg}) to the group $SU(3)_c\times SU(2)_L\times SU(2)_R\times U(1)_Y$. The usual right-handed down-like quarks are $SU(2)_R$ singlets in this case.  The $c\to u$ transitions are driven via the loop diagrams,  where the $W_R$ boson and the exotic quark run in the loop. The parameters that enter the amplitude (the $W_R$ mass, the masses and mixings of exotic quarks) are not strongly constrained by the experimental data on the FCN transitions among down-like quarks. They are most severely constrained by the experimental upper bound on $\Delta m_D$. By saturating this bound via the $W_R$ exchange mechanism, the following upper bounds on the effects of this scenario are obtained \cite{eeg} 
\begin{eqnarray*}
Br(D^0\to\mu^+\mu^-)^R&<&10^{-10}~,\nonumber\\
B(c\to u\gamma)^R&<&10^{-9}~.\nonumber
\end{eqnarray*}
 This mechanism is hardly observable in $D^0\to \mu^+\mu^-$ decay and it is unobservable in $c\to u\gamma$ decay.

\chapter{Short and long distance contributions to $\boldsymbol{|\Delta c|=1}$ meson decays}

The flavour changing neutral  $c\to u\gamma$ and $c\to ul^+l^-$ decays were proposed as the interesting probes for possible physics beyond the standard model in the previous chapter. These quark processes have to be experimentally explored in the hadron processes. A hadron decay of interest is induced by the quark decay $c\to u\gamma(l^+l^-)$ at short distances and the corresponding part is called the short distance contribution. In addition, there are other mechanisms that can induce the same hadron decay and these give rise to the so-called long distance contribution. The dominant long distance contribution is due to the $W$ exchange between four quarks accompanied by the emission of the real photon $\gamma$ or virtual photon $\gamma^*\to l^+l^-$. A certain hadronic decay can serve as a probe for the  $c\to u\gamma(l^+l^-)$ transition at short distances only if the long distance contribution does not dominate over the short distance contribution in this decay. 
In this chapter, I present the general tools for  short and long distance contributions to the exclusive hadron decays.  First the various  contributions to the hadron decays of interest are 
 schematically sketched and the general expressions for the amplitudes are given. In the second section the short distance amplitudes are given. In the third section,   
 the formal framework  for treating the long distance contributions is presented. For this purpose the $W$ exchange among four quarks is expressed in the form of the effective non-leptonic weak Lagrangian that already accounts for the strong interactions among quarks at short distances.    

\vspace{0.2cm}

The $c\to u\gamma$ and $c\to ul^+l^-$ are rare weak decays and can be experimentally searched for  in the decays of the hadronic states that can decay only weakly. The strong and the electromagnetic channels would overshadow the weak channels in the experiment otherwise. The meson states with no strong or electromagnetic decay channels are the pseudoscalar bound states of a quark and an anti-quark with two different flavours.

\subsubsection{Different contributions to the meson decays}

The transition $\boldsymbol{c\to u\gamma}$ can be experimentally probed in meson decays with the flavour content $c\bar
q\to u\bar q\gamma$. Among these, the radiative two body exclusive decay channels of $c\bar q$ pseudoscalar states $P$ are the most suitable. The flavour $q$ can be chosen among  $u,~d,~s,~c$ or $b$ so that the long distance contribution is as small as possible. For the case of the real and transversely polarized photon in the final state $u\bar q$ must be a vector meson $V$ \footnote{The angular momentum of the $P\to M\gamma$ final state must be $|\vec J|=0$. If $M$ is a pseudoscalar state, the orbital momentum of the final state must be $|\vec L|=1$. In the  center of mass frame, the component of the orbital momentum along the momentum of the final particles vanishes. This means that the component of the photon spin along its momentum must also vanish. The final $M$ state is accompanied by a purely longitudinal photon, which is possible only of photon is virtual.}. Let me  first illustrate various contributions to $P\to V\gamma$ decays with   the  categorization  similar to the one used in \cite{GP,DHT,BGHP,KSW,FPS1,FS,FPS3}). 

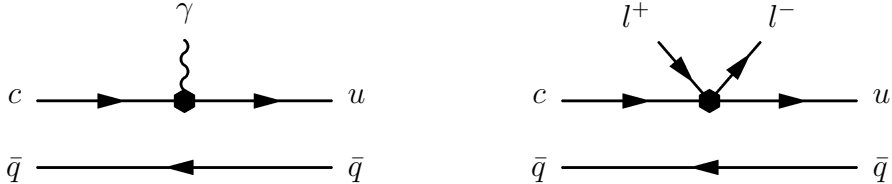
\begin{figure}[h]

\centering
\mbox{
\subfigure[The short distance contribution arising from the transition $c\to u\gamma$. The hexagon denotes the action of the effective local Lagrangian ${\cal L}^{c\to u\gamma}$  (\ref{o7}) with $c_7^{eff}$, given by  (\ref{c7}) in the standard model.]
{
\begin{fmffile}{f17a}
  \fmfframe(8,0)(8,0){
  \begin{fmfgraph*}(40,25)
  \fmfpen{thin}
  \fmfleft{p1,l1,l2,p3}\fmfright{p4,r1,r2,p5}\fmftop{t1}
  \fmf{fermion,tension=1}{l2,v1}\fmf{fermion}{v1,r2} 
  \fmf{fermion}{r1,l1}
  \fmffreeze
  \fmf{boson}{v1,t1}
  \fmfv{de.sh=hexagon,de.filled=full,decor.size=4thick}{v1}
   \fmflabel{$\bar q$}{l1}\fmflabel{$\bar q$}{r1}\fmflabel{$\gamma$}{t1}
   \fmflabel{$c$}{l2}\fmflabel{$u$}{r2}
  \end{fmfgraph*} }
\end{fmffile}
}
\quad
\subfigure[The short distance contribution arising from the transition $c\to ul^+l^-$. The hexagon denotes the action of the effective Lagrangian ${\cal L}^{c\to ul^+l^-}$ (\ref{o9}) with $c_9^{eff}$, given by  (\ref{c9}) in the standard model.]
{
\begin{fmffile}{f17b}
  \fmfframe(8,0)(8,0){
  \begin{fmfgraph*}(40,25)
  \fmfpen{thin}
  \fmfleft{p1,l1,l2,p3}\fmfright{p4,r1,r2,p5}\fmftop{m1,t1,t2,m2}
  \fmf{fermion,tension=1}{l2,v1}\fmf{fermion}{v1,r2} 
  \fmf{fermion}{r1,l1}
  \fmffreeze
  \fmf{fermion}{t1,v1,t2}
  \fmfv{de.sh=hexagon,de.filled=full,decor.size=4thick}{v1}
   \fmflabel{$\bar q$}{l1}\fmflabel{$\bar q$}{r1}
   \fmflabel{$l^+$}{t1}\fmflabel{$l^-$}{t2}
   \fmflabel{$c$}{l2}\fmflabel{$u$}{r2}
  \end{fmfgraph*} }
\end{fmffile}
}
     }
\caption{The sketch of the short distance contributions to the meson decays (a)  $M_1\to M_2\gamma$ and (b) $M_1\to M_2l^+l^-$.}   
\label{fig17}
\end{figure}

{\bf The short
distance contribution} is induced by the $c\to u\gamma$ decay and $\bar q$ is merely a spectator. It is illustrated in Fig. \ref{fig17}a, where the hexagon denotes the action of the effective local Lagrangian ${\cal L}^{c\to u\gamma}$  (\ref{o7}) with $c_7^{eff}$, given by (\ref{c7}) in the standard model.

There are two types
of  long distance contributions. The most serious is {\bf the long
distance weak annihilation contribution} illustrated in Fig. \ref{fig2}. The transition between the initial and final meson is induced by the $W$ exchange and
the photon is emitted from the initial or the final meson. There are three cases of interest. (i) The quark $q=d,~s$ or $b$: The $W$ exchange $c\bar q\to u\bar q$ in the ``$s$'' channel 
is  illustrated in Fig. \ref{fig2}a and is proportional to $V_{cq}^*V_{uq}$. Since $|V_{cd}^*V_{ud}|\simeq |V_{cs}^*V_{us}|\simeq 0.22$ is relatively large, this contribution is found to be dominant in $D_s^+\to K^+\gamma$ and $D^+\to \rho^+\gamma$ decays studied in Chapter 5. Due to the smallness of  $V_{cb}^*V_{ub}$, the long distance weak annihilation contribution is much smaller in $B_c\to B_u^*\gamma$ decay studied in Chapter 4. (ii) The quark 
$q=u$: The $W$ exchange $c\bar u\to d\bar d$ in the ``$t$'' channel 
is  illustrated in Fig. \ref{fig2}b and is proportional to a relatively large factor $V_{cd}^*V_{ud}$. This mechanism  turns out to be dominant in  $c\bar u\to u\bar u\gamma$ decays $D^0\to \rho^0\gamma$ and $D^0\to \omega\gamma$ since the final mesons $\rho^0$ and $\omega$ are the  mixtures of the $u\bar u$ and $d\bar d$ states. (iii) The quark $q=c$: This case  is not of interest since $c\bar c$ states can decay electromagnetically or strongly.

The
second is {\bf the long distance penguin contribution}
sketched in Fig. \ref{fig3}. The $W$ exchange induces the $c\to ud\bar
d$ or $c\to us\bar s$ transition and the $\bar dd$, $\bar ss$ quark pairs subsequently annihilate to a real photon. The quark structure of this contribution is similar to the short distance contribution in Fig. \ref{fig1}a, but the two contributions incorporate two different regimes of the strong interactions. The short distance contribution in Fig. \ref{fig1} involves only the perturbative strong interactions among the quarks and we have explicitly accounted only for the one gluon exchanges in Section 2.1.1. In addition there are the nonperturbative strong interactions among quarks and  these are especially effective among  $\bar dd$ and $\bar ss$  quark-antiquark pairs in Fig. \ref{fig1}a as these can hadronize to the 
  virtual $\rho^0$, $\omega$ and $\phi$ mesons and finally convert to a real 
photon. This mechanism is incorporated into  the long distance penguin contribution in Fig. \ref{fig3} and is reasonably approximated by summing over the virtual vector mesons $\rho^0$, $\omega$ and $\phi$. This approximation is based on the vector meson dominance hypothesis according to which the electromagnetic interactions of hadrons are mediated via the neutral vector mesons. The idea was proposed long time ago \cite{VMD} and  still proves to be useful in describing the electromagnetic interactions of hadrons at low energies (more recent review is given in  \cite{lichard} and references therein). Let me note at this point that 
the long distance penguin contribution given in Fig. \ref{fig3}  turns out to be relatively small  since it vanishes in the exact $SU(3)$ flavour limit. This is analogous to the GIM suppression in the similar diagram on Fig. \ref{fig1}a.   

\vspace{0.2cm}

The $\boldsymbol{c\to ul^+l^-}$ transition can be experimentally probed in meson decays with the flavour content $c\bar
q\to u\bar ql^+l^-$. The three body exclusive decay channels of the pseudoscalar states $P\sim c\bar q$ are the most appropriate. The final state $M\sim u\bar q$ can be a pseudoscalar $P^{\prime}$ or vector $V$ meson\footnote{  In the case of $P\to Ml^+l^-$ decay, the pseudoscalar state $M=P^\prime$ is allowed since the virtual photon in   $P\to P^\prime\gamma^* \to P^\prime l^+l^-$ can be purely longitudinal.}.

The short distance contribution induced by the $c\to ul^+l^-$ decay is sketched in Fig. \ref{fig17}b and the hexagon denotes the action of the effective Lagrangian ${\cal L}^{c\to ul^+l^-}$ (\ref{o9}) with $c_9^{eff}$ equal to (\ref{c9}) in the standard model.

The long distance contributions to $P\to Ml^+l^-$ can be divided to the long distance penguin and weak annihilation contributions. They arise via the mechanism illustrated in Figs. \ref{fig2} and \ref{fig3} where the real photon is replaced by the virtual photon and converts to a charged lepton pair.  

\subsubsection{Different contributions to the baryon decays}

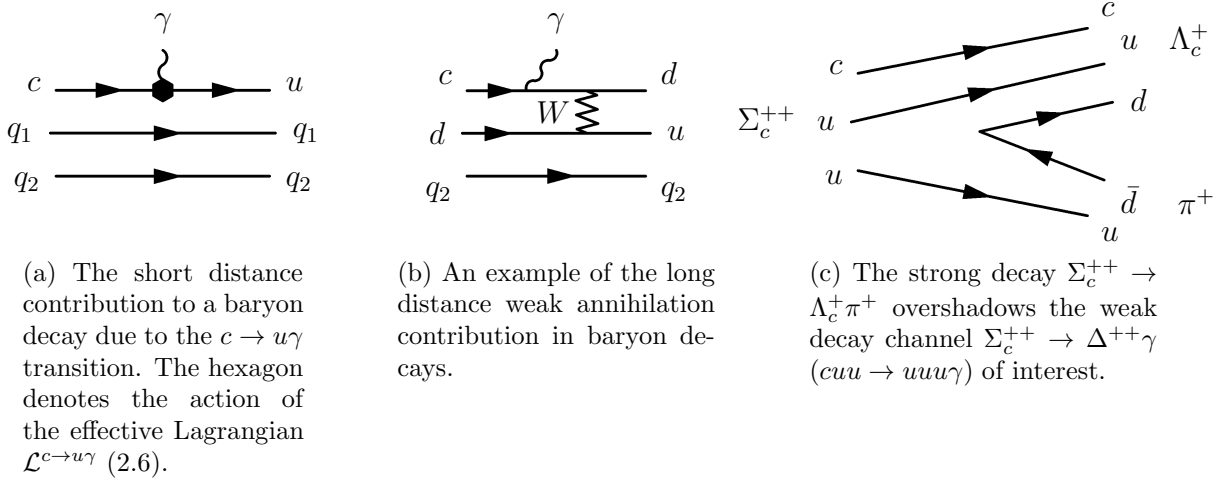
\begin{figure}[h]

\centering
\mbox{
\subfigure[The short distance contribution to a baryon decay due to the $c\to u\gamma$ transition. The hexagon denotes the action of the effective  Lagrangian ${\cal L}^{c\to u\gamma}$  (\ref{o7}).]
{
\begin{fmffile}{f18annn}
  \fmfframe(3,0)(3,0){
  \begin{fmfgraph*}(30,22)
  \fmfpen{thin}
  \fmfleft{p1,l1,l2,l3,p3}\fmfright{p4,r1,r2,r3,p5}\fmftop{t1}
  \fmf{fermion,tension=1}{l3,v1}\fmf{fermion}{v1,r3} 
  \fmf{fermion}{l1,r1}\fmf{fermion}{l2,r2}
  \fmffreeze
  \fmf{boson}{v1,t1}
  \fmfv{de.sh=hexagon,de.filled=full,decor.size=4thick}{v1}
   \fmflabel{$ q_2$}{l1}\fmflabel{$ q_2$}{r1}
  \fmflabel{$ q_1$}{l2}\fmflabel{$ q_1$}{r2}
  \fmflabel{$\gamma$}{t1}
   \fmflabel{$c$}{l3}\fmflabel{$u$}{r3}
  \end{fmfgraph*} }
\end{fmffile}
}
\quad
\subfigure[An example of the long distance weak annihilation contribution in baryon decays. ]
{
\begin{fmffile}{f18bnnn}
  \fmfframe(8,0)(8,0){
  \begin{fmfgraph*}(25,22)
  \fmfpen{thin}
  \fmfleft{p1,l1,l2,l3,p3}\fmfright{p4,r1,r2,r3,p5}\fmftop{t1}
  \fmf{fermion}{l3,v1}\fmf{plain}{v1,m1,r3} 
  \fmf{fermion}{l2,v2}\fmf{plain}{v2,m2,r2} 
  \fmf{fermion}{l1,r1}
  \fmffreeze
  \fmf{boson}{v1,t1}
  \fmf{zigzag,label=$W$}{m1,m2}
   \fmflabel{$ q_2$}{l1}\fmflabel{$ q_2$}{r1}
  \fmflabel{$ d$}{l2}\fmflabel{$ u$}{r2}
  \fmflabel{$\gamma$}{t1}
   \fmflabel{$c$}{l3}\fmflabel{$d$}{r3}
  \end{fmfgraph*} }
\end{fmffile}
}
\quad
\subfigure[The strong decay  $\Sigma_c^{++}\to \Lambda_c^+\pi^+$  overshadows the weak decay channel $\Sigma_c^{++}\to \Delta^{++}\gamma$ ($cuu\to uuu\gamma$) of interest.  ]
{
\begin{fmffile}{f18cn}
  \fmfframe(5,0)(5,0){
  \begin{fmfgraph*}(35,25)
  \fmfpen{thin}
  \fmfleft{p1,l1,l2,l3,p3}\fmfright{r1,q1,q,q2,r2,r3}
  \fmf{fermion}{l3,r3} 
  \fmf{fermion}{l2,r2} 
  \fmf{fermion}{l1,r1}
  \fmf{phantom,tension=2}{l2,v}\fmf{fermion}{q1,v,q2}
   \fmflabel{$ u$}{l1}\fmflabel{$ u$}{r1}  
   \fmflabel{$c$}{l3}\fmflabel{$c$}{r3}
    \fmflabel{$d$}{q2}
   \fmfv{label=$\Sigma_c^{++}~~u$,la.d=3thick}{l2}
   \fmfv{label=$ u~~~\Lambda_c^+$,la.d=3thick}{r2} 
   \fmfv{label=$\bar d~~~~\pi^+$,la.d=3thick}{q1}
  \end{fmfgraph*} }
\end{fmffile}
}

     }
\caption{The sketch of  (a) the short distance and (b) the long distance contributions to the baryon decays.  } 
\label{fig18}  
\end{figure}

The transitions $c\to u\gamma$ and $c\to ul^+l^-$ can be searched for in the baryon decays with the flavour content $cq_1q_2\to uq_1q_2\gamma$ and  $cq_1q_2\to uq_1q_2l^+l^-$, respectively. Only the baryon decays  where the long distance  contributions do  not dominate over the short distance contributions are suitable for this purpose in the question is if any baryon decays of this kind exist. The short distance part is sketched in Fig. \ref{fig18}a. The long distance weak annihilation part via $W$ exchange in ``$t$'' channel is shown in Fig. \ref{fig18}b. The disturbing  long distance weak annihilation contribution is absent only when all valence quarks have equal charge. The least exotic decay to look for $c\to u\gamma$ transition, which is not expected to be dominated by the long distance contributions, is  $cuu\to uuu\gamma$ decay 
$\Sigma_c^{++}\to \Delta^{++}\gamma$. Unfortunately $\Sigma_c^{++}$ can decay strongly and the decay channel $\Sigma_c^{++}\to \Lambda_c^+\pi^+$ shown in Fig. \ref{fig18}c completely overshadows the weak decay channel of interest. We are left only with more exotic decays $\Xi_{cc}^{++}\to 
\Sigma_c^{++}\gamma$  ($ccu\to cuu\gamma$) and $\Omega_{ccc}^{++}\to \Xi_{cc}^{++}\gamma$ ($ccc\to ccu\gamma$), which are expected to be suitable as  probes for  the $c\to u\gamma$ transition \cite{moriond,singer.baryon}. As these two decays are to exotic for the experimental investigation at present, I stop the discussion on the baryon decays at this point and consider only meson decays in what follows.

\section{General expressions for the amplitudes}

In this section the general forms of the amplitudes for the decays $P\to V\gamma$, $P\to Vl^+l^-$ and $P\to P^{\prime}l^+l^-$ are presented.  The amplitudes have to be invariant under the electromagnetic gauge transformation implemented by $\epsilon^{\mu}\to \epsilon^{\mu}+Cq^{\mu}$ for  real or virtual photon with momentum $q$ and polarization $\epsilon$, where $C$ is a general constant. The first principle calculation based on the standard model renders the gauge invariant amplitudes automatically. The gauge invariance may not be automatic when a model is used instead and one has to take special care to get the amplitudes of the form required by the gauge invariance. 

\subsubsection{The amplitudes for $\boldsymbol{P\to V\gamma}$ and $\boldsymbol{P\to V\gamma^*\to Vl^+l^-}$ decays}

First I consider the decay  $\boldsymbol{P(p)\to V(p^{\prime},\epsilon^{\prime})\gamma(q,\epsilon)}$ with the real photon in the final state. The amplitude is linear in polarizations $\epsilon$ and $\epsilon^{\prime}$. Since $V$ and $\gamma$ are real, $\epsilon q=0$ and $\epsilon^{\prime}p^{\prime}=0$, and the general Lorentz invariant amplitude is given by
$${\cal A}[P(p)\to V(p^{\prime},\epsilon^{\prime})\gamma(q,\epsilon)]=\epsilon_{\mu}^*\epsilon_{\nu}^{*\prime}[A_1p^{\mu}p^{\nu}+A_2g^{\mu\nu}+A_3\epsilon^{\mu\nu\alpha\beta}p_{\alpha}q_{\beta}]$$
with the  Lorentz invariant form factors $A_i$. 
The electromagnetic gauge invariance imposes $A_2=-A_1p\cdot q $ and the amplitude can be generally written as
\begin{equation}
\label{amp.gamma}
 {\cal A}[P(p)\to V(p^{\prime},\epsilon^{\prime})\gamma(q,\epsilon)]=\epsilon_{\mu}^*\epsilon_{\nu}^{*\prime}[iA_{PV}(p^{\mu}q^{\nu}-g^{\mu\nu}p\cdot q)+A_{PC}\epsilon^{\mu\nu\alpha\beta}p_{\alpha}q_{\beta}]~.
\end{equation}

Now I turn to  the amplitude for $\boldsymbol{P\to V\gamma^*\to Vl^+l^-}$ decay. In this case $\epsilon q\not=0$ and  
$${\cal A}[P(p)\to V(p^{\prime},\epsilon^{\prime})\gamma^*(q,\epsilon)]=\epsilon_{\mu}^*\epsilon_{\nu}^{*\prime}[(A_1p^{\mu}+A_4q^{\mu})q^{\nu}+A_2g^{\mu\nu}+A_3\epsilon^{\mu\nu\alpha\beta}p_{\alpha}q_{\beta}]~.$$
The gauge invariance forces $A_2=-A_1p\cdot q-q^2A_4$ and $D\to V\gamma^*$ amplitude  has a general form
\begin{align}
\label{amp.gammas}
{\cal A}&[P(p)\to V(p^{\prime},\epsilon^{\prime})\gamma^*(q,\epsilon)]=\\
&=\epsilon_{\mu}^*\epsilon_{\nu}^{*\prime}[iA_{PV}(p^{\mu}q^{\nu}-p\cdot q g^{\mu\nu})-iA_{PV}^\prime(q^{\mu}q^{\nu}-q^2g^{\mu\nu})+A_{PC}\epsilon^{\mu\nu\alpha\beta}p_{\alpha}q_{\beta}]~.\nonumber
\end{align}
The photon polarization  $\epsilon_\mu$ is replaced by $e_0\bar v(p_+)\gamma^\mu u(p_-)/q^2$ in the amplitude for the charged lepton pair in the final state and the term proportional to $q^{\mu}$ does not contribute as $\bar v\!\!\not{\!q} u=0$, 
\begin{align}
\label{amp.ll}
 {\cal A}&[P(p)\to V(p^{\prime},\epsilon^{\prime})l^+(p_+)l^-(p_-)]=\nonumber\\
&=\tfrac{e_0}{q^2}\bar v(p_+)\gamma_\mu u(p_-)\epsilon_{\nu}^{*\prime}[iA_{PV}(p^{\mu}q^{\nu}-g^{\mu\nu}p\cdot q)+iA_{PV}^{\prime}q^2g^{\mu\nu}+A_{PC}\epsilon^{\mu\nu\alpha\beta}p_{\alpha}q_{\beta}]~.
\end{align}

The orbital momentum of the $P\to V\gamma^*$ final state can be $L=0,1,2$. The amplitudes (\ref{amp.gamma}) and (\ref{amp.ll}) involve the parity conserving part given by $A_{PC}$  ($V\gamma^*$ are in P-wave) and the parity violating part given by $A_{PV}$  ($V\gamma^*$ are in S-wave or D-wave).

\subsubsection{The amplitude for $\boldsymbol{P\to P^{\prime}\gamma^*\to P^{\prime}l^+l^-}$ decay}

The general Lorentz decomposition of the amplitude for $P\to P^\prime\gamma^*$ decay is given by
$${\cal A}[P(p)\to P^\prime(p^{\prime})\gamma^*(q,\epsilon)]=\epsilon_{\mu}^*[B_1(p+p^\prime)^{\mu}+B_2q^{\mu}]$$
and the gauge invariance requires $B_1=-q^2B_2/(m_P^2-m_{P^\prime}^2)$, so
\begin{equation}
\label{amp.pgammas}
{\cal A}[P(p)\to P^\prime(p^{\prime})\gamma^*(q,\epsilon)]=B_{PC}~\epsilon_{\mu}^*[q^2(p+p^\prime)^{\mu}-(m_P^2-m_{P^\prime}^2)q^{\mu}]~.
\end{equation}
In the case of the real photon in the final state, $q^2=0$ and $\epsilon q=0$, and the amplitude vanishes.  The photon polarization  $\epsilon_\mu$ is replaced by $e_0\bar v(p_+)\gamma_\mu u(p_-)/q^2$ for the case of the lepton pair in the final state and the term proportional to $q^{\mu}$ does not contribute as $\bar v\!\not{\! q} u=0$ 
\begin{equation}
\label{amp.pll}
{\cal A}[P(p)\to P^\prime(p^{\prime})l^+(p_+)l^-(p_-)]=2e_0~B_{PC}~\bar v(p_+)\gamma_{\mu}u(p_-)~p^\mu.
\end{equation}
The factor $q^2$ in (\ref{amp.pgammas}) has canceled with photon  propagator $1/q^2$ as it should in order to maintain the gauge invariance.  
The amplitude involves only one Lorentz invariant function $B_{PC}$. 
The orbital momentum of the $P\to P^\prime\gamma^*$ final state is $L=1$ and the parity is conserved in this process.  

\vspace{0.2cm}

Now I turn to the discussion of the general theoretical tools for the calculation of the short and long distance parts of the amplitudes.

\section{Short distance contributions}
 
The short distance contributions to the hadronic decay $i\to f$  are induced by $c\to u\gamma$ or $c\to ul^+l^-$ quark decays and their amplitudes are given by
\begin{equation}
\label{asd}
{\cal A}_{SD}=\langle f|:i{\cal L}^{c\to u\gamma}:|i\rangle\qquad {\rm or} \qquad {\cal A}_{SD}=\langle f|:i{\cal L}^{c\to ul^+l^-}:|i\rangle~.
\end{equation}
The effective Lagrangian ${\cal L}^{c\to u\gamma}$  is given in (\ref{o7}) and depends on the coefficient $c_7^{eff}$, whose standard model value is given in (\ref{c7}). The value of $c_7^{eff}$  alters in different scenarios of physics beyond the standard model. The effective Lagrangian ${\cal L}^{c\to ul^+l^-}$ as predicted by the standard model is given in (\ref{o9}) with the coefficient $c_9^{eff}$ given in (\ref{c9}). Possible physics beyond the standard model can change the value of $c_9^{eff}$ as well as the form of the effective  Lagrangian ${\cal L}^{c\to ul^+l^-}$. 

\section{Long distance contributions}

\subsection{Effective nonleptonic weak Lagrangian}

The dominant long distance contributions to the meson decays of interest are induced by the $W$ exchange between four quark states, which is responsible for the transition between the initial and final meson states. In this section the hard gluon corrections to the tree level $W$ exchange are incorporated. Then the resulting four-quark interaction is conveniently expressed as a product of two colour singlet quark currents, which has a suitable form to study the transitions among the colour-singlet meson states. 

\vspace{0.2cm}

The low energy effective Lagrangian for the tree level $W$ exchange between the charm  and the three light quarks is given by
\begin{equation}
\label{3.1}
{\cal L}^{|\Delta c|=1}=-{G_F\over \sqrt{2}}V_{cq_j}^*V_{uq_i}O_1^{ij}\ ,\qquad O_1^{ij}=\bar u^{\alpha}\gamma^{\mu}(1-\gamma_5)q_i^{\alpha}~\bar q_j^{\beta}\gamma_{\mu}(1-\gamma_5)c^{\beta}~,
\end{equation}
where $q_{i,j}$ represent the fields of $d$ or $s$ quarks. 

Strong interactions affect this simple picture in a two-fold way. Hard gluon exchanges can be accounted for by the perturbative methods and renormalization-group techniques. They give rise to new effective weak vertices. Long-range strong confinement forces are responsible for the binding of quarks inside the asymptotic hadronic states. The basic tool in the calculation of the amplitudes among the hadron states is to separate the two regimes by means of the operator product expansion ${\cal L}\propto C_i(\mu)O_i(\mu)$ \cite{willson}. The strong corrections, arising from the exchange of the hard gluons with the virtualities between $m_W$ and some hadronic scale $\mu$, are incorporated into the Wilson coefficients $C_i(\mu)$ and these are independent on the particular decay channel. All the long-range QCD effects are incorporated in the hadronic matrix elements $\langle f|O_i(\mu)|i\rangle$ of local four-quark operators $O_i(\mu)$.

The strong one-loop corrections to the tree level Lagrangian (\ref{3.1}) are shown in Fig. \ref{fig7} given in Appendix A. The diagrams have infinite amplitudes and the Lagrangian (\ref{3.1}) needs counter-terms that renormalize the interactions at the scale $\mu$. The renormalized Lagrangian is calculated in  Appendix A. It  involves the operator $O_1^{ij}$ (\ref{3.1}) and the new operator 
\begin{equation}
\label{3.2}
O_2^{ij}=\bar q_j^{\alpha}\gamma^{\mu}(1-\gamma_5)q_i^{\alpha}~\bar u^{\beta}\gamma_{\mu}(1-\gamma_5)c^{\beta}~
\end{equation}
and depends on the renormalization scale $\mu$.
As the bare Lagrangian should not depend on $\mu$, the operators $O_1(\mu)$ and $O_2(\mu)$ are multiplied by the $\mu$-dependent coefficients $C_1(\mu)$ and $C_2(\mu)$ \footnote{In contrast to the  Wilson coefficients $c_1(\mu)$ and $c_2(\mu)$ used in the Chapter 2, the Wilson coefficients $C_1(\mu)$ and $C_2(\mu)$ defined here do not include the CKM factors.} 
\begin{equation}
\label{3.3}
{\cal L}^{|\Delta c|=1}_W=-{G_F\over \sqrt{2}}V_{cq_j}^*V_{uq_i}[C_1(\mu)O_1^{ij}(\mu)+C_2(\mu)O_2^{ij}(\mu)]~.
\end{equation}
 The  coefficients $C_i(\mu)$  are determined in  Appendix A by imposing that the $\mu$-dependencies of $C_i(\mu)$ and $O_i(\mu)$ cancel. In the leading logarithmic approximation, discussed in Section 2.1.1, the evolution is given by 
\begin{equation*}
[\delta_{ji}~\mu{d\over d\mu}-{g^2\over 8\pi^2}b_{ij}(\mu)]~C_i(\mu)=0\qquad 
{\rm with}\qquad b=\begin{pmatrix}-1&3\\
                                 3&-1\\\end{pmatrix}
\end{equation*}
and the solution  is given in  Eq. (\ref{2.14}). The boundary condition $C_1(m_W)=1$ and $C_2(m_W)=0$  is obtained from the tree level effective Lagrangian (\ref{3.1}). In the following chapters I will study the $D$ and the $B_c$  meson decays and the suitable renormalization scales  are $\mu={\cal O}(m_c)$ and $\mu={\cal O}(m_b)$, respectively. The evolution from $\mu=m_W$ to $m_b$ with five active quark flavours and $m_b=5$ GeV, $\alpha_s(m_Z)\simeq 0.12$ \cite{PDG} gives 
\begin{align}
\label{3.4}
C_1(m_b)&={1\over 2}\biggl[{\alpha_s(m_W)\over \alpha_s(m_b)}\biggr]^{{6\over 23}}+{1\over 2} \biggl[{\alpha_s(m_W)\over \alpha_s(m_b)}\biggr]^{-{12\over 23}}\simeq 1.1~,\nonumber\\
 C_2(m_b)&={1\over 2}\biggl[{\alpha_s(m_W)\over \alpha_s(m_b)}\biggr]^{{6\over 23}}-{1\over 2} \biggl[{\alpha_s(m_W)\over \alpha_s(m_b)}\biggr]^{-{12\over 23}}\simeq -0.22~.
\end{align}
Further evolution from $\mu=m_b$ to $\mu=m_c$ with four active quark flavours and $m_c=1.25$ GeV gives
\begin{align}
\label{3.5}
C_1(m_c)&={1\over 2}\biggl[{\alpha_s(m_W)\over \alpha_s(m_b)}\biggr]^{{6\over 23}}\biggl[{\alpha_s(m_b)\over \alpha_s(m_c)}\biggr]^{{6\over 25}}+{1\over 2} \biggl[{\alpha_s(m_W)\over \alpha_s(m_b)}\biggr]^{-{12\over 23}}\biggl[{\alpha_s(m_b)\over \alpha_s(m_c)}\biggr]^{-{12\over 25}}\!\simeq 1.21~,\nonumber\\
C_2(m_c)&={1\over 2}\biggl[{\alpha_s(m_W)\over \alpha_s(m_b)}\biggr]^{{6\over 23}}\biggl[{\alpha_s(m_b)\over \alpha_s(m_c)}\biggr]^{{6\over 25}}-{1\over 2} \biggl[{\alpha_s(m_W)\over \alpha_s(m_b)}\biggr]^{-{12\over 23}}\biggl[{\alpha_s(m_b)\over \alpha_s(m_c)}\biggr]^{-{12\over 25}}\!\simeq-0.43~.
\end{align}  

In the next-to-leading logarithmic approximation the authors of \cite{buras:neubert} get the following values
\begin{alignat}{2}
\label{ci}
C_1(m_b)&=1.13\pm 0.01~,& \qquad C_2(m_b)&=-0.285\pm 0.015~,\\
C_1(m_c)&=1.35\pm 0.04~,& \qquad C_2(m_c)&=-0.63\pm 0.05\nonumber
\end{alignat}
for $\alpha_s(m_Z)=0.118\pm 0.003$, $m_b=4.8$ GeV and $m_c=1.4$ GeV.

\vspace{0.2cm}

Now I turn  to the amplitude for a given  hadronic decay $i\to f$ 
\begin{equation}
{\cal A}=\langle f|:e^{i\int d^4 x[{\cal L}_W(x)+{\cal L}_{em}(x)+{\cal L}_{strong}(x)]}:|i\rangle~.
\end{equation}
The initial state $i$ of interest is a pseudoscalar meson $P$. The relevant final state $f$ is composed of two states $f_1$ and $f_2$: in the case of $
P\to V\gamma$ one of them is a vector meson and the other is a photon; in the case of $P\to M\gamma^*\to Ml^+l^-$ one of them is a vector or a pseudoscalar meson and the other is a virtual photon or lepton pair \footnote{The emission of a real or virtual photon from the light quark can be successfully described through the emission of a neutral vector meson $V^0$ accompanied by the $V^0\to \gamma$ conversion. In this framework $f_1$ and $f_2$ present mesons, one of them being transferred to a photon.}. The effective weak Lagrangian ${\cal L}_W$ is given in (\ref{3.3}) and all the strong interactions, which are not included into ${\cal L}_W$, are incorporated in the strong Lagrangian ${\cal L}_{strong}$. The electromagnetic Lagrangian ${\cal L}_{em}$ is responsible for the  emission of the real or virtual photon in $P\to V\gamma$ or $P\to M\gamma^*$ decays. I will omit  ${\cal L}_{em}$ and ${\cal L}_{strong}$ in the matrix elements bellow, but their presence is always implicit. In the first order of the weak interactions, the amplitude is given by
\begin{equation}
\label{3.6}
{\cal A}=-i~{G_F\over \sqrt{2}}~V_{cq_j}^*V_{uq_i}~\{~C_1(\mu)~\langle f_1f_2 |:O_1^{ij}(\mu):|i\rangle+C_2(\mu)~\langle f_1f_2 |:O_2^{ij}(\mu):|i\rangle~\}~.
\end{equation}
When $\mu$ is taken at the hadronic scale, the Wilson coefficients $C_{1,2}(\mu)$  incorporate the short-range QCD effects and the matrix elements incorporate the long-range nonperturbative QCD effects. 

The time ordered product places all the annihilation operators to the right and all the creation operators to the left, so the amplitude (\ref{3.6}) can be expressed as
\begin{align}
\label{3.7}
{\cal A}_{LD}=\!-i\tfrac{G_F}{\sqrt{2}}V_{cq_j}^*V_{uq_i}\{
  & C_1(\mu)~\langle 0|
[a_{f_2},\bar u^{\alpha}\Gamma^{\mu}q_i^{\alpha}][a_{f_1},\bar q^{\beta}_j]\Gamma_{\mu}[c^{\beta},a_P^{\dagger}]\\
&\qquad +[a_{f_1},\bar u^{\alpha}\Gamma^{\mu}q_i^{\alpha}][a_{f_2},\bar q^{\beta}_j]\Gamma_{\mu}[c^{\beta},a_P^{\dagger}]+[a_{f_1}a_{f_2},\bar u^{\alpha}\Gamma^{\mu}q_i^{\alpha}][\bar q^{\beta}_j\Gamma_{\mu}c^{\beta},a_P^{\dagger}]|0\rangle\nonumber\\
+~& C_2(\mu)~\langle 0|
[a_{f_2},\bar u^{\beta}\Gamma^{\mu}q_i^{\alpha}][a_{f_1},\bar q^{\alpha}_j]\Gamma_{\mu}[c^{\beta},a_P^{\dagger}]\nonumber\\
&\qquad +[a_{f_1},\bar u^{\beta}\Gamma^{\mu}q_i^{\alpha}][a_{f_2},\bar q^{\alpha}_j]\Gamma_{\mu}[c^{\beta},a_P^{\dagger}]+[a_{f_1}a_{f_2},\bar u^{\beta}\Gamma^{\mu}q_i^{\alpha}][\bar q^{\alpha}_j\Gamma_{\mu}c^{\beta},a_P^{\dagger}]|0\rangle\}~,\nonumber
\end{align}
where $\Gamma^{\mu}=\gamma^{\mu}(1-\gamma_5)$ and the Fierz rearrangement was used for the part proportional to $C_2(\mu)$. The matrix elements involve the commutators of the quark and meson fields and are hardly calculable from the first principles. 

\subsection{Factorization approximation}

The calculation of the matrix element in (\ref{3.7}) is greatly simplified in the factorization approximation, where the sum $I=\sum_n|n\rangle\langle n|$ over all states is inserted between the first and the second commutator, and then all the intermediate states except for the vacuum state $|0\rangle\langle 0|$ are omitted
\begin{align}
\label{3.8}
{\cal A}_{LD}=\!-i\tfrac{G_F}{\sqrt{2}}V_{cq_j}^*V_{uq_i}\{
  & C_1(\mu)[\langle f_2|\bar u^{\alpha}\Gamma^{\mu}q_i^{\alpha}|0\rangle\langle f_1|\bar q^{\beta}_j\Gamma_{\mu}c^{\beta}|P\rangle\\
&\qquad+\langle f_1|\bar u^{\alpha}\Gamma^{\mu}q_i^{\alpha}|0\rangle\langle f_2|\bar q^{\beta}_j\Gamma_{\mu}c^{\beta}|P\rangle+\langle f_1f_2|\bar u^{\alpha}\Gamma^{\mu}q_i^{\alpha}|0\rangle\langle 0|\bar q^{\beta}_j\Gamma_{\mu}c^{\beta}|P\rangle]\nonumber\\
+~& C_2(\mu)[\langle f_2|\bar u^{\beta}\Gamma^{\mu}q_i^{\alpha}|0\rangle\langle f_1|\bar q^{\alpha}_j\Gamma_{\mu}c^{\beta}|P\rangle\nonumber\\
&\qquad+\langle f_1|\bar u^{\beta}\Gamma^{\mu}q_i^{\alpha}|0\rangle\langle f_2|\bar q^{\alpha}_j\Gamma_{\mu}c^{\beta}|P\rangle+\langle f_1f_2|\bar u^{\beta}\Gamma^{\mu}q_i^{\alpha}|0\rangle\langle 0|\bar q^{\alpha}_j\Gamma_{\mu}c^{\beta}|P\rangle]\}\nonumber~.
\end{align} 
The initial $i$ and the final  $f_1$, $f_2$ states are colour singlets  and  for any two colour-singlet states $s_1$, $s_2$
$$\langle s_2|\bar q_i^{\alpha}\Gamma^{\mu}q_j^{\beta}|s_1\rangle={1\over N_c}\sum_{\gamma}\langle s_2|\bar q_i^{\gamma}\Gamma^{\mu}q_j^{\gamma}|s_1\rangle\delta^{\alpha\beta}~$$
for $N_c$ colours, giving\footnote{Beyond the factorization approximation also the matrix elements $\tfrac{1}{2}\langle f_1 f_2|\bar u^{\alpha}\lambda^{a}_{\alpha\beta}\Gamma^{\mu}q_i^{\beta}\bar q^{\gamma}_j\lambda^{a}_{\gamma\delta}\Gamma_{\mu}c^{\delta}|P\rangle$ and $\tfrac{1}{2}\langle f_1 f_2|\bar q_j^{\alpha}\lambda^{a}_{\alpha\beta}\Gamma^{\mu}q_i^{\beta}\bar u^{\gamma}\lambda^{a}_{\gamma\delta}\Gamma_{\mu}c^{\delta}|P\rangle$ enter at this stage. They arise via the relation $\lambda^a_{\alpha\beta}\lambda^a_{\gamma\delta}=-2/3\delta_{\alpha\beta}\delta_{\gamma\delta}+2\delta_{\alpha\delta}\delta_{\beta\gamma}$.}  
\begin{align}
\label{3.9}
{\cal A}_{LD}=\!-i{G_F\over \sqrt{2}}V_{cq_j}^*V_{uq_i}\bigl(\{&C_1(\mu)+{1\over N_c}C_2(\mu)\}[\langle f_2|\bar u^{\alpha}\Gamma^{\mu}q_i^{\alpha}|0\rangle\langle f_1|\bar q^{\beta}_j\Gamma_{\mu}c^{\beta}|P\rangle\\
&\quad+\langle f_1|\bar u^{\alpha}\Gamma^{\mu}q_i^{\alpha}|0\rangle\langle f_2|\bar q^{\beta}_j\Gamma_{\mu}c^{\beta}|P\rangle+\langle f_1f_2|\bar u^{\alpha}\Gamma^{\mu}q_i^{\alpha}|0\rangle\langle 0|\bar q^{\beta}_j\Gamma_{\mu}c^{\beta}|P\rangle]\nonumber\\
+~\{&C_2(\mu)+{1\over N_c}C_1(\mu)\}[\langle f_2|\bar q_j^{\alpha}\Gamma^{\mu}q_i^{\alpha}|0\rangle\langle f_1|\bar u^{\beta}\Gamma_{\mu}c^{\beta}|P\rangle\nonumber\\
&\quad+\langle f_1|\bar q_j^{\alpha}\Gamma^{\mu}q_i^{\alpha}|0\rangle\langle f_2|\bar u^{\beta}\Gamma_{\mu}c^{\beta}|P\rangle
+\langle f_1f_2|\bar q_j^{\alpha}\Gamma^{\mu}q_i^{\alpha}|0\rangle\langle 0|\bar u^{\beta}\Gamma_{\mu}c^{\beta}|P\rangle]\bigr)\nonumber.
\end{align} 
The term proportional to $C_1+C_2/N_c$ follows from (\ref{3.7}, \ref{3.8}). The nonzero matrix elements in this term arise only if the currents $\bar u\Gamma^{\mu}q_i$ and $\bar q_j\Gamma_{\mu}c$ have the right flavour structure for a given $P\to f_1f_2$ decay. In the opposite case, the matrix elements of currents $\bar u\Gamma^{\mu}c$ and $\bar q_j\Gamma_{\mu}q_i$ in the term proportional to $C_2+C_1/N_c$ can contribute. The  term proportional to $C_2+C_1/N_c$ follows from (\ref{3.7}, \ref{3.8}) if the Fierz rearrangement in (\ref{3.7}) was  performed on the term proportional to $C_1$. 

The renormalization scale dependence of the matrix elements $\langle f|O_i(\mu)|i\rangle$ and the Wilson coefficients $C_i(\mu)$ should cancel in the bare Lagrangian (\ref{3.6}). The  matrix elements of the currents  (\ref{3.9}), resulting from the factorization, are not scale dependent due to the Ward identities, as explicitly shown in Appendix A.1.  The $\mu$ dependences of the matrix element and the Wilson coefficient can not cancel in the factorization approximation and this approximation can be at best correct at a single value of $\mu$, the so-called factorization scale $\mu_F$ \cite{buras1,buras:neubert}. In order to set a reasonable choice for $\mu_F$, one has to be aware that  the factorization approximation amounts to neglecting any interaction among the states that take part in the first and the second matrix element. In particular, the soft gluon exchanges with virtualities bellow $\mu$  among these states are neglected in this approximation (the hard gluon exchanges have been incorporated in the Wilson coefficients). Because the factorized hadronic matrix elements can only account for the interaction between quarks remaining in the same hadron, the Wilson coefficients should contain those gluon effects which redistribute the quarks. A suitable choice of $\mu_F$ is made when the soft gluons with virtualities bellow $\mu_F$ are no longer very effective in rearranging the quarks grouped into the colour-singlet pairs, i.e. $\mu_F\sim  {\cal O}(m_c)$ and $\mu_F\sim{\cal O}(m_b)$ for $D$ and $B$ meson decays, respectively \cite{buras:neubert}. At $\mu_F$, the factorized amplitude (\ref{3.9}) is expected to present a reasonable  approximation and it is suitable to define  $\mu$-independent coefficients $a_1$ and $a_2$ for $D$ and $B$ meson decays \cite{BSW}
\begin{equation}
\label{3.13}
a_1=C_1(\mu_F)+\tfrac{1}{N_c}C_2(\mu_F)~,\qquad a_2=C_2(\mu_F)+\tfrac{1}{N_c}C_1(\mu_F)~.
\end{equation}
 With $N_c=3$ and $C_{1,2}(m_{c,b})$ given in (\ref{ci}) this 
amounts to 
\begin{alignat}{2}
\label{3.10}
a_1^b&=1.03\pm 0.01~,&\quad a_2^b= 0.091\pm 0.015~, \\
a_1^c&=1.14\pm 0.04 ~,&\quad a_2^c=-0.18\pm 0.05~. \nonumber
\end{alignat}

On the other hand, the coefficients $a_1$ and $a_2$ can be employed as free parameters in the study of the nonleptonic $D$ and $B$ decays based on the factorization and can be fitted from the experimental data. I denote the measured coefficients based on the factorized amplitudes  by $a_{1,2}^{eff}$. The first extensive analysis of this kind was performed in \cite{BSW}. More recent analysis give 
\begin{alignat}{3}
\label{ai}
(a_1^b)^{eff}&=1.08\pm 0.10~&,\qquad (a_2^b)^{eff}&=0.21\pm 0.06~&,\quad  &\cite{buras1,buras:neubert,neubert2,aiphen}\\
(a_1^c)^{eff}&=1.2\pm 0.1~&,\qquad (a_2^c)^{eff}&=-0.5\pm 0.1~. &\quad   &\cite{NRSX}~\nonumber
\end{alignat}
The coefficients $a_1^{eff}$ and $a_2^{eff}$ are expected to match the predicted coefficients $a_1$ and $a_2$ (\ref{3.10}, \ref{ai}) if the factorization approximation is a good. The coefficients $a_1$ and $a_1^{eff}$ indeed agree well. The discrepancy in the coefficients $a_2$ and $a_2^{eff}$ indicates that nonfactorizable contributions must play an important role, especially in $D$ decays which occur at the energies where many hadronic resonances are present. 
 Formally, the measured coefficients $a_1^{eff}$ and $a_2^{eff}$ contain also the nonfactorizable part and can be generally parameterized in terms of $\epsilon_1(\mu_F)$ and $\epsilon_8(\mu_F)$ \cite{buras:neubert,neubert2} 
\begin{equation*}
a_1^{eff}=a_1[1+\epsilon_1(\mu_F)]+C_2(\mu_F)\epsilon_8(\mu_F)~,\qquad a_2^{eff}=a_2[1+\epsilon_1(\mu_F)]+C_1(\mu_F)\epsilon_8(\mu_F)
\end{equation*}
with $\epsilon_{1,8}(\mu_F)\to 0$ when the factorization is good at $\mu_F$.  The nonfactorizable parts $\epsilon_{1}(\mu_F)$ and $\epsilon_{8}(\mu_F)$ can be fitted from the experimental data. In $1/N_c$ expansion, only the quantity $\zeta=1/N_c+\epsilon_8(\mu_F)$ parameterizes the nonfactorizable contributions in $a_{1,2}^{eff}$  \cite{buras:neubert,neubert2} \footnote{The large $N_c$ counting rules of QCD imply $C_1=1+{\cal O}(1/N_c^2)$, $C_2={\cal O}(1/N_c)$, $\epsilon_1={\cal O}(1/N_c^2)$ and $\epsilon_8={\cal O}(1/N_c)$ \cite{buras:neubert,neubert2}} 
\begin{equation*}
a_1^{eff}=C_1(\mu_F)+{\cal O}(1/N_c^2)~,\qquad a_2^{eff}=C_2(\mu_F)+\zeta ~C_1(\mu_F)~.
\end{equation*}
In the case of the exact factorization $\zeta=1/N_c=1/3$. The measured coefficient $(a_{1}^b)^{eff}$ (\ref{ai}) gives  $\zeta^b=0.45\pm 0.05$ \cite{buras:neubert,neubert2} close to $1/3$. The coefficient $(a_{2}^c)^{eff}$ (\ref{ai}) gives a small value $\zeta^c=0.09\pm 0.1$ and indicates that the factorization should be accompanied by  $N_c^{eff}=1/\zeta\to\infty$ in $D$ meson decays. Having this in mind, I will employ the effective coefficients $(a_{1,2}^{c,b})^{eff}$ given in (\ref{ai}), which take into account the factorizable and also some nonfactrizable contributions, and I drop the superscript $^{eff}$ from now on.

The factorized nonleptonic amplitude (\ref{3.9}) is finally given by 
\begin{align}
\label{factor}
{\cal A}&(P\to f_1f_2)=\!-i\tfrac{G_F}{\sqrt{2}}V_{cq_j}^*V_{uq_i}\\
\times\bigl(~&a_1[\langle f_2|\bar u\Gamma^{\mu}q_i|0\rangle\langle f_1|\bar q_j\Gamma_{\mu}c|P\rangle+\langle f_1|\bar u\Gamma^{\mu}q_i|0\rangle\langle f_2|\bar q_j\Gamma_{\mu}c|P\rangle+\langle f_1f_2|\bar u\Gamma^{\mu}q_i|0\rangle\langle 0|\bar q_j\Gamma_{\mu}c|P\rangle]\nonumber\\
+&a_2[\langle f_2|\bar q_j\Gamma^{\mu}q_i|0\rangle\langle f_1|\bar u\Gamma_{\mu}c|P\rangle+\langle f_1|\bar q_j\Gamma^{\mu}q_i|0\rangle\langle f_2|\bar u\Gamma_{\mu}c|P\rangle
+\langle f_1f_2|\bar q_j\Gamma^{\mu}q_i|0\rangle\langle 0|\bar u\Gamma_{\mu}c|P\rangle]~\bigr)~\nonumber
\end{align} 
 with $\Gamma^{\mu}=\gamma^{\mu}(1-\gamma_5)$ and $a_{1,2}$ given in (\ref{ai}). This amplitude can be expressed  in terms of {\bf the weak  nonleptonic effective Lagrangian}, which is  a product of the colour singlet currents responsible for  the transitions among the hadronic states
\begin{equation}
\label{eff}
{\cal L}^{|\Delta c|=1}_{eff}\!=\!-\frac{G_F}{\sqrt{2}}V_{cq_j}^*V_{uq_i}\bigl[a_1 ~\bar u\gamma^{\mu}(1-\gamma_5)q_i~\bar q_j\gamma_{\mu}(1-\gamma_5)c
+a_2~\bar q_j\gamma_{\mu}(1-\gamma_5)q_i~\bar u\gamma^{\mu}(1-\gamma_5)c\bigr]~,
\end{equation}
where $q_{i,j}$ denotes the $d$ or $s$ quark fields. 
The amplitude for long distance contribution to the $P\to f_1f_2$ decay can be therefore written as 
\begin{align}
\label{ald}
{\cal A}_{LD}=\langle f_1f_2|:i{\cal L}^{|\Delta c|=1}_{eff}:|P\rangle~.
\end{align} 

\subsection{Bremsstrahlung  and  gauge invariance}

In this section I discuss the general features of the bremsstrahlung part of amplitude for $P\to V\gamma^*$ and $P\to P^\prime\gamma$ decays, where  $\gamma^*$ denotes a virtual or real meson. 
In particular, I derive a  procedure for the calculation of bremsstrahlung amplitude which automatically leads to the gauge invariant results. Bremsstrahlung contribution is  a specific part of the {\bf long distance weak annihilation contribution}.  The long distance weak annihilation contribution is induced by the effective nonleptonic Lagrangian (\ref{eff}), where one of the currents has the flavour of the initial meson $P$, the other has the flavour of the final meson $M$ and the photon is emitted before or after the weak vertex.  The bremsstrahlung contribution has the following additional properties: (i) The probability for  the emission of a photon from a meson  is proportional to the net charge of the meson. The bremsstrahlung part of $P\to M\gamma^*$ amplitude is nonzero only when $P$ and $M$ are charged. (ii)  The emission of the photon does not alter the quantum numbers of a meson. 

The bremsstrahlung amplitude has to be invariant under the electromagnetic gauge transformation and this issue is studied in detail. 
The first principle calculation based on the standard model renders the gauge invariance automatically, but when an effective model is employed special care has to be taken in this respect. I derive a general effective Lagrangian, intuitively expressed in terms of the mesonic fields, and show that it leads to the gauge invariant amplitude. The resulting bremsstrahlung amplitude for $P\to V\gamma$ decays turns out to be equal zero. The bremsstrahlung amplitudes for $P\to Vl^+l^-$ and $ P\to P^\prime l^+l^-$ decays depend on the shape of the corresponding electromagnetic form factors.  A more profound effective Lagrangian will be employed when in the charm meson decays are studied in Chapter 5, leading to the same form of the bremsstrahlung amplitudes as derived in this section bellow\footnote{In the  hybrid model for the charm meson decays, which is employed in Chapter 5,  the bremsstrahlung amplitudes for  $D\to Pl^+l^-$ decays vanish in the limit $m_P^2\ll m_D^2$. The bremsstrahlung amplitudes for  $D\to Vl^+l^-$ decays vanish in the exact $SU(3)$ flavour limit.}.

I consider an effective Lagrangian for the weak decays $P\to V\gamma^*$ and $P\to P^\prime\gamma^*$, which proceed via the $W$ boson exchange in the ``s'' channel as shown in Fig. \ref{fig2}a. The strong corrections to the $W$ exchange are neglected throughout this discussion for the reasons of clarity. Defining the decay constants $f_P$ and $g_V$ as
\begin{equation}
\label{2.21}
\langle P(p)|j^{\mu}_W|0\rangle=if_Pp^{\mu}~,\qquad \langle V(p,\epsilon)|j^{\mu}_W|0\rangle=g_V\epsilon^{\mu}~,
\end{equation}
the weak current can be effectively expressed with the $P(x)$ and $V(x)$ fields as $j_W^{\mu}(x)=\tfrac{g_2}{2\sqrt{2}}[f_P\partial^{\mu}P(x)+g_VV^{\mu}(x)]$. The free and the weak  part of the effective Lagrangian for $P$, $V$ and $P^\prime$ fields is  given by
$${\cal L}_0=\partial ^{\mu}P^\dagger\partial_{\mu}P+\partial ^{\mu}P^{\prime\dagger}\partial_{\mu}P^\prime-\tfrac{1}{4}\vec F^{\mu\nu}\vec F_{\mu\nu}-\bigl[i\tfrac{g_2}{2\sqrt{2}}V_{\!C\!K\!M}W_{\mu}^\dagger(f_P\partial^{\mu}P+f_{P^\prime}\partial^{\mu}P^\prime+g_VV^{\mu})+h.c.\bigr]$$
with 
$$ \vec F^{\mu\nu}=\partial^{\mu}\vec V^\nu-\partial^\nu \vec V^\mu+c~\vec V^\mu\times \vec V^{\nu}~,\qquad \vec V=\begin{pmatrix}\tfrac{1}{\sqrt{2}}(V^\dagger+V)\\\tfrac{1}{\sqrt{2}}i(V^\dagger-V)\\V^0\end{pmatrix}~$$
 and $\vec V$ can be thought as a triplet of $\rho$ meson fields. The corresponding Feynman rules are given in  Fig. \ref{fig23}a.

\begin{figure}[h]
\centering
\mbox{
\subfigure[]
{
\begin{fmffile}{f23a1n}
\fmfframe(10,0)(10,0){
  \begin{fmfgraph*}(20,10)
  \fmfpen{thin}  
  \fmfleftn{l}{1} \fmfrightn{r}{1}
  \fmf{dashes}{v,l1}\fmf{zigzag}{v,r1}
  \fmfv{decor.size=1.2thick,decor.shape=circle,decor.filled=full,label=$
  -ig_2\frac{1}{2\sqrt{2}}f_PV_{CKM}p^\mu$,la.d=10thick,la.a=90}{v}
  \fmflabel{$P(p)$}{l1}
  \fmflabel{$W^\mu$}{r1}
  \end{fmfgraph*} }
\end{fmffile}
\quad
\begin{fmffile}{f23a2n}
\fmfframe(10,0)(10,0){
  \begin{fmfgraph*}(20,10)
  \fmfpen{thin}  
  \fmfleftn{l}{1} \fmfrightn{r}{1}
  \fmf{zigzag}{v,l1}\fmf{dashes}{v,r1}
  \fmfv{decor.size=1.2thick,decor.shape=circle,decor.filled=full,label=$
  -ig_2\frac{1}{2\sqrt{2}}f_{P^\prime}V_{CKM}p^{\prime\mu}$,la.d=10thick,la.a=90}{v}
  \fmflabel{$P^\prime(p^\prime)$}{r1}
  \fmflabel{$W^\mu$}{l1}
  \end{fmfgraph*} }
\end{fmffile}
\quad
\begin{fmffile}{f23a3n}
\fmfframe(10,0)(10,0){
  \begin{fmfgraph*}(20,10)
  \fmfpen{thin}  
  \fmfleftn{l}{1} \fmfrightn{r}{1}
  \fmf{dashes}{v,r1}\fmf{zigzag}{v,l1}
  \fmfv{decor.size=1.2thick,decor.shape=circle,decor.filled=full,label=$
  g_2\frac{1}{2\sqrt{2}}g_VV_{CKM}$,la.d=10thick,la.a=90}{v}
  \fmflabel{$W$}{l1}
  \fmflabel{$V$}{r1}
  \end{fmfgraph*} }
\end{fmffile}
}
     }
\mbox{
\subfigure[]
{
\begin{fmffile}{f23bn}
\fmfframe(10,5)(10,5){
  \begin{fmfgraph*}(20,20)
  \fmfpen{thin}  
  \fmfleftn{l}{1} \fmfrightn{r}{1} \fmfbottom{b}
  \fmf{dashes}{v,l1}\fmf{dashes}{v,r1}
  \fmffreeze
  \fmf{boson,label=$\gamma$}{v,b}
  \fmfv{decor.size=1.2thick,decor.shape=circle,decor.filled=full,label=$
  -ie(p+p^\prime)_\nu G_P^{\mu\nu}(q^2)$,la.d=10thick,la.a=90}{v}
  \fmflabel{$P^+(p^\prime)$}{r1}
  \fmflabel{$P^+(p)$}{l1}\fmflabel{$q,\mu$}{b}
  \end{fmfgraph*} }
\end{fmffile}
\quad
\begin{fmffile}{f23b2n}
\fmfframe(10,5)(10,5){
  \begin{fmfgraph*}(20,20)
  \fmfpen{thin}  
  \fmfleftn{l}{1} \fmfrightn{r}{1} \fmfbottom{b}
  \fmf{zigzag}{v,l1}\fmf{dashes}{v,r1}
  \fmffreeze
  \fmf{boson}{v,b}
  \fmfv{decor.size=1.2thick,decor.shape=circle,decor.filled=full,label=$
  ieg_2\frac{1}{2\sqrt{2}}f_PV_{CKM}G_{Pd}^{\mu\nu}(q^2)$,la.d=10thick,la.a=90}{v}
  \fmflabel{$W^\nu$}{l1}
  \fmflabel{$P^+(p)$}{r1}\fmflabel{$q,\mu$}{b}
  \end{fmfgraph*} }
\end{fmffile}
\quad
\begin{fmffile}{f23b3n}
\fmfframe(10,5)(10,5){
  \begin{fmfgraph*}(20,20)
  \fmfpen{thin}  
  \fmfleftn{r}{1} \fmfrightn{l}{1} \fmfbottom{b}
  \fmf{zigzag}{v,l1}\fmf{dashes}{v,r1}
  \fmffreeze
  \fmf{boson}{v,b}
  \fmfv{decor.size=1.2thick,decor.shape=circle,decor.filled=full,label=$
  ieg_2\frac{1}{2\sqrt{2}}f_PV_{CKM}G_{Pd}^{\mu\nu}(q^2)$,la.d=10thick,la.a=90}{v}
  \fmflabel{$W^\nu$}{l1}
  \fmflabel{$P^+(p)$}{r1}\fmflabel{$q,\mu$}{b}
  \end{fmfgraph*} }
\end{fmffile}
}
    }
\mbox{
\subfigure[]
{
\begin{fmffile}{f23cnn}
\fmfframe(10,5)(10,5){
  \begin{fmfgraph*}(20,20)
  \fmfpen{thin}  
  \fmfleftn{l}{1} \fmfrightn{r}{1} \fmfbottom{b}
  \fmf{dashes}{v,l1}\fmf{dashes}{v,r1}
  \fmffreeze
  \fmf{boson}{v,b}
  \fmfv{decor.size=1.2thick,decor.shape=circle,decor.filled=full,label=$
  ieG_V^{\mu\delta}(q^2)~[g_{\delta\nu}(q-p^\prime)_\beta+g_{\nu\beta}
  (p+p^\prime)_\delta-g_{\delta\beta}(p+q)_\nu]$,la.d=10thick,la.a=90}{v}
  \fmflabel{$V_\nu^+(p^\prime)$}{r1}
  \fmflabel{$V_\beta^+(p)$}{l1}\fmflabel{$q,\mu$}{b}
  \end{fmfgraph*} }
\end{fmffile}
}
    }    
\caption{The Feynman rules given by the effective Lagrangian (\ref{3.22}).} 
\label{fig23}
\end{figure}

The electromagnetic interactions of the pseudoscalars with the charge $e$ are introduced by replacing the partial derivative by the covariant derivative $\partial^\mu P\to (\partial^\mu+ieA^\mu)P$. The interactions of the form $A^\mu A_\mu PP^\dagger$ are not if interest and will be omitted. The photon emission from the charged meson $V$ is implemented by the transition $V\to VV^0$, contained in the $\vec F^{\mu\nu}\vec F_{\mu\nu}$ term, followed by the transition $V^0\to \gamma$ via the vector meson dominance. The free and electro-weak part of the Lagrangian is then given by
\begin{align}
\label{3.22}
{\cal L}&=\partial ^{\mu}P^\dagger\partial_{\mu}P+\partial ^{\mu}P^{\prime\dagger}\partial_{\mu}P^\prime-\tfrac{1}{4}\vec F_{\gamma}^{\mu\nu}\vec F_{\gamma,\mu\nu}-\!\bigl[i\tfrac{g_2}{2\sqrt{2}}V_{C\!K\!M}W_{\mu}^\dagger(f_P\partial^{\mu}P+f_{P^\prime}\partial^{\mu}P^\prime+g_VV^{\mu})+h.c.\bigr]\!+\nonumber\\
&-\tfrac{1}{4}F^{\mu\nu}F_{\mu\nu}+ieA^\mu(\partial_\mu P^\dagger~P-P^\dagger~\partial_\mu P)+e\tfrac{g_2}{2\sqrt{2}}V_{C\!K\!M}A^\mu W_\mu^\dagger(P+P^\dagger)
\end{align}
with 
$$F^{\mu\nu}=\partial^{\mu}A^\nu-\partial^\nu A^\mu~,\qquad\vec F_{\gamma}^{\mu\nu}=\partial^{\mu}\vec V^\nu-\partial^\nu \vec V^\mu+e\vec V^\mu\times \vec V^{\nu}~,\qquad \vec V=\begin{pmatrix}\tfrac{1}{\sqrt{2}}(V^\dagger+V)\\\tfrac{1}{\sqrt{2}}i(V^\dagger-V)\\\gamma\end{pmatrix}~.$$
Similar effective Lagrangian, but only for the pseudoscalar field, has been considered in \cite{lichard.brem}. 
The Feynman rules for the electromagnetic interactions, given in the second line of the Lagrangian (\ref{3.22}), are shown in Figs. \ref{fig23}b and  \ref{fig23}c. The mesons are not elementary particles, they have an internal structure and I deliberately multiply the vertices by the unknown form factors $G_P^{\mu\nu}(q^2)$, $G_V^{\mu\nu}(q^2)$ and $G_{Pd}^{\mu\nu}(q^2)$ (the subscript $d$ denotes a form factor connected with a direct $\gamma\!-\!W\!-\!P$ vertex) subject to the condition  
\begin{equation}
\label{3.23}
G_P^{\mu\nu}(0)=G^{\mu\nu}_V(0)=G_{Pd}^{\mu\nu}(0)=1~. 
\end{equation}
The form factors are expected to  have either (i) the polar shape given by the propagator of a neutral vector meson $V^0$, (ii) the flat shape or (iii) the linear combination of these  
\begin{align}
\label{3.123}
G^{\mu\nu}_{\!(i)}(q^2)&=G_{\!(i)}(q^2)g^{\mu\nu}=g^{\mu\nu}~,~~\quad G^{\mu\nu}_{\!(ii)}(q^2)=G_{\!(ii)}(q^2)\biggl[g^{\mu\nu}-\frac{q^\mu q^\nu}{m_{V^0}^2}\biggr]=\frac{m_{V^0}^2}{m_{V^0}^2-q^2}\biggl[g^{\mu\nu}-\frac{q^\mu q^\nu}{m_{V^0}^2}\biggr]~,\nonumber\\
  G^{\mu\nu}_{\!(iii)}(q^2)&=K_1g^{\mu\nu}+\sum_{i=2}^N K_i\biggl[g^{\mu\nu}-\frac{q^\mu q^\nu}{m_{V^0_i}^2}\biggr]\frac{m_{V^0_i}^2}{m_{V^0_i}^2-q^2}\quad{\rm with}\quad\sum_{i=1}^N K_i=1 
\end{align} 
for the photon with momentum $q$. Let me point out that for any of these shapes
\begin{equation}
\label{3.23a}
q_\mu G^{\mu\nu}(q^2)=q^\nu~.
\end{equation}

\begin{figure}[h]
\centering
\mbox{
  \begin{fmffile}{f24o1n}
  \fmfframe(3,3)(3,3){
  \begin{fmfgraph*}(20,20)
  \fmfpen{thin}  
  \fmfleftn{l}{1} \fmfrightn{r}{1}\fmftopn{t}{4}
  \fmf{dashes}{l1,v1}\fmf{dashes,label=$P$,tension=0.7}{v1,v2}
  \fmf{zigzag,label=$W$}{v2,v3}\fmf{dashes}{v3,r1}
  \fmffreeze
  \fmf{boson}{v1,t2}\fmfdot{v1}
  \fmflabel{$P$}{l1}\fmflabel{$P^\prime$}{r1}
  \fmflabel{$\gamma^*$}{t2}
  \end{fmfgraph*} }
\end{fmffile}
\quad
\begin{fmffile}{f24o2n}
  \fmfframe(3,3)(3,3){
  \begin{fmfgraph*}(20,20)
  \fmfpen{thin}  
  \fmfleftn{l}{1} \fmfrightn{r}{1}\fmftopn{t}{4}
  \fmf{dashes}{l1,v1}
  \fmf{zigzag,label=$W$}{v1,v3}\fmf{dashes}{v3,r1}
  \fmffreeze
  \fmf{boson}{v1,t2}\fmfdot{v1}
  \fmflabel{$P$}{l1}\fmflabel{$P^\prime$}{r1}
  \fmflabel{$\gamma^*$}{t2}
  \end{fmfgraph*} }
\end{fmffile}
\quad
\begin{fmffile}{f24o3n}
  \fmfframe(3,3)(3,3){
  \begin{fmfgraph*}(20,20)
  \fmfpen{thin}  
  \fmfleftn{l}{1} \fmfrightn{r}{1}\fmftopn{t}{4}
  \fmf{dashes}{l1,v1}\fmf{zigzag,label=$W$}{v1,v2}
  \fmf{dashes,label=$P^\prime$,tension=0.7}{v2,v3}
  \fmf{dashes}{v3,r1}
  \fmffreeze
  \fmf{boson}{v3,t3}\fmfdot{v3}
  \fmflabel{$P$}{l1}\fmflabel{$P^\prime$}{r1}
  \fmflabel{$\gamma^*$}{t3}
  \end{fmfgraph*} }
\end{fmffile}
\quad
\begin{fmffile}{f24o4n}
  \fmfframe(3,3)(3,3){
  \begin{fmfgraph*}(20,20)
  \fmfpen{thin}  
  \fmfleftn{l}{1} \fmfrightn{r}{1}\fmftopn{t}{4}
  \fmf{dashes}{l1,v1}\fmf{zigzag,label=$W$}{v1,v3}
  \fmf{dashes}{v3,r1}
  \fmffreeze
  \fmf{boson}{v3,t3}\fmfdot{v3}
  \fmflabel{$P$}{l1}\fmflabel{$P^\prime$}{r1}
  \fmflabel{$\gamma^*$}{t3}
  \end{fmfgraph*} }
\end{fmffile}

    }    
\caption{The bremsstrahlung diagrams for $P\to P^\prime\gamma^*$ decay as given by the
effective Lagrangian  (\ref{3.22}). The corresponding Feynman rules are given in Fig. \ref{fig23}.} \label{fig24}
\end{figure}

The bremsstrahlung amplitude for the $\boldsymbol{P\to P^\prime \gamma^*}$ decay is given by the diagrams  in Fig. \ref{fig24} and can be calculated by applying the Feynman rules in Fig. \ref{fig23}
\begin{align}
\label{3.125}
|{\cal A}&[P(p)\to P^\prime\gamma^*(q,\epsilon)]|=\frac{G_F}{\sqrt{2}}eV_{\!C\!K\!M}V_{\!C\!K\!M}^\prime f_Pf_{P^\prime}\nonumber\\
&\times\epsilon_\mu^*\bigl[\frac{G_P^{\mu\nu}(q^2)m_{P^\prime}^2-G_{P^\prime}^{\mu\nu}(q^2)m_P^2}{m_P^2-m_{P^\prime}^2}(p+p^\prime)_\nu+G_{Pd}^{\mu\nu}(q^2)p^\prime_\nu+G_{P^\prime d}^{\mu\nu}(q^2)p_\nu\bigr]~.
\end{align}
The amplitude is invariant under the gauge transformation $\epsilon\to \epsilon +Cq$ since the term in parenthesis vanishes at $q^2=0$ and since $q_\mu G^{\mu\nu}(q^2)=q^\nu$ (\ref{3.23a}). The amplitude can be cast into a manifestly gauge invariant form (\ref{amp.pgammas}) by introducing 
$$q^2B_{PC}(q^2)\equiv \frac{G_P(q^2)m_{P^\prime}^2-G_{P^\prime}(q^2)m_P^2}{m_P^2-m_{P^\prime}^2}+G_{Pd}(q^2)+G_{P^\prime d}(q^2)~.$$
The function $B_{PC}$ is regular at $q^2\!=\!0$ due to the condition (\ref{3.23})  and enters the gauge invariant amplitude (\ref{amp.pgammas}) via 
\begin{equation*}
|{\cal A}[P(p)\to P^\prime\gamma^*(q,\epsilon)]|=\tfrac{G_F}{\sqrt{2}}eV_{\!C\!K\!M}V_{\!C\!K\!M}^\prime f_Pf_{P^\prime}B_{PC}(q^2) \epsilon_\mu^*[q^2(p+p^\prime)^{\mu}-(m_P^2-m_{P^\prime}^2)q^{\mu}]~.
\end{equation*}
 Note that the diagrams with $\gamma\!-\! W\!-\! P$ vertices in Fig. \ref{fig24} are essential in maintaining the gauge invariance.

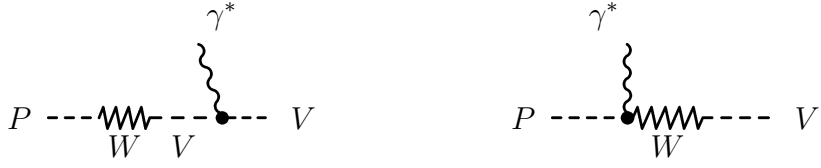
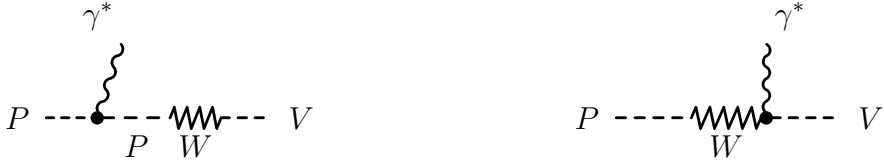
\begin{figure}[h]
\centering
\mbox{
\subfigure[The bremsstrahlung diagrams for $P\to V\gamma^*$ decay.]
{
\begin{fmffile}{f25o3mm}
  \fmfframe(13,3)(13,3){
  \begin{fmfgraph*}(30,20)
  \fmfpen{thin}  
  \fmfleftn{l}{1} \fmfrightn{r}{1}\fmftopn{t}{4}
  \fmf{dashes}{l1,v1}\fmf{zigzag,label=$W$}{v1,v2}
  \fmf{dashes,label=$V$,tension=0.7}{v2,v3}
  \fmf{dashes}{v3,r1}
  \fmffreeze
  \fmf{boson}{v3,t3}\fmfdot{v3}
  \fmflabel{$P$}{l1}\fmflabel{$V$}{r1}
  \fmflabel{$\gamma^*$}{t3}
  \end{fmfgraph*} }
\end{fmffile}
\quad
\begin{fmffile}{f25o2mm}
  \fmfframe(13,3)(13,3){
  \begin{fmfgraph*}(30,20)
  \fmfpen{thin}  
  \fmfleftn{l}{1} \fmfrightn{r}{1}\fmftopn{t}{4}
  \fmf{dashes}{l1,v1}
  \fmf{zigzag,label=$W$}{v1,v3}\fmf{dashes}{v3,r1}
  \fmffreeze
  \fmf{boson}{v1,t2}\fmfdot{v1}
  \fmflabel{$P$}{l1}\fmflabel{$V$}{r1}
  \fmflabel{$\gamma^*$}{t2}
  \end{fmfgraph*} }
\end{fmffile}
}
    }
\mbox{
\subfigure[Kinematicaly forbidden diagram.]
{    
  \begin{fmffile}{f25o1mm}
  \fmfframe(16,3)(16,3){
  \begin{fmfgraph*}(30,20)
  \fmfpen{thin}  
  \fmfleftn{l}{1} \fmfrightn{r}{1}\fmftopn{t}{4}
  \fmf{dashes}{l1,v1}\fmf{dashes,label=$P$,tension=0.7}{v1,v2}
  \fmf{zigzag,label=$W$}{v2,v3}\fmf{dashes}{v3,r1}
  \fmffreeze
  \fmf{boson}{v1,t2}\fmfdot{v1}
  \fmflabel{$P$}{l1}\fmflabel{$V$}{r1}
  \fmflabel{$\gamma^*$}{t2}
  \end{fmfgraph*} }
\end{fmffile}
}
\quad
\subfigure[This diagram does not exist since there are no $\gamma-W-V$ vertices in the
effective Lagrangian (\ref{3.22}).]
{
\begin{fmffile}{f25o4mm}
  \fmfframe(16,3)(16,3){
  \begin{fmfgraph*}(30,20)
  \fmfpen{thin}  
  \fmfleftn{l}{1} \fmfrightn{r}{1}\fmftopn{t}{4}
  \fmf{dashes}{l1,v1}\fmf{zigzag,label=$W$}{v1,v3}
  \fmf{dashes}{v3,r1}
  \fmffreeze
  \fmf{boson}{v3,t3}\fmfdot{v3}
  \fmflabel{$P$}{l1}\fmflabel{$V$}{r1}
  \fmflabel{$\gamma^*$}{t3}
  \end{fmfgraph*} }
\end{fmffile}
}
    }    
\caption{The bremsstrahlung diagrams for $P\to V\gamma^*$ decay as given by the
effective Lagrangian  (\ref{3.22}) are given in Fig. (a). 
The corresponding Feynman rules are given in Fig. \ref{fig23}.}
\label{fig25} 
\end{figure}

The bremsstrahlung amplitude for the $\boldsymbol{P\to V\gamma^*}$ decay is given by the diagrams  in Fig. \ref{fig25}a. Note that the diagram in Fig. \ref{fig25}b is prohibited since a virtual pseudoscalar can not transform to a real vector (a real pseudoscalar can on the other hand transform to a virtual vector, like in the decay $\pi^+\to W^+\to \mu^+\nu_\mu$). The diagram in Fig. \ref{fig25}c does not exist since there is no $\gamma\!-\! W\!-\!V$ vertex in the Lagrangian (\ref{3.22}). The amplitude for the diagrams in Fig. \ref{fig25}a is
\begin{align}
\label{3.126}
|{\cal A}&[P(p)\to V(p^\prime,\epsilon^\prime)\gamma^*(q,\epsilon)]|=\frac{G_F}{\sqrt{2}}eV_{C\!K\!M}V_{C\!K\!M}^\prime f_Pg_V\nonumber\\
&\times \epsilon^{*}_{\mu}\epsilon^{\prime *}_{\nu}\biggl(G_{Pd}^{\mu\nu}(q^2)-G_V^{\mu\delta}(q^2)~p_\alpha~\frac{g^{\alpha\beta}-\tfrac{p^\alpha p^\beta}{m_V^2}}{m_P^2-m_V^2} ~\bigl[g_{\delta\nu}(q-p^\prime)_\beta+g_{\nu\beta}(p^\prime+p)_\delta-g_{\delta\beta}(p+q)_\nu\bigr]\biggr)\nonumber\\
&=\frac{G_F}{\sqrt{2}}eV_{C\!K\!M}V_{C\!K\!M}^\prime f_Pg_V\epsilon^{*}_{\mu}\epsilon^{\prime *}_{\nu}\biggl([G_{Pd}^{\mu\nu}(q^2)-G_V^{\mu\nu}(q^2)]-\frac{G_V^{\mu\delta}(q^2)}{m_V^2}[q_\delta q_\nu-q^2g_{\delta\nu}]\biggr)~.
\end{align}
The amplitude is invariant under the gauge transformation $\epsilon\to \epsilon +Cq$ since $q_\mu G^{\mu\nu}(q^2)=q^\nu$ (\ref{3.23a}) and can be cast to a general form (\ref{amp.gammas}) by introducing
$$q^2A_{PV}^{\prime}(q^2)=G_{Pd}(q^2)-G_V(q^2)+\frac{q^2}{m_V^2}G_V(q^2)~.$$
The function $A_{PV}^{\prime}(q^2)$ is regular at $q^2\!=\!0$ due to the condition (\ref{3.23}) and enters the gauge invariant amplitude (\ref{amp.gammas}) via
\begin{equation*}
|{\cal A}[P(p)\to V(p^\prime,\epsilon^\prime)\gamma^*(q,\epsilon)]|=\frac{G_F}{\sqrt{2}}eV_{C\!K\!M}V_{C\!K\!M}^\prime f_Pg_V ~A_{PV}^{\prime}(q^2)~\epsilon^{*}_{\mu}\epsilon^{\prime *}_{\nu}[q^2g^{\mu\nu}-q^\mu q^\nu]~.
\end{equation*}
In the case of the decay $P\to V\gamma$ with the real photon in the final state,  $q^2=0$ and $\epsilon\cdot q=0$, so 
\begin{equation}
\label{3.brem}
{\cal A}[P\to V\gamma]=0~
\end{equation}
and the bremsstrahlung amplitude based on the general effective Lagrangian (\ref{3.22}) vanishes.

\chapter{The $\boldsymbol{c\to u\gamma}$ transition  in  $\boldsymbol{B_c\to B_u^*\gamma}$ decay}

In this chapter I propose and explore the unique opportunity to observe the flavour changing neutral transition $c\to u\gamma$  in the beauty conserving decay $B_c\to B_u^*\gamma$ with flavour content $c\bar b\to u\bar b\gamma$. This possibility was originally suggested and studied in \cite{FPS3} and subsequently discussed in \cite{moriond,southampton,cracow,AS}.

\section{The short distance contribution}

The amplitude for the short distance contribution is given in (\ref{asd})
\begin{align*}
{\cal A}_{SD}&=-i\frac{G_F}{\sqrt{2}}{e_0\over 8\pi^2}c_7^{eff}m_c\langle \gamma(q,\epsilon)B_u^*|\bar u\sigma^{\mu\nu}(1+\gamma_5)cF_{\mu\nu}|B_c\rangle\\
&=-\frac{G_F}{\sqrt{2}}{e_0\over 4\pi^2}c_7^{eff}m_c\epsilon^*_{\mu}q_{\nu}\langle B_u^*|\bar u\sigma^{\mu\nu}(1+\gamma_5)c|B_c\rangle~
\end{align*}
with $c_7^{eff}=-(1.5+4.4i)[1\pm 0.2] 10^{-3}$ (\ref{c7}) \cite{GHMW} in the standard model. The matrix element $q_{\nu}\langle B_u^*|\bar 
u\sigma^{\mu\nu}(1+\gamma_5)c|B_c\rangle$ can be generally expressed as\footnote{The definitions of the form factors in this chapter are taken from  \cite{Soares}.}
\begin{align}
\label{4.2}
\langle B_u^*(p^{\prime},\epsilon^{\prime})|\bar 
u\sigma^{\mu\nu}q_{\nu}c|B_c(p)\rangle&=\epsilon^{\mu\alpha\beta\gamma}\epsilon_{\alpha}^{*\prime}p_{\beta}^{\prime}p_{\gamma}F_1(q^2)~
,\nonumber\\
~\nonumber\\
\langle B_u^*(p^{\prime},\epsilon^{\prime})|\bar 
ui\sigma^{\mu\nu}q_{\nu}\gamma_5c|B_c(p)\rangle&=[(m_{B_c}^2-m_{B_u^*}^
2)\epsilon^{*\prime\mu}-\epsilon^{*\prime}\cdot q(p+p^{\prime})^{\mu}]F_2(q^2)\nonumber\\
&+\epsilon^{*\prime}\cdot q~[q^{\mu}-{q^2\over m_{B_c}^2-m_{B_u^*}^2}(p+p^{\prime})^{\mu}]F_3(q^2)~.
\end{align} 
For real photon with $q^2=0$ and $\epsilon\cdot q=0$ only $F_1(0)$ and $F_2(0)$ form factors contribute and due to 
$$\sigma_{\mu\nu}\gamma_5=-\tfrac{i}{2}\epsilon_{\mu\nu\alpha\beta}\sigma^{\alpha\beta}$$
they are related via $F_2(0)=\tfrac{1}{2}F_1(0)$. The resulting 
amplitude $A_{SD}$ is automatically invariant under the electromagnetic gauge transformation $\epsilon\to \epsilon+Cq$ \cite{FPS3}
\begin{align}
\label{4.1}
{\cal A}_{SD}&=-\frac{G_F}{\sqrt{2}}{e_0\over 4\pi^2}c_7^{eff}m_c\bigl[\epsilon^{\mu\alpha\beta\gamma}\epsilon_{\mu}^*\epsilon_{\alpha}^{*\prime}p_{\beta}^{\prime}p_{\gamma}-i[(p\cdot q)(\epsilon^*\cdot\epsilon^{*\prime})-(\epsilon^{*\prime}\cdot q)(p\cdot 
\epsilon^{*})]\bigr]F_1(0).
\end{align}
 The form factor $F_1(0)$ defined in (\ref{4.2}) will be  calculated at the end of the chapter using the Isgur-Scora-Grinstein-Wise (ISGW) constituent quark model \cite{ISGW}, which is expected to reasonably account for the strong interactions in the heavy mesons $B_c$ and $B_u^*$. 

\section{The long distance penguin contribution}

The long distance penguin contribution is illustrated in Fig. \ref{fig3}. The intermediate $\bar dd$ and $\bar ss$ quark-antiquark pairs hadronize into the vector mesons $\rho^0$, $\omega$ and $\phi$ and finally convert to a photon. The contribution of $\bar bb$ is neglected in view of the large mass of $\Upsilon$. The corresponding  amplitude (\ref{ald}) is induced by the part of the nonleptonic effective Lagrangian (\ref{eff}) proportional to $a_2$
$${\cal A}_{LD}^{peng}=-i{G_F\over \sqrt{2}}\sum_{q=d,s}V_{cq}^*V_{uq}a_2\langle B_u^*\gamma|\bar q\gamma_{\mu}(1-\gamma_5)q~\bar u\gamma^{\mu}(1-\gamma_5)c|B_c\rangle~.$$
The appropriate renormalization scale for $a_2$ is $\mu_F\!\simeq\! m_c$ as $\bar b$ is merely spectator in this contribution, giving  $a_2\!=\!a_2^c\!=\!-0.5$ (\ref{ai}). 
Using the factorization approximation (\ref{factor}) and $V_{cd}^*V_{ud}\simeq -V_{cs}^*V_{us}$,
$${\cal A}_{LD}^{peng}=-i{G_F\over \sqrt{2}}V_{cs}^*V_{us}a_2^c\langle B_u^*|\bar u\gamma^{\mu}(1-\gamma_5)c|B_c\rangle\langle\gamma|\bar s\gamma_{\mu}s-\bar d\gamma_{\mu}d|0\rangle$$
and the electromagnetic interaction in the second matrix element is implicit. 
In the vector meson dominance approximation the matrix element
$$
\langle\gamma|\bar s\gamma_{\mu}s-\bar d\gamma_{\mu}d|0\rangle=\langle \gamma|-ie_0(\tfrac{2}{3}\bar u\gamma^{\nu}u-\tfrac{1}{3}\bar d\gamma^{\nu}d-\tfrac{1}{3}\bar s\gamma^{\nu}s)A_{\nu}|V^0\rangle\langle V^0|\bar s\gamma_{\mu}s-\bar d\gamma_{\mu}d|0\rangle
$$
 can be expressed in terms of the couplings $g_V$ defined via
\begin{equation*}
\langle V(q,\epsilon)|j_V^{\mu}|0\rangle=g_V(q^2)\epsilon^{*\mu}~,
\end{equation*}
where $j_V^{\mu}$ is properly normalized vector current\footnote{ In the case of $\rho^0$, for example, $j_V^{\mu}=\tfrac{1}{\sqrt{2}}(\bar u\gamma^{\mu}u-\bar d\gamma^{\mu}d)$.} with the same flavour structure as the vector meson $V$,
\begin{equation}
\label{cvmd1}
\langle\gamma|\bar s\gamma_{\mu}s-\bar d\gamma_{\mu}d|0\rangle=e_0~C_{V\!M\!D}~\epsilon_{\mu}^*\qquad{\rm with }\qquad
 C_{V\!M\!D}\equiv {g_{\rho}^2(0)\over  2m_{\rho}^2}-{g_{\omega}^2(0)\over  
6m_{\omega}^2}-{g_{\phi}^2(0)\over  3m_{\phi}^2}~.
\end{equation}
The  $g_{V^0}(m_{V^0}^2)$ couplings  can determined phenomenologically from the experimental data on $V^0\to \gamma^*\to e^+e^-$ decays. Taking the central values and the errors on $g_V(m_V)$ from \cite{PDG} and assuming $g_{V^0}(0)\simeq g_{V^0}(m_{V^0}^2)$, which is a reasonable approximation for light vector mesons, the amplitude is given by 
\begin{align}
\label{cvmd}
{\cal A}_{LD}^{peng}&=-i{G_F\over \sqrt{2}}V_{cs}^*V_{us}~a_2^c~e_0~C_{V\!M\!D}~\epsilon^*_{\mu}\langle B_u^*|\bar u\gamma^{\mu}(1-\gamma_5)c|B_c\rangle~ \quad {\rm with }\nonumber\\
 C_{V\!M\!D}&=(-1.2\pm1.2)\cdot 10^{-3}~{\rm GeV}^2~. 
\end{align}
The three terms in $C_{V\!M\!D}$ almost exactly cancel since  $\langle 0|\bar s\gamma_{\mu}s-\bar d\gamma_{\mu}d|0\rangle$ vanishes in the exact $SU(3)$ flavour limit, so the long distance penguin contribution is relatively small. Due to the cancelations in $C_{V\!M\!D}$, the magnitude of this contribution is fairly uncertain. The uncertainties in (\ref{cvmd}) arise from to the errors in the experimental data on $V^0\to e^+e^-$ rates. 

The matrix element $\langle B_u^*|\bar u\gamma^{\mu}(1-\gamma_5)c|B_c\rangle$ can be generally expressed as 
\begin{align}
\label{4.1a} 
&\langle B_u^*(p^\prime,\epsilon^\prime)|\bar 
u\gamma^{\mu}(1-\gamma_5)c|B_c(p)\rangle=i\biggl\{\frac{2}{
 m_{B_c}+m_{B_u^*}}\epsilon^{\mu\alpha\beta\gamma}\epsilon_{\alpha}^{*\prime}p_{\beta}p_{\gamma}^\prime V(q^2)+i2m_{B_u^*}{\epsilon^{*\prime}\cdot q\over q^2}q^{\mu}A_0(q^2)\nonumber\\
&+i(m_{B_c}+m_{B_u^*})\bigl[\epsilon^{\mu *\prime}-{\epsilon^{*\prime} \cdot q\over q^2}q^{\mu}\bigr]A_1(q^2)
-i{\epsilon^{*\prime}\cdot q\over 
m_{B_c}+m_{B_u^*}}\bigl[(p+p^{\prime})^{\mu}-{m_{B_c}^2-m_{B_u ^*}^2\over q^2}q^{\mu}\bigr]A_2(q^2)q\biggr\}~.
\end{align}
The matrix element is finite at $q^2=0$ and the equality $2m_{B_u^*}A_0(0)= (m_{B_c}+m_{B_u^*})A_1(0)-(m_{B_c}-m_{B_u^*})A_2(0)$ must hold \cite{BSW}. The amplitude (\ref{cvmd}) at $q^2=0$ and $\epsilon\cdot q=0$ is  given by
\begin{align}
\label{4.4}
{\cal A}_{LD}^{peng}&=-i{G_F\over \sqrt{2}}{V_{cs}^*V_{us}a_2^ce_0C_{V\!M\!D}\over m_{B_c}+m_{B_u^*}}
\biggl[
-2i\epsilon^{\mu\alpha\beta\gamma}\epsilon^*_{\mu}\epsilon_{\alpha}^{*\prime}p_{\beta}^{\prime}p_{\gamma}V(0)\nonumber\\
&-(m_{B_c}+m_{B_u^*})^2\epsilon^{*}\cdot\epsilon^{*\prime}A_1(0)
+2(\epsilon^*\cdot p^{\prime})(\epsilon^{*\prime}\cdot q)A_2(0)\biggr]~.
\end{align}

\begin{figure}[h]

\centering
\mbox{
\subfigure[ $H_{++}$]
{
\begin{fmffile}{f21a}
  \fmfframe(8,0)(8,0){
  \begin{fmfgraph*}(25,8)
  \fmfpen{thin}
  \fmfleft{l1}\fmfright{r1}
   \fmf{fermion,tension=1,label=$\Leftarrow$,la.si=right,la.d=30}{v,l1}
   \fmf{fermion,tension=1,label=$\Rightarrow$,la.si=left,la.d=30}{v,r1}
   \fmflabel{$V_1(\vec p)$}{l1}\fmflabel{$V_2(-\vec p)$}{r1}
   \fmfv{label=$P$,la.d=2thick,la.a=90,de.sh=circle,
    dec.siz=1thick,de.fil=shaded}{v}
  \end{fmfgraph*} }
\end{fmffile}
}
\quad
\subfigure[ $H_{--}$]
{
\begin{fmffile}{f21b}
  \fmfframe(5,0)(5,0){
  \begin{fmfgraph*}(25,8)
  \fmfpen{thin}
  \fmfleft{l1}\fmfright{r1}
   \fmf{fermion,tension=1,label=$\Rightarrow$,la.si=right,la.d=30}{v,l1}
   \fmf{fermion,tension=1,label=$\Leftarrow$,la.si=left,la.d=30}{v,r1}
   \fmflabel{$V_1$}{l1}\fmflabel{$V_2$}{r1}
   \fmfv{label=$P$,la.d=2thick,la.a=90,de.sh=circle,dec.siz=1thick,de.fil=shaded}{v}
  \end{fmfgraph*} }
\end{fmffile}
}
\quad
\subfigure[ $H_{00}$]
{
\begin{fmffile}{f21c}
  \fmfframe(5,0)(5,0){
  \begin{fmfgraph*}(25,8)
  \fmfpen{thin}
  \fmfleft{l1}\fmfright{r1}
   \fmf{fermion,tension=1,label=$\Uparrow$,la.si=right,la.d=30}{v,l1}
   \fmf{fermion,tension=1,label=$\Downarrow$,la.si=left,la.d=30}{v,r1}
   \fmflabel{$V_1$}{l1}\fmflabel{$V_2$}{r1}
   \fmfv{label=$P$,la.d=3thick,la.a=90,de.sh=circle,dec.siz=1thick,de.fil=shaded}{v}
  \end{fmfgraph*} }
\end{fmffile}
}
     }
\caption{Helicity states in the decay $P\to V_1V_2$.} 
\label{fig21}
\end{figure}

The electromagnetic gauge invariance under $\epsilon^\mu\to \epsilon^\mu+Cq^\mu$ imposes the relation
\begin{equation}
\label{4.5}
A_2(0)- {(m_{B_c}+m_{B_u^*})^2\over m_{B_c}^2-m_{B_u^*}^2}A_1(0)=0~.
\end{equation}
The physical significance of this condition was studied for other decays in \cite{GP,FPS1,FS} and can be easily understood by decomposing the amplitude (\ref{4.4}) in terms of different  helicity states. For a moment we pretend that the photon has finite mass $m_{\gamma}$ and we eventually set it to zero. In this case the vector meson and the massive photon can have helicities $+$, $-$, $0$, whereas the amplitude (\ref{4.4}) is decomposed to the three helicity amplitudes sketched in Fig.  \ref{fig21} \cite{GP} \footnote{These expressions are obtained by taking the polarizations of the particles as $\epsilon^\mu(k,\pm)=(0,\vec e_{\pm})$ with $\vec e_{\pm}\vec k=0$ and $\epsilon^{\mu}(k,0)=(|\vec k|/m,\vec kk^0/|\vec k|m)$.}
 \begin{align}
\label{4.4a}
{\cal A}_{LD}^{peng}&=-i{G_F\over \sqrt{2}}{V_{cs}^*V_{us}a_2^ce_0C_{V\!M\!D}\over m_{B_c}+m_{B_u^*}}(A_{++}+A_{--}+A_{00})\\
A_{++}&\stackrel{m_{\gamma}\to 0}{\longrightarrow}~ -(m_{B_c}+m_{B_u^*})^2A_1(0)-2p^{\prime}\cdot q V(0)\nonumber\\
A_{--}&\stackrel{m_{\gamma}\to 0}{\longrightarrow}~-(m_{B_c}+m_{B_u^*})^2A_1(0)+2p^{\prime}\cdot q V(0)\nonumber\\
A_{00}&\stackrel{m_{\gamma}\to 0}{\longrightarrow}~\frac{1}{m_{\gamma}}~{m_{B_c}+m_{B_u^*}\over m_{B_u^*}}~p^{\prime}\cdot q~\bigl[(m_{B_c}+m_{B_u^*})A_1(0)-(m_{B_c}-m_{B_u^*})A_2(0)\bigr]~.\nonumber
\end{align}
The photon is massless and can not have helicity zero. The helicity amplitude $A_{00}$ proportional to the left hand side of (\ref{4.5}) must be discarded and the helicity amplitudes $A_{\pm\pm}$ are retained. Since $A_2(0)$ is contained only in $A_{00}$, while $A_1(0)$ is contained also in $A_{\pm\pm}$, the amplitude $A_{00}$ is discarded  by expressing $A_2(0)$ via  
\begin{equation}
\label{4.5a}
A_2(0)\to {(m_{B_c}+m_{B_u^*})^2\over m_{B_c}^2-m_{B_u^*}^2}A_1(0)
\end{equation}
giving
\begin{align}
\label{4.6}
{\cal A}_{LD}^{peng}=-i\frac{G_F}{\sqrt{2}}V_{cs}^*V_{us}a_2^ce_0C_{V\!M\!D} 
\bigl[&
-2i\epsilon^{\mu\alpha\beta\gamma}\epsilon^*_{\mu}\epsilon_{\alpha}^{*\prime}p_{\beta}^{\prime}p_{\gamma}\frac{V(0)}{m_{B_c}+m_{B_u^*}}\nonumber\\
&+2[(\epsilon^*\cdot p^{\prime})(\epsilon^{*\prime}\cdot q)-(\epsilon^{*}\cdot\epsilon^{*\prime})(p\cdot q)]\frac{A_1(0)}{m_{B_c}-m_{B_u^*}}\bigr]~.
\end{align}
The form factors $A_1(0)$ and $A_2(0)$ will be determined using the ISGW model bellow.

\section{The long distance weak annihilation contribution}

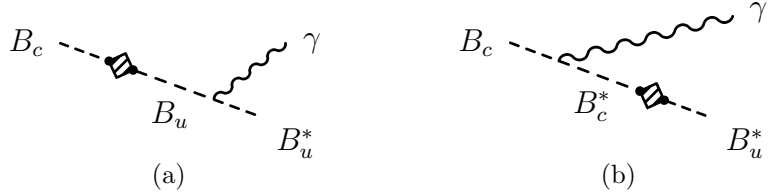
\begin{figure}[h]
\centering
\mbox{
\subfigure[]
{
\begin{fmffile}{f19a}
\fmfframe(8,0)(8,0){
  \begin{fmfgraph*}(30,20)
  \fmfpen{thin}  
  \fmfleftn{l}{1} \fmfrightn{r}{3}
  \fmfrpolyn{shaded,tension=1}{k}{4}   
  \fmf{dashes}{l1,k1} 
  \fmf{dashes,label=$B_u$,tension=0.6}{k3,v}\fmf{dashes}{v,r1} 
  \fmffreeze
  \fmf{boson,tension=1}{v,r2}
  \fmfv{decor.size=1.2thick,decor.shape=circle,decor.filled=full}{k1,k3}
  \fmflabel{$\gamma$}{r2}
  \fmflabel{$B_c$}{l1}
  \fmflabel{$B_u^*$}{r1}
  \end{fmfgraph*} }
\end{fmffile}
}
\quad
\subfigure[]
{
\begin{fmffile}{f19b}
\fmfframe(8,0)(8,0){
  \begin{fmfgraph*}(30,20)
  \fmfpen{thin}  
  \fmfleftn{l}{1} \fmfrightn{r}{4}
  \fmfrpolyn{shaded,tension=1}{k}{4}   
  \fmf{dashes}{l1,v} 
  \fmf{dashes,label=$B_c^*$,tension=0.6}{v,k1}\fmf{dashes}{k3,r1}
  \fmffreeze 
  \fmf{boson,tension=1}{v,r3}
  \fmfv{decor.size=1.2thick,decor.shape=circle,decor.filled=full}{k1,k3}
  \fmflabel{$\gamma$}{r3}
  \fmflabel{$B_c$}{l1}
  \fmflabel{$B_u^*$}{r1}
  \end{fmfgraph*} }
\end{fmffile}
}
    }
\caption{Long distance weak annihilation contribution to $B_c\to B_u^*\gamma$
decay. The box denotes the action 
of the nonleptonic effective Lagrangian (\ref{4.7}) and the two dots  denote the weak currents.}
\label{fig19}
\end{figure}

\begin{figure}[h]
\centering
\mbox{
\subfigure[]
{
\begin{fmffile}{f34a}
\fmfframe(8,0)(8,0){
  \begin{fmfgraph*}(30,20)
  \fmfpen{thin}  
  \fmfleftn{l}{1} \fmfrightn{r}{4}
  \fmfrpolyn{shaded,tension=1}{k}{4}   
  \fmf{dashes}{l1,v} 
  \fmf{dashes,label=$B_c$,tension=0.6}{v,k1}\fmf{dashes}{k3,r1}
  \fmffreeze 
  \fmf{boson,tension=1}{v,r3}
  \fmfv{decor.size=1.2thick,decor.shape=circle,decor.filled=full}{k1,k3}
  \fmflabel{$\gamma$}{r3}
  \fmflabel{$B_c$}{l1}
  \fmflabel{$B_u^*$}{r1}
  \end{fmfgraph*} }
\end{fmffile}
}
\quad
\subfigure[]
{
\begin{fmffile}{f34b}
\fmfframe(8,0)(8,0){
  \begin{fmfgraph*}(30,20)
  \fmfpen{thin}  
  \fmfleftn{l}{1} \fmfrightn{r}{3}
  \fmfrpolyn{shaded,tension=1}{k}{4}   
  \fmf{dashes}{l1,k1} 
  \fmf{dashes,label=$B_u^*$,tension=0.6}{k3,v}\fmf{dashes}{v,r1} 
  \fmffreeze
  \fmf{boson,tension=1}{v,r2}
  \fmfv{decor.size=1.2thick,decor.shape=circle,decor.filled=full}{k1,k3}
  \fmflabel{$\gamma$}{r2}
  \fmflabel{$B_c$}{l1}
  \fmflabel{$B_u^*$}{r1}
  \end{fmfgraph*} }
\end{fmffile}
}
    }
\caption{The diagram (a) is kinematicaly forbidden. The diagram (b) is one of
the bremsstrahlung diagrams and  exactly cancels with the other
bremsstrahlung diagrams. The box denotes the action 
of the nonleptonic effective Lagrangian (\ref{4.7}) and the two dots  denote the weak currents.} 
\label{fig34}
\end{figure}
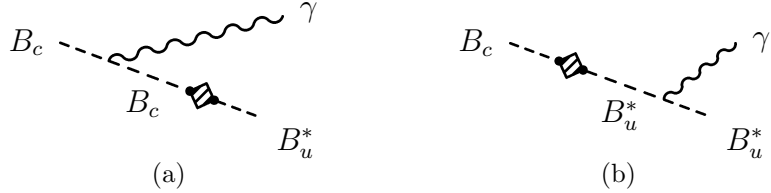

The long distance weak annihilation contribution for $B_c\to B_u^*\gamma$ decay in terms of quark degrees of freedom is illustrated in Fig. \ref{fig2}a with $\bar q=\bar b$. The diagrams in terms of pseudoscalar and vector mesons are shown in Fig. \ref{fig19} and the box represents the action of the relevant part of the nonleptonic Lagrangian (\ref{eff})
\begin{equation}
\label{4.7}
{\cal L}^{WA}=-\tfrac{G_F}{\sqrt{2}}a_1V_{cb}^*V_{ub}~\bar 
u\gamma^{\mu}(1-\gamma_5)b~\bar b\gamma_{\mu}(1-\gamma_5)c~.
\end{equation}
The coefficient $a_1$ is taken at the factorization scale $\mu_F\!\simeq\! {\cal O}(m_b)$ giving $a_1\!=\!a_1^b\!=\!1.08$ (\ref{ai})\footnote{The difference in $a_1^b=1.08$ and $a_1^c=1.2$ (\ref{ai}) is not essential  and the choice the renormalization scale $\mu_F$ does not lead to the sizable uncertainties.}.   
The diagram in Fig. \ref{fig34}a is not considered since it is not allowed kinematically. The diagram in Fig. \ref{fig34}b is one of the bremsstrahlung diagrams given in Fig. \ref{fig25}a and is not included since the bremsstrahlung amplitude for $P\to V\gamma$ decay vanishes, as shown in Section 3.3.3. The contributions with axial and scalar poles are neglected and the weak annihilation amplitude will involve only the parity conserving part. 
Note that the long distance weak annihilation contribution is relatively small due to the factor 
$V_{cb}^*V_{ub}$ in (\ref{4.7}), which makes $B_c\to B_u^*\gamma$ decay interesting for observing $c\to u\gamma$ transition. The amplitude for the  diagrams in Fig.  \ref{fig19} can be expressed in terms of the decay constants and the magnetic moments defined by      
\begin{eqnarray}
\label{4.8}
\langle 0|A_{\mu}|P\rangle &=&f_Pp_{\mu}~,\\
\langle V|V_{\mu}|0\rangle &=&g_V\epsilon_{\mu}^*\nonumber~,\\
{\cal A}(P(p)\to V(p^{\prime},\epsilon^{\prime})\gamma(\epsilon 
))&=&\mu_Pe\epsilon^{\mu\nu\alpha\beta}\epsilon_{\mu}^*\epsilon_{\nu}^{*\prime}p
_{\alpha}p_{\beta}^{\prime}~\nonumber
\end{eqnarray}
and  
 $\mu_{B_c}$, $\mu_{B_u}$, $f_{B_c}$, $f_{B_u}$, $g_{B_c^*}$, 
$g_{B_u^*}$ will be determined using ISGW model.\\

\section{The amplitude}

The total amplitude for the $B_c\to B_u^*\gamma$ decay is given by the sum of the short distance amplitude (\ref{4.1}), the long distance penguin amplitude (\ref{4.6}) and the long distance weak annihilation amplitude coming from (\ref{4.7}, \ref{4.8}). It has  the gauge invariant form (\ref{amp.gamma})
\begin{equation}
\label{4.amp} 
{\cal A}[P(p)\to V(p^{\prime},\epsilon^{\prime})\gamma(q,\epsilon)]=\epsilon_{\mu}^*\epsilon_{\nu}^{*\prime}[iA_{PV}(p^{\mu}q^{\nu}-g^{\mu\nu}p\cdot q)+A_{PC}\epsilon^{\mu\nu\alpha\beta}p_{\alpha}q_{\beta}]~
\end{equation}
with 
\begin{eqnarray}
\label{4.9}
A_{PV}&=&-{G_F\over \sqrt{2}}~ e_0\biggl(V_{cs}^*V_{ud}\biggl[{c_7^{eff}\over 4\pi^2}~m_c~F_1(0)+2a_2^c~C_{\!V\!M\!D}~{A_1(0)\over 
m_{B_c}-m_{B_u^*}}\biggr]\biggr)~,\nonumber\\
A_{PC}&=&-{G_F\over \sqrt{2}}~ e_0\biggl(V_{cs}^*V_{ud}\biggl[{c_7^{eff}\over 4\pi^2}~m_c~F_1(0)+2a_2^c~C_{\!V\!M\!D}{V(0)\over 
m_{B_c}+m_{B_u^*}}\biggr]\nonumber\\
&+&V_{cb}^*V_{ub}a_1^b\biggl[{\mu_{B_c}g_{B_c^*}g_{B_u^*}\over 
m_{B_c^*}^2-m_{B_u^*}^2}+{\mu_{B_u}m_{B_c}^2f_{B_c}f_{B_u}\over 
m_{B_c}^2-m_{B_u}^2}\biggr]
\biggr)~
\end{eqnarray}
and the corresponding decay rate is given by
$$
\Gamma=\frac{1}{4\pi}\biggl(\frac{m_{B_c}^2-m_{B_u^*}^2}{2m_{B_c}}\biggr)^3(|A_{PC}|^2+|A_{PV}|^2)~.
$$

\section{The model}

The strong dynamics within the $B_c$ and $B_u^*$ mesons  has to be incorporated by relaying on one of the available approaches listed in the introduction. 
The heavy quark symmetry is based on the static approximation for the heavy quark in a meson with a single heavy quark. 
The $B_c$ meson is composed of two heavy quarks  and their masses do not permit both valence quarks to be at rest at the same time, so the heavy quark symmetry does not apply to this case in this sense\footnote{The form factors $F_1$, $V$ and $A_1$  
that parameterize the short and long distance amplitudes  cannot be safely related using the 
Isgur-Wise relations \cite{IW1} frequently used for the mesons containing one heavy quark.}. In \cite{FPS3} we have decided to use nonrelativistic constituent Isgur-Wise-Scora-Grinstein (ISGW) quark model \cite{ISGW} which is considered  to be reliable for the states composed of two heavy quarks and is therefore suitable for treating the $B_c$ meson; in addition the velocity of $B_u^*$ in the rest 
frame of $B_c$ is to a fair approximation nonrelativistic. 

The ISGW constituent quark model \cite{ISGW,ISGW2} describes the heavy meson in terms of two constituent quarks with mass $M$ that  move under the influence of the effective 
potential $V(r)=-4\alpha_s/(3r)+c+br$, $c=-0.81$ GeV, $b=0.18$ GeV$^2$  
\cite{ISGW2}. Instead of the accurate solutions of the Schr\"odinger equation, the variational solutions 
$$\psi(\vec r)=\pi^{-{3\over 4}}\beta^{{3\over 2}}e^{-{\beta^2r^2\over 
2}}~~~{\rm or}~~~ \psi(\vec k)=\pi^{-{3\over 4}}\beta^{-{3\over 2}}e^{-{k^2\over 
2\beta^2}}~~~~~~{\rm for~~ S~~ state}$$
are used and $\beta$ is employed as the variational parameter. The meson 
state, composed of the constituent quarks $q_1$ and $\bar q_2$, is given by   
\begin{equation}
\label{4.meson}
|M(p)\rangle\!=\!\!\!\sum_{C,s1,s2}\!\!{1\over \sqrt{3}}\sqrt{{2E\over (2\pi)^3}}\int\!\! d\vec 
k\psi(\vec k)\sqrt{{M_1\over E_1}}\sqrt{{M_2\over 
E_2}}f_{s2,s1}\delta(p\!-\!p_1\!-\!p_2)b_1^\dagger(\vec p_1,s_1,C)d_2^\dagger(\vec 
p_2,s_2,\bar C)|0\rangle~,
\end{equation} 
where $\vec k$ is the momentum of the constituents in the meson rest frame, $C$ 
denotes the colour, while 
$f_{s2,s1}=(\bar\uparrow\downarrow+\bar\downarrow\uparrow)/\sqrt{2}$ for 
pseudoscalar and 
$f_{s2,s1}=(\bar\uparrow\downarrow-\bar\downarrow\uparrow)/\sqrt{2},\bar\uparrow
\uparrow,\bar\downarrow\downarrow$ for vector mesons\footnote{The spinors are normalized as in \cite{itzykson}.}. In the non-relativistic limit the following expressions are obtained \cite{FPS3}\footnote{Similar, but semi-relativistic, quark model has been presented in \cite{Soares} and  can serve as a suitable cross check for the derived form factors.} 
\begin{eqnarray}
\label{4.9a}
V(q^2)&=&{m_{B_c}+m_{B_u^*}\over2}\biggl[{1\over 
M_u}-{M_b(M_c-M_u)\beta_{B_c}^2\over 
M_cM_um_{B_u^*}(\beta_{B_c}^2+\beta_{B_u^*}^2)}\biggr]F_3(q^2)~,\nonumber\\
A_1(q^2)&=&={2m_{B_c}\over m_{B_c}+m_{B_u^*}}F_3(q^2)~,\nonumber\\
F_1(q^2)&=&2\biggl[1+(m_{B_c}-m_{B_u^*})\biggl({1\over 
2M_u}-{M_b(M_c+M_u)\beta_{B_c}^2\over 
2M_cM_um_{B_u^*}(\beta_{B_c}^2+\beta_{B_u^*}^2)}\biggr)\biggr]F_3(q^2)~,\nonumber\\
\mu_{B_c}&=&\sqrt{{m_{B_c^*}\over 
m_{B_c}}}\biggl({2\beta_{B_c}\beta_{B_c^*}\over 
\beta_{B_c}^2+\beta_{B_c^*}^2}\biggr)^{3/ 2}\biggl[{2\over 3M_c}-{1\over 
3M_b}\biggr]~,\nonumber\\
f_{B_c}&=&{2\sqrt{3}\beta_{B_c}^{3/2}\over \pi^{3/4}\sqrt{m_{B_c}}}~,\nonumber\\
g_{B_c^*}&=&m_{B_c^*}{2\sqrt{3}\beta_{B_c^*}^{3/ 2}\over \pi^{3/ 4}\sqrt{m_{B_c^*}}}
\end{eqnarray}
and analogously for $\mu_{B_u}$, $f_{B_u}$ and $g_{B_u^*}$. Here 
$$F_3(q^2)=\sqrt{m_{B_u^*}\over m_{B_c}}\biggl({2\beta_{B_c}\beta_{B_u^*}\over 
\beta_{B_c}^2+\beta_{B_u^*}^2}\biggr)^{3/2}\exp\biggl(-{M_b^2\over 
2m_{B_c}m_{B_u^*}} {[(m_{B_c}-m_{B_u^*})^2-q^2]\over 
\kappa^2(\beta_{B_c}^2+\beta_{B_u^*}^2)}\biggr)~.$$
The results for $V(q^2)$ and $A_1(q^2)$ 
reproduce the results of \cite{ISGW}, while $F_1(q^2)$ represents, 
to my knowledge, the new result within the ISGW model. 
The parameter $\kappa$ is introduced in order to allow the computed shapes of the form factors to be modified by a common factor \cite{ISGW}. The value of $\kappa$ is taken be unique  for all the meson states \cite{ISGW} and is equal to $\kappa=1$ for the case of the meson wave function given in (\ref{4.meson}). In \cite{ISGW} the multiplicative constant $\kappa=0.7$ is chosen in order to give the calculated electromagnetic form factor for the pion in the better agreement with the experimental data. 

Using $\kappa=0.7$, the constituent quark masses 
$M_u=0.33$ GeV, $M_c=1.82$ GeV, $M_b=5.2$ GeV \cite{ISGW2} and the parameters $\beta$ 
\cite{ISGW2} and meson masses given in Table \ref{4.tab.beta},   I get \cite{FPS3}
\begin{align}
\label{4.10}
f_{B_u}&=0.18~{\rm GeV}~,~~g_{B_u^*}=0.86~{\rm GeV}^2~,~~\mu_{B_u}=1.81~{\rm GeV}^{-1}~,\\
f_{B_c}&=0.51~{\rm GeV}~,~~g_{B_c^*}=2.41~{\rm GeV}^2~,~~\mu_{B_c}=0.28~{\rm GeV}^{-1}~,\nonumber
\end{align}
while the values of the $B_c\to B_u^*$ form factors at $q^2=0$ are \cite{FPS3}
\begin{equation}
\label{4.form}
A_1(0)=0.24~,\qquad V(0)=1.3~\quad{\rm and}\quad F_1(0)=0.48~.
\end{equation}

\begin{table}[h]
\begin{center}
\begin{tabular}{|c||c|c|c|c|}
\hline
& $B_c$&$B_c^*$&$B_u$&$B_u^*$\\
\hline
$m[{\rm GeV}]$&6.40 \cite{CDF}&6.42 \cite{ISGW2}&5.28 \cite{PDG}&5.325 \cite{PDG}\\
\hline
$\beta[{\rm GeV}]$&0.92&0.75&0.43&0.40\\
\hline
\end{tabular}
\caption{Parameters $\beta$ taken from \cite{ISGW2}  and masses of 
pseudoscalar and vector mesons.  }
\label{4.tab.beta}
\end{center}
\end{table}

\section{The results}

\subsubsection{The standard model predictions}

The amplitude for the $B_c\to B_u^*\gamma$ decay in the standard model is given by the expression (\ref{4.amp}) with $c_7^{eff}=-(1.5+4.4i)[1\pm 0.2]10^{-3}$ (\ref{c7}) \cite{GHMW}.
 I use  the central value of the current quark mass 
$m_c=1.25$ GeV from \cite{PDG} and $V_{cb}=0.04$, $V_{ub}=0.0035$ together with the results of the ISGW  model. The short distance, the long distance penguin 
and the long distance weak annihilation parts  of amplitudes $A_{PC}$ and $A_{PV}$ (\ref{4.9}) needed to 
compute the amplitude (\ref{4.amp})  are given in 
Table \ref{4.tab.amp}. The error bars in the table arise from the parameter 
$C_{V\!M\!D}$ (\ref{cvmd}), which has the largest uncertainty. In Table \ref{4.2tab} I present the total branching ratio ($Br^{tot}$)
and separately also the short distance ($Br^{SD}$) and long distance ($Br^{LD}$) parts of the branching ratio for $B_c\to 
B_u^*\gamma$ decay. Here, $\tau(B_c)=0.46{+0.18\atop 
-0.16}\pm0.03$ ps is taken as measured by CDF Collaboration recently \cite{CDF}. Note 
that short and long distance contributions  give branching ratios of comparable size $\sim 
10^{-8}$, which in principle allows to use the $B_c\to B_u^*\gamma$ decay for probing the $c\to u\gamma$ transition in the standard model.

\begin{table}[h]
\begin{center}
\begin{tabular}{|c||c|c|c||c|c|c|}
\hline
& $A^{SD}(PV)$ & $A^{LD}_{peng.}(PV)$ & $A^{LD}_{annih.}(PV)$ & $A^{SD}(PC)$ & 
$A^{LD}_{peng.}(PC)$ & $A^{LD}_{annih.}(PC)$ \\
\hline
$B_c\to B_u^*\gamma$&$5.7+17~i$&$-14\pm 14$&$0$&$5.7+17~i$&$-7.3\pm 7.3$&$-21$\\
\hline 
\end{tabular}
\caption{ The standard model predictions for the amplitudes corresponding to different contributions in $B_c\to B_u^*\gamma$ decay. The parity violating  and conserving amplitudes $A_{PV,PC}$, defined in (\ref{4.amp}), for short distance ($A^{SD}$), long distance penguin ($A^{LD}_{peng.}$) and long distance weak annihilation ($A^{LD}_{annih.}$) contributions  are given in units of $10^{-11}$ GeV$^{-1}$ . 
The error-bars are due to the uncertainty in 
$C_{V\!M\!D}=(1.2\pm 1.2)~10^{-3}$ GeV$^2$ (\ref{cvmd}).}
\label{4.tab.amp}
\end{center}
\end{table}

\begin{table}[h]
\begin{center}
\begin{tabular}{|c||c|c|c|}
\hline
& $Br^{SD}$ & $Br^{LD}$ & $Br^{tot}$ \\
\hline
$B_c\to B_u^*\gamma$& $4.7\cdot 10^{-9}$ & $(7.5 {+7.7\atop -4.3})\cdot 10^{-9}$ 
& $(8.5 {+5.8 \atop -2.5})\cdot 10^{-9}$\\
\hline
\end{tabular}
\caption{ The standard model predictions for the $B_c\to B_u^*\gamma$ branching ratios  as given by the ISGW model: the short distance part $Br^{SD}$, the long distance part  $Br^{LD}$ and the total branching ratio $Br^{tot}$.  The 
error-bars  are due to the uncertainty in 
$C_{V\!M\!D}=(1.2\pm 1.2)~10^{-3}$ GeV$^2$ (\ref{cvmd}).}
\label{4.2tab}
\end{center}
\end{table} 

\vspace{0.1cm}

I conclude this section by a  qualitative comparison of  the standard model predictions based on the ISGW model with the predictions based on some other models. 

The form factor $F_1(0)$ entering the short distance contribution to $B_c\to B_u^*\gamma$ has been reexamined using the QCD sum rules approach in \cite{AS} with the result $F_1(0)=0.9\pm 0.1$. This result is almost two times bigger than ISGW result $F_1(0)=0.48$ (\ref{4.form}) and renders the short distance part of the $B_c\to B_u^*\gamma$ branching ratio of the order of $2\cdot 10^{-8}$. 

I have qualitatively estimated  the long distance weak annihilation contribution using the QCD sum rule and vector meson dominance approaches. The QCD sum rules analogue of the diagram in Fig. \ref{fig19}b (only the part of the diagram left of the  box) was considered in the study of the decays $B_c^*\to B_c\gamma$ \cite{AIP}, $B_c\to \rho^+\gamma$ and $B_c\to K^{*+}\gamma$ \cite{AS1}. The corresponding results are in a reasonable agreement the results of the ISGW model, which reflects the fact that the photon interaction with the $B_c$ meson composed of two heavy quarks is well understood.  The amplitude for the diagram in Fig. \ref{fig19}a depends on the magnetic moment  of $B_u$ and the ISGW prediction on $\mu_{B_u}$ (\ref{4.10}) is about two times bigger than  vector meson dominance prediction presented in Table 8 of \cite{casalbuoni}. The  overestimation of the $B_u$ magnetic moment in the ISGW model is connected with the general failure of the constituent quark models in reproducing  the magnetic moments connected with the light quark.  
In the vector meson dominance approach, the light quark emits a photon through the vector meson exchange and seems to give a more accurate phenomenologically description. 
A possible overestimation of the long distance weak annihilation contribution based on the ISGW model indicates that the long distance background to $B_c\to B_u^*\gamma$ decay may be smaller than indicated in Table \ref{4.2tab}.

\subsubsection{The signatures of the physics beyond the standard model}

The standard model prediction for  $B_c\to B_u^*\gamma$ branching ratio is of the order of $10^{-8}$ and the 
experimental detection of this decay 
at the branching ratio well above $10^{-8}$ would clearly indicate a signal of  
physics beyond the standard model. This decay is especially sensitive to the scenarios that could significantly enhance the $c\to u\gamma$ rate compared to the standard model predictions. The long distance contributions are not expected to be significantly effected by  possible scenarios of new physics. 

The supersymmetric models were studied in Section 2.2.2. It was argued that the minimal supersymmetric model with a plausible mechanism of supersymmetry breaking renders the $c\to u\gamma$ rate comparable to that in the standard model. Given the uncertainties in the long distance contribution in the $B_c\to B_u^*\gamma$ decay, it would be difficult to distinguish the effect of the minimal supersymmetric model from the standard model contribution in this decay. The $Br(c\to u\gamma)$ can be enhanced up to $1.2\cdot 10^{-5}$ (\ref{brsusy}) in some versions of the nonminimal supersymmetric model (by adding a pair of additional Higgs doublets to the minimal model for example \cite{BGM}). This corresponds to $c_7^{eff}\le 0.14$ (\ref{c7susy}) and the short distance part of $Br(B_c\to B_u^*\gamma)$ could be as large as $4\cdot 10^{-6}$, which would be clearly  observable when the experiments reach the corresponding sensitivity. 

The effect of the fourth generation on the $c\to u\gamma$ rate was discussed in Section 2.2.4. The heavy $\hat b$ quark could enhance the coefficient $c_7^{eff}$ up to $2.8\cdot 10^{-2}$ and the short distance part of the $Br(B_c\to B_u^*\gamma)$ decay could be as large as $3\cdot 10^{-7}$. This could be distinguished from the standard model contribution, which is an order of magnitude smaller. 

The effects of the extended Higgs sector and that of the left-right symmetry were studied in Sections 2.2.1 and 2.2.4, respectively, and have been shown to give the negligible contributions to the $c\to u\gamma$ decay.      

\subsubsection{Experimental status}

Needless to say that the observation of the unique $B_c\to B_u^*\gamma$ channel for probing the flavour changing neutral transition $c\to u\gamma$  is experimentally very challenging. The $B_c$ meson has recently been observed by the CDF collaboration \cite{CDF} and only a handful of $B_c$ mesons have been detected by now. Apart from the aspect described in this work, the $B_c$ meson is an interesting state as the charm quark and beauty quark weak decay channels are of comparable importance in  $B_c$ meson decays. The standard model predictions for various channels have been extensively studied in the literature and are expected to be searched for at the Tevatron, B-factories and Large Hadron Collider. The $B_c$ production at various experimental facilities have been studied in  \cite{cheung}. The Tevatron and B factories will not produce enough of $B_c$ to make the $B_c\to B_u^*\gamma$ decay observable. The Large Hadron Collider is expected to produce $2.1\cdot 10^{8}$ $B_c$ mesons with $p_T(B_c)>20$ GeV at the integrated luminosity $100$ fb$^{-1}$. By searching for the decay channel $B_c\to B_u^*\gamma$  at The Large Hadron Collider one could probe  any enhancement of the $c\to u\gamma$ rate arising from new physics.

\chapter{Weak decays of charmed mesons}

The approximate symmetries of quantum chromodynamics in the infinite quark mass limit for the heavy quarks ($Q=c,b$, $~m_Q\!\to\! \infty$) and in the chiral limit for the light quarks ($q=u,d,s$, $~m_q\!\to \!0$) can be used together to build up an effective chiral Lagrangian for heavy and light mesons. This Lagrangian describes the strong interaction among the effective meson fields and their couplings to electromagnetic and weak currents together with the relevant symmetry breaking terms. In Section 1 the heavy quark and the chiral symmetries are introduced and  combined in the heavy chiral Lagrangian \cite{heavy.chiral}. The light vector mesons are incorporated using the hidden symmetry approach \cite{hidden}. A specific model, the so called {\it hybrid model}, for the weak currents, the shapes of the form factors and the $SU(3)$ flavour breaking is proposed \cite{FPS1,FPS2,BFO1,BFOP}. In Section 2 the relevant free parameters of the effective theory are determined and the model is applied to the semileptonic decays \cite{BFO1}. In Section 3 the two-body nonleptonic exclusive charm meson decays are studied \cite{BFOP}. The understanding of nonleptonic decays is necessary in order to develop a model for the long distance contributions to  the  rare charm meson decays. In the last two sections the hybrid model is adapted  for rare charm meson decays, which are interesting for probing the flavour changing neutral currents.   The $D\to V\gamma$ \cite{FPS1,FS} and $D\to Vl^+l^-$ \cite{FPS2} decays  are studied in Section 4, while  $D\to Pl^+l^-$ decays are studied in Section 5 ($P$ and $V$ denote light pseudoscalar and vector mesons and $l$ denotes a charged lepton).

\section{Heavy meson chiral Lagrangian\\
 for light and heavy pseudoscalar and vector mesons}

\subsection{Heavy quark symmetry and heavy mesons}

The heavy quark symmetry was introduced in \cite{HQET1,HQET2} and extensively studied afterwards. A good review with many applications and  relevant references is given in \cite{neubert}.

The heavy quark symmetry  applies to a heavy meson, moving with velocity $v$, composed of a heavy quark $Q$ and the light degrees of freedom.  The  heavy quark inside the meson moves essentially with the velocity  $v^{\mu}$ and is almost on-shell. Its momentum can be decomposed as $p_Q=m_Qv^\mu+k^\mu$, where  $k$ is much smaller than $m_Qv$. Interactions of the heavy quark with the light degrees of freedom change this residual momentum by an amount of the order of $\Delta k\!\sim\!\Lambda_{Q\!C\!D}$ and the corresponding changes in the heavy quark velocity vanish as $\Lambda_{Q\!C\!D}/m_Q\!\to \! 0$. For this physical system  it is suitable to project out the ``large'' $h_v$ and ``small'' $H_v$ components of the heavy quark field $Q$ \cite{neubert}
\begin{equation}
\label{5.2}
h_v(x)\equiv\exp(im_Qv\cdot x)\tfrac{1}{2}(1+\!\!\not{\! v})Q(x)~,\qquad  H_v(x)\equiv\exp(im_Qv\cdot x)\tfrac{1}{2}(1-\!\!\not{\! v})Q(x)~,
\end{equation}
so that the ``small'' component $H_v$ is equal to zero if the heavy quark moves exactly with the meson's velocity $v$ and 
\begin{equation}
\label{5.3}
Q(x)=\exp(-im_Qv\cdot x)[h_v(x)+H_v(x)]~.
\end{equation}
In terms of these fields, the heavy quark part of the Lagrangian is given by \cite{neubert}
\begin{equation}
\label{5.1}
{\cal L}_Q=Q(i\!\!\not{\!D}-m_Q)Q=\bar h_viv\cdot Dh_v-\bar H_v(iv\cdot D+2m_Q)H_v+\bar h_vi\!\!\not{\! D_\bot}H_v+\bar H_vi\!\!\not{\! D_\bot}h_v
\end{equation}
with $D^\mu=\partial^\mu-\tfrac{1}{2}ig_s\lambda_aG_\mu^a$ and $D_\bot^\mu=D^\mu-v^\mu v\cdot D$. Since the main $x$-dependence of the field $h_v(x)$  (\ref{5.2}) has been factored out, the field  $h_v$ corresponds to the massless degrees of freedom (\ref{5.1}), whereas $H_v$ corresponds to fluctuations with twice the heavy quark mass (\ref{5.1}). The heavy degrees of freedom, represented by $H_v$, can be eliminated  on the classical level by inserting the expression (\ref{5.3}) to the Dirac equation of motion $(i\!\!\not{\! D}-m_Q)Q=0$ 
\begin{equation}
\label{5.1a}
i\!\!\not{\! D}h_v+(i\!\!\not{\! D}-2m_Q)H_v =0~.
\end{equation} 
Multiplying this by $P_\pm$, two equations are obtained 
$$ -iv\cdot Dh_v=i\!\!\not{\! D}_\bot H_v~,\qquad (iv\cdot D+2m_Q)H_v=i\!\!\not{\!D}_\bot h_v~.$$
The second can be solved to give the small component in terms of the large component
$$H_v=(iv\cdot D+2m_Q-i\epsilon)^{-1}i\!\!\not{\!D}_\bot h_v~.$$
The small component, which is of the order of $1/m_Q$ is inserted back to the Dirac equation (\ref{5.1a})  and the equation of motion for $h_v$ is obtained. By the variational principle this equation of motion follows from the Lagrangian of the heavy quark effective theory (HQET) 
\begin{align}
\label{5.4}
{\cal L}_{eff}&=\bar h_viv\cdot Dh_v+\bar H_vi\!\!\not{\! D}_\bot(iv\cdot D+2m_Q-i\epsilon)^{-1}i\!\!\not{\! D}_\bot h_v\nonumber\\
&=\bar h_v iv\cdot Dh_v+\frac{1}{2m_Q}h_v(iD_\bot)^2h_v+\frac{g}{4m_Q}\bar h_v\sigma_{\alpha\beta}G^{\alpha\beta}h_v+{\cal O}(1/m_Q^2)~.
\end{align}
The second expression is  derived from the first one by expanding in $D/m_Q$, which converges since the phase $\exp(im_Qv\cdot x)$ has been factored out in (\ref{5.2}). Derivatives acting on $h_v$ produce powers of the residual momenta $k$, which is much smaller than $m_Q$. 

In the $m_Q\to \infty$ limit only the first term  ${\cal L}_{eff}^0= \bar h_viv\cdot Dh_v $  in the Lagrangian (\ref{5.4}) remains.  The strong interaction with the light degrees of freedom do not alter the velocity of the heavy quark in this limit. Since there are no Dirac matrices in ${\cal L}_{eff}^0$, the interactions with gluons do not have effect on the heavy quark spin and the Lagrangian ${\cal L}_{eff}^0$ is invariant under the $SU(2)$ spin transformations. The Lagrangian ${\cal L}_{eff}^0$ does not depend on the mass of the heavy quark $m_Q$.  The combined spin-flavour symmetry under the $SU(2N_h)$ transformations at the  leading $(\Delta k/m_Q)^0$ order  of the theory with $N_f$ heavy flavours is called {\bf the heavy quark symmetry}. This symmetry is lost in at the  order $\Delta k/m_Q$. The most successful application of the heavy quark symmetry is to $B\to D^*l^-\bar \nu_l$ and $B\to Dl^-\bar \nu_l$ decays, where it relates the form factors in the kinematic region where the $b$ and $c$ quark have equal velocities \cite{HQET1,neubert,HQET2}. In this work, the heavy quark symmetry will be employed to study the decays of the heavy mesons to the light mesons, where its implications are not so powerful.

\vspace{0.1cm}

Now we need to choose the suitable fields to build the effective field theory for the heavy mesons, which is based on the heavy quark symmetry. We consider the ground state pseudoscalar and  vector heavy mesons containing a given heavy quark $Q$. The light degrees of freedom carry the spin $1/2$. The pseudoscalar  and vector meson differ in the direction of the heavy quark spin. The rotations of the heavy quark spin present the symmetry of the strong Lagrangian in the limit $m_Q\to \infty$  and pseudoscalar and  vector mesons have the same mass at this order. They are both  dynamical at the same energy scale and they have to be introduced in an effective theory together. The terms containing  the pseudoscalar and vector fields  should be related by the heavy quark symmetry. It is convenient to gather the pseudoscalar field  and a vector meson field in a common filed $H(x)$ in such a way that the spin symmetry is rendered automatically. For this purpose we study the Lorentz structure of an object composed of a heavy quark with spin $|\vec s_Q|=1/2$ and the light degrees of freedom with spin $|\vec s_l|=1/2$. An important implication of the heavy quark spin symmetry is that the spin of the heavy quark $\vec s_Q$ is conserved during the free propagation of the meson. Since the spin of the meson is conserved, the spin of the light degrees of freedom $\vec s_l$ must be conserved as well.  The Lorentz structure of this composed object can be represented by the bi-spinor $u_Q(v)\bar v_l(v)$. Under the heavy quark spin rotation the bi-spinor transforms as
$$ u_Q(v)\bar v_l(v)\to \bigl(S u_Q(v)\bigr)v_l(v)~,\qquad S\in SU(2)~.$$
The bi-spinor ``$0^-$'' for a pseudoscalar meson in its  rest frame $v=(1,\vec 0)$  is given by\footnote{The Dirac representation of the $\gamma^\mu$ matrices is used as in \cite{neubert}.}
$$\tfrac{1}{\sqrt{2}}[u(\Uparrow) \bar v(\downarrow)\!+\!u(\Downarrow)\bar v(\uparrow)]\!=\!-\tfrac{1}{\sqrt{2}}\begin{pmatrix}0&I\\
                                  0&0\end{pmatrix}\!=\!-\tfrac{1}{\sqrt{2}}\tfrac{1}{2}(1+\gamma^0)\gamma_5~,\quad {\rm so} \quad 0^-(\vec v=0)\sim-\tfrac{1}{\sqrt{2}}\tfrac{1}{2}(1+\gamma^0)\gamma_5~,$$
where $\Uparrow$ and $\uparrow$ denote the third spin components of the heavy quark and the light degrees of freedom, respectively. The bi-spinor ``$1^-$'' for the vector meson is
$$u( \Uparrow)\bar v(\uparrow)\!=\!-\tfrac{1}{2}\begin{pmatrix}0&\sigma_1+i\sigma_2\\
                                         0&0\end{pmatrix}~,\quad u(\Downarrow)\bar v(\downarrow)\!=\!-\tfrac{1}{2}\begin{pmatrix}0&\sigma_1-i\sigma_2\\
                                         0&0\end{pmatrix}~,$$
$$\tfrac{1}{\sqrt{2}}[u(\Uparrow)\bar v(\downarrow)\!-\!u(\Downarrow)\bar v(\uparrow)]\!=\!-\tfrac{1}{\sqrt{2}}\begin{pmatrix}0&\sigma_3\\ 
                                              0&0\end{pmatrix}~,\quad\qquad
{\rm so}\quad 1^-(\vec v=0,\epsilon)\sim \tfrac{1}{\sqrt{2}}\tfrac{1}{2}(1+\gamma^0)\!\!\not{\! \epsilon}$$
with $\epsilon_\pm^\mu=\tfrac{1}{\sqrt{2}}(0,1,\pm i,0)$ and $\epsilon_3=(0,0,0,1)$. For a general velocity $v$ this is easily generalized to 
$$0^-(v)\sim-\tfrac{1}{\sqrt{2}}\tfrac{1}{2}(1+\!\!\not{\! v})\gamma_5~,\qquad  1^-(v,\epsilon)\sim \tfrac{1}{\sqrt{2}}\tfrac{1}{2}(1+\!\!\not{\! v})\!\!\not{\! \epsilon}~.$$
We gather the pseudoscalar and vector fields together with their bi-spinor Lorentz structure in the filed $H(x)$. In the case of charm mesons, which are of interest in this chapter, the basic field of the effective Lagrangian is \cite{neubert,heavy.chiral}
\begin{equation}
\label{ha}
H_a(x)=\tfrac{1}{2}(1+\!\!\not{\! v})[-D^v_a(x)~\gamma_5+D^{*v}_{a\mu}(x)~\gamma^\mu]~
\end{equation}
with $v^\mu D^{*v}_\mu=0$. The fields $D_a^v$ and $D^{*v}_{a\mu}$ annihilate the pseudoscalar and vector mesons with flavour $c\bar q_a$, respectively, and $q_1=u$, $q_2=d$, $q_3=s$.  
The fields $D^v(x)$ and $D^{*v}(x)$ are defined so that the free heavy meson Lagrangian, given in (\ref{5.5}) bellow, is independent on the heavy meson mass in the limit $m_H\to \infty$
\begin{equation}
\label{5.105}
D^v(x)=\sqrt{m_H}~\exp(im_Hv\cdot x)D(x)~,\qquad D^{*v}_{\mu}(x)=\sqrt{m_H}~\exp(im_Hv\cdot x)D^{*}_\mu(x)~.
\end{equation}
The fields $D^v$ and $D^{*v}$ have mass dimension $3/2$. The main $x$-dependence in the field $H(x)$ is factored out and the terms with derivatives on $H(x)$ are suppressed by $\Delta k/m_H$, where $k=p_H-m_Hv$ denotes the residual momentum. In the lowest order in $\Delta k/m_H$, 
the free heavy meson Lagrangian is given by the expression invariant under the Lorentz,  heavy quark spin and flavour symmetry and with the lowest number of the derivatives 
\begin{equation}
\label{5.5}
{\cal L}_H=iTr[H_av_\mu\partial^\mu\bar H_a]~,
\end{equation}
where the trace is taken in the $4\times 4$ space of Dirac indices and the heavy meson creation operators are contained in
\begin{equation}
\bar H_a=\gamma^0H^\dagger_a\gamma^0=[D_a^{v\dagger}(x)~\gamma_5+D_{a\mu}^{*v\dagger}(x)~\gamma^\mu]\tfrac{1}{2}(1+\!\!\not{\! v})~.
\end{equation}
The Lagrangian (\ref{5.5}) is invariant under the heavy quark spin transformation 
$$H_a\to SH_a\quad{\rm and}\quad\bar H_a\to \bar H_aS^\dagger~,\qquad S\in SU(2)$$
since $SS^\dagger=I$ and $Tr[AB]=Tr[BA]$. It is also invariant under the heavy quark flavour transformation $c\to b$ as it does not depend on the heavy meson mass. When we express the Lagrangian ${\cal L}_H$ (\ref{5.5}) in terms of the fields $D^v$ and $D^{*v}$, we get  nothing but the free heavy meson Lagrangian in the lowest order in $\Delta k/m_H$ \cite{SS}
\begin{align*} 
{\cal L}_H&=\partial^\mu D\partial_\mu D^{\dagger}-m_H^2DD^{\dagger}-\tfrac{1}{2}(\partial_\mu D_{\nu}^*-\partial_\nu D_{\mu}^*)(\partial^\mu  D^{*\dagger\nu}-\partial^\nu  D^{*\dagger\mu})-m_H^2D^*_\mu D^{*\dagger\mu}\\
&=-2iD^vv_\mu\partial^\mu D^{v\dagger}+2iD^{v*}
_\nu v_\mu\partial^\mu D^{v*\dagger\nu}+{\cal O}(1/m_H)~.
\end{align*}
This result is expected even if we do not work with the field $H(x)$. The full power of the formalism with  the effective field $H(x)$ becomes apparent when the interactions of the heavy mesons are studied.

\subsection{Chiral symmetry and light pseudoscalar mesons}

In the limit of massless light quarks $u$, $d$ and $s$, the quantum chromodynamics is described in terms of the Lagrangian 
$${\cal L}_{QCD}=\bar qi\!\!\not{\! D}q-\tfrac{1}{4}G_{\mu\nu}^aG^{\mu\nu a}+{\cal L}^{{heavy \atop quarks}}=\bar{q_L}i\!\!\not{\! D}q_L+\bar{q_R}i\!\!\not{\! D}q_R-\tfrac{1}{4}G_{\mu\nu}^aG^{\mu\nu a}+{\cal L}^{{heavy \atop quarks}}$$
with
$$q=(u,d,s)^T~,\quad q_L=\tfrac{1}{2}(1-\gamma_5)q~,\quad  q_R=\tfrac{1}{2}(1+\gamma_5)q$$
and is invariant under the global transformation
$$SU(3)_L\times SU(3)_R\times U(1)_V~.$$  
 The absence of the parity doublets in the hadron spectrum indicates that the chiral symmetry $G=SU(3)_L\times SU(3)_R$ is spontaneously broken to a vector subgroup $H=SU(3)_V$. The eight pseudoscalar particles are identified with the Goldstone bosons corresponding to the eight broken generators $A_{i}$ of the coset space $G/H$. The general element of the coset space $G/H$ can be expressed in terms of the generators $A_i$ and the parameters  $a_i$ as $\exp(ia_iA_i)$ or equivalently as 
\begin{equation}
\label{xi}
\xi(x)=e^{i\Pi(x)/f}~ \qquad{\rm with}\qquad \Pi=\begin{pmatrix}\tfrac{\pi^0}{\sqrt{2}}+\tfrac{\eta_8}{\sqrt{6}}&\pi^+&K^+\\
\pi^-&-\tfrac{\pi^0}{\sqrt{2}}+\tfrac{\eta_8}{\sqrt{6}}&K^0\\ K^-&\bar K^0&-\tfrac{2\eta_8}{\sqrt{6}}~\end{pmatrix}~.
\end{equation}
 The value of parameter $f$ will be given when the weak interactions are discussed. The element of coset space (\ref{xi})  transforms under the chiral symmetry  as \cite{CWZ} \footnote{A general group element $g\in G$ can be uniquely decomposed into the product
$g=e^{ia_iA_i}e^{iv_iV_i}$ (${i=1,..,8}$),
where $a_i$ and $v_i$ are the parameters and $A_i$ and $V_i$ are the generators of $G/H$ and $H$, respectively \cite{CWZ}. An element $g_L\in G$ can be decomposed to 
$$g_Le^{ia_iA_i}=e^{ia_i^\prime A_i}e^{iv_i^\prime V_i}\quad {\rm or}\quad e^{ia_i^\prime A_i}=g_Le^{ia_iA_i}e^{-iv_i^\prime V_i}.$$
Under parity $g_L\!\to\! g_R~$, $A_i\!\to\! -A_i$ and $V_i\!\to\! V_i$ so $e^{-ia_i^\prime A_i}=g_Re^{-ia_iA_i}e^{-iv_i^\prime V_i}$ or \cite{CWZ}
$$e^{ia_i^\prime A_i}=g_Le^{ia_iA_i}e^{-iv_i^\prime V_i}=e^{iv_i^\prime V_i}e^{ia_iA_i}g_R^+~,$$
which is equivalent to the chiral transformation (\ref{5.6}) via the correspondence $\xi=e^{ia_iA_i}\in G/H$ and $U=e^{iv_i^{\prime} V_i}\in H$.}
\begin{equation}
\label{5.6}
\xi(x)\to g_L\xi(x)U^\dagger(x)=U(x)\xi(x)g_R^+~,\qquad U(x)\in SU(3)_V, ~~g_{L,R}\in SU(3)_{L,R}~.
\end{equation}
The space time dependent matrix $U(x)\in SU(3)$ is defined by the equation  $g_L\xi(x)U^\dagger(x)=U(x)\xi(x)g_R^+$ above and is a complicated non-linear function of $g_L$, $g_R$ and the coset field $\xi(x)$. For the case of the chiral transformation in the  subgroup $SU(3)_V$, $U\!=\!g_L\!=\!g_R$. The chiral transformation on $\xi$ (\ref{5.6}) implies that the field $\Sigma=\xi^2$ transforms linearly 
\begin{equation}
\label{5.9}
\Sigma(x)\to g_L\Sigma(x)g_R^\dagger~, \qquad\qquad \Sigma(x)=\xi(x)^2~.
\end{equation}

The chiral perturbation theory \cite{chiral} exploits the spontaneous breaking of the chiral symmetry in order to describe  the interactions of the low energy pseudoscalar mesons. It is  an effective field theory and employs the most general Lagrangian, expressed in terms of mesonic fields contained in $\Sigma(x)$, invariant under the symmetries of QCD: Lorentz and chiral symmetry, parity, charge conjugation and time reversal. The infinite number of terms consistent with this condition can be grouped according to the number of the derivatives. The terms with derivatives on the pseudoscalar fields are suppressed by the powers of $E/\Lambda_\chi$ when  the low energy pseudoscalars with the energy $E$ are involved. The chiral-symmetry breaking scale $\Lambda_\chi$ is evaluated to be of the order of $1$ GeV \cite{chiral}.  The interactions of light pseudoscalars in the lowest order ${\cal O}(E^2)$  are given by \cite{chiral} \footnote{There are no nontrivial terms of the order of ${\cal O}(E^0)$ and ${\cal O}(E)$  invariant under the chiral symmetry.}
\begin{equation}
\label{chiral}
{\cal L}=\tfrac{1}{8}f^2tr[\partial^\mu\Sigma\partial_\mu\Sigma^\dagger]+{\cal O}(E^4)~,
\end{equation}
where $tr$ denotes a trace in a three dimensional flavour space and the constant $f^2/8$ has been chosen as to get a canonical kinetic term for the mesonic fields.

Chiral symmetry is not exact symmetry of QCD. It is explicitly broken by the quark mass term
\begin{equation}
\label{5.9a}
{\cal L}_M=-\bar q \hat mq\quad{\rm with} \qquad q=\begin{pmatrix} u\\d\\s\end{pmatrix}~,\quad\hat m=\begin{pmatrix} m_u&0&0\\0&m_d&0\\0&0&m_s\end{pmatrix}~,
\end{equation}
which transforms as $(\bar 3_L,3_R)\oplus (3_L,\bar 3_R)$ under the chiral transformation. At the first order of the quark masses, this breaking is taken into account by adding a term transforming exactly in the same way \cite{chiral}
\begin{equation}
\label{chiral.breaking}
{\cal L}_M^{breaking}=\lambda_0~tr[\hat m\Sigma+\Sigma^\dagger \hat m]~.
\end{equation} 
This term introduces the masses of the Goldstone bosons and indicates that quark masses are of the order of ${\cal O}(E^2)$.

\subsubsection{The $\eta$ and $\eta^\prime$ mesons}


The octet of the pseudoscalar Goldstone fields incorporates the pions, kaons and the flavor octet state $\eta_8$ (\ref{xi}).  The physical meson $\eta$ is composed mostly of the flavour octet state $\eta_8$ with a small admixture of the singlet state $\eta_0$. The physical meson $\eta^\prime$ is mostly the flavour singlet state $\eta_0$ with a small admixture of the octet state $\eta_8$.  The singlet state $\eta_0$ is not the Goldstone boson since $U(1)_A$ is not the symmetry at the quantum level. The properties of the meson $\eta^\prime$ are quite sensitive on the way $\eta_0$ state is incorporated in to the theory and  I will not consider the processes  with an external $\eta^\prime$ meson in this work. The presence of  $\eta^\prime$ as the intermediate state can sometimes not be avoided and I will include it where necessary\footnote{In this work  $\eta^\prime$ meson is present only in the diagrams for the decays $D^0\to \rho^0\gamma$ and $D^0\to \omega\gamma$ on Fig. \ref{fig33}b.}. The properties of the meson $\eta$ are rather insensitive on the way $\eta_0$ state is incorporated in to the theory and I will consider the processes  with an external $\eta$ meson. I use a simple scheme of $\eta-\eta^\prime$ mixing given by \cite{PDG,chiral,eta}
\begin{equation}
\label{thetap}
\eta_8=\cos \theta_P\eta+\sin\theta_P\eta^\prime~,\quad  \eta_0=-\sin \theta_P\eta+\cos\theta_P\eta^\prime\quad{\rm with}\quad \theta_P=(-20\pm 5)^0~. 
\end{equation}
The flavour singlet state $\eta_0$ is incorporated in the matrix $\Pi$ (\ref{xi}) as if  $U(1)_A$  was not anomalous
\begin{align*}
\Pi&=\begin{pmatrix}\tfrac{\pi^0}{\sqrt{2}}+\tfrac{\eta_8}{\sqrt{6}}+\tfrac{\eta_0}{\sqrt{3}}&\pi^+&K^+\\
\pi^-&-\tfrac{\pi^0}{\sqrt{2}}+\tfrac{\eta_8}{\sqrt{6}}+\tfrac{\eta_0}{\sqrt{3}}&K^0\\ K^-&\bar K^0&-\tfrac{2\eta_8}{\sqrt{6}}+\tfrac{\eta_0}{\sqrt{3}}\end{pmatrix}\\
&=\begin{pmatrix}\tfrac{\pi^0}{\sqrt{2}}+K^d_\eta\eta+K^d_{\eta^\prime}\eta^\prime&\pi^+&K^+\\
\pi^-&-\tfrac{\pi^0}{\sqrt{2}}+K^d_\eta\eta+K^d_{\eta^\prime}\eta^\prime&K^0\\ K^-&\bar K^0&K^s_\eta\eta+K^s_{\eta^\prime}\eta^\prime\end{pmatrix}~
\end{align*}
 and the coefficients $K^{d,s}_{\eta,\eta^\prime}$ are functions of the $\eta-\eta^\prime$ mixing angle $\theta_P$
 \begin{alignat}{2}
\label{mixing}
K_\eta^d&=\frac{\cos\theta_P}{\sqrt{6}}-\frac{\sin\theta_P}{\sqrt{3}}~&,\qquad K_\eta^s&=-\frac{2\cos\theta_P}{\sqrt{6}}-\frac{\sin\theta_P}{\sqrt{3}}~,\\
K_{\eta^\prime}^d&=\frac{\sin\theta_P}{\sqrt{6}}+\frac{\cos\theta_P}{\sqrt{3}}~&,\qquad K_{\eta^\prime}^s&=-\frac{2\sin\theta_P}{\sqrt{6}}+\frac{\cos\theta_P}{\sqrt{3}}~.\nonumber
\end{alignat}

\subsection{Strong interactions of heavy mesons and light pseudoscalars}

The strong interactions of  heavy and light mesons  in the limit $m_Q\to \infty$ and $m_q\to 0$ can be studied using the effective field theory combining heavy quark and chiral symmetry. The strong interactions are given by the most general Lagrangian invariant under the heavy quark, chiral, Lorentz, C, P and T transformations. In the kinematical region, where the energies of the light pseudoscalars are small and the velocity of the heavy quark remains practically the same, the  terms are grouped according to the number of derivatives on the heavy and light meson fields. Only the lowest orders in the chiral  $E/\Lambda_\chi$ expansion and the heavy quark $\Delta k/m_H$ expansion are important in this kinematical region. 
The heavy meson chiral Lagrangians were introduced in \cite{heavy.chiral,heavy.chiral2,heavy.chiral3} and the more recent review on their  applications is given in \cite{casalbuoni}. The applications to charm and beauty meson decays have different aspects of advantages and drawbacks. In the case of the beauty  meson decays, the expansion in $\Delta k/m_B$ converges quickly. The chiral expansion can be however problematic in the case when beauty meson is decaying only to the light particles. The light pseudoscalars can be rather energetic in this case and the kinematical region, where chiral expansion is meaningful, may not exist. In  charm meson decays, the kinematical regions with a good chiral expansion cover a bigger fraction of the phase space. The heavy quark expansion in the powers of $\Delta k/m_D$ does not converge so quickly, however.      

\vspace{0.1cm}

In order to write the Lagrangian invariant under the heavy quark and chiral transformations, 
we must establish how the field $H^a$ (\ref{ha}) transforms under the chiral transformation  and how the field $\xi(x)$ (\ref{xi}) transforms under the heavy quark transformation. The heavy quark transformation does not have any effect on the field $\xi$ as this field contains only the light degrees of freedom. The heavy meson field with flavour $Q\bar q_a$ transforms  according to the representation $\bar 3$  under the unbroken $SU(3)_V$ group. It is suitable to define the  heavy meson field $H_a$ (\ref{ha}) as  a singlet under the transformations of the  coset space $G/H$. In this case it transforms under the chiral transformation  according to representation $\bar 3$ of $SU(3)_V$ \cite{heavy.chiral,wise1} \footnote{The chiral transformation of $H_a$ could also be defined as $H_a\to H_b(g_{L}^\dagger)_{ba}$. In this case  $PH_aP^{-1}$ should transform under the chiral transformation as $PH_aP^{-1}\to PH_bP^{-1}(g_{R}^\dagger)_{ba}$, which implies an awkward definition of parity $PH_a(\vec x,t)P^{-1}=\gamma^0 H_b(-\vec x,t)\gamma^0\Sigma_{ba}(-\vec x,t)$. If the chiral transformation is defined as $H_a\to H_bU^{\dagger}_{ ba}(x)$ (\ref{5.15}),  the parity transformation has simpler form $PH_a(\vec x,t)P^{-1}=\gamma^0 H_a(-\vec x,t)\gamma^0$ \cite{wise1}.}
\begin{equation}
\label{5.15}
H_a\to H_bU^{\dagger}_ {ba}(x)~.
\end{equation} 

The strong interactions of the heavy and light mesons are given by the most general Lagrangian invariant under the Lorentz, parity, heavy quark and chiral symmetries. The transformation properties of the fields are gathered in Appendix E. For the reasons of the predictability I keep only the lowest orders on the chiral and the heavy quark expansions. 
The lowest order interaction terms of the heavy and light degrees of freedom are given at the order $E/\Lambda_{\chi}$ and $(\Delta k/m_H)^0$. They are represented by terms with no derivatives on heavy field $H(x)$ and one derivative on light field $\xi(x)$.   In addition the interactions of light degrees of freedom are given at the order $(E/\Lambda_{\chi})^2$ and the 
kinetic term for the heavy mesons is given at the order $(\Delta k/m_H)^0$  \cite{heavy.chiral}
\begin{equation}
\label{5.8}
{\cal L}= iTr[H_av_\mu(\partial^\mu+{\cal V}^\mu)\bar H_a]+igTr[H_b\gamma_\mu\gamma_5{\cal A}^\mu_{ba}\bar H_a]+\tfrac{1}{8}f^2tr[\partial^\mu\Sigma\partial_\mu\Sigma^\dagger]~.
\end{equation}
The fields ${\cal V}$ and ${\cal A}$ are defined as
\begin{equation}
\label{5.18}
{\cal A}_\mu=\tfrac{1}{2}(\xi^\dagger\partial_\mu\xi-\xi\partial_\mu\xi^\dagger)~,\qquad {\cal V}_\mu=\tfrac{1}{2}(\xi^\dagger\partial_\mu\xi+\xi\partial_\mu\xi^\dagger)
\end{equation}
and thir chiral transformations are given by (\ref{5.6}) 
\begin{equation}
{\cal A}_\mu\to U{\cal A}_\mu U^\dagger~, \qquad {\cal V}_\mu\to U{\cal V}_\mu U^\dagger+U\partial_\mu U^\dagger~.
\end{equation}
The first term is the chiral invariant generalization of the heavy meson free Lagrangian (\ref{5.5}). It describes the interactions among the heavy mesons and an even number of light pseudoscalars contained in ${\cal V}_\mu$.  The second term describes the interactions of the heavy mesons with an odd number of light pseudoscalars contained in  ${\cal A}_\mu$. These interactions are given  in terms of a single dimensionless coupling $g$, 
 which is a  free parameter of the effective field theory. Its value can not be determined using the symmetry arguments and will be discussed in  Section 5.2.  

\subsection{Hidden symmetry and light vector mesons}

The aim of this chapter is to study the decays of charm mesons to the final states that include the light vector mesons. For this purpose the octet of the light vector mesons $\rho$, $\phi$, $\omega$ and $K^*$ has to be incorporated into the effective field theory. I am going to follow the idea of the hidden symmetry approach, which was originally introduced to study the interactions of the light pseudoscalar and vector mesons in \cite{hidden}. The idea was then extended to the interactions among the light and heavy, pseudoscalar and vector mesons in \cite{hidden1}. The subsequent applications are reviewed in \cite{casalbuoni}. 

In the standard model all the elementary vector particles are introduced as the gauge bosons of a local gauge symmetry. The vector mesons are not elementary particles in the quantum chromodynamics, but they can effectively behave as elementary vector particles when they are introduced  as the gauge bosons of some local gauge symmetry. For this purpose a new local gauge symmetry must be introduced in a way that it renders the interactions of the light vector fields in a phenomenologically viable way.  Successful phenomenological applications are obtained if  $SU(3_V)$ is promoted from a global to a local symmetry and the vector mesons are the corresponding gauge bosons \cite{hidden}. The effective field theory based on the group $G=[SU(3)_L\times SU(3)_R]_{global}$ spontaneously broken to $H=[SU(3)_V]_{global}$ (\ref{5.8}) is replaced by the effective field theory based on the group $G^\prime=[SU(3)_L\times SU(3)_L]_{global}\times [SU(3)_V]_{local}$  \cite{hidden}. One is free to fix the gauge of $[SU(3)_V]_{local}$  and the two theories are equivalent until the terms with the derivatives on light vector fields are introduced. 

\vspace{0.1cm}

First let us rewrite the Lagrangian (\ref{5.8}) to a from invariant under the extended group $G^\prime=[SU(3)_L\times SU(3)_R]_{global}\times [SU(3)_V]_{local}$. This is achieved by introducing two new fields $\xi_L,\xi_R\in SU(3)_L\times SU(3)_R$ that transform under $G^\prime$ as 
\begin{equation}
\label{5.10}
\xi_L(x)\to g_L\xi_L(x)h^\dagger(x)~, \quad \xi_R(x)\to g_R\xi_R(x)h^\dagger(x)\qquad \qquad h(x)\in [SU(3)_V]_{local}~,
\end{equation}
so that the field 
$$\hat \Sigma(x)=\xi_L^\dagger(x)\xi_R(x)$$
 transforms under $G^\prime$ just like the field $\Sigma=\xi^2$ transforms under $G$, namely $\hat\Sigma\to g_L\hat\Sigma g_R^\dagger$. The field $\hat \Sigma$ is a singlet under the transformation $[SU(3)_V]_{local}$ and so $[SU(3)_V]_{local}$ is called  {\bf the hidden symmetry}. The field $H_a\sim Q\bar q_a$ transforms under $G^\prime$ according the representation $\bar 3$ of $SU(3)_V$
$$H_a\to H_bh^\dagger_{ab}(x)~.$$
The Lagrangian analogous to (\ref{5.8}), but invariant to $G^\prime$ instead to $G$, is 
 \begin{equation}
\label{5.11}
{\cal L}= iTr[H_av_\mu(\partial^\mu+\hat {\cal V}^\mu)\bar H_a]+igTr[H_b\gamma_\mu\gamma_5\hat{\cal A}^\mu_{ba}\bar H_a]+\tfrac{1}{8}f^2tr[\partial^\mu\hat\Sigma\partial_\mu\hat\Sigma^\dagger]
\end{equation}
with $\hat {\cal V}$ and $\hat {\cal A}$ defined as
\begin{equation}
\label{5.215}
\hat {\cal A}_\mu=\tfrac{1}{2}(\xi_L^\dagger\partial_\mu\xi_L-\xi_R\partial_\mu\xi_R^\dagger)~,\qquad \hat{\cal V}_\mu=\tfrac{1}{2}(\xi_L^\dagger\partial_\mu\xi_L+\xi_R\partial_\mu\xi_R^\dagger)
\end{equation}
and their transformations under $G^\prime$ are given by (\ref{5.10}) 
\begin{equation}
\hat {\cal A}_\mu\to h(x)\hat {\cal A}_\mu h^\dagger(x)~, \qquad \hat{\cal V}_\mu\to h(x)\hat{\cal V}_\mu h^\dagger(x)+h(x)\partial_\mu h^\dagger(x)~.
\end{equation}
By fixing the gauge of  $h(x)\in [SU(3)_V]_{local}$ so that $\xi_L^\dagger=\xi_R\equiv \xi$ (\ref{5.10}), the Lagrangian (\ref{5.11}) based on the group $G^\prime$ is equivalent to the Lagrangian (\ref{5.8}) based on the group $G$.

\vspace{0.1cm}  

Now we introduce the light vector mesons as the gauge bosons  of the local group $[SU(3)_V]_{local}$. They span the octet representation of $SU(3)_V$ and are incorporated in the field
 $\rho$
\begin{equation}
\label{rho}
\rho_\mu=i\tfrac{\tilde g_V}{\sqrt{2}}\begin{pmatrix}\tfrac{\rho^0_\mu+\omega_\mu}{\sqrt{2}}&\rho^+_\mu&K^{*+}_\mu\\
\rho^-_\mu&\tfrac{-\rho^0_\mu+\omega_\mu}{\sqrt{2}}&K^{*0}_\mu\\
K^{*-}_\mu&\bar K^{*0}_\mu&\phi_\mu\end{pmatrix}
\end{equation}
with a new free parameter $\tilde g_V$ to be discussed later. The field $\rho_\mu$ transforms as the gauge field under the gauge group $G^\prime$
\begin{equation}
\label{5.16}
\rho_\mu\to h(x)\rho_\mu h^\dagger(x)+h(x)\partial_\mu h^\dagger(x)\qquad\qquad h\in [SU(3)_V]_{local}~.
\end{equation}
The strong Lagrangian for the heavy and light, pseudoscalar and vector  mesons is given by the most general expression invariant under the group $G^\prime$ and under  Lorentz, parity and  heavy quark transformations. The transformation properties of the fields are gathered in  Appendix E.  In addition to the terms given in (\ref{5.11}), the presence  of the gauge field $\rho_\mu$ allows for  new terms. 
The lowest order interaction terms of the heavy and light degrees of freedom are given at the order $E/\Lambda_{\chi}$ and $(\Delta k/m_H)^0$. They are represented by terms with no derivatives on heavy fields $H(x)$ and one derivative on light fields $\xi_{L,R}(x)$.   The interactions of light degrees of freedom are given at the order $(E/\Lambda_{\chi})^2$. In addition, the kinetic term for the light vector mesons and the heavy mesons are kept \cite{hidden,hidden1,casalbuoni}
 \begin{align}
\label{5.12}
&{\cal L}_1= \tfrac{1}{8}f^2tr[\partial^\mu\hat\Sigma\partial_\mu\hat\Sigma^\dagger]-a\tfrac{f^2}{2}tr[(\hat{\cal V}_\mu-\rho_\mu)^2]+\tfrac{1}{2\tilde g_V^2}tr[F_{\mu\nu}(\rho)F^{\mu\nu}(\rho)]\\
&+iTr[H_av_\mu\{\partial^\mu+\hat {\cal V}^\mu+\kappa(\hat {\cal V}^\mu-\rho^\mu)\}\bar H_a]+igTr[H_b\gamma_\mu\gamma_5\hat{\cal A}^\mu_{ba}\bar H_a]+i\beta Tr[H_bv_\mu(\hat {\cal V}^\mu-\rho^\mu)_{ba}\bar H_a]~,\nonumber
\end{align}  
where $F^{\mu\nu}(\rho)\!=\!\partial^\mu\rho^\nu\!-\!\partial^\nu\rho^\mu\!+\![\rho^\mu,\rho^\nu]$. The $\rho$ meson field is accompanied by the vector current $\hat {\cal V}$  so that the combination $\rho-\hat {\cal V}$ is invariant under the transformation of $G^\prime$.  The current $\hat {\cal V}$ (\ref{5.215}) is of the order of ${\cal O}(E)$ in the chiral expansion  and so  the vector meson field $\rho$ is of the order of ${\cal O}(E)$ as well \cite{hidden}. 

The Lagrangian (\ref{5.12}) does not incorporate the interactions of two vector mesons  and a pseudoscalar meson as these interactions are of the higher order in the chiral and heavy quark expansion. However, these vertices are essential for the dynamics of the decays studied in this chapter\footnote{These terms are particularly important for the mechanisms $D\to VV^0\to V\gamma$ and $D\to VV^0\to Vl^+l^-$ discussed in Section 5.4. } and the lowest order interaction terms of this kind are added to the Lagrangian ${\cal L}_1$ (\ref{5.12}) \cite{FPS1,FS,FPS2,casalbuoni,hidden,VVP,CFN,BFO2}
\begin{align}
\label{5.12a}
{\cal L}&={\cal L}_1+{\cal L}_2\nonumber\\
{\cal L}_2&=-4\tfrac{C_{VV\Pi}}{f}\epsilon^{\mu\nu\alpha\beta}tr[\partial_\mu\rho_\nu\partial_\alpha\rho_\beta \Pi]+i\lambda Tr[H_a\sigma_{\mu\nu}F^{\mu\nu}(\rho)_{ab}\bar H_b]~.
\end{align}
The first term is responsible the interactions of two light vectors and a light pseudoscalar and is of the order of $(E/\Lambda_\chi)^4$ \cite{hidden,VVP,BFO2}. The second term gives the interaction of a heavy vector, heavy pseudoscalar and light vector mesons and is of the order of $(E/\Lambda_\chi)^2$ \cite{BFO1,casalbuoni,CFN,BFO2}. In contrast to the Lagrangian ${\cal L}_1$ (\ref{5.12}), the Lagrangian ${\cal L}_2$ (\ref{5.12a}) in not invariant under the ``naive parity''
\begin{equation}
\label{5.12b}
{\cal P}(t,\vec x)=(t,-\vec x)~,\quad
H_a(x)\to H_a({\cal P}x) ~,\quad
\xi_{L,R}(x)\to \xi_{L,R}({\cal P}x)~,\quad \rho^\mu(x)\to {\cal P}^\mu_{~\nu}\rho^\nu({\cal P}x)~,
\end{equation}
which is not the symmetry of QCD. 
The Lagrangian is of course invariant under the parity transformation
\begin{equation}
\label{5.12c}
{\cal P}(t,\vec x)=(t,-\vec x)~,\quad
H_a(x)\to \gamma^0 H_a({\cal P}x)\gamma^0 ~,\quad
\xi_{L,R}(x)\to \xi^\dagger_{L,R}({\cal P}x)~,\quad \rho^\mu(x)\to {\cal P}^\mu_{~\nu}\rho^\nu({\cal P}x)~.
\end{equation}
This kind of terms are called the anomalous terms and were originally introduced by Wess, Zumino and Witten \cite{WZ,witten} in order to incorporate the processes like $\pi^0\to \gamma\gamma$ and $K^+K^-\to \pi^+\pi^0\pi^-$. 

We are free to fix the gauge  of the local symmetry $h(x)\in [SU(3)_V]_{local}$ in the Lagrangian ${\cal L}_1$ (\ref{5.12})   so that $\xi_L^\dagger=\xi_R\equiv \xi$ (\ref{5.10}) and $h=U$ (\ref{5.6}). In this gauge, the fields $\hat{ \cal A}$, $\hat {\cal V}$ in (\ref{5.12}) are replaced by ${ \cal A}$, ${\cal V}$ \footnote{The relation $\tfrac{1}{8}f^2tr[\partial^\mu\hat\Sigma\partial_\mu\hat\Sigma^\dagger]=-\tfrac{f^2}{2}tr[\hat {\cal A}_{\mu}\hat {\cal A}^\mu]$ is used.} and the complete strong Lagrangian ${\cal L}={\cal L}_1+{\cal L}_2$ (\ref{5.12}, \ref{5.12a}) is given by
\begin{align}
\label{5.13}
&{\cal L}= -\tfrac{f^2}{2}\{tr[{\cal A}_{\mu}{\cal A}^\mu]+a~tr[({\cal V}_\mu-\rho_\mu)^2]\}+\tfrac{1}{2\tilde g_V^2}tr[F_{\mu\nu}(\rho)F^{\mu\nu}(\rho)]-4\tfrac{C_{VV\Pi}}{f}\epsilon^{\mu\nu\alpha\beta}tr[\partial_\mu\rho_\nu\partial_\alpha\rho_\beta \Pi]\nonumber\\
&+iTr[H_av_\mu\{\partial^\mu+{\cal V}^\mu+\kappa({\cal V}^\mu-\rho^\mu)\}\bar H_a]+igTr[H_b\gamma_\mu\gamma_5{\cal A}^\mu_{ba}\bar H_a]\nonumber\\
&+i\beta Tr[H_bv_\mu({\cal V}^\mu-\rho^\mu)_{ba}\bar H_a]+i\lambda Tr[H_a\sigma_{\mu\nu}F^{\mu\nu}(\rho)_{ab}\bar H_b]~.
\end{align}  
If the terms with the derivatives of the field $\rho$ were not present, the equation of motion for $\rho_\mu$ would be $\rho_\mu={\cal V}_\mu$. In this case all the terms of the form ${\cal V}^\mu-\rho^\mu$ would vanish and the Lagrangian (\ref{5.13}) based on the group $G^\prime$ would exactly match the Lagrangian (\ref{5.8}) based on the group $G$.


\subsection{Electromagnetic interactions}

We have considered only the strong interactions among the hadronic states up to now. 
In order to calculate the amplitudes for the processes with real or virtual photons in the final state, the electromagnetic interactions have to be incorporated. First we will establish the electromagnetic gauge transformations of the fields.  By imposing the invariance under the local transformations, the interactions of the hadronic fields and  the electromagnetic field $A_\mu$ will arise\footnote{Note the difference in the notation of the electromagnetic field $A_\mu$ and the field ${\cal A}_\mu$ (\ref{5.18}) containing the pseudoscalar mesons.}.

The transformation of the fields $\xi$, $H$ and $\rho$ under the chiral transformation is given in (\ref{5.6}), ({\ref{5.15}) and (\ref{5.16}), respectively. We have to determine the chiral transformations $g_L$, $g_R$, and $U$ for the particular case of the electromagnetic gauge transformation. These are determined by considering electromagnetic gauge transformations of the quark field $q$
$$q\to e^{ie_0{\cal Q}\alpha(x)}q\qquad {\rm with}\quad {\cal  Q}=\begin{pmatrix}2/3&0&0\\0&-1/3&0\\0&0&-1/3\end{pmatrix} ~,\quad q=\begin{pmatrix}u\\d\\s\end{pmatrix}~.$$
The $g_L$ and $g_R$ are defined as the chiral transformations of the quark fields $q_{L,R}\to g_{L,R}~q_{L,R}^\dagger$, while  $U(x)$ is defined by the equation  $g_L\xi U^\dagger=U\xi g_R$ (\ref{5.6}). For the case of the  electromagnetic local gauge transformation they  read
\begin{equation}
\label{5.17}
g_L(x)=g_R(x)=U(x)\equiv g_0(x)=e^{ie_0{\cal Q}\alpha(x)}\qquad{\rm with}\quad {\cal Q}=\begin{pmatrix}2/3&0&0\\0&-1/3&0\\0&0&-1/3\end{pmatrix} ~.
\end{equation}
The local electromagnetic transformations of the basic hadronic fields and the photon field $A_\mu$ are then given as
\begin{alignat}{2}
\label{5.19}
\xi&\to g_0(x)\xi g_0^\dagger(x)~,&\qquad\xi^\dagger&\to g_0(x)\xi^\dagger g_0^\dagger(x)~,\nonumber\\
H_a&\to e^{ie_0{\cal Q}^\prime \alpha(x)}H_bg_{0ba}^{\dagger }(x)~,&\qquad\bar H_a&\to g_{0ab}(x)\bar H_b e^{-ie_0{\cal Q}^\prime \alpha(x)}~,\nonumber\\
\rho_\mu&\to g_0(x)\rho_\mu g_0^\dagger(x)+g_0(x)\partial_\mu g_0^\dagger(x)~,&\qquad A_\mu&\to g_0(x)A_\mu g_0^\dagger(x)+g_0(x)\partial_\mu g_0^\dagger(x)~,
\end{alignat}
where ${\cal Q}^\prime$ is the charge of the heavy quark, i.e. ${\cal Q}^\prime=2/3$ for $c$ quark.  The transformation of the field $H_a\sim Q\bar q_a$ involves $g_0$ for the light antiquark and  $\exp(\pm ie_0{\cal Q}^\prime\alpha)$ for the heavy quark. 
 
The interactions of the hadronic degrees of freedom and  the electromagnetic field $A_{\mu}$ are obtained by imposing the invariance under the local gauge transformations (\ref{5.19}). This is achieved by replacing the partial derivatives $\partial^\mu\xi$ and $\partial^\mu H_a$ by the covariant derivatives $D^\mu$ in the following way
\begin{equation}
\label{5.221}
D_\mu\xi=(\partial_\mu+ie_0{\cal Q} A_\mu)\xi~,\quad D_\mu\xi^\dagger=(\partial_\mu+ie_0{\cal Q} A_\mu)\xi^\dagger~,\quad D_\mu \bar H^a= (\partial_\mu-ie_0{\cal Q}^\prime A_\mu)\bar H^a~.
\end{equation} 
The heavy chiral Lagrangian for strong and electromagnetic interactions is given in terms of 
\begin{equation}
\label{5.20}
{\cal A}^D_\mu=\tfrac{1}{2}(\xi^\dagger D_\mu\xi-\xi D_\mu\xi^\dagger)~,\qquad {\cal V}^D_\mu=\tfrac{1}{2}(\xi^\dagger D_\mu\xi+\xi D_\mu\xi^\dagger)
\end{equation}
as \cite{FPS1,FPS2,FS,BFO2}
\begin{align}
\label{hybrid}
{\cal L}&= {\cal L}^{light}+{\cal L}^{heavy}\\
{\cal L}^{light}&=
-\tfrac{f^2}{2}\{tr[{\cal A}_{\mu}^D{\cal A}^{D\mu}]+a~tr[({\cal V}^D_\mu-\rho_\mu)^2]\}\nonumber\\
&+\tfrac{1}{2\tilde g_V^2}~tr[F_{\mu\nu}(\rho)F^{\mu\nu}(\rho)]-4\tfrac{C_{VV\Pi}}{f}~\epsilon^{\mu\nu\alpha\beta}~tr[\partial_\mu\rho_\nu\partial_\alpha\rho_\beta \Pi]\nonumber\\
{\cal L}^{heavy}&=iTr[H_av^\mu\{\partial_\mu-ie_0Q^\prime A^D_\mu+{\cal V}^D_\mu+\kappa({\cal V}^D_\mu-\rho_\mu)\}\bar H_a]+igTr[H_b\gamma_\mu\gamma_5{\cal A}^{D\mu}_{ba}\bar H_a]\nonumber\\
&+i\beta Tr[H_bv^\mu({\cal V}_\mu^D-\rho_\mu)_{ba}\bar H_a]+i\lambda Tr[H_a\sigma_{\mu\nu}F^{\mu\nu}(\rho)_{ab}\bar H_b]-\lambda^{\prime}~ Tr[H_a\sigma_{\mu\nu}F^{\mu\nu}(A)\bar H_a]~\nonumber
\end{align}
with $F^{\mu\nu}(A)\!=\!\partial^\mu A^\nu\!-\!\partial^\nu A^\mu$.
The last term proportional to the parameter $\lambda^\prime$ does not appear as a result of the covariant derivative and is gauge invariant by itself. It is introduced in order to incorporate the interactions of  the heavy pseudoscalar, the heavy vector and the photon, which are important for the dynamics of the radiative meson decays studied in this chapter. This term  is of the order of $(E/\Lambda_\chi)^2$ in the chiral expansion  and is of the anomalous kind just like the interaction terms  for two vectors and a pseudoscalar in (\ref{5.12a}). 

The values of the free parameters of the Lagrangian (\ref{hybrid}) will be discussed in the next section. At this point  I discuss only the value of the parameter $a$, which appears in the second term of ${\cal L}_{light}$.  It is fixed by the vector meson dominance hypothesis \cite{VMD,hidden}, which assumes that there are no direct vertices between the photon and the two pseudoscalars. The pseudoscalars interact with the photon only through the vector mesons, which is achieved by fixing
\begin{equation}
\label{a}
a=2~. 
\end{equation}

\subsection{The weak currents}

Finally we have to incorporate the weak interactions of the hadrons. The weak charm meson decays of interest are induced by the effective nonleptonic weak Lagrangian (\ref{eff})
\begin{align*}
{\cal L}^{|\Delta c|=1}_{eff}=-\tfrac{G_F}{\sqrt{2}}V_{cq_j}^*V_{uq_i}[~&a_1 ~\bar u\gamma^{\mu}(1-\gamma_5)q_i~\bar q_j\gamma_{\mu}(1-\gamma_5)c\nonumber\\
+~&a_2~\bar q_j\gamma_{\mu}(1-\gamma_5)q_i~\bar u\gamma^{\mu}(1-\gamma_5)c~]\quad{\rm with} \quad q_{i,j}=d,s
\end{align*}
extensively discussed in Section 3.3. It incorporates $W$ exchange between four quarks and the hard gluon exchanges. It is expressed in terms of the colour singlet weak currents and has a suitable form to study the weak interactions among the colour singlet mesonic states. I study the light-to-light $\bar q_a\gamma^\mu(1-\gamma_5)q_b$ and heavy-to-light $\bar q_a\gamma^\mu(1-\gamma_5)c$ currents separately.

\subsubsection{Weak current for light quarks}

The weak current $\bar q_a\gamma^\mu(1-\gamma_5)q_b$ is incorporated along the similar line as the electromagnetic interaction of the light quarks was incorporated in the previous section (although the notion of the currents instead of the covariant derivatives will be used here).

We have to determine the transformations $g_L$, $g_R$ and $U$ for the weak transformation with the corresponding Noether current 
$$(J_W^{light})_{ba}^\mu=\bar q_a\gamma^\mu(1-\gamma_5)q_b=2\bar q_L\hat T_{ab}\gamma^\mu q_L~,\qquad q=(u,~d,~s)^T~.$$
Here $\hat T^{ab}$ denotes a $3\times 3$ matrix with the entry $1$ in the $a$-th row and the $b$-th column and other entries equal to zero. The gauge transformation responsible for this current is 
$$q_L\to e^{i\hat T^{ab}\beta(x)}q_L~, \qquad q_R\to q_R, $$ 
so the corresponding chiral transformations $g_{L}$ and $g_R$ for the weak transformations read 
\begin{equation}
\label{5.23}
 g_L= e^{i\hat T^{ab}\beta(x)}~,\qquad g_R=I~.
\end{equation}
The transformation $U(x)$ is defined by the equation $\exp(i\hat T^{ab}\beta)~\xi U^\dagger\!=\!U\xi$ (\ref{5.6}) and does not need be explicitly calculated as the  Lagrangian (\ref{hybrid}) is already invariant under a general $U(x)$. By applying the weak transformation  (\ref{5.23}) to the fields
\begin{align}
\label{5.24}
\xi&\to e^{i\hat T^{ab}\beta(x)}\xi U(x)^\dagger=U(x)\xi~,\quad\qquad \xi^\dagger\to U(x)\xi ^{\dagger}e^{-i\hat T^{ab}\beta(x)}=U(x)\xi~,\\
\rho_\mu&\to U(x)\rho_\mu U^\dagger(x)+U(x)\partial_\mu U^\dagger(x)\nonumber
\end{align}
in ${\cal L}^{light}$ (\ref{hybrid})\footnote{The transformation (\ref{5.24}) on ${\cal L}^{heavy}$ renders the effective weak current $\bar q_a\gamma^\mu(1-\gamma_5)q_b$ that involves a pair of the heavy fields $H$ and $\bar H$. As the processes of interest only involve the light mesons in the final state, this contribution can be safely neglected.}, the weak current $(J_W^{light})_{ba}$ is defined via
$${\cal L}^{light}\to {\cal L}^{light}-\tfrac{1}{2}(J_W^{light})_{ba}^\mu\partial_\mu\beta~,$$
 so
\begin{equation}
\label{current.light}
\bar q_a\gamma^\mu(1-\gamma_5)q_b\simeq (J_W^{light})_{ba}^\mu\simeq \bigl(if^2\xi [{\cal A}^{D\mu}+a({\cal V}^D-\rho^D)^\mu]\xi^\dagger\bigr)_{ba}~.
\end{equation}
      
\subsubsection{Weak current for heavy quark and light antiquark}


The  weak current for heavy quark and light antiquark $(J_W^{heavy})_a^\mu=\bar q_a\gamma^\mu(1-\gamma_5)c$ is not so restricted by the heavy quark and chiral symmetry as the weak current for light quarks $J_W^{light}$. The reason for this is, that the transformation among a light and a heavy quark is not a symmetry of the theory. The chiral symmetry implies only the symmetries for the transformations among the three light quarks and so it restricts the effective weak current for the light quarks. The heavy quark symmetry implies the symmetries for the transformations among  the $b$ and $c$ quarks and so it restricts the effective weak current $\bar c\gamma^\mu(1-\gamma_5)b$. In spite of the fact, that the derivation of the effective current $J_W^{heavy}$ can not follow the line of the derivation for $J_W^{light}$ above, the chiral and heavy quark symmetries are also useful in this case. The heavy quark spin symmetry relates the effective currents involving a pseudoscalar $D$ and a vector $D^*$ meson. The chiral symmetry relates the effective currents for different light mesons. 

The most general effective current $(J_W^{heavy})_a^\mu=\bar q_a\gamma^\mu(1-\gamma_5)c$ can be derived based on the fact, that
 it transforms as $(\bar 3_L,1_R)$ under the chiral transformation $SU(3)_L\times SU(3)_R$. It is expressed in terms of the field $H_a$ so, that the heavy quark symmetry is manifest. The current $J_W^{heavy}$ should be linear in the heavy meson field $H_a$ as the current $\bar q_a\gamma^\mu(1-\gamma_5)c$ is linear in the field $c$. In the kinematical region, where light degrees of freedom have small momentum and  the heavy quark is almost on-shell, only few terms in the expansions $E/\Lambda_\chi$ and $\Delta k/m_H$ are important. The current  for  heavy mesons and light pseudoscalars at the order $(E/\Lambda_\chi)^0$ and $(\Delta k/m_H)^0$  was originally proposed in \cite{heavy.chiral}. In order to incorporate the light vector meson field $\rho$, which is of the order of $E/\Lambda_\chi$, the terms up to the order  $E/\Lambda_\chi$ in chiral expansion have to be considered. The most general current with $V-A$ structure at the order $E/\Lambda_\chi$ and $(\Delta k/m_H)^0$ is derived in Appendix F following \cite{BFO1,pl4}. Using the shorthand notation  $H=(H_1,H_2,H_3)$ the result is given by
\begin{align}
\label{current.heavy1}
(J_W^{heavy})_\mu&=\tfrac{1}{2}i\alpha Tr[\gamma_\mu(1-\gamma_5)H]\xi^\dagger\\
&-\alpha_1 Tr[(1-\gamma_5)H](\rho-{\cal V})_\mu\xi^\dagger-\alpha_2Tr[\gamma_\mu(1-\gamma_5)H]v^\alpha (\rho-{\cal V})_\alpha \xi^\dagger\nonumber\\
&+\alpha_3Tr[(1-\gamma_5)H]{\cal A}_{\mu}\xi^\dagger+\alpha_4Tr[\gamma_\mu(1-\gamma_5)H]v^\alpha{\cal A}_\alpha \xi^\dagger\nonumber\\
&+Tr[\gamma^{\delta}(1-\gamma_5)H]\bigl(g_{\delta\mu}v_\alpha-g_{\delta\alpha}v_\mu-ig_{\delta\gamma}\epsilon^{\gamma}_{~\mu\alpha\beta}v^\beta\bigr)\{\alpha_1(\rho-{\cal V})_\alpha-\alpha_3{\cal A}_{\alpha}\}\xi^\dagger~.\nonumber
\end{align}
In the calculation of the charm meson decays to the leading order in the heavy and chiral expansion, we will need only the current proportional to $D$, $DP$ or $D^*$ at the order $(E/\Lambda_\chi)^0$ and the current proportional to $DV$ at the order of $E/\Lambda_\chi$ ($P$ and $V$ denote light pseudoscalar and vector meson). The last term in (\ref{current.heavy1}) contains at least one light meson field and does not contain the heavy pseudoscalar and it is not of interest for this work. The terms proportional to $\alpha_3$ and $\alpha_4$ give the sub-leading contributions to  current $D\to P$ and are not  of interest as well. The  free parameters $\alpha$, $\alpha_1$ and $\alpha_2$ for the relevant terms will be determined in the next section.  

The  electromagnetic interaction in the sector of the light quarks are  incorporated by replacing the partial derivative $\partial ^\mu\xi$ with the covariant derivative  $(\partial^\mu+ie_0{\cal Q})\xi$ (\ref{5.221}) in the expression for the weak current (\ref{current.heavy1}). 

The weak current of the order of $(\Delta k/m_H)^0$ does not lead to the gauge invariant  bremss\-trahlung amplitudes for $D\to P\gamma^*$ and $D\to V\gamma^*$ decays.   The gauge invariant amplitudes were obtained in Section 3.3.3 by considering the general effective model, which gives rise to the diagrams in Figs. \ref{fig24} and \ref{fig25}a. The model for the charm meson decays, presented here, contains all diagrams in Figs. \ref{fig24} and \ref{fig25}a, except for the second diagram in both figures. It is induced by the weak current, which contains $D$ meson and the photon field. This current was introduced in the effective theory presented in Section 3.3.3,  when the partial derivative  in $j_W^\mu=\tfrac{g_2}{2\sqrt{2}}f_D\partial^\mu D$ was replaced by the covariant derivative $j_W^\mu=\tfrac{g_2}{2\sqrt{2}}f_D(\partial^\mu+ie_DA^\mu)D$ (\ref{3.22}). The current $j_W^\mu\propto e_DA^\mu D$ is not generated at the order $(\Delta k/m_H)^0$, since there are no derivatives on the heavy meson field $H$ at this order. This current is, however, required at the order $\Delta k/m_H$ by the so called velocity reparametrization invariance (VRI) \cite{VRI,grinstein1}. The velocity reparametrization exploits the fact, that the velocity $v$ in the definitions of the fields $D^v$ and $D^{*v}$ (\ref{5.105}) need not be the velocity of the heavy meson. It can be thought as a parameter as long as $p_H=m_Hv+k$ is the momentum of the heavy meson. The Lagrangian must be invariant under the reparametrization of the velocity \cite{VRI}
$$v\to v+q/m_H~,\qquad k\to k-q~.$$
This is achieved when  the velocity and the derivatives on the heavy meson fields appear in the combination $v^\mu+i{\cal D}^\mu/M$  and one uses the field $\tilde H$
$$\tilde H=H+\tfrac{i}{2m_H}{\cal D}_\mu[\gamma^\mu,H] $$
instead of the field $H$  \cite{casalbuoni,VRI,grinstein1}. Here ${\cal D}$ denotes the covariant derivative ${\cal D}^\mu=\partial^\mu+ie_DA^\mu$ and $e_D$ is the charge of the heavy field. The VRI requirement introduces only the higher order term in the expansion $\Delta k/m_H$,  which are systematically neglected in this work. In order to get the  gauge invariant  bremsstrahlung amplitudes for $D\to P\gamma^*$ and $D\to V\gamma^*$ decays,  I include only the term imposed by the VRI corresponding to the first term in (\ref{current.heavy1})
\begin{equation}
\label{current.VRI}
(J_W^{heavy})_\mu^{VRI}=\tfrac{1}{2}i\alpha~ Tr\bigl[\gamma_\mu(1-\gamma_5)~\bigl(H+\tfrac{i}{2m_H}{\cal D}^\nu[\gamma_\nu,H]\bigr)\bigr]\xi^\dagger~.\nonumber
\end{equation}
The weak vertex $DW\gamma$ arising from this current is given in Fig.  \ref{fig23}b. Together with other diagrams in Figs. \ref{fig24} and \ref{fig25}a, this diagram ensures the gauge invariance of $D\to P\gamma^*$ and $D\to V\gamma^*$ bremsstrahlung amplitudes. 

Finally, I collect the relevant terms for the heavy-to-light weak current including the electromagnetic interactions
\begin{align}
\label{current.heavy}
(J_W^{heavy})_\mu&=\tfrac{1}{2}i\alpha~ Tr\bigl[\gamma_\mu(1-\gamma_5)~\bigl(H+\tfrac{i}{2m_H}{\cal D}^\nu[\gamma_\nu,H]\bigr)\bigr]\xi^\dagger~\\
&-\alpha_1 Tr\bigl[(1-\gamma_5)H\bigr](\rho-{\cal V}^D)_\mu\xi^\dagger-\alpha_2Tr\bigl[\gamma_\mu(1-\gamma_5)H\bigr]v^\alpha (\rho-{\cal V}^D)_\alpha \xi^\dagger~.\nonumber
\end{align}

\subsection{Extrapolation away from the kinematical point  where chiral and heavy quark expansions are valid}

The chiral and the heavy quark expansions are valid in the kinematical region where the light mesons have small momentum and the velocity of the heavy quark does not change drastically. In this kinematical region the physical process is well described by the strong and electromagnetic Lagrangian (\ref{hybrid}) and the weak currents (\ref{current.light}, \ref{current.heavy}). In the case of the heavy-to-light  transition $\langle M(p^\prime)|\bar q\gamma^\mu(1-\gamma_5)c|D(p)\rangle$, represented by the diagrams in Fig. \ref{fig40},  the chiral and the heavy quark expansions are meaningful in the region near  $q^2_{max}=(p-p^\prime)^2=(m_D-m_M)^2$. 
The velocity of the light meson $M$ in the $D$ meson rest frame  is equal to zero at $q_{max}^2$ and the chiral expansion is valid; at the same time the virtuality $k=-m_Pv$ of the intermediate heavy meson $D^\prime$ is small and the heavy quark expansion is valid. The problem is how to extrapolate the amplitude from the zero recoil point to the rest of the allowed kinematical region. Following the idea proposed in \cite{BFO1} and applied in \cite{FPS2,BFOP,BFO2}, I shall make a very simple, physically motivated assumption: the vertices do not change significantly, while the propagators of the off-shell heavy mesons are given by the full propagators\footnote{The additional factors $m_D$ and $m_{D^*}$ are present due to the normalization of the fields $D^v$ and $D^{*v}$ in (\ref{5.105}).}
\begin{equation}
\label{5.106}
\frac{im_D}{p^2-m_D^2}\quad {\rm for} \quad D\qquad\qquad{\rm and }\qquad\qquad-im_{D^*}~\frac{g^{\mu\nu}-\tfrac{p^\mu p^\nu}{m_{D^*}^2}}{p^2-m_{D^*}^2}\quad {\rm for}\quad D^*
\end{equation}
instead of the propagators given by the heavy quark effective field theory ${\cal L}_H=iTr[H_av_\mu\partial^\mu\bar H_a]$ (\ref{5.5})
\begin{equation}
\label{5.106a}
\frac{i}{2v\cdot k}\quad {\rm for} \quad D\qquad\qquad{\rm and }\qquad\qquad -i~\frac{g^{\mu\nu}-v^\mu v^\nu}{2v\cdot k}\quad {\rm for}\quad D^*~.
\end{equation}
These assumptions imply, for example, that the heavy-to-light matrix element $\langle M(p^\prime)|\bar q\gamma^\mu(1-\gamma_5)c|D(p)\rangle$  has a polar part, arising from the diagram in Fig. \ref{fig40}a,  and a flat part, arising from the diagram in Fig.  \ref{fig40}b.
With these assumptions, the following general features are incorporated: (i) similar prediction at $q^2_{max}$ as given by the  HQET propagators (\ref{5.106a}); (ii) natural explanation of the polar shape of form factors when appropriate\footnote{A more popular approach for the extrapolation  assumes the polar behavior of all the form factors 
\cite{BSW,casalbuoni1}.}; (iii) natural explanation of a flat $q^2$ behavior of the $D\to V$ form factors $A_1$ and $A_2$ (\ref{5.55}, \ref{formV}), which has been indicated by   QCD sum-rule and lattice results \cite{burchat};
 (iv) the relations (\ref{5.35}, \ref{5.35a}) among  $D\to M$ form factors at $q^2=0$ are satisfied automatically.  This assumption, used together with the Lagrangian (\ref{hybrid}), the currents (\ref{current.light},
\ref{current.heavy}) and the flavour $SU(3)$ breaking effects proposed in the next section, will be referred to as the {\bf hybrid model}.

\begin{figure}[h]
\centering
\mbox{
\subfigure[]
{
\begin{fmffile}{f40o1n}
\fmfframe(15,0)(15,0){
  \begin{fmfgraph*}(25,15)
  \fmfpen{thin}  
  \fmfleftn{l}{1} \fmfrightn{r}{1} \fmftopn{t}{1}
  \fmf{dashes}{l1,v1} 
  \fmf{dashes,label=$D^\prime$}{v1,r1}
  \fmffreeze 
  \fmf{dashes}{v1,t1}
  \fmfv{decor.size=1.2thick,decor.shape=circle,decor.filled=full}{r1}
  \fmflabel{$j^\mu=\bar q\gamma^\mu(1-\gamma_5)c$}{r1}
  \fmflabel{$D$}{l1}
  \fmflabel{$M$}{t1}
  \end{fmfgraph*} }
\end{fmffile}
}
\qquad\qquad
\subfigure[]
{
\begin{fmffile}{f40o2n}
\fmfframe(15,0)(15,0){
  \begin{fmfgraph*}(10,15)
  \fmfpen{thin}  
  \fmfleftn{l}{1} \fmfrightn{r}{1} \fmftop{t,tt,t1}
  \fmf{dashes}{l1,r1} 
  \fmffreeze 
  \fmf{dashes}{r1,t1}
  \fmfv{decor.size=1.2thick,decor.shape=circle,decor.filled=full}{r1}
  \fmflabel{$j^\mu=\bar q\gamma^\mu(1-\gamma_5)c$}{r1}
  \fmflabel{$D$}{l1}
  \fmflabel{$M$}{t1}
  \end{fmfgraph*} }
\end{fmffile}
}
    }
\caption{The diagrams corresponding to the matrix element of the
current $\langle M|\bar 
q\gamma^\mu(1-\gamma_5)c|D\rangle$ in the hybrid model. Here $D^\prime$ denotes
a pseudoscalar or a vector charmed  meson.}
\label{fig40}
\end{figure}
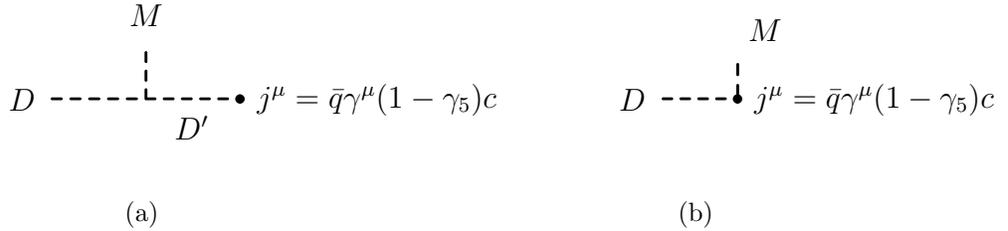

\subsection{ The $SU(3)_V$ flavour breaking}

The heavy meson chiral Lagrangian for strong and electromagnetic interactions (\ref{hybrid}) and the weak currents (\ref{current.light}, \ref{current.heavy}) are based on the $SU(3)_L\times SU(3)_L$ chiral symmetry spontaneously broken to the diagonal $SU(3)_V$ symmetry. This renders the  physical observables that respect the exact  $SU(3)_V$ flavour symmetry.   
 The $SU(3)_V$ flavour symmetry is broken in QCD by the quark mass term  (\ref{5.9a}). Systematic incorporation of the $SU(3)_V$ breaking effects requires the addition off all possible terms that transform under the chiral symmetry in the same way as the quark mass term (\ref{5.9a}). These terms are formally suppressed in the chiral expansion, since they involve the quark mass matrix $\hat  m$ (\ref{5.9a}) of the order of $(E/\Lambda_\chi)^2$. The systematic incorporation of the $SU(3)_V$ breaking terms, as well as the other higher order terms in the chiral expansion, introduces a number a new free parameters. In order to retain a certain level of predictability, the symmetry breaking terms are not incorporated to the Lagrangian. The $SU(3)_V$ breaking effects are incorporated in to the hybrid model by using the physical values of the meson masses and the decay constants in the final formulae for the amplitudes.  

\subsubsection{Meson decay constants}

The decay constants for the {\bf light} pseudoscalar $P$ and vector $V$ mesons are defined as 
\begin{equation}
\label{decay.constants}
\langle 0|j_{V-A}^{\mu}|P(p)\rangle =if_P p^{\mu}~,\qquad \langle 0|j_{V-A}^\mu|V(\epsilon,p)\rangle=g_V\epsilon^\mu~,
\end{equation}
where $j_{V-A}^\mu$ are properly normalized currents of the form $\bar q_2\gamma^\mu(1-\gamma_5)q_1$ with the same flavour structure as the corresponding meson. 
The decay constants $f_P$ and $g_V$ are given by the weak current (\ref{current.light}): $f_{\pi^+}=f$, $g_{\rho^+}=af^2\tilde g_V /\sqrt{2}$ and $SU(3)$ related expressions for other light mesons.  As anticipated above, I prefer to use the measured  values for $f_P$ and $g_V$. The decay constants $f_P$ and $g_V$ are measured in the leptonic decays $P^+\to \mu^+ \nu_\mu$ and  $V^0\to e^+e^-$, respectively. Their values based on the data from \cite{PDG} and are collected in Table \ref{const.tab}. At this point we can set the value of the constant $f$ to the measured value of the pion decay constant
\begin{equation}
\label{f}
f=132~{\rm MeV}~.
\end{equation}
With $a=2$ (\ref{a}) and $\tilde g_V=5.8$ (to be determined in (\ref{gv}) bellow),  the prediction for $g_{\rho^+}$ is $g_{\rho^+}=af^2\tilde g_V /\sqrt{2}=0.15$ GeV$^2$, which is close to the measured value given in Table \ref{const.tab}. 

\vspace{0.1cm}

The  decay constants for the {\bf heavy} pseudoscalar $f_D$ and vector $f_{D^*}$ mesons are defined  as
\begin{equation*}
\langle 0|q_a\gamma^\mu(1-\gamma_5)c|D_a(p)\rangle\! =\!-if_{D_a} p^{\mu},\qquad \langle 0|q_a\gamma^\mu(1-\gamma_5)c|P(\epsilon,p)\rangle\!=\! if_{D_a^*}m_{D_a^*}\epsilon^\mu.
\end{equation*}
 The  pseudoscalar and vector decays constants  are equal in the heavy quark limit $m_Q\to \infty$. They are given $f_D=f_{D^*}=\alpha/\sqrt{m_D}$ by the first term of the weak current (\ref{current.heavy}). The physical values of the heavy meson decay constants are uncertain at present. The value of $f_{D_s^+}$ based on the observation of  $D^+_s\to \mu^+\nu_\mu$ and $D_s^+\to \tau^+\nu_\tau$ channels has been reported by several experiments and extends over a wide range $f_{D_s^+}=194-430$ MeV. For $f_{D^+}$ only the experimental upper bound  is available at present \cite{PDG}. In absence of the reliable experimental data, one is forced to use the theoretical estimates of the decay constants. I will relay on the heavy quark predictions $f_{D^*}\!=\!f_D$, $f_{D_s^*}\!=\!f_{D_s}$ and take $f_{D}$ and  $f_{D_s}$, given  in Table \ref{const.tab}, from the recent lattice QCD results \cite{f.lattice}.  Using the value of decay constant $f_D$, the parameter $\alpha$ can be set to $\alpha=f_D m_D\simeq 0.29$ GeV$^2$.  

\subsubsection{Meson masses}

The measured masses for the light and the heavy mesons, that are used in this chapter, are gathered  in Table \ref{const.tab}. 

Let me briefly comment on the meson masses that arise from the heavy meson chiral Lagrangian given above.  After the explicit breaking of the chiral symmetry, the square masses of the light pseudoscalars are given in terms of the quark masses and the parameter $\lambda_0$  (\ref{chiral.breaking}). The masses of the light vector mesons are given by the term proportional to the parameter $a$ in ${\cal L}^{light}$ (\ref{hybrid}) as $m_V^2=\tfrac{1}{2}a\tilde g_V^2f^2$. This relation fixes \cite{casalbuoni,hidden} the parameter $\tilde g_V$  to \begin{equation}
\label{gv}
\tilde g_V=5.8~
\end{equation} 
 for  $a=2$ (\ref{a}), $f=132$ MeV (\ref{f}) and $m_\rho=0.77$ GeV.

The masses of the heavy mesons do not explicitly appear in 
the Lagrangian (\ref{hybrid}) and the weak currents (\ref{current.light}, \ref{current.heavy}) due to the  heavy quark flavour symmetry in the limit $m_Q\to \infty$. The masses enter the amplitude only due to the factor $\sqrt{m_H}$, which  corresponds to an  external heavy meson. This factor  is due to the 
normalization of the field operators $D^v(x)$ and $D^{*v}(x)$ in (\ref{5.105}). In the hybrid model additional dependence on the heavy meson masses arises because  the full heavy meson propagators (\ref{5.106}) are used instead of the HQET propagators (\ref{5.106a}).   

\begin{table}[ht]
\begin{center}
\begin{tabular}{|c|c|c||c|c|c||c|c|c|c|}
\hline
$H$ & $m_H$ & $f_H$ & $P$ & $m_P$ & $f_P$ & $V$ & $m_V$ & $g_V$ & $\Gamma_V$ \\
&[GeV] & [GeV] & & [GeV] & [GeV] &  & [GeV] & [GeV$^2$] & [GeV]\\
\hline
\hline
$D$ & $1.87$ & $0.21 \pm 0.04~$ & $\pi$ & $0.14$ & $0.135$ & $\rho$ & $0.77$ & 
$0.17$ & $0.15$ \\
$D_s$ & $1.97$ & $0.24 \pm  0.04~$ & $K$ & $0.50$ & $0.16$ & $K^*$ & $0.89$ & $0.19$ 
& $0.050$ \\
$D^*$ & $2.01$ & $0.21 \pm  0.04~$ & $\eta$ & $0.55$ & $0.13$ & 
$\omega$ & $0.78$ & $0.15$&$0.0084$  \\
$D_s^*$ & $2.11$ & $0.24 \pm  0.04~$ & $\eta'$ & $0.96$ & $0.11$ & 
$\phi$ & $1.02$ & $0.24$&$0.0044$\\
\hline
\end{tabular}
\end{center}
\caption{The measured values of the meson messes, decay constants and decay widths \cite{PDG}. The measured decay constants  $f_D$ and $f_{D^*}$ have sizable uncertainties and the values are taken from lattice QCD results \cite{f.lattice}. }
\label{const.tab}
\end{table}

\section{The parameters of the heavy meson chiral 
         Lagrangian and application to the 
         charm  meson semileptonic decays}

It this section I discuss the values of the parameters $g$, $C_{VV\Pi}$, $\lambda$, $\lambda^\prime$ present in the Lagrangian (\ref{hybrid}) and the values of the parameters $\alpha_1$, $\alpha_2$ present in the weak current (\ref{current.heavy}). The value of the parameter $\beta$ (\ref{hybrid}) will be discussed in the study of nonleptonic decays in Section 5.3. The value of the parameter $\kappa$ (\ref{hybrid}) will not be discussed in this work as it does not enter any decay amplitudes of interest\footnote{The parameter $\kappa$ is responsible for the shape of the bremsstrahlung  $D\to D\gamma^*$. Such bremsstrahlung contribution  is kinematically forbidden in $D\to V\gamma$ and $D\to Vl^+l^-$ decays. In the $D\to P l^+l^-$ decays, this particular bremsstrahlung contribution is proportional to $m_P^2$ and can be neglected in comparison with the terms proportional to $m_D^2$.}. Except for the parameter $\lambda^\prime$, these parameters can not be determined by exploiting the chiral and heavy quark symmetries further. They are free parameters of the effective field theory and have to be determined either by the measurement or by using  some other  model of the strong interaction. 

In this section I demonstrate also for the applicability of the hybrid model by predicting the  charm meson semileptonic decay rates, as done in \cite{BFO1}.

\subsubsection{The parameters $\boldsymbol{\lambda}$, $\boldsymbol{\alpha_1}$ and $\boldsymbol{\alpha_2}$ and semileptonic $\boldsymbol{D\to Vl^+\mu_l}$ decays} 
 
The amplitudes for the semileptonic decays $D\to Vl^+\nu_l$ with a light vector meson $V$ in the final state depend on the parameters $\lambda$, $\alpha_1$, $\alpha_2$  of the hybrid model. The same parameters appear also in the amplitudes for the corresponding $B$ meson semileptonic decays. The three parameters $\lambda$, $\alpha_1$ and $\alpha_2$ are fitted from the experimental data on $D^+\to \bar K^{*0}e^+\nu_e$ decay, as proposed in \cite{BFO1}.

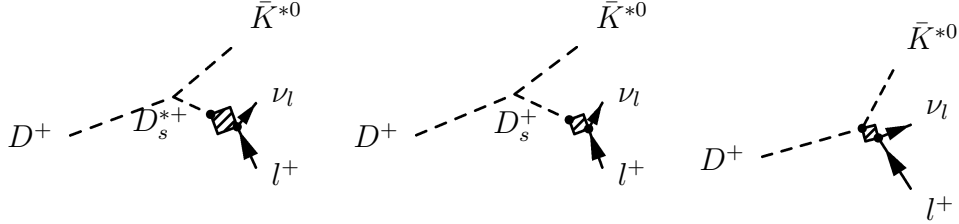
\begin{figure}[h]
\centering
\mbox{
\begin{fmffile}{f26o1}
\fmfframe(5,3)(5,3){
  \begin{fmfgraph*}(25,25)
  \fmfpen{thin}  
  \fmfleftn{l}{1} \fmfright{p1,r1,r2,r3}
  \fmfrpolyn{shaded,tension=0.7}{k}{4} 
  \fmfv{decor.size=1.2thick,decor.shape=circle,decor.filled=full}{k1,k3}
  \fmf{dashes}{l1,v1} 
  \fmf{dashes,label=$D_s^{*+}$,tension=1,la.d=10}{v1,k1}
  \fmf{dashes}{v1,r3} 
  \fmf{fermion}{r1,k3,r2}
  \fmflabel{$l^+$}{r1}\fmflabel{$\bar K^{*0}$}{r3}
  \fmflabel{$D^+$}{l1}
  \fmflabel{$\nu_l$}{r2}
  \end{fmfgraph*} }
\end{fmffile}
\quad
\begin{fmffile}{f26o2}
\fmfframe(5,3)(5,3){
  \begin{fmfgraph*}(25,25)
  \fmfpen{thin}  
  \fmfleftn{l}{1} \fmfright{p1,r1,r2,r3}
  \fmfrpolyn{shaded,tension=0.7}{k}{4} 
  \fmfv{decor.size=1.2thick,decor.shape=circle,decor.filled=full}{k1,k3}
  \fmf{dashes}{l1,v1} 
  \fmf{dashes,label=$D_s^{+}$,tension=0.5,la.d=10}{v1,k1}
  \fmf{dashes}{v1,r3} 
  \fmf{fermion}{r1,k3,r2}
  \fmflabel{$l^+$}{r1}\fmflabel{$\bar K^{*0}$}{r3}
  \fmflabel{$D^+$}{l1}
  \fmflabel{$\nu_l$}{r2}
  \end{fmfgraph*} }
\end{fmffile}
\quad
\begin{fmffile}{f26o3}
\fmfframe(5,0)(5,0){
  \begin{fmfgraph*}(20,25)
  \fmfpen{thin}  
  \fmfleftn{l}{1} \fmfright{p1,r1,r2,r3}
  \fmfrpolyn{shaded,tension=2}{k}{4} 
  \fmfv{decor.size=1.2thick,decor.shape=circle,decor.filled=full}{k1,k3}
  \fmf{dashes}{l1,k1}
  \fmf{dashes}{k1,r3} 
  \fmf{fermion}{r1,k3,r2}
  \fmflabel{$l^+$}{r1}\fmflabel{$\bar K^{*0}$}{r3}
  \fmflabel{$D^+$}{l1}
  \fmflabel{$\nu_l$}{r2}
  \end{fmfgraph*} }
\end{fmffile}
    }
\caption{The diagrams for the semileptonic decay $D^+\to \bar K^{*0}l^+\nu_l$ in
the hybrid model.  
The box denotes the action of the nonleptonic effective Lagrangian (\ref{eff}). The two dots in the box denote the two weak currents.}
\label{fig26}
\end{figure}

The Feynman diagrams for $D^+\to \bar K^{*0}e^+\nu_e$ decay in the hybrid model are shown in Fig.  \ref{fig26}. Analogous diagrams contribute to other semileptonic decays. The matrix element of the weak current is expressed in terms of the form factors as  \cite{BSW,casalbuoni} \footnote{There was an additional factor $i$ in the definition (\ref{4.1a}), since the mesons had different time reversal and charge conjugation assignments in the quark model. The signs of the form factors $V$ and $A_0$ are reversed compared to \cite{BFO1}.} 
\begin{align}
\label{5.55} 
\langle V&(p^\prime,\epsilon^\prime)|\bar 
q\gamma^{\mu}(1-\gamma_5)c|D(p)\rangle=\frac{2}{
 m_{D}+m_{V}}\epsilon^{\mu\alpha\beta\gamma}\epsilon_{\alpha}^{*\prime}p_{\beta}p_{\gamma}^\prime V(q^2)+i2m_{V}{\epsilon^{*\prime}\cdot q\over q^2}q^{\mu}A_0(q^2)\\
&+i(m_{D}+m_{V})\bigl[\epsilon^{\mu *\prime}-{\epsilon^{*\prime} \cdot q\over q^2}q^{\mu}\bigr]A_1(q^2)-i{\epsilon^{*\prime}\cdot q\over 
m_{D}+m_{V}}\bigl[(p+p^{\prime})^{\mu}
-{m_{D}^2-m_{V}^2\over q^2}q^{\mu}\bigr]A_2(q^2)~\nonumber
\end{align}
with $\epsilon^{0123}\equiv 1$. The matrix element  is finite at $q^2=0$ and the form factors must satisfy the relation
\begin{equation}
\label{5.35}
2m_{V}A_0(0)= (m_{D}+m_{V})A_1(0)-(m_{D}-m_{V})A_2(0)~.
\end{equation}
 In the hybrid model the form factors are calculated from the diagrams in Fig.  \ref{fig26}  \cite{BFO1} \footnote{The expression for the form factor $A_0$ is corrected in comparison with \cite{BFO1}. The semileptonic rates as predicted by \cite{BFO1} do not change, however, as these do not depend on the form factor $A_0$ due to the small mass of the leptons $\mu$ and $e$.}
\begin{align}
\label{formV}
V(q^2)&=-K_{DV}\bigl[2(m_D+m_V)\sqrt{\frac {m_{D^{\prime *}}}{m_D}}
\frac{ m_{D^{\prime *}}}{q^2-m_{D^{\prime *}}^2}f_{D^{\prime *}}\lambda\bigr]
\frac{\tilde g_V}{\sqrt{2}}~,\nonumber\\
A_0(q^2)&=-K_{DV}\biggl[\frac{1}{m_V}\sqrt{\frac {m_{D^{\prime *}}}{m_D}}
\frac{q^2}{q^2-m_{D^\prime}^2}f_{D^\prime}\beta+\frac{\sqrt{m_D}}{m_V}\alpha_1-\frac{q^2+m_D^2-m_V^2}{2m_D^2}\frac{\sqrt{m_D}}{m_V}\alpha_2\biggr]\frac{\tilde g_V}{\sqrt{2}}~,\nonumber\\
A_1(q^2)&=K_{DV}\bigl[-\frac {2\sqrt{m_D}}{m_D+m_V}\alpha_1\bigr]\frac{\tilde g_V}{\sqrt{2}}~,\nonumber\\
A_2(q^2)&=K_{DV}\bigl[-\frac {m_D+m_V}{m_D\sqrt{m_D}}\alpha_2\bigr]\frac{\tilde g_V}{\sqrt{2}}~.
\end{align}
The poles $D^{\prime *}$ and the constants $K_{DV}$  are given in Table \ref{5.1.tab}. The form factors as predicted by  the hybrid model automatically satisfy the relation (\ref{5.35}). In the case of the pole assumption for   form factors, on the other hand, the pole masses have to be related in order to satisfy (\ref{5.35}); it is unreasonable to assume such a relation, since the pole masses are taken from the measurement and are not free parameters. 
The contribution of the form factor $A_0$ to semileptonic rate is negligible due to the small masses of the muon and the electron. The form factors $V$, $A_1$ and $A_2$ depend on the parameters $\lambda$, $\alpha_1$ and $\alpha_2$, respectively. The three parameters are fixed by using the experimental data on $\boldsymbol{D^+\to \bar K^{*0}e^+\nu_e}$ decay, which has been most precisely measured. By assuming the pole shape for each of the form factors, the experimenters have fitted the data to obtain $Br=4.8\pm 0.5\%$, $V(0)/A_1(0)=1.85\pm 0.12$ and $A_2(0)/A_1(0)=0.72\pm 0.09$, taken from the Particle Group Average of data from different experiments \cite{PDG}. In the hybrid model, the form factor $V$ has the polar shape, while $A_1$ and $A_2$ are flat, so it is not appropriate to fit theoretical form factors to the experimental ones at $q^2\!=\!0$.  The three observables  $\Gamma_L/\Gamma_T=1.23\pm 0.13$, $\Gamma_+/\Gamma_-=0.16\pm 0.04$ and $Br=4.7\pm 0.4\%$ \cite{PDG} are used instead. They give eight sets of the solutions for the three parameters $\lambda$, $\alpha_1$ and $\alpha_2$. The quark model calculation indicates that $\lambda<0$ \footnote{The parameter $\lambda$ is responsible for the magnitude of the $D\to D^*V^0\to D^*\gamma$ coupling and can be related to the magnetic moment of the light antiquark $\mu_a=e_a/M_a$ in the heavy meson $Q\bar q_a$. Here $M_a$ is interpreted as the mass parameter in the constituent quark model. In the hybrid model and in the exact $SU(3)_V$ flavour limit $M_a^{-1}=-4\lambda$, so $\lambda<0$ if $M_a$ is  a mass parameter \cite{casalbuoni}.}. Among the four sets with $\lambda<0$ given in Table \ref{set.tab}, I choose the set 2  \footnote{ The updated values for the heavy meson decays constants are used here and the calculated $\lambda$, $\alpha_1$ and $\alpha_2$ are different to the ones in \cite{BFO1}.}
\begin{equation}
\label{lambda}
\lambda\!=\!-0.38\pm 0.07~{\rm GeV}^{-1}~,~~~~ \alpha_1\!=\!0.14\pm 0.01~{\rm GeV}^{1/2}~,~~~~ \alpha_2\!=\!0.10\pm 0.03~{\rm GeV}^{1/2}~.
\end{equation}
This set gives $V(0)\!=\!-1.0\pm 0.2$, $A_1(0)\!=\!-0.55\pm 0.05$, $A_2(0)\!=\!-0.43\pm 0.13$ and the corresponding ratios  $V(0)/A_1(0)=1.8\pm 0.2$, $A_2(0)/A_1(0)=0.8\pm 0.2$ are close to the experimental data. Other three sets with $\lambda<0$ give the ratios far from the experimental data and can be excluded since the ratios are not very strongly dependent on the  shapes of the form factors.

\begin{table}[ht]
\begin{center}
\begin{tabular}{|c|c|c|c|}\hline
& $\lambda$ [GeV$^{-1}$]
& $\alpha_1$ [GeV$^{1/2}$]
& $\alpha_2$ [GeV$^{1/2}$]  \\ \hline
set 1 & $-0.38 \pm 0.07$ & $0.14 \pm 0.01$ & $0.83 \pm 0.04$\\
set 2 & $-0.38 \pm 0.07$ & $0.14 \pm 0.01$ & $0.10 \pm 0.03$\\
set 3 & $-0.82 \pm 0.14$ & $0.064 \pm 0.007$ & $0.60 \pm 0.03$\\
set 4 & $-0.82 \pm 0.14$ & $0.064 \pm 0.007$ & $-0.18 \pm 0.03$\\ \hline
\end{tabular}
\caption{Four solutions for the model parameters
as determined from the experimental data on decay $D^+\to\bar{K}^{*0}l^+\nu_l$.}
\label{set.tab}
\end{center}
\end{table}

The hybrid model predictions for the remaining {\bf $\boldsymbol{D\to Ve^+\nu_e}$ semileptonic decays} are given  in Table \ref{5.1.tab} and agree well with the available experimental data. The quoted errors do not include any systematic errors related to the validity of the model and are due only to uncertainties of the input parameters. The corrections due to the chiral and $1/m_c$ expansion are expected to effect the result and the combined error is of the order of $30~\%$. 
   
\begin{table}[ht]
\begin{center}
\begin{tabular}{|c||c|c|c||c|c|c|}
\hline
decay&$D^{\prime *}$&$D^\prime$&$K_{DV}V_{CKM}$& $Br[\%]$&$\tfrac{\Gamma_L}{\Gamma_T}$&$\tfrac{\Gamma_+}{\Gamma_-}$\\
\hline
\hline
$D^+\to\bar K^{*0}$&$D_s^{*+}$&$D_s^+$&$\cos\theta_C$&*&*&*\\
\hline
$D^0\to K^{*-}$&$D_s^{*+}$&$D_s^+$&$\cos\theta_C$&$1.9\pm 0.2$&$1.23\pm 0.13$&$0.16\pm 0.04$\\
                &                  &              & &$[2.02\pm 0.33]^{exp}$& & \\
\hline
$D^+_s\to\phi$&$D_s^{*+}$&$D_s^+$&$\cos\theta_C$&$1.7\pm 0.1$&$1.2\pm 0.1$&$0.16\pm 0.04$\\
              &          &       &               &$[2.0\pm 0.5]^{exp}$&$[0.72\pm 0.18]^{exp}$& \\
\hline
$D^0\to\rho^-$&$D^{*+}$&$D^+$&$\sin\theta_c$&$0.17\pm 0.02$&$1.4\pm 0.2$&$0.15\pm 0.10$\\
\hline
$D^+\to\rho^0$&$D^{*+}$&$D^+$&$-\sin\theta_C/\sqrt{2}$&$0.22\pm 0.02$&$1.4\pm 0.2$&$0.15\pm 0.10$\\
              &        &     &                                  &$[0.22\pm 0.08]^{exp}$& & \\
 \hline
$D^+\to\omega$&$D^{*+}$&$D^+$&$\sin\theta_C/\sqrt{2}$&$0.21\pm 0.02$&$1.4\pm 0.2$&$0.16\pm 0.10$\\
\hline
$D_s^+\to K^{*0}$&$D^{*+}$&$D^+$&$\sin\theta_C$&$0.17\pm 0.02$&$1.3\pm 0.2$&$0.15\pm 0.10$\\
\hline
\end{tabular}
\end{center}
\caption{ The hybrid model predictions for $Br$, $\Gamma_L/\Gamma_T$ and $\Gamma_+/\Gamma_-$  are given in the last three columns together with the available experimental data \cite{PDG} in the brackets. The $*$ denotes that the experimental data has been used to fix the free parameters. The quoted errors take into account only the experimental uncertainties in the input parameters, but not the validity of the model. The second and third column give the resonance states $D^\prime$ and $D^{\prime *}$ for $D\to Ve^+\nu_e$ diagrams in Fig.  \ref{fig26}.}
\label{5.1.tab}
\end{table}

\subsubsection{The parameter $\boldsymbol{\lambda^\prime}$ and the heavy quark symmetry}   

The parameter $\lambda^\prime$ is the only parameter of this effective field theory that can be determined by exploiting the symmetries further. The $\lambda^\prime$ term in ${\cal L}^{heavy}$ (\ref{hybrid}) gives rise to the $D^*D\gamma$ coupling and describes the contribution of the magnetic moment of the {\it  heavy quark} in a heavy meson. The magnetic moment of the {\it light quark} is incorporated by the vector meson dominance mechanism $D\to D^*V^0$ followed by transition $V^0\to \gamma$. The $D\to D^*V^0$ coupling is given by the $\lambda$ term in (\ref{hybrid}), while  $V^0\to \gamma$ is given by the vector meson dominance (\ref{VMD}) bellow.
Assuming that  $\lambda^\prime$ term accounts only for the photon emission from the charm quark in the charm meson 
\begin{equation}
\label{5.107}
{\cal A}[D^*(v,\eta)\to D\gamma(q,\epsilon)]^{{emission\atop from~c}}=4\lambda^\prime e_0\epsilon_{\mu\nu\alpha\beta}\epsilon^\mu\eta^\nu v^\alpha q^{\beta} \sqrt{m_D m_{D^*}}\!=\!-ie_c\epsilon^\mu\langle D|\bar c\gamma_\mu c|D^*\rangle~.
\end{equation}
The quantity on the right hand side can be expressed in terms of the well-known Isgur-Wise function \cite{neubert,HQET1,casalbuoni}\footnote{The matrix elements related by the heavy quark spin or flavour symmetry, i.e. $\langle B(v^\prime)|h^{(b)}_{v^\prime}\Gamma h_v^{(b)}|B(v)\rangle$, $\langle D(v^\prime)|h^{(c)}_{v^\prime}\Gamma h_v^{(b)}|B(v)\rangle$, $\langle D^*(v^\prime)|h^{(c)}_{v^\prime}\Gamma h_v^{(b)}|B(v)\rangle$ and $\langle D(v^\prime)|h^{(c)}_{v^\prime}\Gamma h_v^{(c)}|D^*(v)\rangle$, can be expressed in terms of a single Isgur-Wise function as $\langle M^\prime(v^\prime)|\bar h^\prime\Gamma h|M(v)\rangle=-\xi (v\cdot v^\prime) Tr[\bar H^\prime (v^\prime)\Gamma H(v)]$ \cite{neubert}. The field $H$ is given in (\ref{ha}) and $\Gamma$ denotes a general combination of the Dirac matrices. The value of the matrix element  $\langle M(v)|\bar h\gamma^0 h|M(v)\rangle$ is given by the normalization of the meson wave functions and renders the Isgur-Wise function at the point of equal velocities $\xi(1)=1$.}
$$-ie_c\epsilon^\mu\langle D(v^\prime)|\bar c\gamma_\mu c|D^*(v,\eta)\rangle=\tfrac{2}{3}e_0\sqrt{m_D m_{D^*}}\xi(v\cdot v^\prime)\epsilon_{\mu\nu\alpha\beta}\epsilon^\mu\eta^\nu v^\alpha v^{\prime\beta}~.$$
The value of the Isgur-Wise function at $v=v^\prime$ is given by the normalization of the meson wave function, $\xi(1)=1$. The value of $\lambda^\prime$ is given by the equation (\ref{5.107}) taken at $v=v^\prime$
\begin{equation}
\label{lambda.prime}
\lambda^\prime=-\frac{1}{6m_D^*}\simeq-0.083~{\rm GeV}^{-1}~. 
\end{equation}

 \subsubsection{Coupling $\boldsymbol{g}$, vertices $\boldsymbol{DD^*\pi}$, $\boldsymbol{DD^*\gamma}$ and  semileptonic decays $\boldsymbol{D\to Pl^+\nu_l^-}$}

In the lowest order of the heavy quark and the chiral expansion, the strong interactions among the heavy mesons and the low energy light pseudoscalars are expressed in terms of a single coupling $g$ introduced in (\ref{5.8}). This coupling  gives the magnitude of $D^*D\Pi$, $D^*D^*\Pi$, $B^*B\Pi$ and  $B^*B^*\Pi$ vertices, where $\Pi$ denotes one or more low energy light pseudoscalars. The theoretical predictions for the value of $g$ range from $0.15-1.0$ \cite{casalbuoni,stewart}. The parameter $g$ would be most directly determined phenomenologically by measuring the $D^*\to D\pi$ rate. Unfortunately,  only the branching fractions for $D^*\to D\pi$ and $D^*\to D\gamma$ have been measured, but the total width of $D^*$ is still unknown. The value of $g$ can be determined from the measured ratios \cite{PDG,stewart}
\begin{align}
\label{5.30}
R_\gamma^0\equiv Br(D^{*0}\to D^0\gamma)/Br(D^{*0}\to D^0\pi^0)&=0.616\pm 0.076~,\\
R_\gamma^+\equiv Br(D^{*+}\to D^+\gamma)/Br(D^{*+}\to D^+\pi^0)&=0.055\pm 0.017~.\nonumber
\end{align}
I will evaluate these ratios in the hybrid model and then use the experimental data to determine $g$. The $D^*\to D\pi$ is induced by the term proportional to $g$ in the ${\cal L}^{heavy}$ (\ref{hybrid}). The photon emission from the heavy quark in the  $D^*\to D\gamma$ decay is given by the $\lambda^\prime$ term in (\ref{hybrid}). The photon emission  from the light degrees of freedom proceeds via the exchange of the light vector meson $D^*\to DV^0\to D\gamma$ and the  coupling $D^*DV$ is given by the $\lambda$ term in (\ref{hybrid}). The $V^0\!\to \!\gamma$ conversion is given by the  term proportional to $a\!=\!2$ (\ref{a})  in ${\cal L}^{light}$ (\ref{hybrid}) 
\begin{equation}
\label{VMDexact}
{\cal L}_{V\!\gamma} = -e_0 {\tilde g}_V f^2 (\rho^{0\mu} + \tfrac{1}{3} 
\omega^{\mu} - \tfrac{{\sqrt 2}}{3} \phi^{\mu})~ A_\mu~.
\end{equation}
Instead of using the exact $SU(3)$ symmetry values $\tilde g_V=5.8$ and 
$f=132$ MeV, I follow the discussion in the section on the flavour  $SU(3)_V$ breaking effects and I express the couplings $V^0\gamma$  in terms of the measurable quantities $g_{\rho}$, $g_{\omega}$, $g_{\phi}$ defined in (\ref{decay.constants}) as in \cite{FPS1,FS,FPS2}
\begin{equation}
\label{VMD}
{\cal L}_{V\gamma} = -\tfrac{e_0}{\sqrt{2}}  (g_{\rho}\rho^{0\mu} + 
\tfrac{g_{\omega}}{3} \omega^{\mu} - \tfrac{\sqrt{2}g_{\phi}}{ 3}\phi^{\mu})~A_{\mu}.
\end{equation}
With these ingredients   
\begin{align*}
R_\gamma^0&=16 e_0^2f^2\biggl(\frac{|\vec p_\gamma|}{|\vec p_\pi|}\biggr)^3\biggl[\frac{\lambda^{\prime}}{g}+\frac{\tilde g_V}{2\sqrt{2}}\bigl(\frac{g_\rho}{m_\rho^2}+\frac{g_\omega}{3m_\omega^2}\bigr)\frac{\lambda}{g}\biggr]^2~,\\
R_\gamma^1&=16 e_0^2 f^2\biggl(\frac{|\vec p_\gamma|}{|\vec p_\pi|}\biggr)^3\biggl[\frac{\lambda^{\prime}}{g}-\frac{\tilde g_V}{2\sqrt{2}}\bigl(\frac{g_\rho}{m_\rho^2}-\frac{g_\omega}{3m_\omega^2}\bigr)\frac{\lambda}{g}\biggr]^2
\end{align*}
and the experimental data in (\ref{5.30}) gives\footnote{In absence of the $SU(3)_V$ flavour violation, the numerical factors $0.77$ and $-0.42$ would be replaced by the charges $2/3$ and $-1/3$, respectively}
$$|\lambda^\prime+0.77\lambda|=(0.84\pm 0.05) |g|\qquad {\rm and}\qquad |\lambda^\prime-0.42\lambda|=(0.214\pm 0.03) |g|~.$$
The values of the parameters $\lambda=-0.38\pm 0.07$ (\ref{lambda}) and $\lambda^\prime=-0.058$ (\ref{lambda.prime}) can be independently checked by evaluating the ratio $|\lambda^\prime+0.77\lambda|/|\lambda^\prime-0.42\lambda|$ giving $5.7\pm 2.5$, compared to the experimental ratio $3.9\pm 0.9$. Considering the simplicity of the model and the quoted errors this is in reasonable agreement. The value of $|g|$ is obtained by taking $\lambda=-0.38\pm 0.07$ (\ref{lambda}) and $\lambda^\prime=-0.058$ (\ref{lambda.prime}) 
\begin{equation}
\label{5.217}
|g|=0.44\pm 0.10\quad {\rm from}~~R_\gamma^0\quad\qquad{\rm and}\quad\qquad |g|=0.36\pm 0.15\quad {\rm from}~~R_\gamma^+~.
\end{equation}
These values are roughly compatible with the value $|g|=0.27\pm 0.1$, which has been obtained  from the experimental data on $R_\gamma^0$ and $R_\gamma^+$ (\ref{5.30}) by using the heavy meson chiral Lagrangian to order $1/m_c$ and $E^2$ \cite{stewart}. The semileptonic decays $D\to Pl^+\bar \nu_l$ studied in the next section favor the value $|g|=0.15$.  I choose to use the value of $g$ obtained from $R_\gamma^0$ and $R_\gamma^+$ in \cite{stewart}
\begin{equation}
\label{gabs}
|g|=0.27\pm 0.1~.
\end{equation}
This value is somewhere in between the hybrid model values  of $|g|=0.4\pm 0.2$, which is favored by $R_\gamma^0$, $R_\gamma^+$ (\ref{5.217}), and $|g|=0.15$ favored by semileptonic decays $D\to Pl^+\nu_l$.

\vspace{0.1cm}

\begin{figure}[h]
\centering
\mbox{
\begin{fmffile}{f27o1}
\fmfframe(5,3)(5,3){
  \begin{fmfgraph*}(25,25)
  \fmfpen{thin}  
  \fmfleftn{l}{1} \fmfright{p1,r1,r2,r3}
  \fmfrpolyn{shaded,tension=0.7}{k}{4} 
  \fmfv{decor.size=1.2thick,decor.shape=circle,decor.filled=full}{k1,k3}
  \fmf{dashes}{l1,v1} 
  \fmf{dashes,label=$D^{*+}$,tension=0.7,la.d=10}{v1,k1}
  \fmf{dashes}{v1,r3} 
  \fmf{fermion}{r1,k3,r2}
  \fmflabel{$l^+$}{r1}\fmflabel{$\pi^-$}{r3}
  \fmflabel{$D^0$}{l1}
  \fmflabel{$\nu_l$}{r2}
  \end{fmfgraph*} }
\end{fmffile}
\qquad\quad
\begin{fmffile}{f27o2}
\fmfframe(5,3)(5,3){
  \begin{fmfgraph*}(20,25)
  \fmfpen{thin}  
  \fmfleftn{l}{1} \fmfright{p1,r1,r2,r3}
  \fmfrpolyn{shaded,tension=1.5}{k}{4} 
  \fmfv{decor.size=1.2thick,decor.shape=circle,decor.filled=full}{k1,k3}
  \fmf{dashes}{l1,k1}
  \fmf{dashes}{k1,r3} 
  \fmf{fermion}{r1,k3,r2}
  \fmflabel{$l^+$}{r1}\fmflabel{$\pi^-$}{r3}
  \fmflabel{$D^0$}{l1}
  \fmflabel{$\nu_l$}{r2}
  \end{fmfgraph*} }
\end{fmffile}
    }
\caption{The diagrams for the semileptonic decay $D^0\to \pi^-l^+\nu_l$ in
the hybrid model.  
The box denotes the action of the nonleptonic effective Lagrangian (\ref{eff}). The two dots in the box denote the two weak currents.}
\label{fig27}
\end{figure}
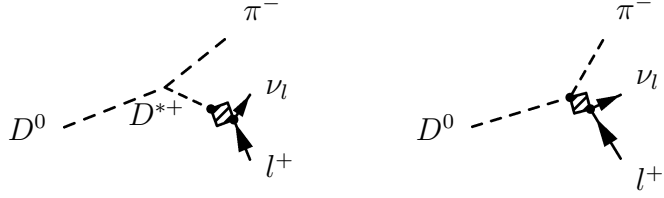

Having chosen the value of $g$, the consistency of the model can be checked by applying it to the {\bf semileptonic $\boldsymbol{D\to Pl^+\nu_l}$ decays}, where $P$ is a light pseudoscalar meson. The diagrams for $D^0\to \pi^-l^+\nu_l$ decay are shown in Fig.  \ref{fig27}. Similar diagrams apply to other decays. The matrix of the weak current is parameterized in terms of the form factors 
\begin{equation}
\label{defP}
\langle P(p^\prime)|\bar q\gamma_\mu(1-\gamma_5)c|D(p)\rangle=[(p+p^\prime)_\mu-\tfrac{m_D^2-m_P^2}{q^2}q_\mu]f_1(q^2)+\tfrac{m_D^2-m_P^2}{q^2}q_\mu f_0(q^2)~.
\end{equation}
The matrix element is finite at $q^2=0$ and imposes the relation
\begin{equation}
\label{5.35a}
f_1(0)=f_0(0)~.
\end{equation}
 In the hybrid model this relation is automatically satisfied and the form factors are given by the diagrams in Fig. \ref{fig27} as \cite{BFO1} \footnote{The expression for $f_0$ in \cite{BFO1} is not correct, but this form factor has negligible contribution to the semileptonic decays.}
\begin{align}
\label{formP}
f_1(q^2)&=K_{DP}\biggl[-\frac{f_D}{2}+gf_{D^{\prime *}}\frac{m_{D^{\prime *}}\sqrt{m_Dm_{D^{\prime *}}}}{q^2-m_{D^{\prime *}}^2}\biggr]~,\\
f_0(q^2)&=K_{DP}\biggl[-\frac{f_D}{2}-gf_{D^{\prime *}}\sqrt{\frac{m_D}{m_{D^{\prime *}}}}+\frac{q^2}{m_D^2-m_P^2}\bigl(-\frac{f_D}{2}+gf_{D^{\prime *}}\sqrt{\frac{m_D}{m_{D^{\prime *}}}}\bigr)\biggr]~\nonumber.
\end{align}
The constants $K_{DP}$ and the corresponding poles $D^{\prime *}$ are given in Table \ref{5.2.tab}.
The contribution of the form factor $f_0$ to the decay rate is negligible due to the small lepton masses.
 By comparing the hybrid model predictions for $g=\pm 0.27\pm 0.1$ with the experimental data, the possibility $g=-0.27$ can be excluded as it gives too small branching ratios for all the measured decays\footnote{The parameters$g=0.27\pm 0.1$ and $g=-0.27\pm 0.1$ predict $Br(D^0\to \bar K^- e^\nu_e)$ at $5.5\pm 1.5~\%$ and $0.24{+0.4\atop -0.2}$, respectively, while the measured branching ratio is $3.66\pm 0.18~\%$.},  so
\begin{equation}
\label{g}
g=0.27\pm 0.1~.
\end{equation}
The positive sign of $g$ is in agreement with the constituent quark model  which gives $g=1$. 

The hybrid model predictions for the  semileptonic branching ratios $Br^{hybrid}$ are given in Table \ref{5.2.tab}. They are slightly to high compared to the experimental data, but the agreement is still reasonable given the simplicity of the model and the uncertainty in the parameter $g=0.27\pm 0.1$ (the semileptonic  data favors the value $g\simeq 0.15$).  The hybrid model results $Br^{hybrid}$ are compared with those obtained by applying the heavy meson chiral Lagrangian  at $q_{max}^2=(m_D-m_P)^2$ \footnote{The velocity of the light pseudoscalar in the $D$ meson rest frame  at $q_{max}^2$ is zero and the chiral expansion is valid; at the same time the virtuality of the intermediate heavy meson $k=-m_Pv$ is small and the heavy quark expansion is valid.} and by assuming the polar behavior of the form factor $f_1$ \cite{heavy.chiral,casalbuoni} 
\begin{equation}
\label{5.30a}
f_1(q^2)=f_1(q_{max}^2)\frac{q^2_{max}-m_{D^{\prime *}}^2}{q^2-m_{D^{\prime *}}^2}=-K_{DP}~g~\frac{f_D}{2}~\frac{m_D-m_P}{m_P+m_{D^{\prime *}}-m_D}~\frac{m_{D^{\prime *}}^2-q^2_{max}}{m_{D^{\prime *}}^2-q^2}~.
\end{equation}
The branching ratios $Br^{pole}$ based on $f_1$ (\ref{5.30a}) and $g=0.27\pm 0.1$ are too small compared to the experimental data and call for higher value of the parameter $g$. This indicated that the $q^2$ shape as suggested by the hybrid model (polar shape for the diagram in Fig.  \ref{fig27}a and flat shape for the diagram in Fig.  \ref{fig27}b) is more suitable than the pole assumption assumed in (\ref{5.30a}).   

\begin{table}[h]
\begin{center}
\begin{tabular}{|c||c|c||c|c||c|}
\hline
$D\to Pe^+\nu_e$& $D^{\prime *}$&$K_{DP}V_{CKM}$&$Br^{hybrid}[\%]$&$Br^{pole}[\%]$&$Br^{exp}[\%]$\\
\hline
$D^0\to K^-$&$D_s^{*+}$&$\cos\theta_C/f_K$&$5.5\pm 1.5$&$0.4\pm 0.2$&$3.66\pm 0.18$\\
$D^+\to \bar K^0$&$D_s^{*+}$&$\cos\theta_C/f_K$&$14\pm 4$&$1.0\pm 0.5$&$6.7\pm 0.9$\\
$D_s^+\to \eta$&$D_s^{*+}$&$K_\eta^s\cos\theta_C/f$&$4.7\pm 1.2$&$0.4\pm 0.2$&$2.5\pm 0.7$\\
$D^0\to \pi^-$&$D^{*+}$ &$\sin\theta_C/f_\pi$&$ 0.68\pm 0.18$&$0.09\pm 0.04$&$0.37\pm0.06$\\
$D^+\to\pi^0$&$D^{*+}$&$-\tfrac{1}{\sqrt{2}}\sin\theta_C/f_\pi$&$0.89\pm 0.24$&$0.12\pm 0.06$&$0.31\pm 0.15$\\
$D^+\to \eta$&$D^{*+}$&$K_\eta^d\sin\theta_C/f$&$0.26\pm 0.07$&$0.02\pm 0.01$&$<0.5$\\
$D^+_s\to K^0$&$D^{*+}$&$\sin\theta_C/f_K$&$0.49\pm 0.12$&$0.06\pm 0.03$&\\
\hline
\end{tabular}
\end{center}
\caption{The semileptonic $D\to Pe^+\nu_e$ decays: the constants $K_{DP}$ and poles $D^{\prime *}$ for the calculation of the form factors (\ref{formP}); the hybrid model predictions $Br^{hybrid}$ based on the form factor $f_1(q^2)$ (\ref{formP}); the strict HQET result extended by the pole assumption $Br^{pole}$ based on the form factor $f_1(q^2)$ (\ref{5.30a}); the experimental data $Br^{exp}$ \cite{PDG}. The quoted errors for the predicted branching ratios arise form the uncertainty in $g=0.27\pm 0.1$ (\ref{gabs}, \ref{g}). In the hybrid model, the value $g\!\simeq\! 0.15$ is favored by the semileptonic data. The constants $K_{\eta}^{d,s}$ are given in (\ref{mixing}).}
\label{5.2.tab}
\end{table}

\subsubsection{The coupling $\boldsymbol{C_{VV\Pi}}$ and decays $\boldsymbol{V\to P\gamma}$}

The coupling among two light vectors and a light pseudoscalar meson $C_{VV\Pi}$ (\ref{hybrid}) can be determined in the case of the exact $SU(3)_V$ flavour symmetry following the hidden symmetry approach of \cite{hidden,witten} and it is found to be $C_{VV\Pi}=3\tilde g_V^2/32\pi^2=0.33$. In this work, the coupling $C_{VV\Pi}$ is determined phenomenologically \cite{FS} from the experimental data on $V\to P\gamma$. This decays in the hybrid model occur via $V\to PV^0\to P\gamma$ and the $SU(3)_V$ flavour breaking is taken into account by using the physical values  of the masses and decay constants: 
$$\Gamma(V\to P\gamma)=\frac{e_0^2 }{96\pi}\frac{(m_V^2-m_P^2)^3}{m_V^3}|g_{VP\gamma}|^2$$
with
\begin{alignat*}{2}
g_{\omega\pi\gamma}&=4\frac{g_\rho}{m_\rho^2}\frac{C_{VV\Pi}}{f_\pi}~,&\qquad g_{K^{*+}K^+\gamma}&=2\bigl(\frac{g_\omega}{3m_\omega^2}+\frac{g_\rho}{m_\rho^2}-\frac{2}{3}\frac{g_\phi}{m_\phi^2}\bigr)\frac{C_{VV\Pi}}{f_K}~,\\
g_{\rho\pi\gamma}&=4\frac{g_\omega}{3m_\omega^2}\frac{C_{VV\Pi}}{f_\pi}~,&\qquad g_{K^{*0}K^0\gamma}&=2\bigl(\frac{g_\omega}{3m_\omega^2}-\frac{g_\rho}{m_\rho^2}-\frac{2}{3}\frac{g_\phi}{m_\phi^2}\bigr)\frac{C_{VV\Pi}}{f_K}~.
\end{alignat*}
I choose the value, which reproduces  the observed width $\Gamma(K^{*+}\to K^+\gamma)=(5.0\pm 0.5)\cdot 10^{-5}$ GeV \cite{PDG96},   
\begin{equation}
\label{cvvp1}
|C_{VV\Pi}|=0.31
\end{equation}
 and gives the reasonable agreement in other channels shown in Table \ref{5.3.tab}. The sign of $C_{VV\Pi}$ will be determined as positive by comparing hybrid model amplitudes for $D\to V\gamma$ decays to the quark model results in Section 5.4 (\ref{cvvp}).

\begin{table}[h]
\begin{center}
\begin{tabular}{|c||c|c|c|}
\hline
& $\Gamma(\omega\to \pi\gamma)$[GeV]&$\Gamma(\rho^+\to\pi^+\gamma)$[GeV]&$\Gamma(K^{*0}\to K^0\gamma)$[GeV]\\
 \hline
 hybrid & $9.8\cdot 10^{-4}$&$7.7\cdot 10^{-5}$&$1.42\cdot 10^{-4}$\\
exp & $(7.25\pm 0.5)\cdot 10^{-4}$&$(6.8\pm 0.6)\cdot 10^{-5}$&$(1.2\pm 0.1)\cdot 10^{-4}$\\
\hline
\end{tabular}
\end{center}
\caption{The comparison of the hybrid model predictions with $|C_{VV\Pi}|=0.31$ and the experimental data \cite{PDG96} on the $V\to P\gamma$ rates.}
\label{5.3.tab}
\end{table}

\section{Nonleptonic two-body charmed meson decays}

Before I turn to the discussion of the rare charm meson decays  $D\to V\gamma$, $D\to Vl^+l^-$ and $D\to Pl^+l^-$ in the next two sections, the hybrid model is applied to the  nonleptonic two-body charmed meson decays \cite{BFOP}. 
The understanding of nonleptonic decays is necessary in order to develop a model for the long distance contributions to  the  rare charm meson decays. The study of the nonleptonic decays leads to  a better insight into the  hybrid model, its applicability  and  the factorization approximation. To my knowledge, this analysis \cite{BFOP} presents the first application of the heavy meson chiral Lagrangian approach to the  nonleptonic two-body charmed meson decays. The exclusive nonleptonic charmed meson decays are challenging to understand theoretically and have been extensively studied using various  approaches  \cite{BSW,nonleptonic}. The agreement of the theoretical predictions with the experimental data is, however,  rather poor at present.

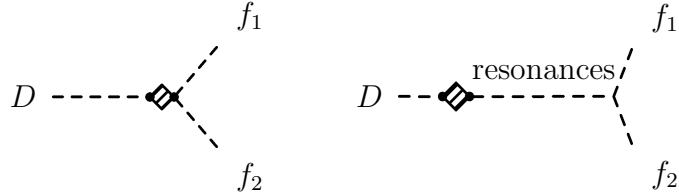
\begin{figure}[h]
\centering
\mbox{
\begin{fmffile}{f28o1n}
\fmfframe(5,2)(5,2){
  \begin{fmfgraph*}(25,15)
  \fmfpen{thin}  
  \fmfleftn{l}{1} \fmfright{r1,r2}
  \fmfrpolyn{shaded,tension=2}{k}{4} 
  \fmfv{decor.size=1.2thick,decor.shape=circle,decor.filled=full}{k1,k3}
  \fmf{dashes}{l1,k1} 
  \fmf{dashes}{k3,r2}\fmf{dashes}{k3,r1} 
  \fmflabel{$f_2$}{r1}
  \fmflabel{$D$}{l1}
  \fmflabel{$f_1$}{r2}
  \end{fmfgraph*} }
\end{fmffile}
\quad
\begin{fmffile}{f28o2a}
\fmfframe(5,2)(5,2){
  \begin{fmfgraph*}(35,15)
  \fmfpen{thin}  
  \fmfleftn{l}{1} \fmfright{r1,r2}
  \fmfrpolyn{shaded,tension=0.8}{k}{4} 
  \fmfv{decor.size=1.2thick,decor.shape=circle,decor.filled=full}{k1,k3}
  \fmf{dashes}{l1,k1}
 \fmf{dashes,label=resonances,la.s=left,tension=0.3}{k3,p}
  \fmf{dashes}{p,r2}\fmf{dashes}{p,r1} 
  \fmflabel{$f_2$}{r1}
  \fmflabel{$D$}{l1}
  \fmflabel{$f_1$}{r2}
  \end{fmfgraph*} }
\end{fmffile}
   }
\caption{The annihilation contribution to the weak nonleptonic decay $D\to f_1f_2$.
The box denotes the action of the nonleptonic effective Lagrangian (\ref{eff})
and the two dots in it denote the two weak currents.}
\label{fig28}
\end{figure}

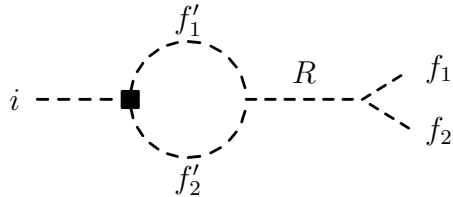
\begin{figure}[h]

\centering
\mbox{
\begin{fmffile}{f29}
  \fmfframe(7,0)(7,0){
  \begin{fmfgraph*}(50,20)
  \fmfpen{thin}
  \fmfleft{l1}\fmfright{p1,r1,r2,p2}
  \fmf{dashes,tension=1}{l1,v1}
  \fmf{dashes,left,label=$f_1^\prime$,la.s=left,la.d=20,tension=0.4}{v1,v2}
  \fmf{dashes,right,label=$f_2^\prime$,la.s=right,la.d=20,tension=0.4}{v1,v2}
  \fmf{dashes,tension=0.8,label=$R$}{v2,v3}\fmf{dashes}{r1,v3,r2}
  \fmflabel{$i$}{l1}\fmflabel{$f_2$}{r1}\fmflabel{$f_1$}{r2}
  \fmfv{de.sh=square,de.si=3thick,de.f=full}{v1}
  \end{fmfgraph*} }
\end{fmffile}
     }
\caption{The strong restcattering via the resonance $R$ in the weak nonleptonic decay $i\to f_1f_2$. 
The box donotes the action of 
the weak nonleptonic effective Lagrangian (\ref{eff}).}
\label{fig29}
\end{figure}

The nonleptonic charm decays are induced by the nonleptonic weak Lagrangian  (\ref{eff}) with the relevant coefficient $a_1^c$ and $a_2^c$ for charm decays given in (\ref{ai}). 
As in most of the studies, this analysis relies on the  factorization approximation (\ref{factor}) and the amplitude for $D\to f_1f_2$ decay is expressed as the product of  two matrix elements of the current
\begin{align*}
\langle f_1&f_2|\bar u^{\alpha}\gamma^{\mu}(1-\gamma_5)q_i^{\alpha}~\bar q^{\beta}_j\gamma_{\mu}(1-\gamma_5)c^{\beta}|D\rangle=
\langle f_2|\bar u\gamma^{\mu}(1-\gamma_5)q_i|0\rangle\langle f_1|\bar q_j\gamma_{\mu}(1-\gamma_5)c|D\rangle\\
&+\langle f_1|\bar u\gamma^{\mu}(1-\gamma_5)q_i|0\rangle\langle f_2|\bar q_j\gamma_{\mu}(1-\gamma_5)c|D\rangle
+\langle f_1f_2|\bar u\gamma^{\mu}(1-\gamma_5)q_i|0\rangle\langle 0|\bar q_j\gamma_{\mu}(1-\gamma_5)c|D\rangle~.
\end{align*}
The first two terms correspond to the spectator contribution.  The last term involves the weak annihilation of the valence quarks of the initial $D$ meson and is called the annihilation contribution. The strong interactions among the two final mesons in the annihilation contribution can be successfully described by the dominance of the light scalar or pseudoscalar  resonances with masses close to $m_D$, as shown in Fig.  \ref{fig28} \cite{nonleptonic}. The hybrid model contains only the light pseudoscalar and vector mesons and is therefore not applicable to the annihilation amplitudes. I have explicitly checked this by applying the hybrid model to   $D^0\to \bar K^0\phi$ decay, which contains only the annihilation contribution, and I found that the calculated branching ratio is much smaller than the measured branching ratio of $(8.6\pm 1.0)\cdot 10^{-3}$ \cite{PDG}. For this reason  the hybrid model is applied only to the {\bf charm meson two body nonleptonic decays, where the annihilation contribution is absent or negligible}. 
Another problem is related to the fact that  the final state mesons $f_1$ and $f_2$  interact strongly also in the spectator contributions \cite{nonleptonic}. In the factorization approximation, the hard gluon exchange interactions among $f_1$ and $f_2$ are     incorporated into the coefficients $a_1^c$ and $a_2^c$ (\ref{3.13}, \ref{3.10}), while the soft gluon exchange interactions are neglected. The later can be incorporated to some extent by using the effective coefficients  $a_1^c$ and $a_2^c$ (\ref{ai}) given by  the global fit to the  charm meson nonleptonic data.  The remaining strong interactions of $f_1$ and $f_2$ can be described by the rescattering $D\to f_1^\prime f_2^\prime\to R\to f_1f_2$ through the resonances $R$, as shown in Fig.  \ref{fig29}.  The rescattering can significantly affect the amplitude for  $D\to f_1f_2$ decay when several intermediate resonances $R_i$ contribute. The corresponding  amplitudes $D\to f_1^\prime f_2^\prime\to R_i\to f_1f_2$ with different magnitudes and phases  interfere. The rescattering effect are not expected to be very significant in the decays $D\to f_1f_2$, where the rescattering  occurs dominantly through a single resonance and so the interference is not important. This is the case when {\bf the final state contains only a single isospin} (for example $D^+\to \bar K^0\pi^+$ with $|I,I_3\rangle=|3/2,3/2\rangle$) and I will focus only on the decays of this type. Within these two limitations, I systematically study all Cabibbo allowed and Cabibbo suppressed two body charm meson nonleptonic decays: 
\begin{alignat}{3}
&D\to P_1P_2&&:&\  &D^+\to\bar K^0\pi^+\\
&D\to PV&&:&\ &D^+\to\bar K^{*0}\pi^+,~D^+\to\bar K^0\rho^+,~D_s^+\to \phi \pi^+,~ D_s^+\to \rho^+\eta,\nonumber\\
&~       && &\ &D^+\to \phi\pi^+,~D^+\to\rho^+\eta,~D^0\to \phi\pi^0(\eta),~D^0\to \omega \eta\nonumber\\
&D\to V_1V_2&&:&\ &D^+\to \bar K^{*0}\rho^+,~D_s^+\to \rho^+\phi,~D^0\to \phi\rho^0(\omega),~D^+\to \rho^+\phi~.\nonumber
\end{alignat}

\begin{figure}[h]
\centering
\mbox{
\subfigure[ The decays $D\to PV$.]
{
\begin{fmffile}{f30a1}
\fmfframe(5,2)(5,2){
  \begin{fmfgraph*}(23,20)
  \fmfpen{thin}  
  \fmfleftn{l}{1} \fmfright{r1,r2}
  \fmfrpolyn{shaded,tension=1}{k}{4} 
  \fmfv{decor.size=1.2thick,decor.shape=circle,decor.filled=full}{k1,k3}
  \fmf{dashes}{l1,v1} \fmf{dashes}{v1,r2}
  \fmf{dashes,label=$D$,tension=1,la.d=10}{v1,k1}
  \fmf{dashes}{k3,r1} 
  \fmflabel{$P$}{r1}\fmflabel{$V$}{r2}\fmflabel{$D$}{l1}
  \end{fmfgraph*} }
\end{fmffile}
\quad
\begin{fmffile}{f30a2}
\fmfframe(5,2)(5,2){
  \begin{fmfgraph*}(23,20)
  \fmfpen{thin}  
  \fmfleftn{l}{1} \fmfright{r1,r2}
  \fmfrpolyn{shaded,tension=1.5}{k}{4} 
  \fmfv{decor.size=1.2thick,decor.shape=circle,decor.filled=full}{k1,k3}
  \fmf{dashes}{l1,k1} \fmf{dashes}{k1,r2}
  \fmf{dashes}{k3,r1} 
  \fmflabel{$P$}{r1}\fmflabel{$V$}{r2}\fmflabel{$D$}{l1}
  \end{fmfgraph*} }
\end{fmffile}
\quad
\begin{fmffile}{f30a3}
\fmfframe(5,2)(5,2){
  \begin{fmfgraph*}(23,20)
  \fmfpen{thin}  
  \fmfleftn{l}{1} \fmfright{r1,r2}
  \fmfrpolyn{shaded,tension=1}{k}{4} 
  \fmfv{decor.size=1.2thick,decor.shape=circle,decor.filled=full}{k1,k3}
  \fmf{dashes}{l1,v1} \fmf{dashes}{v1,r2}
  \fmf{dashes,label=$D^*$,tension=1,la.d=10}{v1,k1}
  \fmf{dashes}{k3,r1} 
  \fmflabel{$V$}{r1}\fmflabel{$P$}{r2}\fmflabel{$D$}{l1}
  \end{fmfgraph*} }
\end{fmffile}
\quad
\begin{fmffile}{f30a4}
\fmfframe(5,2)(5,2){
  \begin{fmfgraph*}(23,20)
  \fmfpen{thin}  
  \fmfleftn{l}{1} \fmfright{r1,r2}
  \fmfrpolyn{shaded,tension=1.5}{k}{4} 
  \fmfv{decor.size=1.2thick,decor.shape=circle,decor.filled=full}{k1,k3}
  \fmf{dashes}{l1,k1} \fmf{dashes}{k1,r2}
  \fmf{dashes}{k3,r1} 
  \fmflabel{$V$}{r1}\fmflabel{$P$}{r2}\fmflabel{$D$}{l1}
  \end{fmfgraph*} }
\end{fmffile}
}
    }
\mbox{
\subfigure[The decays $D\to P_1P_2$]
{
\begin{fmffile}{f30b1}
\fmfframe(5,2)(5,2){
  \begin{fmfgraph*}(20,20)
  \fmfpen{thin}  
  \fmfleftn{l}{1} \fmfright{r1,r2}
  \fmfrpolyn{shaded,tension=1}{k}{4} 
  \fmfv{decor.size=1.2thick,decor.shape=circle,decor.filled=full}{k1,k3}
  \fmf{dashes}{l1,v1} \fmf{dashes}{v1,r2}
  \fmf{dashes,label=$D^*$,tension=1,la.d=10}{v1,k1}
  \fmf{dashes}{k3,r1} 
  \fmflabel{$P_1(P_2)$}{r1}\fmflabel{$P_2(P_1)$}{r2}\fmflabel{$D$}{l1}
  \end{fmfgraph*} }
\end{fmffile}
\quad
\begin{fmffile}{f30b2}
\fmfframe(5,2)(5,2){
  \begin{fmfgraph*}(20,20)
  \fmfpen{thin}  
  \fmfleftn{l}{1} \fmfright{r1,r2}
  \fmfrpolyn{shaded,tension=1.5}{k}{4} 
  \fmfv{decor.size=1.2thick,decor.shape=circle,decor.filled=full}{k1,k3}
  \fmf{dashes}{l1,k1} \fmf{dashes}{k1,r2}
  \fmf{dashes}{k3,r1} 
  \fmflabel{$P_1(P_2)$}{r1}\fmflabel{$P_2(P_1)$}{r2}\fmflabel{$D$}{l1}
  \end{fmfgraph*} }
\end{fmffile}
}
\quad
\subfigure[The decays $D\to V_1V_2$]
{
\begin{fmffile}{f30c1}
\fmfframe(5,2)(5,2){
  \begin{fmfgraph*}(20,20)
  \fmfpen{thin}  
  \fmfleftn{l}{1} \fmfright{r1,r2}
  \fmfrpolyn{shaded,tension=1}{k}{4} 
  \fmfv{decor.size=1.2thick,decor.shape=circle,decor.filled=full}{k1,k3}
  \fmf{dashes}{l1,v1} \fmf{dashes}{v1,r2}
  \fmf{dashes,label=$D^*$,tension=1,la.d=10}{v1,k1}
  \fmf{dashes}{k3,r1} 
  \fmflabel{$V_1(V_2)$}{r1}\fmflabel{$V_2(V_1)$}{r2}\fmflabel{$D$}{l1}
  \end{fmfgraph*} }
\end{fmffile}
\quad
\begin{fmffile}{f30c2}
\fmfframe(5,2)(5,2){
  \begin{fmfgraph*}(20,20)
  \fmfpen{thin}  
  \fmfleftn{l}{1} \fmfright{r1,r2}
  \fmfrpolyn{shaded,tension=1.5}{k}{4} 
  \fmfv{decor.size=1.2thick,decor.shape=circle,decor.filled=full}{k1,k3}
  \fmf{dashes}{l1,k1} \fmf{dashes}{k1,r2}
  \fmf{dashes}{k3,r1} 
  \fmflabel{$V_1(V_2)$}{r1}\fmflabel{$V_2(V_1)$}{r2}\fmflabel{$D$}{l1}
  \end{fmfgraph*} }
\end{fmffile}
}
   }
\caption{The diagrams for two body nonleptonic charm meson decays in the hybrid
model. 
The box denotes the action of the weak nonleptonic effective Lagrangian (\ref{eff}).
The factorization approximation is used and 
the two dots in the box denote the two weak currents in the Lagrangian
(\ref{eff}).}
\label{fig30}
\end{figure}
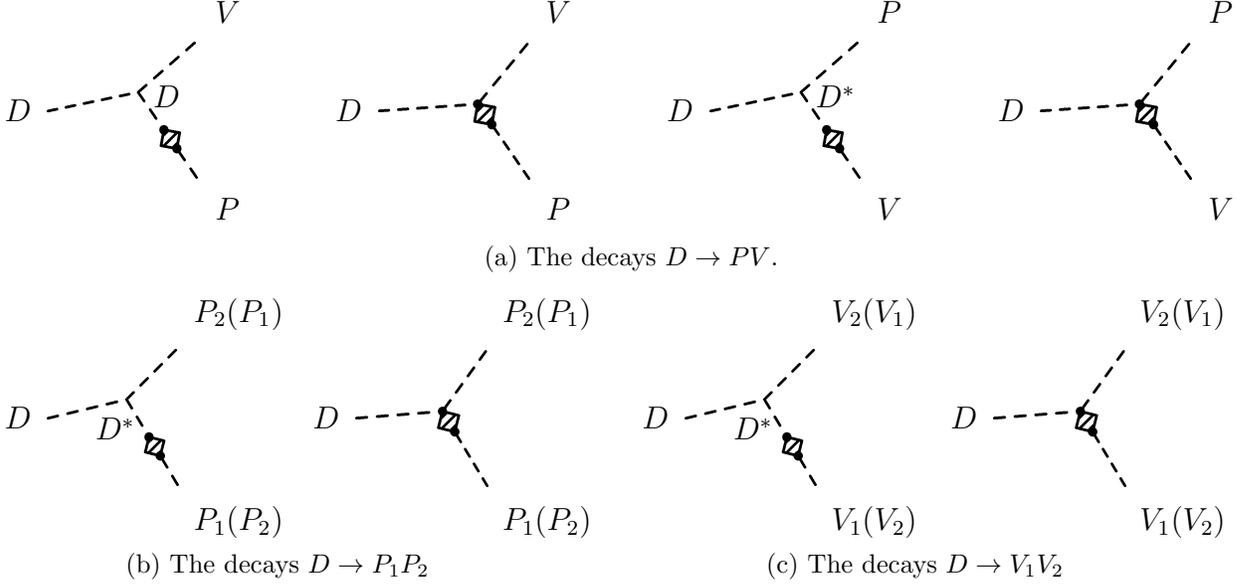

The calculation of the amplitudes using the hybrid model is now strait-forward. In the factorization approximation they are expressed in terms od the form factors $f_0$, $f_1$ (\ref{formP}) and $V$, $A_0$, $A_1$, $A_2$ (\ref{formV}). 

The relevant diagrams for $D\to P_1P_2$ decays are shown in Fig.  \ref{fig30}a and the amplitude is given by \cite{BFOP}
\begin{equation}
\label{5.40}
{\cal A}[D (p) \to P V(\epsilon^*) ] = 
\frac{G_F}{{\sqrt 2}} ~\epsilon^* \cdot p~2 
[-m_Vw_V K_{V}~ f_P ~A_0(m_P^2)+ w_P K_{P}~ g_V ~f_1(m_V^2)]~.
\end{equation}
The factors $w_V$, $w_P$, $K_{V}$ and $K_{P}$ 
are  given in Table \ref{tabpv}. 
In the decays with $\eta$ in the final state,  
the factors $K_V$ and $K_{P}$ depend on the $\eta -\eta 
'$ mixing angle $\theta_P$ through the  functions $K^{d,s}_{\eta}$ given in (\ref{mixing}).

\begin{table}[h]
\begin{center}
\begin{tabular}{|c|c|c|c|c|c|c|c|c|}
\hline
$D\to V P$ & $D^{\prime}$ & $D^{\prime*}$ & $w_V$ & $K_{V}$ & $w_P$ & $K_{P}$ \\
\hline
\hline
$D^+\to\bar K^{*0}\pi^+$ & $D_s^+$ & $D^{*0}$ & $a_1 c^2$ & $1$ &  $a_2 
c^2$ & $1$ \\
\hline
$D^+\to\rho ^+\bar K^0$ & $D^0$ & $D_s^{*+}$ & $a_2 c^2$ & $1$ &  $a_1 
c^2$ & $1$ \\
\hline
$D_s^+\to\phi\pi^+$ & $D_s^+$ & $$ & $a_1 c^2$ & $1$ &  $0$ & $0$ \\
\hline
$D^+\to\phi\pi^+$ & $$ & $D^{*0}$ & $0$ & $0$ &  $a_2 s c$ & $1$ \\
\hline 
$D^0\to\phi\pi^0$ & $$ & $D^{*0}$ & $0$ & $0$ &  $a_2 sc$ & $1/\sqrt{2}$ 
\\
\hline
$D_s^+\to\rho^+\eta$ & $$ & $D_s^{*+}$ & $0$ & $0$ &  $a_1 c^2 $ & 
$K^s_\eta$ \\

\hline
$D^+\to\rho^+\eta$ & $D^0$ & $D^{*+}$ &  $a_2 s c(K^s_\eta-K^d_{\eta})$ & 
$1$ &  $-a_1 s c$ & $K^d_\eta$ 
\\
\hline
$D^0\to\phi\eta$ & $$ & $D^{*0}$ & $0$ & $0$ & $a_2 s c$ & $K^d_\eta$ \\
\hline
$D^0\to\omega\eta$ & $D^0$ & $D^{*0}$ & $a_2 s c(K^d_\eta-K^s_\eta)$ & 
$1/\sqrt{2}$ & $a_2 s c$ & $K^d_\eta/\sqrt{2}$ \\
\hline
\end{tabular} 
\end{center}
\caption{The pole mesons and the constants $w_V$, $K_{V}$, $w_P$ and $K_{P}$ 
for the Cabibbo allowed and
Cabibbo suppressed $D\to VP$ decays. Here $c=\cos\theta_C$, $s=\sin\theta_C$ 
and  $\theta_C$ is the Cabibbo angle. The $K_{\eta}^{d,s}$ are functions of the $\eta$-$\eta '$ mixing angle $\theta_P$  given in (\ref{mixing}).}
\label{tabpv}
\end{table}

The contributions to the $D\to P_1P_2$ decay are shown in Fig.  \ref{fig30}b and the amplitude is given by
\begin{align*}
{\cal A}[D (p) \to P_{(1)}P_{(2)}] = 
\frac{G_F}{{\sqrt 2}} 
[&-i w_1~K_{P(1)}~ f_{P(2)}~ (m_D^2-m_{P(1)}^2) ~F^{(1)}_{0}(m_{P(2)}^2) \\
&-i w_2 ~ K_{P(2)}~ f_{P(1)}~(m_D^2-m_{P(2)}^2)~ F^{(2)}_{0}(m_{P(1)}^2)]
\end{align*}
with $w_1$, $w_2$, $K_{P(1)}$ and $K_{P(2)}$ presented in Table \ref{tabpp}.

\begin{table}[h]
\begin{center}
\begin{tabular}{|c|c|c|c|c|c|c|c|c|}
\hline
$D\to P_{(1)}P_{(2)}$ & $D^{\prime *}_{1}$ & $D^{\prime *}_{2}$ & $w_1$ & $K_{P(1)}$ & 
$w_2$ & $K_{P(2)}$ \\
\hline
\hline
$D^+\to\bar K^0\pi^+$ & $D^{*+}_{s}$ & $D^{*0}$ & $a_1 c^2$ &  $1$ & $a_2 
c^2$ & $1$ \\
\hline
\end{tabular} 
\end{center}
\caption{The pole mesons and the constants $w_1$, $K_{P(1)}$, $w_2$ and 
$K_{P(2)}$ 
for the $D\to P_{1}P_{2}$ decay. Here $c=\cos\theta_C$, $s=\sin\theta_C$ 
and  $\theta_C$ is the Cabibbo angle.}
\label{tabpp}
\end{table}

Finally, the diagrams for $D \to V_{1}V_{2}$ decay are given in Fig.  \ref{fig30}c and the amplitude is 
\begin{align*}
{\cal A}[D(p) &\to  V_{(1)}(p_1,\epsilon_1),V_{(2)}(p_2,\epsilon_2)] =\\
=\frac{G_F } {{\sqrt 2}}  
\biggl(&w_1 K_{V(1)}~ g_{V(2)}~ \epsilon_{2\mu}\biggl[
-{2V^{(1)}(m_{V(2)}^2)\over m_D+m_{V(1)}}\varepsilon^{\mu\nu\alpha\beta}
~\epsilon_{1\nu}^*~p_{\alpha}~p_{1\beta}
+ i (m_D+m_{V(1)})~A^{(1)}_{1}m_{V(2)}^2~\epsilon_1^{\mu 
*}\\
&\qquad\qquad\qquad\qquad-i{A^{(1)}_{2}(m_{V(2)}^2) \over m_D+m_{V(1)}}~\epsilon_1^* \cdot p_{2}~ 
(p+p_{1})^{\mu}\biggr] \\ 
+  &w_2 K_{V(2)} ~g_{V(1)}~ 
\epsilon_{1\mu}\biggl[
-{2V^{(2)}(m_{V(1)}^2)\over m_D+m_{V(2)}}\varepsilon^{\mu\nu\alpha\beta}
~\epsilon_{2\nu}^*~p_{\alpha}~p_{2\beta} +  i (m_D+m_{V(2)})~A^{(2)}_{1}(m_{V(1)}^2)~\epsilon_2^{\mu *}\\
&\qquad\qquad\qquad\qquad-  
i{A^{(2)}_{2}(m_{V(1)}^2) \over m_D+m_{V(2)}}~\epsilon_2^* \cdot p_{1}~ 
(p+p_{2})^{\mu}\biggr] \biggr) ~.
\end{align*}
The factors $w_1$, $w_2$, $K_{V(1)}$ and $K_{V(2)}$ for $D\to 
V_{(1)}V_{(2)}$ processes are given in Table \ref{tabvv}.

\begin{table}[h]
\begin{center}
\begin{tabular}{|c|c|c|c|c|c|c|c|c|}
\hline
$D\to V_{(1)}V_{(2)}$ & $D^{\prime *}_{1}$ & $D^{\prime *}_{2}$ & $w_1$ & $K_{V(1)}$ & $w_2$ & 
$K_{V(2)}$ \\
\hline
\hline
$D^+\to\bar K^{*0}\rho^+$ & $D^{*+}_{s}$ & $D^{*0}$ & $a_1 c^2$ & $1$ & 
$a_2 c^2$ & $1$ \\
\hline
$D_s^+\to \rho^+\phi$ & $D^{*+}_{s}$ & $$ & $a_1 c^2$ &  $1$ & $0$ & 
$0$\\\hline 
$D^0\to\rho^0\phi$ & $D^{*0}$ & $$ & $a_2 sc$ &  $1/\sqrt{2}$ & $0$ & $0$ 
\\
\hline 
$D^+\to\rho^+\phi$ & $D^{*0}$ & $$ & $a_2 sc$ & $1$ & $0$ & $0$\\
\hline 
$D^0\to\omega\phi$ & $D^{*0}$ & $$ & $a_2 sc$ & $1/\sqrt{2}$ & $0$ & 
$0$\\
\hline  
\end{tabular} 
\end{center}
\caption{The pole mesons and the constants $w_1$, $K_{V(1)}$, $w_2$ and 
$K_{V(2)}$ 
for the Cabibbo allowed and
Cabibbo suppressed $D\to V_{1}V_{2}$ decays. Here $c=\cos\theta_C$, $s=\sin\theta_C$ and  $\theta_C$ is the Cabibbo angle.}
\label{tabvv}
\end{table} 

First, I present the results 
for the decay amplitudes, which depend only on the form factors $f_0$ and 
$f_1$ and consequently only on the parameter $g$. These are decays 
 $D^+ \to \bar K^0 \pi^+$, 
$D^+\to \phi \pi^+$, $D_s^+\to \rho^+\eta $, $D^0\to \phi\eta $ and 
$D^0\to \phi\pi^0$. 
The predicted branching ratios for $g=0.27\pm 0.1$ (\ref{g}) are compared to the experimental data \cite{PDG} in Table \ref{5.10.tab} on top. The quoted errors are due to the uncertainties in $g=0.27\pm 0.1$ and $\theta_P=(-20\pm 5)^o$. The $\eta-\eta^\prime$ mixing angle  $\theta_P$ enters the $D_s^+\to\rho^+\eta$ and $D^0\to\phi\eta$ decays through the coefficients $K^{s,d}_\eta$ given in (\ref{mixing}).

Further, there are five decays $D\to V_1V_2$ that depend only on the form factors $V$, $A_1$ and $A_2$ and consequently on the parameters $\lambda$, $\alpha_1$ and $\alpha_2$. With the values of these parameters taken from (\ref{lambda}), the predicted branching ratios are given in Table \ref{5.10.tab} in the middle. 

The remaining $D\to PV$  decay rates depend also on the form factor $A_0(m_P^2)$ (\ref{5.40}) and consequently on the parameter $\beta$ (\ref{formV}), which gives the magnitude of the $DDV$ and $BBV$ coupling in (\ref{hybrid}). The parameter $\beta$ has been left undetermined up to now:
the semileptonic $D\to Vl^+\nu_l$ rates are not sensitive to $A_0$ due to the small mass of the lepton;  the $D^*\to D\gamma$ rates are insensitive to $\beta$ as well. The nonleptonic $D\to PV$ decays depend on $\beta$ through $A_0(m_P^2)$ and are rather insensitive to value of $\beta$ as this parameter is multiplied with a small factor $m_P^2$ in $A_0(m_P^2)$ (\ref{formV}). Among the observed $D_s^+\to \phi\pi^+$, $D^+\to \bar K^{*0}\pi^+$ and $D^+\to\bar K^0\rho^+$ decays, the last decay should be the most sensitive to $\beta$ since $m_K^2>m_\pi^2$. The predictions for $\beta=(-10,-5,0,5,10)$, $g=0.27\pm 0.1$ and $\lambda$, $\alpha_1$ and $\alpha_2$ (\ref{lambda}) are compared to the experimental data in Table \ref{beta.tab}. Although it is difficult to decide between the various values of $\beta$, it seems that negative values of $\beta$ are preferred 
\begin{equation}
\label{beta}
\beta=-2.5\pm 2.5~.
\end{equation}
The branching ratios for the remaining $D\to PV$ decays are presented in Table \ref{5.10.tab} at the bottom.

\begin{table}[!htb]
\begin{center}
\begin{tabular}{|c|c|c|}
\hline
$$ & $Br^{hybrid}[\%]$ & $Br^{exp}[\%]$\\
\hline
\hline
$D^+ \to \phi \pi^+$ & $0.6\pm 0.2$  & $0.61\pm 0.06$\\
\hline
$D_s^+ \to \rho^+\eta$ & $13\pm 5$ & $10.3\pm 3.2$\\
\hline
$D^+ \to \bar K^0 \pi^+$ & $3.4\pm 1.1$ & $2.89\pm 0.26$\\
\hline
$D^0 \to \phi \eta$ & $0.030\pm 0.012$ & $ <0.28$\\
\hline
$D^0 \to \phi \pi^0$ & $0.10\pm 0.03$ & $ <0.14$\\
\hline 
\hline
$D_s^+ \to \phi \rho^+ $ & $7.5\pm 1.0$  & $6.7\pm 2.3$\\\hline
$D^0 \to \phi \rho^0$ & $0.038\pm 0.007$ &  $0.06\pm 0.03$\\\hline
$D^+ \to \bar K^{*0}\rho^+$ & $5.2\pm 0.7$ & $2.1\pm 1.3$\\\hline
$D^+ \to \phi \rho^+$ & $0.19\pm 0.03$ &  $<1.4$\\\hline
$D^0 \to \phi \omega $ & $0.036\pm 0.004$ &  $<0.21$\\
\hline
\hline
$D_s^+\to \phi\pi^+$&$2.3\pm 0.2$&$3.6\pm 0.9$\\
\hline
$D^+\to \bar K^{*0}\pi^+$&$0.8{+1.1\atop -0.6}$&   $1.90\pm 0.19$\\
\hline
$D^+\to \rho^+\bar K^0$&$21\pm 10$&$6.6\pm 2.5$\\
\hline
$D^+\to \rho^+\eta$&$0.31\pm 0.15$&$<1.2$\\
\hline
$D^0\to\omega\eta$&$0.07\pm 0.04$&$-$\\
\hline
\end{tabular} 
\end{center}
\caption{The hybrid model predictions \cite{BFOP} and the experimental data \cite{PDG} for the nonleptonic charmed two body decays. The quoted errors in the theoretical predictions take into account only the uncertainties of the model parameters $g$ (\ref{g}), $\lambda$, $\alpha_1$, $\alpha_2$ (\ref{lambda}), $\beta$ (\ref{beta}) and $\theta_P$ (\ref{thetap}), but not the validity of the model. }
\label{5.10.tab}
\end{table}

Among all the observed charm decays discussed so far, only $D^+\to \bar K^{*0}\pi^+$ and $D^+\to\bar K^0\rho^+$ are sensitive to the relative sign of $g$ and $(\alpha_1,\alpha_2)$. In these decays, we can independently check the positive relative sign  of $g$ and $(\alpha_1,\alpha_2)$ advocated above (\ref{lambda}, \ref{g}). The positive relative sign  $g\!=\!0.27$, $\alpha_1\!=\!0.14$, $\alpha_2\!=\!0.10$
gives $Br^{hyb}(D^+\to \bar K^{*0}\pi^+)=0.8{+1.1\atop -0.6}\%$ (see Table \ref{5.10.tab}) and agrees with $Br^{exp}(D^+\to \bar K^{*0}\pi^+)=1.0\pm 0.19\%$. The negative  relative sign   $g\!=\!0.27$, $\alpha_1\!=\!-0.14$, $\alpha_2\!=\!-0.10$ gives $Br^{hyb}(D^+\to \bar K^{*0}\pi^+)=34\pm 6\%$ and is disfavored by the experimental data.

\begin{table}[!htb]
\begin{center}
\begin{tabular}{|c|c|c|c|c|c||c|}
\hline
$D\to VP$ & $Br^{hyb}[\%]$ & $Br^{hyb}[\%]$ & $Br^{hyb}[\%]$ & $Br^{hyb}[\%]$ & $Br^{hyb}[\%]$ & $Br^{exp}[\%]$\\
 & $\beta=-10$ & $\beta=-5$ & $\beta=0$ & $\beta=5$ & $\beta=10$ & \\
\hline
$D_s^+\to \phi\pi^+$&$2.6\pm 0.1$&$2.4\pm 0.1$&$2.2\pm 0.1$&$2.0\pm 0.1$&$1.8\pm 0.1$&$3.6\pm 0.9$\\
\hline
$D^+\to \bar K^{*0}\pi^+$&$0.54{+1.0\atop -0.5}$& $0.71{+1.0\atop -0.5}$& $0.91{+1.2\atop -0.7}$&$1.1{+1.3\atop -0.8}$&$1.4{+1.4\atop -1.0}$&  $1.90\pm 0.19$\\
\hline
$D^+\to \rho^+\bar K^0$&$12\pm 6$&$18\pm 7$&$25\pm 8$&$33\pm 9$&$43\pm 11$&$6.6\pm 2.5$\\
\hline  
\end{tabular} 
\end{center}
\caption{The branching ratios three observed decays $D\to PV$ that depend on the parameter $\beta$ in the form factor $A_0$ (\ref{5.40}). The theoretical errors are due to  uncertainties in $g$ (\ref{g}) and $\lambda$, $\alpha_1$, $\alpha_2$ (\ref{lambda}).}
\label{beta.tab}
\end{table}

The predicted nonleptonic decay rates in Table \ref{5.10.tab} do agree with the experimental data given the fact that the theoretical and experimental uncertainties are large. The nonleptonic decays, where the annihilation contribution is absent or negligible and the rescattering is not seizable, are reasonably well understood in terms of the hybrid model accompanied by the factorization approximation. 
 
\vspace{0.4cm}

\newpage

In the $D\to V_1V_2$ decays, the additional experimental information comes from the partial wave analysis of the final state. The final vector mesons are produced in three helicity states $++$, $--$, $00$ (see Fig. \ref{fig21}) or equivalently in the three angular momentum states $S$, $P$, $D$. The authors of \cite{kamal} have analyzed the $D\to \bar K^*\rho$ decays for which the experimental data on the total  and partial wave branching ratios exists. They applied the factorization approximation, they neglected the annihilation contribution (this is reliable for $D^+\to\bar K^{*0}\rho^+$ decay studied above, but not for $D^0\to \rho^0\bar K^{*0}$ and $D^0\to \bar K^{*-}\rho^+$ decays) and predicted the experimental observables using several models for the $D\to V$ form factors, among others the hybrid model based on \cite{BFOP}. The hybrid model, like every other model employed in \cite{kamal}, predicts the ratio  $|Br^{S~wave}|/|Br^{D~wave}|(D^+\to \bar K^{*0}\rho^+)$ which is about $5-8$ times bigger than the measured ratio $1.3\pm 0.8$ extracted from \cite{PDG}. The authors of \cite{kamal} claim the inconsistency of the experimental partial wave data for  $D^+\to \bar K^{*0}\rho^+$ in \cite{PDG}.   

\section{The $\boldsymbol{D\to V\gamma}$ and $\boldsymbol{D\to Vl^+l^-}$ decays}

Finally, with all these ingredients I can turn to the study of the rare charm meson decays $D\to V\gamma$ and $D\to Vl^+l^-$ with a light vector meson $V$ and a photon or a charged lepton $l=e,\mu$ in the final state.  The  Cabibbo suppressed decays have the flavour structure $c\bar q\to u\bar q\gamma$  and $c\bar q\to u\bar ql^+l^-$, respectively. They are induced by the flavour changing neutral transitions $c\to u\gamma$ and $c\to ul^+l^-$ at short distances and by the long distance mechanisms.  Different mechanisms were sketched in Chapter 3 and the   long distance contribution was divided to the weak annihilation (Fig.  \ref{fig2}) and long distance penguin (Fig.  \ref{fig3}) part.
In addition to Cabibbo suppressed decays, I systematically study all the Cabibbo allowed and Cabibbo doubly suppressed $D\to V\gamma$ and $D\to Vl^+l^-$ decays.   These can not proceed through a flavour changing neutral transition of a single quark and are induced only through the long distance mechanisms. 

None of these decays has been observed so far. The upper limits on $D^0\to V^0\gamma$ branching ratios  are in the range $10^{-4}$ \cite{radiative.exp1}, while there are no experimental results on the $D^+$ and $D_s^+$ decay channels yet.  As a result of an ongoing experimental efforts \cite{radiative.exp2}, it is reasonable to expect that the  data on $D\to V\gamma$ decays  forthcoming during the next few years. The present upper limits on $D\to Vl^+l^-$ branching ratios are in the  range $10^{-4}-10^{-3}$ \cite{PDG,vll.exp}. Let me point out that there no available upper bounds  for the Cabibbo allowed channels $D_s^+\to \rho^+\gamma$ and $D_s^+\to \rho^+l^+l^-$, which are predicted at the highest rates in this work (with branching ratios of the order of $10^{-4}$ and $10^{-5}$, respectively) and have the best chances for detection.

The theoretical treatment of these charmed meson decays faces different situation than encountered in the kaon  and beauty meson decays. The charmed meson decays turn out to be dominated by the long distance contributions. The most frequent beauty meson decays of this type, i.e. $B\to K^*\gamma$ and $B\to K^*l^+l^-$, are  driven by short distance contribution via $b\to s\gamma$ and $b\to sl^+l^-$ decays and the long distance contributions are relatively small, as discussed in the introduction. The kaon decays of this type are not possible as the kaon is lighter than the lightest vector meson.

\vspace{0.1cm}

The short distance contributions to $D\to Vl^+l^-$ decays have been studied in different scenarios beyond the standard model \cite{pakvasa,CMM,BHLP,schwartz} and have been reviewed in Chapter 2. In order to perform  search for new physics in these decays, one should have a good control over the long distance contributions. These have not been predicted so far.
The first theoretical study of the long distance contributions to $D\to Vl^+l^-$ decays is presented  in this work and follows the original presentation in \cite{FPS2,genova2}. 

The short distance contributions to $D\to V\gamma$ decays have been studied in different scenarios beyond the standard model \cite{pakvasa,CNS,BGM,BHLP,hewett,eeg} and have been reviewed in Chapter 2.
The long distance contributions to $D\to V\gamma$ decays have been studied in \cite{BGHP,KSW,BFO2,gamma.quark,BFO3}. The theoretical predictions are still rudimentary at present and various models do not always lead to compatible results.   
The first comprehensive theoretical analysis of all $D\to V\gamma$ decays has been presented in \cite{BGHP}.  The authors of \cite{BGHP} have divided the long distance contributions in $D^0\to \bar K^{*0}\gamma$ decay, for example,  to (i) $D^0\to D^{*0}\gamma$ followed by the weak transition $D^{*0}\to \bar K^{*0}$, (ii)  the weak transition $D^0\to \bar K^0$ followed  by $\bar K^0\to \bar K^{*0}\gamma$ and (iii)  the weak decay $D^0\to \rho^0\bar K^{*0}$ followed by  $\rho^0\to \gamma$. The couplings in analysis \cite{BGHP} have been extracted from the available experimental data and from the predictions of the other models. The approach of \cite{BGHP} may have problems with possible double counting, as the contribution shown in Fig.  \ref{fig31}a is accounted by the mechanism (i) and (iii); similarly the contribution shown in Fig.  \ref{fig31}b is accounted by the mechanisms (ii) and (iii). 
The weak annihilation long distance contribution for specific  $D\to V\gamma$ channels has been studied also with the quark model  \cite{gamma.quark} and QCD sum rules \cite{KSW}. The QCD sum rules analysis \cite{KSW} incorporates only the weak annihilation mechanism in which the photon is emitted before the weak transition.

\begin{figure}[h]
\centering
\mbox{
\subfigure[]
{
\begin{fmffile}{f31a}
\fmfframe(8,0)(8,0){
  \begin{fmfgraph*}(30,17)
  \fmfpen{thin}  
  \fmfleftn{l}{1} \fmfrightn{r}{4}
  \fmfrpolyn{shaded,tension=1}{k}{4}   
  \fmf{dashes}{l1,v} 
  \fmf{dashes,label=$D^{*0}$,tension=0.6,la.s=right}{v,k1}\fmf{dashes}{k3,r1}
  \fmffreeze 
  \fmf{dashes,label=$\rho^0$,tension=0.6,la.s=left}{v,v1}
  \fmf{boson,tension=1}{v1,r3}
  \fmfv{decor.size=1.2thick,decor.shape=circle,decor.filled=full}{k1,k3}
  \fmflabel{$\gamma$}{r3}
  \fmflabel{$D^0$}{l1}
  \fmflabel{$\bar K^{*0}$}{r1}
  \end{fmfgraph*} }
\end{fmffile}
}
\quad
\subfigure[]
{
\begin{fmffile}{f31b}
\fmfframe(8,0)(8,0){
  \begin{fmfgraph*}(30,17)
  \fmfpen{thin}  
  \fmfleftn{l}{1} \fmfrightn{r}{3}
  \fmfrpolyn{shaded,tension=1}{k}{4}   
  \fmf{dashes}{l1,k1} 
  \fmf{dashes,label=$\bar K^0$,tension=0.6,la.s=right}{k3,v}\fmf{dashes}{v,r1} 
  \fmffreeze
   \fmf{dashes,label=$\rho^0$,tension=0.6,la.s=left,la.d=1}{v,v1}
  \fmf{boson,tension=1}{v1,r2}
  \fmfv{decor.size=1.2thick,decor.shape=circle,decor.filled=full}{k1,k3}
  \fmflabel{$\gamma$}{r2}
  \fmflabel{$D^0$}{l1}
  \fmflabel{$\bar K^{*0}$}{r1}
  \end{fmfgraph*} }
\end{fmffile}
}
    }
\caption{Comment on the approach of Ref. [21] considering the weak radiative decays of the charmed mesons. }
\label{fig31}
\end{figure}
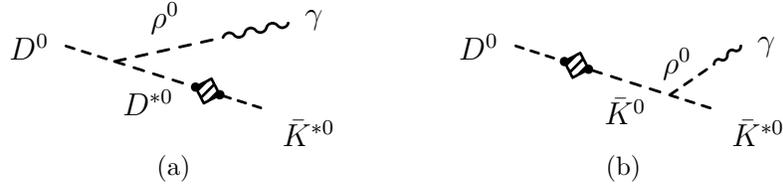

The heavy meson chiral Lagrangian has been applied for the parity conserving part of the weak annihilation contribution. Cabibbo allowed $D\to V\gamma$ decays have been studied in  \cite{BFO2}. The ratios $Br(D_s^+\to \bar K^{*+}\gamma)/Br(D_s^+\to \rho^+\gamma)$ and $Br(D^0\to \rho^0\gamma)/Br(D^0\to \bar K^{*0}\gamma)$ were proposed as possible tests for new physics in \cite{BFO3}. In the present work, the parity violating part of the weak annihilation contribution and the long distance penguin contribution are added \cite{FPS1,FS}. All the decays $D\to V\gamma$  are systematically studied \cite{FPS1,FS}.  
 
\vspace{0.1cm}

The $D\to V\gamma$ and $D\to Vl^+l^-$ decays are studied by adapting the heavy meson chiral Lagrangian approach, more specifically the hybrid model as described above.

\subsection{Long distance contribution}

First I turn to the calculation of the long distance contributions to $D\to V\gamma$ and $D\to Vl^+l^-$ decays. The long distance mechanism in $D\to Vl^+l^-$ decay is due to the exchange of the intermediate photon via the channel $D\to V\gamma^*\to Vl^+l^-$. 
The  $D\to V\gamma^*$ decay involves $S$, $P$ or $D$ orbital momentum final states.  The parity conserving part of the amplitude corresponds to the $P$ wave state, while the parity violating part corresponds to  the $S$ and $D$ wave state.  

The general framework for the long distance contributions is presented in Chapter 3. They are induced by the effective nonleptonic Lagrangian (\ref{eff})
\begin{equation}
\label{eff1}
{\cal L}^{|\Delta c|=1}_{eff}=-\tfrac{G_F}{\sqrt{2}}V_{cq_j}^*V_{uq_i}[a_1^c~ \bar u\gamma^{\mu}(1-\gamma_5)q_i~\bar q_j\gamma_{\mu}(1-\gamma_5)c
+a_2^c~\bar q_j\gamma_{\mu}(1-\gamma_5)q_i~\bar u\gamma^{\mu}(1-\gamma_5)c~]
\end{equation}
with $a_1^c\simeq 1.26$, $a_2^c\simeq-0.55$ (\ref{ai}) and $q_{i,j}=s,d$. In addition the photon is emitted. 
The amplitudes (\ref{ald}) 
$${\cal A}_{LD}(D\to V\gamma)=\langle \gamma V|:i{\cal L}^{|\Delta c|=1}_{eff}:|P\rangle~~,\quad {\cal A}_{LD}(D\to Vl^+l^-)=\langle l^+l^- V|:i{\cal L}^{|\Delta c|=1}_{eff}:|P\rangle$$
are calculated by employing the factorization approximation (\ref{factor}).

The Feynman diagrams are given in terms of the hadronic degrees of freedom contained in the hybrid model: heavy pseudoscalar ($D$) and vector ($D^*$) mesons and light pseudoscalar ($P$) and vector ($V$) mesons. The relevant diagrams 
for $D\to V\gamma$ in the hybrid model  are given in Figs.  \ref{fig32} and  \ref{fig33}. In the case of $D\to Vl^+l^-$ decays, the lepton pair $l^+l^-$  is attached to the virtual photon. The boxes in the diagrams denote the action of the effective nonleptonic weak Lagrangian (\ref{eff1}). This Lagrangian contains a product of two left handed quark currents, each denoted by a dot in a box. 
 Let me comment on different contributions.

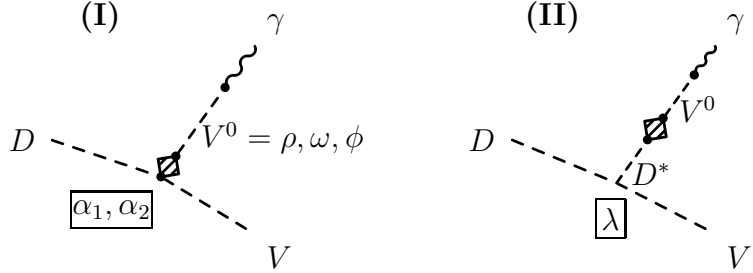
\begin{figure}[h]
\centering
\mbox{
\begin{fmffile}{f32o1n}
\fmfframe(8,5)(8,5){
  \begin{fmfgraph*}(30,25)
  \fmfpen{thin}  
  \fmfleft{l1} \fmfrightn{r}{2}\fmftopn{t}{4}\fmflabel{\bf{(I)}}{t2}
  \fmfrpolyn{shaded,tension=0.5}{k}{4}   
  \fmf{dashes}{l1,k1} 
  \fmf{dashes}{k1,r1} 
  \fmf{dashes,label=$V^0=\rho,,\omega,,\phi$,la.d=0.6,tension=0.3}{k3,v1}
  \fmf{boson,tension=0.5}{v1,r2}
  \fmfv{decor.size=1.2thick,decor.shape=circle,decor.filled=full}{k3,v1}
  \fmfv{decor.size=1.2thick,decor.shape=circle,decor.filled=full,label=\framebox[11mm]{$\alpha_1,,\alpha_2$},la.d=3thick,la.a=-120}{k1}
  \fmflabel{$\gamma$}{r2}
  \fmflabel{$D$}{l1}
  \fmflabel{$V$}{r1}
  \end{fmfgraph*} }
\end{fmffile}
\qquad
\begin{fmffile}{f32o2n}
\fmfframe(8,5)(8,5){
  \begin{fmfgraph*}(30,25)
  \fmfpen{thin}  
  \fmfleftn{l}{1} \fmfrightn{r}{2}\fmftopn{t}{4}\fmflabel{\bf{(II)}}{t2}
  \fmfrpolyn{shaded,tension=0.2}{k}{4}   
  \fmf{dashes,tension=1.1}{l1,v2} 
  \fmf{dashes,tension=1.1}{v2,r1} 
  \fmf{dashes,label=$D^*$,la.d=0.1,tension=0.2,la.si=right}{v2,k1}
  \fmf{dashes,label=$V^0$,la.d=0.1,tension=0.2}{k3,v1}
  \fmf{boson,tension=0.3}{v1,r2}
   \fmfv{decor.size=1.2thick,decor.shape=circle,decor.filled=full}{k1,k3,v1}
  \fmfv{label=\framebox[4mm]{$\lambda$},la.d=3thick,la.a=-110}{v2}   
  \fmflabel{$\gamma$}{r2}
  \fmflabel{$D$}{l1}
  \fmflabel{$V$}{r1}
  \end{fmfgraph*} }
\end{fmffile}
    }
\caption{Long distance penguin diagrams for $D\to V\gamma$ decays. For the case of $D\to Vl^+l^-$ decay, the real photon has to be replaced with the virtual photon and the lines for the charged lepton pair $l^+l^-$ have to be attached.  The parameters,  given in the frames   by the
verteces, indicate which terms in the Lagrangian (\ref{hybrid}) and weak current
(\ref{current.heavy}) are responsible for the couplings. The box denotes the action of the weak nonleptonic effective Lagrangian (\ref{eff}). The box contains two dots each denoting a weak current in the Lagrangian (\ref{eff}).}
\label{fig32}
\end{figure}

\begin{itemize}
\item The {\bf long distance penguin} contribution is presented in Fig.  \ref{fig32} and is induced by the part of the Lagrangian (\ref{eff1}) proportional to $a_2$. The current $\bar u\gamma^\mu(1-\gamma_5)c$ annihilates the initial $D$ meson and creates the final $V$ meson. The other current $\sum_{d,s}V_{cq}^*V_{uq} \bar q\gamma^\mu(1-\gamma_5)q\simeq V_{cs}^*V_{us}[s\gamma^\mu(1-\gamma_5)s-\bar d\gamma^\mu(1-\gamma_5)d]$ is proportional to the $SU(3)$ flavour breaking and creates a photon or a lepton-antilepton pair via the intermediate neutral vector meson $\rho^0$, $\omega$ and $\phi$. The short lifetime of the vector mesons $V^0$ is accounted for by the Breit-Wigner form of the propagator $-i(g^{\mu\nu}-q^\mu q^\nu/m_{V^0}^2)/(q^2-m_{V^0}^2+i\Gamma_{V^0}m_{V^0})$ with decay widths  $\Gamma_{V^0}$  given in Table \ref{const.tab}. This renders the  resonant shape for $D\to Vl^+l^-$ spectrum in terms of  di-lepton mass $m_{ll}=\sqrt{q^2}$. In the regions of $m_{ll}$ far from $m_{V^0}$, the amplitude is given solely by the tail of the Breit-Wigner vector 
meson propagator.
\item The {\bf long distance weak annihilation} is presented in Fig.  \ref{fig33}. One weak current has the flavour of the initial  meson $D$, while  the other has the flavour of the final  meson $V$. The bremsstrahlung diagrams are given in Fig.  \ref{fig33}c and are nonzero only when $D$ and $V$ are charged. The diagrams, where the photon is emitted before and after the weak transition and  do not correspond to bremsstrahlung, are gathered in Figs.  \ref{fig33}a and  \ref{fig33}b, respectively. The hybrid model does not contain the axial heavy mesons and the excited light mesons and their contributions to the mechanisms in Figs.  \ref{fig33}a and  \ref{fig33}b are neglected.  
 The diagrams $III$, $V$, $VI$ and $VIII$ correspond to the emission of the photon from the light quark and give the resonant shape of the $D\to Vl^+l^-$ spectrum with  peaks at $m_{ll}=m_\rho,m_\omega,m_\phi$. The diagram $IV$ corresponds to the emission of the photon from the charm  quark and  gives the nonresonant shape  of the $D\to Vl^+l^-$ spectrum.  The bremsstrahlung diagram $VII$ is imposed by the velocity reparametrization invariance (\ref{current.VRI}) and gives the nonresonant shape of the spectrum. 
\item Note that the parity violating contribution, given by the diagrams $I$ and $V$,  is incorporated solely via the vector meson dominance mechanism, in which the nonleptonic decay $D\to VV^0$ with $V^0=\rho,\omega,\phi$ is followed by the conversion $V^0\to \gamma$. This observation will be important when the gauge invariance is studied bellow. 
\end{itemize}

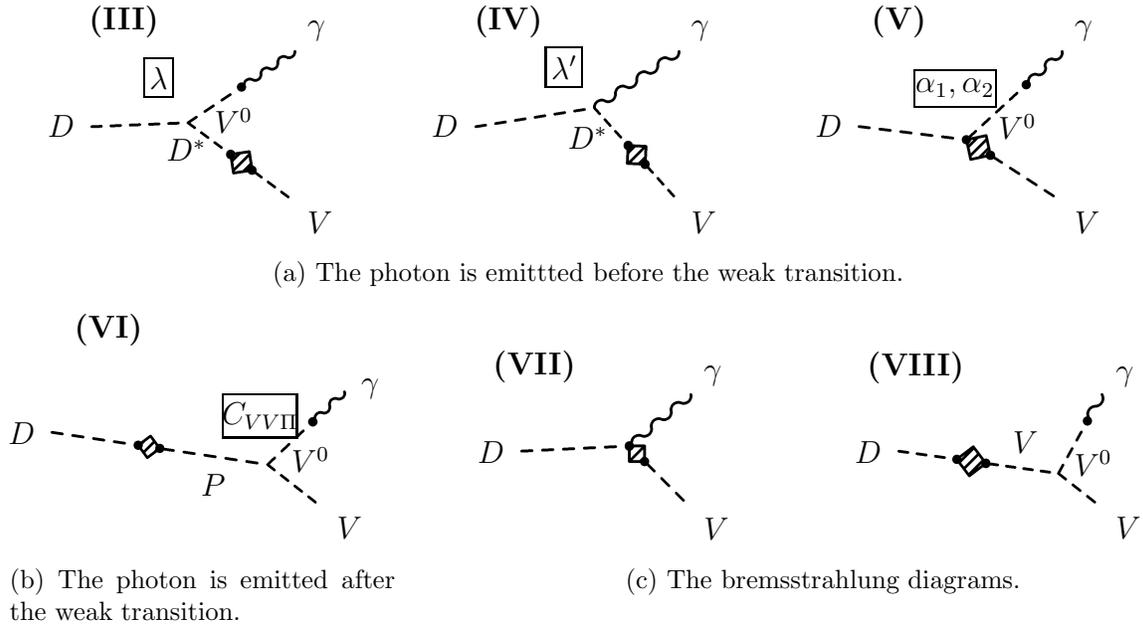
\begin{figure}[htb]
\centering
\mbox{
\subfigure[The photon is emittted before the weak transition.]
{
\begin{fmffile}{f33a1}
\fmfframe(5,2)(5,2){
  \begin{fmfgraph*}(30,20)
  \fmfpen{thin}
  \fmfleftn{l}{1} \fmfrightn{r}{2}\fmftopn{t}{4}\fmflabel{\bf{(III)}}{t2}
  \fmfrpolyn{shaded}{k}{4}  
  \fmfv{decor.size=1.2thick,decor.shape=circle,decor.filled=full}{k1,k3,v2} 
  \fmf{dashes}{l1,v1} 
  \fmf{dashes,label=$V^0$,la.s=right,la.d=10}{v1,v2}
  \fmf{boson}{v2,r2}
  \fmf{dashes,label=$D^*$,la.s=right,la.d=10}{v1,k1} 
    \fmf{dashes}{k3,r1}
    \fmflabel{$D$}{l1}\fmflabel{$V$}{r1}\fmflabel{$\gamma$}{r2}
    \fmfv{label=\framebox[4mm]{$\lambda$},la.d=5thick,la.a=120}{v1} 
  \end{fmfgraph*} }
\end{fmffile}
\quad
\begin{fmffile}{f33a2}
\fmfframe(5,2)(5,2){
  \begin{fmfgraph*}(30,20)
  \fmfpen{thin}
  \fmfleftn{l}{1} \fmfrightn{r}{2}\fmftopn{t}{4}\fmflabel{\bf{(IV)}}{t2}
  \fmfrpolyn{shaded}{k}{4}   
  \fmfv{decor.size=1.2thick,decor.shape=circle,decor.filled=full}{k1,k3} 
  \fmf{dashes}{l1,v1} 
  \fmf{boson}{v1,r2}
  \fmf{dashes,label=$D^*$,la.s=right,la.d=10}{v1,k1} 
  \fmf{dashes}{k3,r1}
  \fmflabel{$D$}{l1}\fmflabel{$V$}{r1}\fmflabel{$\gamma$}{r2}
  \fmfv{label=\framebox[5mm]{$\lambda^\prime$},la.d=4thick,la.a=120}{v1}
  \end{fmfgraph*} }
\end{fmffile}
\quad
\begin{fmffile}{f33a3n}
\fmfframe(5,2)(5,2){
  \begin{fmfgraph*}(30,20)
  \fmfpen{thin}
  \fmfleftn{l}{1} \fmfrightn{r}{2}\fmftopn{t}{4}\fmflabel{\bf{(V)}}{t2}
  \fmfrpolyn{shaded,tension=1.5}{k}{4}   
  \fmfv{decor.size=1.2thick,decor.shape=circle,decor.filled=full}{k3,v1}
  \fmfv{decor.size=1.2thick,decor.shape=circle,decor.filled=full,label=\framebox[11mm]{$
  \alpha_1,,\alpha_2$},la.d=6thick,la.a=110}{k1}
  \fmf{dashes}{l1,k1}\fmf{dashes,label=$V^0$,la.s=right,la.d=10,tension=0.6}{k1,v1} 
  \fmf{boson}{v1,r2}
  \fmf{dashes}{k3,r1}
  \fmflabel{$D$}{l1}\fmflabel{$V$}{r1}\fmflabel{$\gamma$}{r2}
  \end{fmfgraph*} }
\end{fmffile}
}
      }
\mbox{
\subfigure[The photon is emitted after the weak transition.]
{
\begin{fmffile}{f33bn}
\fmfframe(5,2)(5,2){
  \begin{fmfgraph*}(40,20)
  \fmfpen{thin}
  \fmfleftn{l}{1} \fmfright{r1,p1,p2,r2,p3}\fmftopn{t}{4}\fmflabel{\bf{(VI)}}{t2}
  \fmfrpolyn{shaded,tension=2}{k}{4}   
  \fmf{dashes}{l1,k1} 
  \fmf{dashes,label=$P$,tension=0.8,la.s=right}{k3,v1}
  \fmf{dashes,label=$V^0$,la.d=10,la.si=left,tension=0.7,la.s=right}{v1,v2}
  \fmf{boson}{v2,r2}
  \fmf{dashes}{v1,r1}
  \fmfv{decor.size=1.2thick,decor.shape=circle,decor.filled=full}{k3,k1,v2} 
  \fmflabel{$D$}{l1}\fmflabel{$V$}{r1}\fmflabel{$\gamma$}{r2}
  \fmfv{label=\framebox[10mm]{$C_{VV\Pi}$},la.d=5thick,la.a=107}{v1}
  \end{fmfgraph*} }
\end{fmffile}
}
    }
\mbox{
\subfigure[The bremsstrahlung diagrams.]
{
\begin{fmffile}{f33c1n}
\fmfframe(5,2)(5,2){
  \begin{fmfgraph*}(25,15)
  \fmfpen{thin}
  \fmfleftn{l}{1} \fmfrightn{r}{2}\fmftopn{t}{4}\fmflabel{\bf{(VII)}}{t2}
  \fmfrpolyn{shaded,tension=1.5}{k}{4}   
  \fmfv{decor.size=1.2thick,decor.shape=circle,decor.filled=full}{k3,k1}
  \fmf{dashes}{l1,k1}
  \fmf{boson}{k1,r2}
  \fmf{dashes}{k3,r1}
  \fmflabel{$D$}{l1}\fmflabel{$V$}{r1}\fmflabel{$\gamma$}{r2}
  \end{fmfgraph*} }
\end{fmffile}
\qquad
\begin{fmffile}{f33c2nn}
\fmfframe(5,2)(5,2){
  \begin{fmfgraph*}(30,15)
  \fmfpen{thin}
  \fmfleftn{l}{1} \fmfrightn{r}{2}\fmftopn{t}{4}\fmflabel{\bf{(VIII)}}{t2}
  \fmfrpolyn{shaded}{k}{4}   
  \fmf{dashes}{l1,k1} 
  \fmf{dashes,label=$V$,tension=0.8}{k3,v1}
  \fmf{dashes,label=$V^0$,la.d=10,la.si=left,tension=0.5,la.s=right}{v1,v2}
  \fmf{boson}{v2,r2}
  \fmf{dashes}{v1,r1}
  \fmfv{decor.size=1.2thick,decor.shape=circle,decor.filled=full}{k3,k1,v2} 
  \fmflabel{$D$}{l1}\fmflabel{$V$}{r1}\fmflabel{$\gamma$}{r2}
  \end{fmfgraph*} }
\end{fmffile}
}
    }    

\caption{Long distance weak annihilation diagrams for $D\to V\gamma$ decays in the hybrid
model. 
For the case of $D\to Vl^+l^-$ decays, the real photon is replaced with 
the virtual photon and the lines for the charged lepton pair $l^+l^-$ are attached.
The bremsstrahlung diagrams are gathered in Fig. (c). The  diagrams, where the photon is
emitted before and after the weak transition, and do not correspond to  bremsstrahlung are
gathered in Figs. (a) and (b), respectively. The parameters,  given in the frames  by the
verteces, indicate which terms in the Lagrangian (\ref{hybrid}) and weak current
(\ref{current.heavy}) are responsible for the couplings.  
The box denotes the action of the weak nonleptonic effective Lagrangian (\ref{eff}). 
The box contains two dots each denoting a weak current in the Lagrangian (\ref{eff}).}
\label{fig33}
\end{figure}

First I turn to the {\bf bremsstrahlung} contribution. 
Following the general discussion in Section 3.3.3, the bremsstrahlung amplitude for $D\to V\gamma^*$ diagrams in Fig.  \ref{fig25}a can be parameterized in terms of the general electromagnetic form factors $G_V$ and $G_{Dd}$ as (\ref{3.126})
\begin{align}
\label{5.126}
{\cal A}[D(p)\to V(p^\prime,\epsilon^\prime)&\gamma^*(q,\epsilon)]=\frac{G_F}{\sqrt{2}}e_0f_{Cabb}^{(i)}f_Pg_V\epsilon^{*}_{\mu}\epsilon^{\prime *}_{\nu}\nonumber\\
&\times\bigl([G_{Dd}^{\mu\nu}(q^2)-G_V^{\mu\nu}(q^2)]-\frac{G_V^{\mu\delta}(q^2)}{m_V^2}[q_\delta q_\nu-q^2g_{\delta\nu}]\bigr)~
\end{align}
with $f_{Cabb}^{(i)}$ given in Table \ref{tab.vll} for different decays.
I have shown that the amplitude  is manifestly gauge invariant for the case of the flat form factors, the polar form factors and their linear combination (\ref{3.123}). The bremsstrahlung diagrams in Fig.  \ref{fig33}c as given by the hybrid model indicate that $G_{Dd}\!=\!1$ is flat  and $G_{V}$ has polar shape:
\begin{align*}
G_{\rho^+}^{\mu\nu}(q^2)&=\frac{m_{\rho}^2~\bigl[g^{\mu\nu}-\frac{q^\mu q^\nu}{m_{\rho}^2}\bigr]}{m_{\rho}^2-q^2-i\Gamma_{\rho}m_{\rho}}~,\\
G_{K^{*+}}^{\mu\nu}(q^2)&=\frac{m_{\rho}^2~\bigl[g^{\mu\nu}-\frac{q^\mu q^\nu}{m_{\rho}^2}\bigr]}{2(m_{\rho}^2-q^2-i\Gamma_{\rho}m_{\rho})}+\frac{m_{\omega}^2~\bigl[g^{\mu\nu}-\frac{q^\mu q^\nu}{m_{\omega}^2}\bigr]}{6(m_{\omega}^2-q^2-i\Gamma_\omega m_\omega)}+\frac{m_{\phi}^2~\bigl[g^{\mu\nu}-\frac{q^\mu q^\nu}{m_{\phi}^2}\bigr]}{3(m_{\phi}^2-q^2-i\Gamma_\phi m_\phi)}~\nonumber
\end{align*}
with $\Gamma_{V^0}(q^2\!=\!0)=0$ and $G(q^2\!=\!0)=1$. Inserting these electromagnetic form factors to the amplitude (\ref{5.126}) and neglecting the contributions proportional to $\Gamma_{V^0}$ I get
 \begin{equation}
\label{5.127}
{\cal A}[D(p)\to V(p^\prime,\epsilon^\prime)\gamma^*(q,\epsilon)]=-\tfrac{G_F}{\sqrt{2}}e_0f_{Cabb}^{(i)}f_Pg_V\epsilon^{*}_{\mu}\epsilon^{\prime *}_{\nu}(q^2g^{\mu\nu}-q^\mu q^\nu)~L_V(q^2)
\end{equation}
with
\begin{align*}
L_{K^{*+}}&=\frac{1}{m_{K^*}^2}\biggl[{m_{K^*}^2-m_{\rho}^2\over 2(q^2-m_{\rho}^2+i\Gamma_{\rho}m_{\rho})}+{m_{K^*}^2-m_{\omega}^2\over 6(q^2-m_{\omega}^2+i\Gamma_{\omega}m_{\omega})}+{m_{K^*}^2-m_\phi^2\over 3(q^2-m_{\phi}^2+i\Gamma_{\phi}m_{\phi})}\biggr]~,\\
L_{\rho^{+}}&=0~.
\end{align*}
The bremsstrahlung amplitude turns out to be proportional to the $SU(3)$ flavour breaking in the hybrid model: it is equal to zero for the case of $\rho^+$ meson in the final state and small, but nonzero, for the case of $K^{*+}$ meson in the final state.  

The {\bf non-bremsstrahlung} part of the long distance contribution is given by the diagrams in Figs.  \ref{fig32},  \ref{fig33}a and  \ref{fig33}b. The calculation of the corresponding amplitude is strait-forward given the Lagrangian (\ref{hybrid}, \ref{eff1}) and weak currents (\ref{current.light}, \ref{current.heavy}). 

The sum of the long distance  amplitudes for the $D\to V\gamma^*$ diagrams, given in Figs.  \ref{fig32} and  \ref{fig33},  is given by the expression  
\begin{equation}
\label{5.amp1}
{\cal A}^{LD}[D(p)\to V(\epsilon^\prime,p^\prime)~\gamma^*(q,\epsilon)]  =  -i\frac{G_F}{{\sqrt 2}}e_0 f_{Cabb}^{(i)}~\epsilon_\mu^*\epsilon^{*\prime}_{\nu} \bigl[iA_{PV}^{LD~\mu\nu}+\epsilon^{\mu \nu \alpha \beta} p_{\alpha}q_\beta A_{PC}^{LD}\bigr] ~.
\end{equation}
 The amplitude for $D\to V\gamma$ decay  is obtained by taking $q^2=0$, while the amplitude for $D\to Vl^+(p_+)l^-(p_-)$ decay is obtained replacing the photon polarization  $\epsilon_\mu$ with $e_0\bar u(p_-)\gamma_{\mu}v(p_+)/q^2$. The indices $(i)=1,..,9$ denote nine decays $D\to V\gamma^*$ and the corresponding Cabibbo factors $f_{Cabb}^{(i)}$ for Cabibbo allowed, suppressed and doubly suppressed decays are given in Table \ref{tab.vll}. The parity conserving and  violating parts of the amplitude are represented by $A_{PC}^{LD}$ and $A_{PV}^{LD~\mu\nu}$, respectively. They get contributions from different diagrams in Figs.  \ref{fig32} and  \ref{fig33} in the hybrid model. It is suitable to express them in terms of amplitudes $A^{I,..,VIII}$, where  the superscript denotes the contributions of different diagrams $I..VIII$ in Figs.  \ref{fig32} and  \ref{fig33},
\begin{equation}
A_{PC}^{LD} = A^{III,IV}_{PC}+A^{VI}_{PC}+A^{II}_{PC}~,\qquad 
A^{LD~\mu\nu}_{PV}=(A^{VII,VIII}_{PV})^{\mu\nu}+(A^{V}_{PV})^{\mu\nu}+(A^{I}_{PV})^{\mu\nu}~.
\end{equation} 
These are in the hybrid model given by
\begin{align}
\label{aa}
A^{III,IV}_{PC}&=4J^{(i)}g_V f_{D*}\sqrt{{m_{D*}\over m_{D}}}~{m_{D*}\over m_V^2-m_{D*}^2}~,\\
~\nonumber\\
 A^{VI}_{PC}&=2K^{(i)}C_{VV\Pi}~f_D~m_D^2~,
\nonumber\\
~\nonumber\\
 A^{II}_{PC}&=-\sqrt{2}~\lambda ~f_{N^{(i)}}N~ {a_2\sin\theta_C\cos\theta_C\over f_{Cabb}^{(i)}}~f_{D^*}\tilde g_V~ \sqrt{{m_{D*}\over m_{D}}}~{m_{D*}\over q^2-m_{D*}^2}~, \nonumber\\
~\nonumber\\
(A^{VII,VIII}_{PV})^{\mu\nu}&=A_{Bremss.}^{\mu\nu}=-f_Dg_V(q^2 g^{\mu\nu}-q^\mu q^\nu) L^{(i)}~,
\nonumber\\
~\nonumber\\
(A^{V}_{PV})^{\mu\nu}&=\frac{1}{\sqrt{2}}~ M^{(i)}\tilde g_V g_V\sqrt{m_D}~\biggl(g^{\mu\delta}-\frac{p^\mu p^\delta}{m_{V^0}^2}\biggr)\biggl[\alpha_1 g_{\delta}^{~\nu}-\alpha_2{p_{\delta}p^{\nu}\over m_D^2}\biggr]~,
\nonumber\\
~\nonumber\\
(A^{I}_{PV})^{\mu\nu}&=\frac{1}{\sqrt{2}}~f_{N^{(i)}}N~{a_2\sin\theta_C\cos\theta_C\over f_{Cabb}^{(i)}}~\tilde g_V\sqrt{m_D}\biggl(g^{\mu\delta}-\frac{p^\mu p^\delta}{m_{V^0}^2}\biggr)\biggl[\alpha_1 g_{\delta}^{~\nu}-\alpha_2{p_{\delta}p^{\nu}\over m_D^2}\biggr]~.\nonumber
\end{align}
Different parameters in the expressions indicate which terms in the Lagrangian are responsible for various parts of the amplitudes. 
The functions  $J^{(i)}(q^2)$, $K^{(i)}(q^2)$, $L^{(i)}(q^2)$, $M^{(i)}(q^2)$ and $f_{N^{(i)}}N(q^2)$ depend on a given $D\to V\gamma^*$ decay. They are given in Table \ref{tab.vll}:  $J^{(i)}$ and $M^{(i)}$ depend on the $D$ meson in the initial state, while  $K^{(i)}$ and $L^{(i)}$ depend on the vector meson $V$ in the final state. The Table \ref{tab.vll} further gives the functions $J^D(q^2)$, $K^V(q^2)$, $L^V(q^2)$, $M^D(q^2)$ and $N(q^2)$ expressed as
\begin{eqnarray}
\label{jklm}
J^{D^0}\!\!\!&=&\!\!\!\lambda '-{\lambda\tilde g_V\over 2\sqrt{2}}\biggl[{g_{\rho}\over q^2-m_{\rho}^2+i\Gamma_{\rho}m_{\rho}}+{g_{\omega}\over 3(q^2-m_{\omega}^2+i\Gamma_{\omega}m_{\omega})}\biggr]~,
\\
J^{D^+}\!\!\!&=&\!\!\!\lambda '-{\lambda\tilde g_V\over 2\sqrt{2}}\biggl[-{g_{\rho}\over q^2-m_{\rho}^2+i\Gamma_{\rho}m_{\rho}}+{g_{\omega}\over 3(q^2-m_{\omega}^2+i\Gamma_{\omega}m_{\omega})}\biggr]~,
\nonumber\\
J^{D_s^+}\!\!\!&=&\!\!\!\lambda '+{\lambda\tilde g_V\over 2\sqrt{2}}~{2g_{\phi}\over 3(q^2-m_{\phi}^2+i\Gamma_{\phi}m_{\phi})}~,
\nonumber
\end{eqnarray}
\begin{eqnarray}
K^{\bar K^{*0}}\!\!\!&=&\!\!\!\biggl[{g_{\rho}\over q^2-m_{\rho}^2+i\Gamma_{\rho}m_{\rho}}-{g_{\omega}\over 3(q^2-m_{\omega}^2+i\Gamma_{\omega}m_{\omega})}+{2g_{\phi}\over 3(q^2-m_{\phi}^2+i\Gamma_{\phi}m_{\phi})}\biggr]{1\over m_D^2-m_K^2}~,
\nonumber\\
K^{K^{*+}}\!\!\!&=&\!\!\!\biggl[-{g_{\rho}\over q^2-m_{\rho}^2+i\Gamma_{\rho}m_{\rho}}-{g_{\omega}\over 3(q^2-m_{\omega}^2+i\Gamma_{\omega}m_{\omega})}+{2g_{\phi}\over 3(q^2-m_{\phi}^2+i\Gamma_{\phi}m_{\phi})}\biggr]{1\over m_D^2-m_K^2}~,
\nonumber\\
K^{\rho^+}\!\!\!&=&\!\!\!-{2g_{\omega}\over 3(q^2-m_{\omega}^2+i\Gamma_{\omega}m_{\omega})}{1\over m_D^2-m_{\pi}^2}~,
\nonumber\\
K^{\rho^0}\!\!\!&=&\!\!\!-2\sqrt{2}{g_{\rho}\over q^2-m_{\rho}^2+i\Gamma_{\rho}m_{\rho}}
\biggl[{K_{\eta}^d(K_{\eta}^d-K_{\eta}^s)\over m_D^2-m_{\eta }^2}+{K_{\eta^\prime}^d(K_{\eta^\prime}^{d}-K_{\eta^\prime}^{s})\over m_D^2-m_{\eta '}^2}\biggr]
~
\nonumber\\
\!\!\!&+&\!\!\!{\sqrt{2}\over 3}{g_{\omega}\over (q^2-m_{\omega}^2+i\Gamma_{\omega}m_{\omega})}{1\over m_D^2-m_{\pi}^2}~,
\nonumber\\
K^{\omega}\!\!\!&=&\!\!\!-{2\sqrt{2}\over 3}{g_{\omega}\over q^2-m_{\omega}^2+i\Gamma_{\omega}m_{\omega}}\biggl[{K_{\eta}^d(K_{\eta}^d-K_{\eta}^s)\over m_D^2-m_{\eta }^2}+{K_{\eta^\prime}^d(K_{\eta^\prime}^{d}-K_{\eta^\prime}^{s})\over m_D^2-m_{\eta '}^2}\biggr]
\nonumber\\
\!\!\!&+&\!\!\!\sqrt{2}{g_{\rho}\over q^2-m_{\rho}^2+i\Gamma_{\rho}m_{\rho}}{1\over m_D^2-m_{\pi}^2}~,
\nonumber\\
K^{\phi}\!\!\!&=&\!\!\!-{4\over 3}{g_{\phi}\over( q^2-m_{\phi}^2+i\Gamma_{\phi}m_{\phi})}\biggl[{K_{\eta}^s(K_{\eta}^d-K_{\eta}^s)\over m_D^2-m_{\eta }^2}+{K_{\eta^\prime}^s(K_{\eta^\prime}^{d}-K_{\eta^\prime}^{s})\over m_D^2-m_{\eta '}^2}\biggr]~,
\nonumber\\
 L^{\rho^+}\!\!\!&=&\!\!\!0~,
\nonumber\\
 L^{K^{*+}}\!\!\!&=&\!\!\!\frac{1}{m_{K^*}^2}\biggl[{m_{K^*}^2-m_{\rho}^2\over 2(q^2-m_{\rho}^2+i\Gamma_{\rho}m_{\rho})}+{m_{K^*}^2-m_{\omega}^2\over 6(q^2-m_{\omega}^2+i\Gamma_{\omega}m_{\omega})}+{m_{K^*}^2-m_\phi^2\over 3(q^2-m_{\phi}^2+i\Gamma_{\phi}m_{\phi})}\biggr]~,
\nonumber\\  
M^{D^0}\!\!\!&=&\!\!\!{g_{\rho}\over q^2-m_{\rho}^2+i\Gamma_{\rho}m_{\rho}}+{g_{\omega}\over 3(q^2-m_{\omega}^2+i\Gamma_{\omega}m_{\omega})}~,
\nonumber\\
M^{D^+}\!\!\!&=&\!\!\!-{g_{\rho}\over q^2-m_{\rho}^2+i\Gamma_{\rho}m_{\rho}}+{g_{\omega}\over 3(q^2-m_{\omega}^2+i\Gamma_{\omega}m_{\omega})}~,
\nonumber\\
M^{D_s^+}\!\!\!&=&\!\!\!-{2g_{\phi}\over 3(q^2-m_{\phi}^2+i\Gamma_{\phi}m_{\phi})}~,
\nonumber\\
N\!\!\!&=&\!\!\!{g_{\rho}^2\over q^2-m_{\rho}^2+i\Gamma_{\rho}m_{\rho}}-{g_{\omega}^2\over 3(q^2-m_{\omega}^2+i\Gamma_{\omega}m_{\omega})}-{2g_{\phi}^2\over 3(q^2-m_{\phi}^2+i\Gamma_{\phi}m_{\phi})}~.\nonumber
\end{eqnarray}
For the case of the real photon in the final state one should take $q^2\!=\!0$ and $\Gamma_{V^0}(q^2\!=\!0)=0$.
The coefficients $K_{\eta,\eta^\prime}^{d,s}$  in the expressions for $K^{\rho^0}$ and $K^{\omega}$ depend on the $\eta-\eta^\prime$ mixing angle $\theta_P$  (\ref{mixing}). The $\eta$ and $\eta^\prime$ mesons enter as the intermediate states in the diagram $IV$ of Fig.  \ref{fig33}. 

\begin{table}[h]
\begin{center}
\begin{tabular}{|c|c||c|c|c|c|c|c|}
\hline
$i$ & $D\to V l^+l^-$ & $f_{Cabb}^{(i)}$  
& $J^{(i)} $ 
& $K^{(i)} $ & $L^{(i)}$ & $M^{(i)}$ & $f_{N^{(i)}}$ \\
 & $D\to V\gamma$ & &&&&&\\
\hline \hline
$1$ & $ D^0 \to {\bar K}^{*0} \gamma^*$ &$ a_2\cos^2\theta_C$ & $J^{D^0}$ & $K^{\bar K^{*0}}$ & $0$ & $M^{D^0}$ & $0$   \\
\hline
$2$ & $ D_s^+ \to \rho^+ \gamma^*$ & $a_1\cos^2\theta_C$ & $J^{D_s^+}$ & $K^{\rho^+}$ & $L^{\rho^+}$ & $M^{D_s^+}$ & $0$  \\
\hline
\hline
$3$ & $ D^0 \to \rho^{0}\gamma^*$ &$ -a_2\sin\theta_C\cos\theta_C$ & $-J^{D^0}/\sqrt{2}$ & $K^{\rho^0}$ & $0$ & $-M^{D^0}/\sqrt{2}$ & $1/\sqrt{2}$ \\
\hline
$4$ & $ D^0 \to \omega \gamma^*$ &$ -a_2\sin\theta_C\cos\theta_C$ & $J^{D^0}/\sqrt{2}$ & $K^{\omega}$ & $0$ &  $M^{D^0}/\sqrt{2}$ & $1/\sqrt{2}$  \\
\hline
$5$ & $ D^0 \to \phi \gamma^*$ &$ a_2\sin\theta_C\cos\theta_C$ & $J^{D^0}$ & $K^{\phi}$ & $0$ & $M^{D^0}$ & $0$  \\
\hline
$6$ & $ D^+ \to \rho^+ \gamma^*$ &$ -a_1\sin\theta_C\cos\theta_C$ & $J^{D^+}$ & $K^{\rho^+}$ & $L^{\rho^+}$ & $M^{D^+}$ & $1$  \\
\hline
$7$ & $ D_s^+ \to K^{*+ }\gamma^*$ &$ a_1\sin\theta_C\cos\theta_C$ & $J^{D_s^+}$ & $K^{K^{*+}}$ & $L^{K^{*+}}$ & $M^{D_s^+}$  & $1$ \\
\hline
\hline 
$8$ & $ D^+ \to K^{*+} \gamma^*$ &$ -a_1\sin^2\theta_C$ & $J^{D^+}$ & $K^{K^{*+}}$ & $L^{K^{*+}}$ & $M^{D^+}$ & $0$ \\
\hline
$ 9$ & $ D^0 \to K^{*0} \gamma^*$ &$ -a_2\sin^2\theta _C$ & $J^{D^0}$ & $K^{\bar K^{*0}}$ & $0$ & $M^{D^0}$ & $0$ \\
\hline
\end{tabular}
\caption{ The Cabibbo factor $f_{Cabb}^{(i)}$ and the functions $J^{(i)}$, $K^{(i)}$, $L^{(i)}$, $M^{(i)}$ and $f_N^{(i)}$ for nine $D\to Vl^+l^-$ and $D\to V\gamma$ decays. They enter the expressions for the amplitudes ${\cal A}$ (\ref{5.amp1}, \ref{aa}, \ref{5.amp2}, \ref{5.amp3}, \ref{aa2}). The functions $J^D$, $K^V$, $L^V$, $M^D$ and $N$ are further given in (\ref{jklm}). 
 }
\label{tab.vll}
\end{center}
\end{table}

\subsubsection{The gauge invariance}

The amplitudes for  $D\to V\gamma^*(\epsilon,q)$ decays have to be invariant under the gauge transformation $\epsilon^\mu\to \epsilon^\mu+Cq^\mu$ and should have the general form given in (\ref{amp.gammas}).
The parity conserving  and the bremsstrahlung parts of the amplitude (\ref{5.amp1}) are gauge invariant. The non-bremsstrahlung part of the parity violating amplitude (\ref{5.amp1}, \ref{aa}), $A_{PV}^V+A_{PV}^I$,   is not in a   manifestly gauge invariant form  yet. The idea how to achieve the gauge invariance  was proposed in \cite{GP,FPS1,FS,FPS3} and has been used also for the case of $B_c\to B_u^*\gamma$ decay in Section 4.2. 
The parity violating amplitude $A_{PV}^V+A_{PV}^I$ is incorporated via the nonleptonic decay $P\to VV^0$ with $V^0=\rho,~\omega,~\phi$ followed by $V^0\to \gamma^*$. The $VV^0$ intermediate state involves three helicity amplitudes $++,~--,~00$. The real photon in the final state can not have longitudinal polarization and the helicity state $00$ has to be discarded when the decay $P\to  V\gamma$ is considered. This is achieved by relating the form factors $A_1(0)$ and $A_2(0)$ (\ref{4.5}, \ref{4.5a}) that parameterize the nonleptonic decay $P\to VV^0$ in the factorization approximation. The gauge invariance in  $D\to V\gamma^*\to Vl^+l^-$ decay is  achieved by relating the form factors $A_1(q^2)$ and $A_2(q^2)$. This idea is now discussed in detail separately for $A_{PV}^V$ and $A_{PV}^I$ amplitudes: 
 \begin{itemize}
\item 
The amplitude $A_{PV}^I$ corresponds to 
the parity violating part of the {\bf long distance penguin contribution} and is given by  the diagram $I$ in Fig.  \ref{fig32}. First I turn to $D\to V\gamma$ decay with  the real photon in the final state.  
 The nonleptonic decay $D\to VV^0$  is followed by the conversion $V^0\to \gamma$ and the real photon couples only to the transverse polarization of $V^0$, as discussed in Section 4.2 for $B_c\to B_u^*\gamma$ decay. This is equivalent to the requirement that  the parity violating amplitude $\epsilon^*_\mu\epsilon_\nu^{*\prime}(A^{I}_{PV})^{\mu\nu}$ (\ref{5.amp1}) must be invariant under $\epsilon^\mu\to \epsilon^\mu+Cq^\mu$, which amounts to 
$$\alpha_2=\frac{2m_D^2}{m_D^2-m_V^2}~\alpha_1\qquad{\rm at}~~q^2=0~.$$
If $\alpha_1$ and $\alpha_2$ are expressed in terms of the form factors $A_1^{D\to V}(0)$ and $A_2^{D\to V}(0)$ for  $\langle V|\bar u\gamma^\mu(1-\gamma_5)c|D\rangle$ (\ref{formV}) this is equivalent to the relation
$$A_2^{D\to V}(0)= {(m_{D}+m_{V})^2\over m_{D}^2-m_{V}^2}~A_1^{D\to V}(0)~,$$
which is exactly the same condition as obtained in (\ref{4.5}).
The helicity analysis of the final state in Section 4.2 showed, that one should express $A_2(0)$ in terms of $A_1(0)$ (\ref{4.5a}) in the expression for the amplitude
\begin{equation}
\label{5.51}
A_2^{D\to V}(0)\to {(m_{D}+m_{V})^2\over m_{D}^2-m_{V}^2}~A_1^{D\to V}(0)~
\end{equation}
or equivalently
\begin{equation}
\label{5.52}
\alpha_2\to\frac{2m_D^2}{m_D^2-m_V^2}~\alpha_1\qquad{\rm at}~~q^2=0~.
\end{equation}
With this the replacement the gauge invariant long distance penguin amplitude for $D\to V\gamma$ is rewritten bellow (\ref{5.amp2}). The coefficient $N$ (\ref{jklm}), present in $A^{I}_{PV}$ and $A^{II}_{PV}$, has to be evaluated at the $q^2=0$  
\begin{equation}
\label{5.53} 
N(q^2\!=\!0)=-2\biggl({g_{\rho}^2\over  2m_{\rho}^2}-{g_{\omega}^2\over  
6m_{\omega}^2}-{g_{\phi}^2\over  3m_{\phi}^2}\biggr)\equiv -2~C_{V\!M\!D}=(2.4\pm2.4)\cdot 10^{-3}~{\rm GeV}^2~.
\end{equation}
 The three terms in (\ref{5.53}) exactly cancel in  the $SU(3)$ flavour limit and one has to use the accurate values for $g_{V^0}$ and $m_{V^0}$ to evaluate  $C_{V\!M\!D}$  (\ref{cvmd1}, \ref{cvmd}), as explained in Section 4.2. The long distance penguin contribution turns out to be much smaller than the weak annihilation contribution in the $D\to V\gamma$ decays due to the smallness of the coefficient $N(q^2\!=\!0)$. 

Now I turn to the case of the lepton pair in the final state. The gauge invariance under $\epsilon^\mu\to \epsilon^\mu+Cq^\mu$ imposes the replacement
\begin{equation}
\label{5.54} 
\alpha_2\to\frac{2m_D^2}{m_D^2-m_V^2+q^2}~\alpha_1
\end{equation}
or in terms of the $D\to V$ form factors 
\begin{equation}
\label{5.154}
A_2^{D\to V}(q^2)\to {(m_{D}+m_{V})^2\over m_{D}^2-m_{V}^2+q^2}~A_1^{D\to V}(q^2)~.
\end{equation}
This is equivalent to (\ref{5.51}, \ref{5.52})  at $q^2\!=\!0$. 
Let me note that the long distance penguin contribution in $D\to Vl^+l^-$ decays is relatively more important than in $D\to V\gamma$ decays since the maximums of the function $N(q^2)$ (\ref{jklm}) at $q^2\!=\!m_\rho^2,m_\omega^2$ and at $q^2\!=\!m_\phi^2$ are well separated and the $SU(3)$ flavour cancellation is not so effective. 

\item The gauge invariance of the {\bf weak annihilation amplitude} $A_{PV}^{V}$ is achieved exactly in the same way, as for the long distance penguin amplitude $A_{PV}^{I}$. The gauge invariance under $\epsilon\to \epsilon+Cq$ implies  the replacement 
\begin{equation*}
\alpha_2\to\frac{2m_D^2}{m_D^2-m_V^2+q^2}~\alpha_1\qquad{\rm for ~general}~~q^2~,
\end{equation*}
exactly as in (\ref{5.54}). In terms of the $\langle V^0|(V-A)^\mu|D\rangle$ form factors this is equivalent to 
$$A_2^{D\to V_0}(q^2)\to {(m_{D}+m_{V_0})^2\over m_{D}^2-m_{V}^2+q^2}~A_1^{D\to V_0}(q^2)~$$
and the mass $m_{V_0}$ enters the expression only due to the definition of the form factors in (\ref{5.55}). 

\end{itemize}

\subsubsection{The amplitudes}

The manifestly gauge invariant amplitudes for $D\to V\gamma$ (\ref{amp.gamma}) and $D\to Vl^+l^-$ (\ref{amp.ll}) decays are given by the amplitudes (\ref{5.amp1}) together with the replacement  (\ref{5.54}) 
\begin{equation}
\label{5.amp2}
{\cal A}^{LD}[D(p)\to V(\epsilon^\prime,p^\prime)~\gamma(q,\epsilon)]  =  -i\frac{G_F}{{\sqrt 2}}e_0 f_{Cabb}^{(i)}~\epsilon_\mu^*\epsilon^{*\prime}_{\nu} \bigl[iA_{PV}^{LD}(p^\mu q^\nu-g^{\mu\nu}p\cdot q)+\epsilon^{\mu \nu \alpha \beta} p_{\alpha}q_\beta A_{PC}^{LD}\bigr] 
\end{equation}
and
\begin{align}
\label{5.amp3}
{\cal A}^{LD}&[D(p)\to V(\epsilon^\prime,p^\prime)~l^+(p_+)l^-(p_-)]  = \nonumber\\
 &-i\frac{G_F}{{\sqrt 2}}e^2_0 f_{Cabb}^{(i)}~\frac{1}{q^2}\bar u(p_-)\gamma_\mu v(p_+)\epsilon^{*\prime}_{\nu} \bigl[iA_{PV}^{LD}(p^\mu q^\nu-g^{\mu\nu}p\cdot q)+\epsilon^{\mu \nu \alpha \beta} p_{\alpha}q_\beta A_{PC}^{LD}\bigr] ~
\end{align}
with
\begin{equation*}
A_{PC}^{LD} = A^{III,IV}_{PC}+A^{VI}_{PC}+A^{II}_{PC}~,\qquad
A_{PV}^{LD}=A^{VII,VIII}_{PV}+A^{V}_{PV}+A^{I}_{PV}~
\end{equation*} 
and
\begin{align}
\label{aa2}
A^{III,IV}_{PC}&=4J^{(i)}g_V f_{D*}\sqrt{{m_{D*}\over m_{D}}}~{m_{D*}\over m_V^2-m_{D*}^2}~,\\
~\nonumber\\
A^{VI}_{PC}&=2K^{(i)}C_{VV\Pi}~f_D~m_D^2~,
\nonumber\\
~\nonumber\\
A^{II}_{PC}&=-\sqrt{2}~\lambda ~f_{N^{(i)}}N~ {a_2\sin\theta_C\cos\theta_C\over f_{Cabb}^{(i)}}~f_{D^*}\tilde g_V~ \sqrt{{m_{D*}\over m_{D}}}~{m_{D*}\over q^2-m_{D*}^2} ~,\nonumber\\
~\nonumber\\
(A^{VII,VIII}_{PV})^{\mu\nu}&=A_{Bremss.}^{\mu\nu}=-f_Dg_V(q^2 g^{\mu\nu}-q^\mu q^\nu) L^{(i)}~,
\nonumber\\
~\nonumber\\
A^{V}_{PV}&=-\frac{1}{\sqrt{2}}~ M^{(i)}\tilde g_V g_V\sqrt{m_D}~\frac{2}{m_D^2-m_V^2+q^2}~\alpha_1~,
\nonumber\\
~\nonumber\\
A^{I}_{PV}&=-\frac{1}{\sqrt{2}}~f_{N^{(i)}}N~{a_2\sin\theta_C\cos\theta_C\over f_{Cabb}^{(i)}}~\tilde g_V\sqrt{m_D}~\frac{2}{m_D^2-m_V^2+q^2}~\alpha_1~.\nonumber
\end{align}
The functions  $J^{(i)}(q^2)$, $K^{(i)}(q^2)$, $L^{(i)}(q^2)$, $M^{(i)}(q^2)$ and $f_{N^{(i)}}N(g^2)$  are given in Table \ref{tab.vll}.  This table further involves the functions $J^D(q^2)$, $K^V(q^2)$, $L^V(q^2)$, $M^D(q^2)$ and $N(q^2)$ given in (\ref{jklm}). For the case of $D\to V\gamma$ decay, one has to take $q^2\!=\!0$, $\Gamma_{V^0}(0)=0$ and $N(0)$ from (\ref{5.53}).

The decay rate for $D\to V\gamma$ is expressed as
\begin{equation}
\label{5.212}
\Gamma=\frac{1}{4\pi}\biggl(\frac{m_D^2-m_V^2}{2m_D}\biggr)^3(|{\cal A}_{PC}|^2+|{\cal A}_{PV}|^2)~,
\end{equation}
where ${\cal A}_{PC,PV}$ are defined in terms of $A_{PC,PV}$ as
\begin{equation}
\label{defcala}
{\cal A}_{PC,PV}=\tfrac{G_F}{{\sqrt 2}}e_0 f_{Cabb}^{(i)} ~A_{PC,PV}~.
\end{equation}

The decay rate for $D\to Vl^+l^-$ is given by the square of the amplitude,  
summed over the polarizations of the three particles in the final state and 
integrated over the three body phase space
$$
\Gamma={1\over 2m_D(2\pi)^5}\sum_{polar.}\int |{\cal 
A}(p^\prime,p_+,p_-)|^2 ~{d^3p^\prime\over 2p^{\prime0}} {d^3p_{+}\over 2p_{+}^0} 
{d^3p_{-}\over 2p_{-}^0}~ \delta(p^\prime+p_++p_--p)~.
$$

\subsubsection{The sign of $\boldsymbol{C_{VV\Pi}}$}

The magnitudes and the relative signs of the parameters appearing the in expression for the amplitudes have been determined above. The sign of the parameter $C_{VV\Pi}$, that is present in the amplitude $A_{PC}^{VI}$, has been left undetermined. The absolute value of this parameter was determined from $V\to P\gamma$ data $|C_{VV\Pi}|=0.31$ in (\ref{cvvp1}). The sign of $C_{VV\Pi}$ can be determined by inspecting the relative sign of amplitudes $A_{PC}^{VI}$ and $A_{PC}^{III,IV}$  (\ref{5.amp2}) for $D\to V\gamma$ decays and comparing it to the quark model results. For this purpose  I rewrite the weak annihilation amplitude for the $B_c\to B_u^*\gamma$ decay (\ref{4.9}), as obtained in the ISGW quark model \cite{ISGW},  for the case of $D^0\to \bar K^{*0}\gamma$ decay 
\begin{equation}
\label{5.57}
A_{PC}^{III,IV}\!+\!A_{PC}^{VI}~\propto~ {\mu_{D^0}g_{D^{0*}}g_{\bar K^{0*}}\over 
m_{D^{0*}}^2-m_{\bar K^{0*}}^2}+{\mu_{\bar K^{0}}m_{D^0}^2f_{D^0}f_{\bar K^{0}}\over 
m_{D^0}^2-m_{\bar K^0}^2}~.
\end{equation}
 The quark model results for decay constants $f$ and $g$, defined in (\ref{4.8}), are positive (\ref{4.9a}). The magnetic moment $\mu$ defined in (\ref{4.8}) is given by (\ref{4.9a}) 
$$\mu_{D^0}\!=\!\sqrt{{m_{D^{*0}}\over 
m_{D^0}}}\biggl({2\beta_{D^0}\beta_{D^{*0}}\over 
\beta_{D^0}^2+\beta_{D^{*0}}^2}\biggr)^{\tfrac{3}{2}}~\biggl[{e_c\over M_c}+{e_u\over M_u}\biggr]~,\quad \mu_{\bar K^0}\!=\!\sqrt{{m_{\bar K^{*0}}\over 
m_{\bar K^0}}}\biggl({2\beta_{\bar K^0}\beta_{\bar K^{*0}}\over \beta_{\bar K^0}^2+\beta_{\bar K^{*0}}^2}\biggr)^{\tfrac{3}{2}}~\biggl[{e_s\over M_s}+{e_d\over M_d}\biggr]$$
with $e_{u,c}=2/3$ and $e_{d,s}=-1/3$. The quark model gives the negative relative sign for the amplitudes $A_{PC}^{VI}$ and $A_{PC}^{III,IV}$ (\ref{5.57}) for the case of $D^0\to \bar K^{*0}\gamma$ decay \cite{GHMW}. The hybrid model expression $A_{PC}^{III,IV}+A_{PC}^{VI}$ (\ref{aa2}) for $D^0\to \bar K^{*0}\gamma$ decay  with $\lambda\!<\!0$ (\ref{lambda}) and $\lambda^\prime\!<\!0$ (\ref{lambda.prime}) indicates that $C_{VV\Pi}\!>\!0$,
namely
\begin{equation}
\label{cvvp}
C_{VV\Pi}=0.31~.
\end{equation}     
The same conclusion is reached by comparing the quark and hybrid model results for any other $D\to V\gamma$ decay \cite{GHMW}.  The positive sign is in  agreement with the hidden symmetry prediction  $C_{VV\Pi}=3\tilde g_V^2/32\pi^2$ \cite{hidden,witten}. 

\subsection{The short distance contribution}

The short distance contribution is present only in the Cabibbo suppressed $D\to V\gamma$ and $D\to Vl^+l^-$ decays (decays 3 to 7 in Table \ref{tab.vll}) and is induced by the flavour changing neutral quark transitions $c\to u\gamma$ and $c\to ul^+l^-$, respectively. In the standard model, the corresponding branching ratios $Br(c\to u\gamma)=(1.3\pm 0.6)\cdot 10^{-8}$ (\ref{2.60}) \cite{GHMW} and 
$Br(c\to ul^+l^-)\sim (1.7{+0.1\atop -0.7})\cdot 10^{-9}$ (\ref{2.33}) are small. Assuming that the exclusive rates of interest amount to about $10\%$ of the inclusive ones, we have
\begin{equation}
\label{5.213}
Br^{SD}(D\to V\gamma)\sim 10^{-9}\quad{\rm and}\quad   Br^{SD}(D\to Vl^+l^-)\sim 10^{-10}~.
\end{equation}
In the standard model the short distance contributions are negligible compared to the corresponding long distance contributions for the Cabibbo suppressed decays given in Tables \ref{tab.gamma2} and \ref{tab.vll2} bellow
$$Br^{LD}(D\to V\gamma)\sim 10^{-5}\quad{\rm and}\quad   Br^{LD}(D\to Vl^+l^-)\sim 10^{-7}~.$$

\vspace{0.1cm}

The exact evaluation of the short distance contributions to $D\to V\gamma$ decays  amounts to the calculation of (\ref{asd}, \ref{o7})
$${\cal A}^{SD}(D\to V\gamma)=\langle \gamma V|:i{\cal L}^{c\to u\gamma}:|D\rangle\propto
c_7^{eff}\epsilon^\mu q^\nu\langle V|\bar u\sigma_{\mu\nu}(1+\gamma_5)c|D\rangle~.$$
 The matrix elements $\langle V|\bar u\sigma_{\mu\nu}(1+\gamma_5)c|D\rangle$  have not been studied  in this work so far. The form factors for $\langle V|\bar u\sigma_{\mu\nu}(1+\gamma_5)c|D\rangle$ can be related to the form factors for  $\langle V|\bar u\gamma_\mu(1-\gamma_5)c|D\rangle$ (\ref{formV}) at $q_{max}^2=(m_D-m_V)^2$ using the heavy quark symmetry \cite{IW1}. Together with an additional assumption for $q^2$ behavior of the form factors $\langle V|\bar u\sigma_{\mu\nu}(1+\gamma_5)c|D\rangle$, the short distance amplitudes for $D\to V\gamma$ decays can be predicted. In view of the fact, that  $Br^{SD}(D\to V\gamma)\ll Br^{LD}(D\to V\gamma)$ in the standard model, I do not proceed with the evaluation of the short distance amplitudes for these decays.

The exact evaluation of the short distance contributions to $D\to Vl^+l^-$ decays  amounts to the determination of (\ref{o9}, \ref{asd})
$${\cal A}^{SD}(D\to Vl^+l^-)\!=\!\langle l^+l^- V|:i{\cal L}^{c\to ul^+l^-}\!\!:|D\rangle\!=\!-i\frac{G_F}{\sqrt{2}}{e^2\over 8\pi^2}c_9^{eff}\langle V|\bar u_{\alpha}\gamma_{\mu}(1-\gamma_5)c_{\alpha}|D\rangle\langle l^+l^-|\bar l\gamma^{\mu}l|0\rangle~.$$
The  expressions for the form factors $\langle V|\bar u_{\alpha}\gamma_{\mu}(1-\gamma_5)c_{\alpha}|D\rangle$ in the hybrid model are given in (\ref{formV}). The calculation of the short distance amplitudes is then straightforward giving
\begin{align}
\label{5.amp4}
{\cal A}^{SD}[D(p)\to V(\epsilon^\prime,p^\prime)&~l^+(p_+)l^-(p_-)]  = \nonumber\\
 &-i\frac{G_F}{{\sqrt 2}}e^2_0 \bar u(p_-)\gamma_\mu v(p_+)\epsilon^{*\prime}_{\nu} \bigl[iA_{PV}^{SD~\mu\nu}+\epsilon^{\mu \nu \alpha \beta} p_{\alpha}q_\beta A_{PC}^{SD}\bigr]\nonumber~, ~
\end{align}
 with
\begin{align*}
A_{PC}^{SD}&=\frac{c_9^{eff}}{8\pi^2}~2\sqrt{2}f_{N^{(i)}}\lambda~f_{D^*}\tilde g_V~ \sqrt{{m_{D*}\over m_{D}}}~{m_{D*}\over q^2-m_{D*}^2}~,\\
A_{PV}^{SD~\mu\nu}&=-\frac{c_9^{eff}}{8\pi^2}~\sqrt{2}f_{N^{(i)}}\sqrt{m_D}~\tilde g_V~\biggl[\alpha_1 g^{\mu\nu}-\alpha_2{p^{\mu}q^{\nu}\over m_D^2}\biggr]~
\end{align*}
and $f_{N^{(i)}}$ is given in Table \ref{tab.vll} for different Cabibbo suppressed decays. 
The amplitude has the proper gauge invariant form (note that there is no  photon propagator $1/q^2$  as in the long distance contribution) and can be cast to the general form given in (\ref{amp.ll}). In this case no relation between the coefficients $\alpha_1$ and $\alpha_2$ need to be imposed and I use the values given in (\ref{lambda}).

\subsection{The results}

\subsubsection{The $\boldsymbol{D\to V\gamma}$ decays}

First I present the results for the {\bf long distance contributions} to the weak radiative decays of the charm mesons. 
The long distance amplitudes for $D\to V\gamma$ decays (\ref{5.amp2}) are expressed in terms of parity conserving and parity violating amplitudes ${\cal A}_{PC,PV}^{I,..,VIII}$ (\ref{aa2}, \ref{defcala}), where the Arabic numbers denote contributions from different diagrams in Figs.  \ref{fig32} and  \ref{fig33}.  The diagrams $VII$ and $VIII$ correspond to the bremsstrahlung and their amplitude (\ref{aa2}) at $q^2\!=\!0$ is equal to zero. The numerical values of the amplitudes ${\cal A}_{PC,PV}^{I-VI}$ (\ref{aa2}, \ref{defcala}) taken at $q^2\!=\!0$ are given in Table \ref{tab.gamma1} for nine decays.  The long distance penguin contribution, given by ${\cal A}_{PV}^I$ and ${\cal A}_{PC}^{II}$, is small since it is proportional to the breaking of the $SU(3)$ flavour symmetry (\ref{5.53}). The remaining parity conserving amplitudes ${\cal A}_{PC}^{III,IV}$ have  negative relative sign. This is a consequence of the arguments on the sign of $C_{VV\Pi}$ given above.  In the case of $D^0$ decays, these two parity conserving contributions tend to have similar magnitude and almost cancel. The predicted rates (\ref{5.212}) are given in the last column of Table \ref{tab.gamma1}. The predicted branching ratios are not strongly dependent on the errors of the input parameters given in Section 5.2. The main uncertainty comes from the validity of the model, which is estimated to be of the order of $50\%$.

\begin{table}[!htb]
\begin{center}
\begin{tabular}{|c||c|c|c|c|c||c|}
\hline
$D\to V \gamma$ & ${\cal A}_{PC}^{III,IV}$  
& ${\cal A}_{PC}^{VI} $ 
& ${\cal A}_{PC}^{II} $
& ${\cal A}_{PV}^{V}$
& ${\cal A}_{PV}^{I} $ &$Br_{LD}^{th}$\\
\hline \hline
$ D^0 \to {\bar K}^{*0} \gamma$ &$ -4.5$ & $5.6$ & $0$ & $-4.8$ & $0$ & $4.6\cdot 10^{-5}$  \\
\hline
$ D_s^+ \to \rho^+ \gamma$ &$-0.96$ & $7.3$ & $0$ & $-3.6$ & $0$ & $1.7\cdot 10^{-4}$  \\
\hline
\hline
$ D^0 \to \rho^{0} \gamma$ &$ -0.61$ & $0.96$ & $0.015$& $-0.65$ & $0.025$ & $1.2\cdot 01^{-6}$ \\
\hline
$ D^0 \to \omega \gamma$ &$ 0.54$ & $-0.96$ & $0.015$& $0.57$ & $0.025$ & $1.2\cdot 10^{-6}$  \\
\hline
$ D^0 \to \Phi \gamma$ &$ -1.4$ & $1.2$ & $0$& $-1.5$ & $0$ & $3.3\cdot 10^{-6}$  \\
\hline
$ D^+ \to \rho^+ \gamma$ &$ 0.42$ & $-1.3$ & $0.022$& $1.2$ & $0.035$ & $1.4\cdot 10^{-5}$  \\
\hline
$ D_s^+ \to K^{*+ }\gamma$ &$ -0.26$ & $2.3$ & $0.021$& $-0.97$ & $0.033$ & $1.4\cdot 10^{-5}$  \\
\hline
\hline 
$ D^+ \to K^{*+} \gamma$ &$ 0.11$ & $-0.41$ &$ 0$ &  $0.33$ & $ 0$ & $9.5\cdot 10^{-7}$  \\
\hline
$ D^0 \to K^{*0} \gamma$ &$ 0.23$ & $-0.29$ &$ 0$ &  $0.25$ & $ 0$  & $1.2\cdot 10^{-7}$\\
\hline
\end{tabular}
\caption{ The long distance amplitudes and branching ratios for $D\to V\gamma$ decays. The parity conserving and parity violating amplitudes ${\cal A}_{PC,PV}^{I,..,VI}$ (\ref{aa2}, \ref{defcala}) for the diagrams $I-VI$ in Figs.  \ref{fig32} and  \ref{fig33} are given in units $10^{-8}$ GeV$^{-1}$. The predicted branching ratios (\ref{5.212}) are given in the last column. First two decays are Cabibbo allowed, the next five are Cabibbo suppressed and the last two are doubly Cabibbo suppressed. }
\label{tab.gamma1}
\end{center}
\end{table}

In Table \ref{tab.gamma2}, our theoretical predictions \cite{FPS1,FS} (second column) are compared with other theoretical predictions \cite{BGHP,KSW,gamma.quark} (columns 3-5) and with the present experimental upper limits \cite{radiative.exp1} (last column). In view of the discussion on other theoretical approaches at the beginning of Section 5.4, the difference in the predicted branching ratios is well understood. The predictions of \cite{BGHP} tend to be higher due to the possible double counting in the phenomenological approach \cite{BGHP}. The QCD sum rules analysis \cite{KSW} did not incorporate the weak annihilation contributions where the photon is emitted after the weak transition. In our approach this contribution corresponds to $A_{PC}^{VI}$ in Table \ref{tab.gamma1}. The amplitudes $A_{PC}^{VI}$ and $A_{PC}^{III,IV}$ nearly cancel in $D^0$ decays, so the QCD sum rules predictions \cite{KSW} are naturally larger. The amplitude $A_{PC}^{VI}$ dominates over $A_{PC}^{III,IV}$ in charged $D$ meson decays  and the QCD sum rules predictions \cite{KSW} are naturally smaller. The experimental upper limit for the channel $D^0\to \bar K^{*0}\gamma$ is only one order of magnitude above the predicted rate and it is expected to be detected soon. There are unfortunately no experimental limits on the channel $D_s^+\to \rho^+\gamma$, which is predicted at the highest rate. 

The {\bf short distance} parts of the branching ratios in the Cabibbo suppressed decays are of the order of $10^{-9}$ in the standard model (\ref{5.213}) \cite{GHMW}. Since they are much smaller than the corresponding long distance contributions in the charm meson decays, they are not explicitly evaluated.

\begin{table}[!htb]
\begin{center}
\begin{tabular}{|c||c||c|c|c||c|}
\hline
$D\to V \gamma$ & $Br_{LD}^{th}$ & $Br_{LD}^{th}$\cite{BGHP} & $Br_{LD}^{th}$\cite{KSW} & $Br_{LD}^{th}$\cite{gamma.quark}&$Br^{exp}$\cite{radiative.exp1}\\
\hline \hline
$ D^0 \to {\bar K}^{*0} \gamma$ & $4.6\cdot 10^{-5}$&$[7-12]\cdot 10^{-5}$&$1.5\cdot 10^{-4}$&$[8-11]\cdot 10^{-5}$&$<7.6\cdot 10^{-4}$ \\
\hline
$ D_s^+ \to \rho^+ \gamma$ & $1.7\cdot 10^{-4}$&$[0.6-3.8]\cdot 10^{-4}$&$2.8\cdot10^{-5}$&$[0.8-2.1]\cdot 10^{-4}$ & \\
\hline
\hline
$ D^0 \to \rho^{0} \gamma$ &$1.2\cdot 10^{-6}$ &$[1-5]\cdot 10^{-6}$&$3.1\cdot 10^{-6}$&&$<2.4\cdot 10^{-4}$\\
\hline
$ D^0 \to \omega \gamma$ & $1.2\cdot 10^{-6}$&$2\cdot 10^{-6}$&&& $<2.4\cdot 10^{-4}$\\
\hline
$ D^0 \to \Phi \gamma$ & $3.3\cdot 10^{-6}$&$[1-34]\cdot 10^{-6}$&&&$<1.9\cdot 10^{-4}$  \\
\hline
$ D^+ \to \rho^+ \gamma$& $1.4\cdot 10^{-5}$ &$[2-6]\cdot 10^{-5}$&$2.7\cdot 10^{-6}$&& \\
\hline
$ D_s^+ \to K^{*+ }\gamma$  & $1.4\cdot 10^{-5}$ &$[0.8-3]\cdot 10^{-5}$&&& \\
\hline
\hline 
$ D^+ \to K^{*+} \gamma$ & $9.5\cdot 10^{-7}$ &&&$6\cdot 10^{-7}$& \\
\hline
$ D^0 \to K^{*0} \gamma$ & $1.2\cdot 10^{-7}$&&&&\\
\hline
\end{tabular}
\caption{The predicted branching ratios for  $D\to V\gamma$ decays as given by the long distance mechanisms: the predictions presented in this work  are given in the column 2; The theoretical predictions presented in \cite{BGHP}, \cite{KSW} and \cite{gamma.quark} are given in columns 3, 4 and 5, respectively (for comparison see the comments in the text). The experimental upper bounds \cite{radiative.exp1} are given in the last column. The short distance parts of the branching ratios for the Cabibbo suppressed decays $\sim 10^{-9}$ (\ref{5.213}) are negligible in the standard model. }
\label{tab.gamma2}
\end{center}
\end{table}

\subsubsection{The $\boldsymbol{D\to Vl^+l^-}$ decays}

\begin{figure}[!htb]
\begin{center}
\includegraphics[scale=.4]{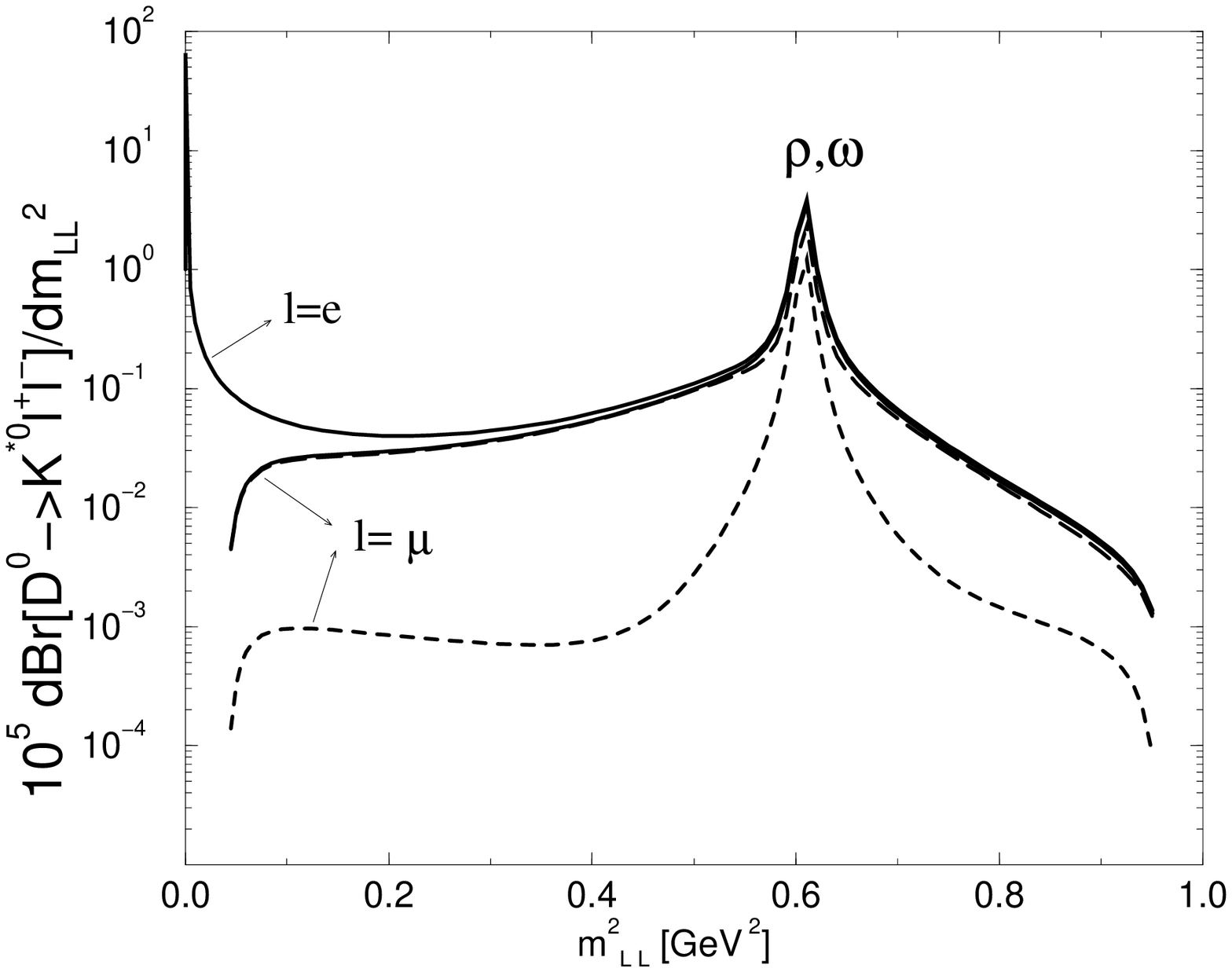} 
\quad
\includegraphics[scale=.4]{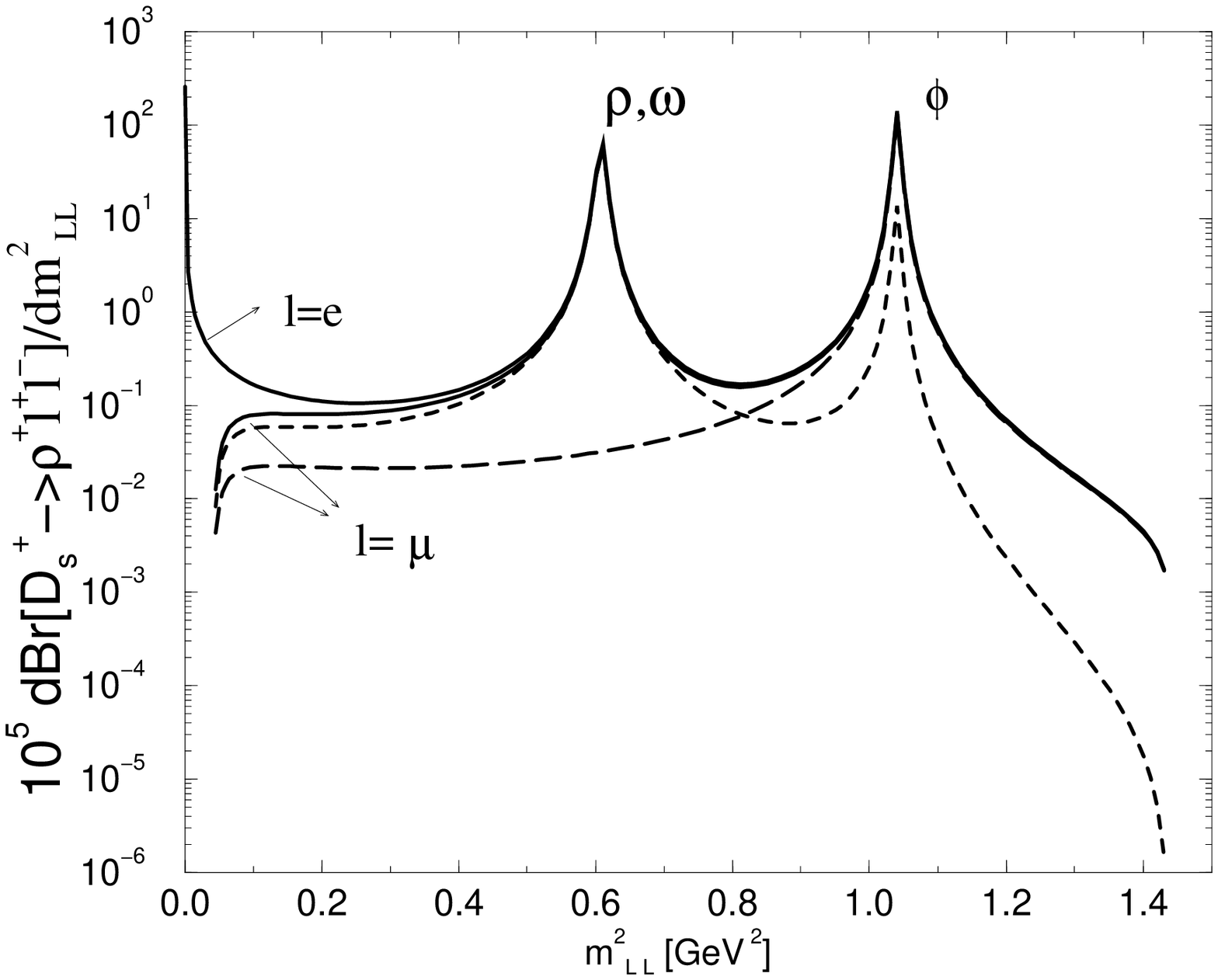} 
\\
\includegraphics[scale=.4]{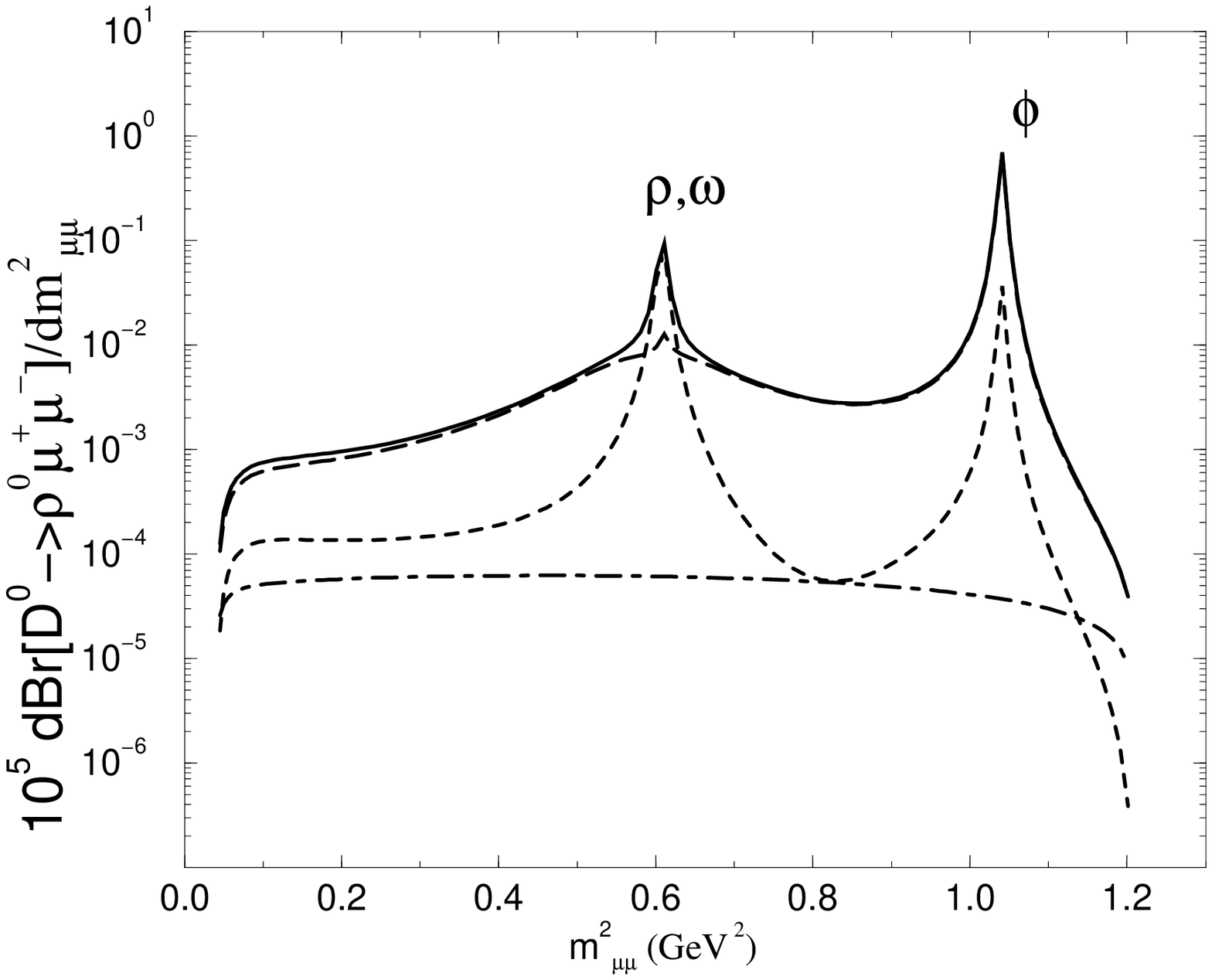} 
\quad
\includegraphics[scale=.4]{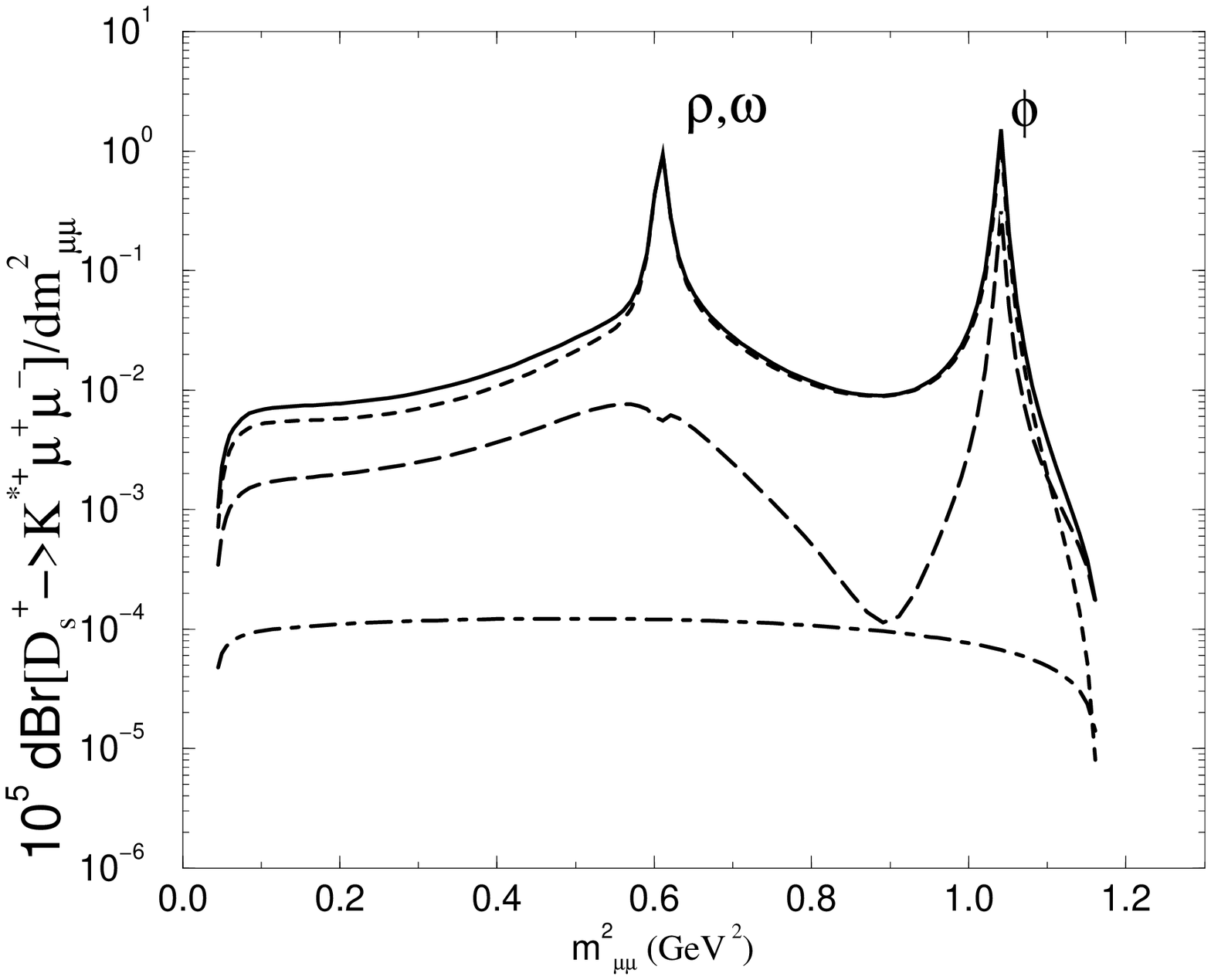} 
 \caption{The differential branching ratios $dBr/dm^2_{ll}$ as a function of the invariant di-lepton mass $m_{ll}^2$ for the Cabibbo allowed decays $D^0\to \bar K^{*0}l^+l^-$, $D_s^+\to \rho^+l^+l^-$ and  Cabibbo suppressed decays $D^0\to \rho^0l^+l^-$, $D_s^+\to K^{*+}l^+l^-$. The solid lines denote the long distance contribution, while the dot-dashed lines denote the short distance contribution. The dotted  and dashed lines indicate the parity conserving and parity violating parts of the long distance contribution, respectively.} 
\label{fig37}
\end{center}
\end{figure}

The allowed kinematical region for the di-lepton mass in the $D\to Vl^+l^-$ decay is $m_{ll}=[2m_l,m_D\!-\!m_V]$. 
The {\bf long distance contribution} has resonant shape with poles at the di-lepton mass $m_{ll}\!=\!m_{\rho^0},m_\omega,m_\phi$. There is another pole at zero di-lepton mass coming from the photon propagator. This pole is significant in $D\to Ve^+e^-$ decays where the kinematically allowed region for the di-lepton mass $m_{ee}$ starts at $2m_e$. The lowest allowed di-lepton mass in $D\to V\mu^+\mu^-$ decays is $m_{\mu\mu}=2m_\mu$ and the contribution due to the pole at zero di-lepton mass is much smaller in this case. In Fig.   \ref{fig37} I plot the differential branching ratios $dBr/dm_{\mu\mu}^2$ in terms of the di-lepton mass squared $m_{\mu\mu}^2$ for the two Cabibbo allowed decays $D^0\to \bar K^{*0}\mu^+\mu^-$ and $D_s^+\to \rho^+\mu^+\mu^-$ and for two Cabibbo suppressed decays $D^0\to \rho^0\mu^+\mu^-$ and $D_s^+\to K^{*+}\mu^+\mu^-$. The full line represents the long distance contribution, while the dotted and dashed lines present the parity conserving and parity violating parts of the long distance contribution, respectively.  In the Cabibbo allowed decays I plot also the long distance contribution for the case of the electron and positron  in the final state.  The two spectrums are essentially identical above $m_{ll}>0.3$ GeV due to the small masses of the electron and muon. 
The predicted branching ratios for eight $D\to V\mu^+\mu^-$ and $D\to Ve^+e^-$ decays are given in Tables \ref{tab.vll2} and \ref{tab.vll3}. The long distance parts of the predicted branching ratios $Br^{th}_{LD}$ are given in the third column.

\begin{table}[!htb]
\begin{center}
\begin{tabular}{|c||c|c|c|c||c|}
\hline
$D\to V \mu^+\mu^-$ & $Br^{th}_{SD}$ & $Br^{th}_{LD}$&$Br^{exp}$\cite{PDG,vll.exp}\\
\hline \hline
$ D^0 \to {\bar K}^{*0} \mu^+\mu^-$ & $0$&$1.2\cdot 10^{-6}$&$<1.18\cdot 10^{-3}$\\
\hline
$ D_s^+ \to \rho^+ \mu^+\mu^-$ & $0$&$3.3\cdot 10^{-5}$&\\
\hline
\hline
$ D^0 \to \rho^{0} \mu^+\mu^-$ & $6.0\cdot 10^{-10}$&$1.5\cdot 10^{-7}$&$<2.3\cdot 10^{-4}$\\
\hline
$ D^0 \to \omega \mu^+\mu^-$ & $5.7\cdot 10^{-10}$&$1.6\cdot 10^{-7}$&$<8.3\cdot 10^{-4}$\\
\hline
$ D^0 \to \phi \mu^+\mu^-$ &  $0$&$4.8\cdot 10^{-8}$&$<4.1\cdot 10^{-4}$ \\
\hline
 $D^+ \to \rho^+ \mu^+\mu^-$& $3.0\cdot 10^{-9}$&$1.4\cdot 10^{-6}$&$<5.6\cdot 10^{-4}$\\
\hline
$ D_s^+ \to K^{*+ }\mu^+\mu^-$  &$1.2\cdot 10^{-9}$&$5.6\cdot 10^{-7}$ &$<1.4\cdot 10^{-3}$ \\
\hline
\hline 
$ D^+ \to K^{*+} \mu^+\mu^-$ & $0$&$2.8\cdot 10^{-8}$&\\
\hline
$ D^0 \to K^{*0} \mu^+\mu^-$ & $0$&$3.1\cdot 10^{-9}$&\\
\hline
\end{tabular}
\caption{The branching ratios for the Cabibbo allowed, suppressed and doubly suppressed $D\to V\mu^+\mu^-$ decays. The predicted short distance parts of the branching ratios in the standard model are given in column two, while the predictions for the long distance parts are given in column three. The present experimental upper limits are gathered in the last column \cite{PDG,vll.exp}. }
\label{tab.vll2}
\end{center}
\end{table}

\begin{table}[!htb]
\begin{center}
\begin{tabular}{|c||c|c|c|c||c|}
\hline
$D\to V e^+e^-$ & $Br^{th}_{SD}$ & $Br^{th}_{LD}$&$Br^{exp}$\cite{PDG,vll.exp}\\
\hline \hline
$ D^0 \to {\bar K}^{*0} e^+e^-$ & $0$&$1.7\cdot 10^{-6}$&$<1.4\cdot 10^{-4}$\\
\hline
$ D_s^+ \to \rho^+ e^+e^-$ & $0$&$3.7\cdot 10^{-5}$&\\
\hline
\hline
$ D^0 \to \rho^{0} e^+e^-$ & $6.6\cdot 10^{-10}$&$1.7\cdot 10^{-7}$&$<1.0\cdot 10^{-4}$\\
\hline
$ D^0 \to \omega e^+e^-$ & $6.3\cdot 10^{-10}$&$1.9\cdot 10^{-7}$&$<1.8\cdot 10^{-4}$\\
\hline
$ D^0 \to \phi e^+e^-$ &  $0$&$8.0\cdot 10^{-8}$&$<5.2\cdot 10^{-5}$ \\
\hline
 $D^+ \to \rho^+ e^+e^-$& $3.0\cdot 10^{-9}$&$1.6\cdot 10^{-6}$&\\
\hline
$ D_s^+ \to K^{*+ }e^+e^-$  &$1.2\cdot 10^{-9}$&$7.3\cdot 10^{-7}$ &\\
\hline
\hline 
$ D^+ \to K^{*+} e^+e^-$ & $0$&$3.9\cdot 10^{-8}$&\\
\hline
$ D^0 \to K^{*0} e^+e^-$ & $0$&$4.4\cdot 10^{-9}$&\\
\hline
\end{tabular}
\caption{ The branching ratios for the Cabibbo allowed, suppressed and doubly suppressed $D\to Ve^+e^-$ decays. 
The predicted short distance parts of the branching ratios in the standard model are given in column two, while the predictions for the long distance parts are given in column three. The present experimental upper limits are gathered in the last column \cite{PDG,vll.exp}.  }
\label{tab.vll3}
\end{center}
\end{table}

The {\bf short distance} contribution is present only in the Cabibbo suppressed decays and is rather flat in terms of the di-lepton mass. The short distance part of $dBr(D^0\to \rho^0\mu^+\mu^-)/dm^2_{\mu\mu}$ and $dBr(D_s^+\to K^{*+}\mu^+\mu^-)/dm^2_{\mu\mu}$ in the standard model  is shown by dashed-dotted line in Fig.  \ref{fig37}. The part induced by the flavour changing neutral transition $c\to ul^+l^-$ is found to be much smaller than the part induced by the long distance mechanisms in the whole region of $m_{ll}^2$. The predictions for short distance parts of the branching ratios are given in Tables \ref{tab.vll2} and \ref{tab.vll3}. They are much smaller than the  long distance parts of the branching ratios and they give negligible contribution to the total rates in the standard model.


The experimental upper bounds \cite{PDG,vll.exp} on the branching ratios are given in the last column of Tables \ref{tab.vll2} and \ref{tab.vll3}. No experimental data is available on $D_s^+\to \rho^+l^+l^-$  decay, which is predicted at the highest rate.

\vspace{0.2cm}

I conclude this section by stressing that the $D\to V\gamma$ and $D\to Vl^+l^-$ decays are not  sensitive to the flavour changing neutral transitions $c\to u\gamma$ and $c\to ul^+l^-$ unless these transitions are significantly enhanced by some mechanism beyond the standard model. The rates are dominated by the long distance contributions.  
Different scenarios of  physics beyond the standard model could alter the short distance contributions, but none of the scenarios discussed in Chapter 2 can not enhance them above the long distance ones. The non-minimal supersymmetric model  could perhaps enhance $Br(c\to u\gamma)$ up to $1.2 \cdot 10^{-5}$  (\ref{brsusy}), which could lead up to $Br^{SD}(D\to V\gamma)\sim 10^{-6}$ for the exclusive decays. 
Such effect would still be difficult to separate from the long distance contributions in the experiment, given the present theoretical and experimental uncertainties. 

The experimental observation of these charm meson decay channels would improve our knowledge on the long distance dynamics in  the heavy meson decays. In $B$ meson decays similar long distance contributions present an important background in the extraction of the short distance contribution. The predicted rates indicate that the  Cabibbo allowed channels $D^0\to \bar K^{*0}\gamma$, $D_s^+\to \rho^+\gamma$, 
$D^0\to \bar K^{*0}l^+l^-$ and $D_s^+\to \rho^+l^+l^-$ will be observed soon.

\section{The $\boldsymbol{D\to Pl^+l^-}$ decays}

I close the chapter on charm meson decays with the study of rare  charm meson decays $D\to Pl^+l^-$ with  the light pseudoscalar $P$ and a charged lepton pair in the final state. The decay $D\to P\gamma$ with a real photon in the final state is forbidden  as the photon would have to be purely longitudinal.  
The Cabibbo suppressed decays have the flavour structure $c\bar q\to u\bar ql^+l^-$ and 
are interesting as possible probes for the flavour changing neutral transition $c\to ul^+l^-$. The short distance contribution, which arises due to the decay $c\to ul^+l^-$ at short distances, is accompanied by the long distance contribution. 
Different mechanisms were sketched in Chapter 3 and the   long distance contribution was divided to the weak annihilation (Fig.  \ref{fig2}) and long distance penguin (Fig.  \ref{fig3}) part. In addition to the Cabibbo suppressed decays, I systematically study all the Cabibbo allowed and Cabibbo doubly suppressed $D\to Pl^+l^-$ decays. These can not proceed through the flavour changing neutral transition of a single quark and are induced only via the long distance mechanisms. 

None of the $D\to Pl^+l^-$ decays have been observed so far and the experimental upper limits on the branching ratios are in the range $10^{-4}-10^{-5}$ at present \cite{PDG,pll.exp}. 

The Cabibbo suppressed decays $D\to Pl^+l^-$ have  been proposed as  possible probes for various scenarios of physics beyond the standard model discussed in Chapter 2 \cite{pakvasa,hewett,BHLP,schwartz}. 
It has been stated in \cite{schwartz} that the ``decays $D\to \pi l^+l^-$, $D_s^+\to K^+ l^+l^-$ constitute a large discovery window and would present a strong evidence for new physics if observed at the branching ratios above $10^{-7}$''. It is therefore of obvious interest to obtain reliable  estimates for the long distance contributions to these modes. One suspects that they are the dominant ones in the standard model and one should have a good control of their estimation in order to perform a meaningful search for new physics. In spite of this,  the long distance contributions to these channels have not been determined  so far with an exception of the channel $D^+\to \pi^+l^+l^-$, which has been studied in \cite{singer}. 
The authors of \cite{singer} have calculated the long distance penguin contribution to $D^+\to \pi^+l^+l^-$ decay via the exchange of the intermediate $\phi$ meson, $D^+\to \pi^+\phi\to\pi^+l^+l^-$ . They have  not accounted for the intermediate mesons $\rho^0$ and $\omega$   and they have neglected the weak annihilation contribution. The results on the $D^+\to \pi^+l^+l^-$ rate within the hybrid model shows, that the long distance penguin contribution via the intermediate meson $\phi$ is indeed dominant for this particular channel.  

The theoretical treatment of $D\to Pl^+l^-$ decays faces different situation than encountered in beauty meson decays. The charm meson decays turn out to be dominated by the long distance contributions, while the most frequent beauty meson decays of this type,  $B\to Kl^+l^-$, arise due to  $b\to sl^+l^-$ decay at short distances.  The kaon decays of this type , $K\to \pi l^+l^-$,  were discussed in  Introduction.  The long distance contribution via the photon exchange is almost absent in the CP violating $K_L\to \pi^0l^+l^-$ channel \cite{burdman1}.   
The  $K^+\to \pi^+l^+l^-$ channel is dominated by the long distance mechanisms and has similar dynamics to that encountered in the $D\to Pl^+l^-$ decays. The $K^+\to \pi^+l^+l^-$ decay has been extensively studied \cite{rare.kaon} and the available literature on the subject gives some feeling for various mechanisms, that have to be considered in the case of charm decays. The energies involved are however different. The chiral perturbation theory gives the reliable predictions  in the kaon decays t,  while  a model for charm decays has to incorporate the light as well as the heavy degrees of freedom.

 In the following I adapt the hybrid model for the calculation of the long  and short distance contributions to all $D\to Pl^+l^-$ decays.

\subsection{Long distance contributions}

The long distance mechanism involves the exchange of a virtual photon $D\to P\gamma^*\to Pl^+l^-$. The pseudoscalar and the virtual photon form the orbital momentum state $ L=1$ and the parity is conserved in the process. The weak transition 
is induced by the effective nonleptonic  Lagrangian (\ref{eff})
\begin{equation}
\label{eff2}
{\cal L}^{|\Delta c|=1}_{eff}\!=\!-\tfrac{G_F}{\sqrt{2}}V_{cq_j}^*V_{uq_i}[a_1^c ~\bar u\gamma^{\mu}(1-\gamma_5)q_i~\bar q_j\gamma_{\mu}(1-\gamma_5)c+a_2^c~\bar q_j\gamma_{\mu}(1-\gamma_5)q_i~\bar u\gamma^{\mu}(1-\gamma_5)c]
\end{equation} 
with $a_1^c\simeq 1.26$, $a_2^c\simeq-0.55$ (\ref{ai}). The amplitudes (\ref{ald}) 
$${\cal A}_{LD}(D\to Pl^+l^-)=\langle l^+l^- P|:i{\cal L}^{|\Delta c|=1}_{eff}:|P\rangle$$
 are calculated by  employing the factorization approximation  (\ref{factor}). 

The Feynman diagrams are given in terms of the hadronic degrees of freedom contained in the hybrid model: heavy pseudoscalar ($D$) and vector ($D^*$) mesons and light pseudoscalar ($P$) and vector ($V$) mesons. The relevant diagrams 
for $D\to Pl^+l^-$ decays in the hybrid model  are given in Figs.  \ref{fig35} and  \ref{fig36}. The boxes in the diagrams denote the action of the weak nonleptonic effective Lagrangian (\ref{eff2}). This Lagrangian contains a product  of two left handed quark currents, each denoted by a dot in a box. 
 Let me comment on different contributions:

\begin{figure}[h]
\centering
\mbox{
\begin{fmffile}{f35o1n}
\fmfframe(3,3)(3,3){
  \begin{fmfgraph*}(26,23)
  \fmfpen{thin}  
  \fmfleftn{l}{1} \fmfright{r1,q1,q2,r2,r3}
  \fmftopn{t}{4}\fmflabel{\bf{(I)}}{t2}
  \fmfrpolyn{shaded,tension=0.4}{k}{4}   
  \fmf{dashes,tension=4}{l1,k1} 
  \fmf{dashes,tension=4}{k1,r1} 
  \fmf{dashes,label=$V^0=\rho^0,,\omega,,\phi$,la.d=0.6,tension=0.3,la.s=right}{k3,v1} 
  \fmf{boson,label=$\gamma$,la.d=20,tension=0.3,la.s=right}{v1,a}
  \fmf{fermion}{r3,a,r2}
  \fmfv{decor.size=1.2thick,decor.shape=circle,decor.filled=full}{k3,v1}
  \fmfv{decor.size=1.2thick,decor.shape=circle,decor.filled=full,label=\framebox[5mm]{$\alpha$},la.d=3thick,la.a=-120}{k1}
  \fmflabel{$D$}{l1}
  \fmflabel{$P$}{r1}
  \fmflabel{$l^+$}{r3}
  \fmflabel{$l^-$}{r2}
  \end{fmfgraph*} }
\end{fmffile}
\quad
\begin{fmffile}{f35o2n}
\fmfframe(8,3)(3,3){
  \begin{fmfgraph*}(26,23)
  \fmfpen{thin}  
  \fmfleftn{l}{1} \fmfright{r1,q1,q2,r2,r3}
  \fmftopn{t}{4}\fmflabel{\bf{(II)}}{t2}
   \fmfrpolyn{shaded,tension=0.05}{k}{4}   
  \fmf{dashes,tension=4}{l1,b} 
  \fmf{dashes,tension=4}{b,r1} 
  \fmf{dashes,label=$D^*$,la.d=0.6,tension=0.05,la.s=left}{b,k1}
  \fmf{dashes,label=$V^0$,la.d=0.6,tension=0.05,la.s=left}{k3,v1}
  \fmf{boson,label=$\gamma$,la.d=20,tension=0.05,la.s=left}{v1,a}
  \fmf{fermion}{r3,a,r2}
  \fmfv{decor.size=1.2thick,decor.shape=circle,decor.filled=full}{k1,k3,v1}
   \fmfv{label=\framebox[4mm]{$g$},la.d=3thick,la.a=-110}{b}  
  \fmflabel{$D$}{l1}
  \fmflabel{$P$}{r1}
  \fmflabel{$l^+$}{r3}
  \fmflabel{$l^-$}{r2}
  \end{fmfgraph*} }
\end{fmffile}
\quad
\begin{fmffile}{f35o3n}
\fmfframe(3,3)(3,3){
  \begin{fmfgraph*}(24,23)
  \fmfpen{thin}  
  \fmfleftn{l}{1} \fmfright{r1,q1,q2,r2,r3}
  \fmftopn{t}{4}\fmflabel{\bf{(I$^D$)}}{t2}
  \fmfrpolyn{shaded,tension=0.4}{k}{4}   
  \fmf{dashes,tension=4}{l1,k1} 
  \fmf{dashes,tension=4}{k1,r1} 
  \fmf{boson,label=$\gamma$,la.d=20,tension=0.3,la.s=left}{k3,a}
  \fmf{fermion}{r3,a,r2}
  \fmfv{decor.size=1.2thick,decor.shape=circle,decor.filled=full}{k3}
    \fmfv{decor.size=1.2thick,decor.shape=circle,decor.filled=full,label=\framebox[5mm]{$\alpha$},la.d=3thick,la.a=-120}{k1}
  \fmflabel{$D$}{l1}
  \fmflabel{$P$}{r1}
  \fmflabel{$l^+$}{r3}
  \fmflabel{$l^-$}{r2}
  \end{fmfgraph*} }
\end{fmffile}
\quad
\begin{fmffile}{f35o4n}
\fmfframe(3,3)(3,3){
  \begin{fmfgraph*}(24,23)
  \fmfpen{thin}  
  \fmfleftn{l}{1} \fmfright{r1,q1,q2,r2,r3}
  \fmftopn{t}{4}\fmflabel{\bf{(II$^D$)}}{t2}
  \fmfrpolyn{shaded,tension=0.05}{k}{4}   
  \fmf{dashes,tension=4}{l1,b} 
  \fmf{dashes,tension=4}{b,r1} 
  \fmf{dashes,label=$D^*$,la.d=0.6,tension=0.05,la.s=left}{b,k1}
  \fmf{boson,label=$\gamma$,la.d=20,tension=0.05,la.s=left}{k3,a}
  \fmf{fermion}{r3,a,r2}
  \fmfv{decor.size=1.2thick,decor.shape=circle,decor.filled=full}{k1,k3}
   \fmfv{label=\framebox[4mm]{$g$},la.d=3thick,la.a=-110}{b} 
  \fmflabel{$D$}{l1}
  \fmflabel{$P$}{r1}
  \fmflabel{$l^+$}{r3}
  \fmflabel{$l^-$}{r2}
  \end{fmfgraph*} }
\end{fmffile}
    }
\caption{Long distance penguin diagrams for $D\to Pl^+l^-$ decays.  The parameters,  given in the frames  by the
verteces, indicate which terms in the Lagrangian (\ref{hybrid}) and weak current
(\ref{current.heavy}) are responsible for the couplings.  The box denotes the action of the weak nonleptonic effective Lagrangian (\ref{eff}). The box contains two dots each denoting a weak current in the Lagrangian (\ref{eff}).}
\label{fig35}
\end{figure}
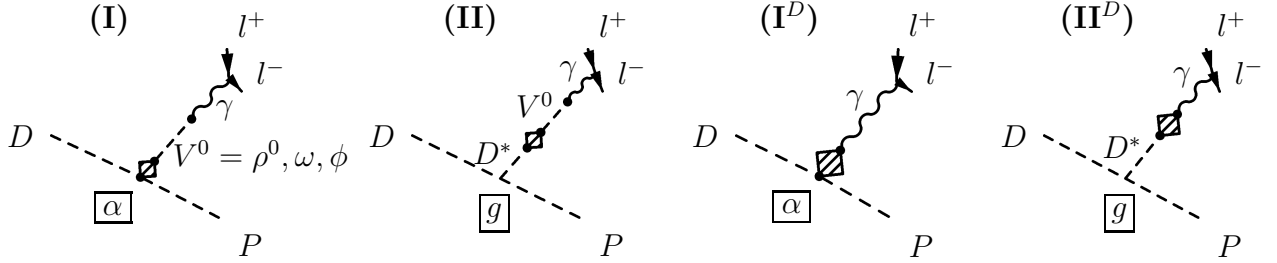

\begin{itemize}
\item
The {\bf long distance penguin} is sketched in Fig.  \ref{fig3} on the quark level. Its amplitude is GIM suppressed
\begin{align*}
{\cal A}_{LD}^{peng}&=-i\tfrac{G_F}{\sqrt{2}}\sum_{q=d,s}V_{cq}^*V_{uq}a_2^c\langle P\gamma^*|\bar q\gamma_{\mu}(1-\gamma_5)q~\bar u\gamma^{\mu}(1-\gamma_5)c|D\rangle\\
&\simeq -i\tfrac{G_F}{\sqrt{2}}V_{cs}^*V_{us}a_2^c\langle P|\bar u\gamma^{\mu}(1-\gamma_5)c|D\rangle\langle\gamma^*|\bar s\gamma_{\mu}s-\bar d\gamma_{\mu}d|0\rangle
\end{align*}
and it is presented by the diagrams in Fig.  \ref{fig35} in the hybrid model. The matrix elements $\langle P|\bar u\gamma^{\mu}(1-\gamma_5)c|D\rangle$ can be expressed in terms of the form factors $f_1$ and $f_0$ (\ref{defP}) and the form factor $f_0$ does not contribute in the decay to a lepton pair. The form factor $f_1$ is given by the expression (\ref{formP}) in the hybrid model. 
The virtual photon is created by the weak current $\bar s(1-\gamma_5)s-\bar d(1-\gamma_5)d$ via the intermediate vector mesons $\rho^0$, $\omega$ and $\phi$ (diagrams $I$ and $II$) or directly (diagrams $I^D$ and $II^D$). The significance of the diagrams, where the photon is directly coupled to the weak current, will be discussed in the connection with the gauge invariance later.

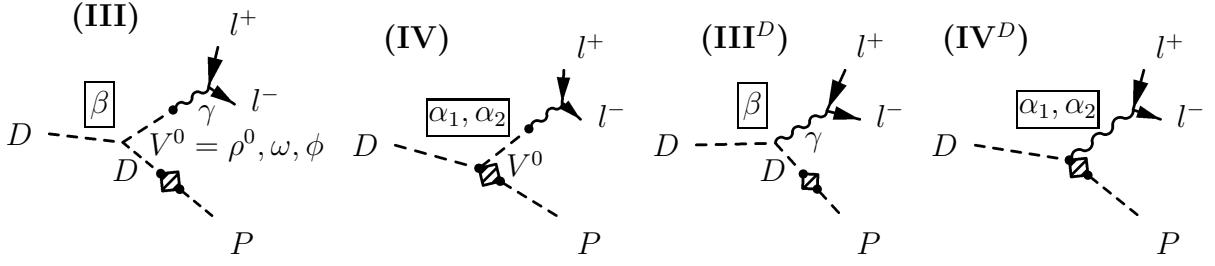
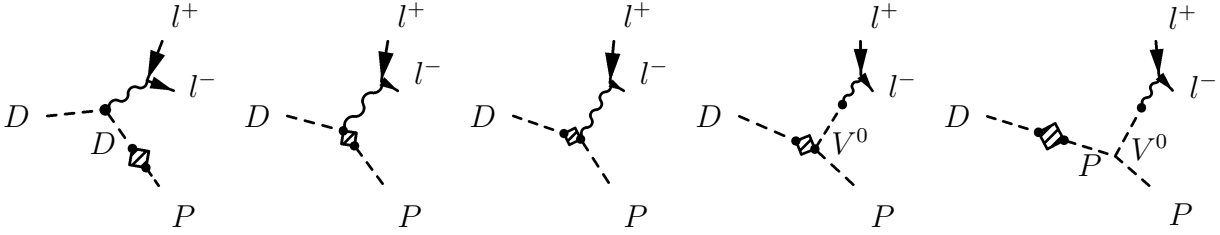
\begin{figure}[h]
\centering
\mbox{
\subfigure[Non-bremsstrahlung diagrams.]
{
\begin{fmffile}{f36a1nn}
\fmfframe(2,2)(2,2){
  \begin{fmfgraph*}(25,23)
  \fmfpen{thin}
  \fmfleftn{l}{1} \fmfrightn{r}{4}\fmftopn{t}{4}\fmflabel{\bf{(III)}}{t2}
  \fmfrpolyn{shaded}{k}{4}  
  \fmfv{decor.size=1.2thick,decor.shape=circle,decor.filled=full}{k1,k3,v2} 
  \fmf{dashes}{l1,v1} 
  \fmf{dashes,label=$V^0=\rho^0,,\omega,,\phi$,la.s=right,la.d=10,tension=0.7}{v1,v2}
  \fmf{boson,label=$\gamma$,la.s=right,la.d=30}{v2,a}\fmf{fermion}{r4,a,r3}
  \fmf{dashes,label=$D$,la.s=right,la.d=10}{v1,k1} 
    \fmf{dashes}{k3,r1}
    \fmflabel{$D$}{l1}\fmflabel{$P$}{r1}\fmflabel{$l^+$}{r4}\fmflabel{$l^-$}{r3}
    \fmfv{label=\framebox[4mm]{$\beta$},la.d=3thick,la.a=120}{v1} 
  \end{fmfgraph*} }
\end{fmffile}
\quad
\begin{fmffile}{f36a2n}
\fmfframe(8,2)(2,2){
  \begin{fmfgraph*}(25,20)
  \fmfpen{thin}
  \fmfleftn{l}{1} \fmfrightn{r}{4}\fmftopn{t}{4}\fmflabel{\bf{(IV)}}{t2}
  \fmfrpolyn{shaded,tension=1.5}{k}{4}   
  \fmfv{decor.size=1.2thick,decor.shape=circle,decor.filled=full}{k3,v1}
  \fmfv{decor.size=1.2thick,decor.shape=circle,decor.filled=full,label=\framebox[11mm]{$
  \alpha_1,,\alpha_2$},la.d=6thick,la.a=110}{k1}
  \fmf{dashes}{l1,k1}\fmf{dashes,label=$V^0$,la.s=right,la.d=10,tension=0.4}{k1,v1} 
  \fmf{boson,tension=0.6}{v1,a}\fmf{fermion}{r4,a,r3}
  \fmf{dashes}{k3,r1}
  \fmflabel{$D$}{l1}\fmflabel{$P$}{r1}
  \fmflabel{$l^+$}{r4}\fmflabel{$l^-$}{r3}
  \end{fmfgraph*} }
\end{fmffile}
\quad
\begin{fmffile}{f36a3nnn}
\fmfframe(2,2)(2,2){
  \begin{fmfgraph*}(22,20)
  \fmfpen{thin}
  \fmfleftn{l}{1} \fmfrightn{r}{4}\fmftopn{t}{4}\fmflabel{\bf{(III$^D$)}}{t2}
  \fmfrpolyn{shaded}{k}{4}  
  \fmfv{decor.size=1.2thick,decor.shape=circle,decor.filled=full}{k1,k3} 
  \fmf{dashes}{l1,v1} 
  \fmf{boson,label=$\gamma$,la.s=right,la.d=20}{v1,a}\fmf{fermion}{r4,a,r3}
  \fmf{dashes,label=$D$,la.s=right,la.d=10}{v1,k1} 
    \fmf{dashes}{k3,r1}
    \fmflabel{$D$}{l1}\fmflabel{$P$}{r1}\fmflabel{$l^+$}{r4}\fmflabel{$l^-$}{r3}
    \fmfv{label=\framebox[4mm]{$\beta$},la.d=3thick,la.a=120}{v1} 
  \end{fmfgraph*} }
\end{fmffile}
\quad
\begin{fmffile}{f36a4nn}
\fmfframe(2,2)(2,2){
  \begin{fmfgraph*}(25,20)
  \fmfpen{thin}
  \fmfleftn{l}{1} \fmfrightn{r}{4}\fmftopn{t}{4}\fmflabel{\bf{(IV$^D$)}}{t2}
  \fmfrpolyn{shaded,tension=1.5}{k}{4}   
  \fmfv{decor.size=1.2thick,decor.shape=circle,decor.filled=full}{k3}
  \fmfv{decor.size=1.2thick,decor.shape=circle,decor.filled=full,label=\framebox[11mm]{$
  \alpha_1,,\alpha_2$},la.d=6thick,la.a=110}{k1} 
  \fmf{dashes}{l1,k1} 
  \fmf{boson,tension=0.6}{k1,a}\fmf{fermion}{r4,a,r3}
  \fmf{dashes}{k3,r1}
  \fmflabel{$D$}{l1}\fmflabel{$P$}{r1}
  \fmflabel{$l^+$}{r4}\fmflabel{$l^-$}{r3}
  \end{fmfgraph*} }
\end{fmffile}
}
    } 
\mbox{
\subfigure[Bremsstrahlung diagrams. The blob in the first diagram indicates that shape
of the electromagnetic form factor $F_D(q^2)$ (given by the parameter $\kappa$
(\ref{hybrid})) is not determined in the hybrid model. The contribution of the
first diagram is fortunately proportional to $m_P^2/m_D^2$ and therefore negligible. ]
{
\begin{fmffile}{f36b1}
\fmfframe(2,2)(2,2){
  \begin{fmfgraph*}(17,20)
  \fmfpen{thin}
  \fmfleftn{l}{1} \fmfrightn{r}{4}
  \fmfrpolyn{shaded,tension=0.6}{k}{4}  
  \fmfv{decor.size=1.2thick,decor.shape=circle,decor.filled=full}{k1,k3} 
  \fmf{dashes}{l1,v1} 
  \fmf{boson}{v1,a}\fmf{fermion}{r4,a,r3}
  \fmf{dashes,label=$D$,la.s=right,la.d=10,tension=0.6}{v1,k1} 
    \fmf{dashes}{k3,r1}
    \fmflabel{$D$}{l1}\fmflabel{$P$}{r1}\fmflabel{$l^+$}{r4}\fmflabel{$l^-$}{r3}
    \fmfv{decor.size=1.7thick,decor.shape=circle,decor.filled=full}{v1} 
  \end{fmfgraph*} }
\end{fmffile}
\quad
\begin{fmffile}{f36b2}
\fmfframe(2,2)(2,2){
  \begin{fmfgraph*}(15,20)
  \fmfpen{thin}
  \fmfleftn{l}{1} \fmfrightn{r}{4}
  \fmfrpolyn{shaded,tension=1.5}{k}{4}   
  \fmfv{decor.size=1.2thick,decor.shape=circle,decor.filled=full}{k3,k1}
  \fmf{dashes}{l1,k1} 
  \fmf{boson,tension=0.6}{k1,a}\fmf{fermion}{r4,a,r3}
  \fmf{dashes}{k3,r1}
  \fmflabel{$D$}{l1}\fmflabel{$P$}{r1}
  \fmflabel{$l^+$}{r4}\fmflabel{$l^-$}{r3}
  \end{fmfgraph*} }
\end{fmffile}
\quad
\begin{fmffile}{f36b3}
\fmfframe(2,2)(2,2){
  \begin{fmfgraph*}(15,20)
  \fmfpen{thin}
  \fmfleftn{l}{1} \fmfrightn{r}{4}
  \fmfrpolyn{shaded,tension=1.5}{k}{4}   
  \fmfv{decor.size=1.2thick,decor.shape=circle,decor.filled=full}{k3,k1}
  \fmf{dashes}{l1,k1} 
  \fmf{boson,tension=0.6}{k3,a}\fmf{fermion}{r4,a,r3}
  \fmf{dashes}{k3,r1}
  \fmflabel{$D$}{l1}\fmflabel{$P$}{r1}
  \fmflabel{$l^+$}{r4}\fmflabel{$l^-$}{r3}
  \end{fmfgraph*} }
\end{fmffile}
\quad
\begin{fmffile}{f36b4}
\fmfframe(2,2)(2,2){
  \begin{fmfgraph*}(18,20)
  \fmfpen{thin}
  \fmfleftn{l}{1} \fmfrightn{r}{4}
  \fmfrpolyn{shaded,tension=1.5}{k}{4}   
  \fmfv{decor.size=1.2thick,decor.shape=circle,decor.filled=full}{k3,v1,k1}
  \fmf{dashes}{l1,k1}\fmf{dashes,label=$V^0$,la.s=right,la.d=10,tension=0.4}{k3,v1} 
  \fmf{boson,tension=0.6}{v1,a}\fmf{fermion}{r4,a,r3}
  \fmf{dashes}{k3,r1}
  \fmflabel{$D$}{l1}\fmflabel{$P$}{r1}
  \fmflabel{$l^+$}{r4}\fmflabel{$l^-$}{r3}
  \end{fmfgraph*} }
\end{fmffile}
\quad
\begin{fmffile}{f36b5}
\fmfframe(2,2)(2,2){
  \begin{fmfgraph*}(25,20)
  \fmfpen{thin}
  \fmfleftn{l}{1} \fmfrightn{r}{4}
  \fmfrpolyn{shaded,tension=1}{k}{4}   
  \fmfv{decor.size=1.2thick,decor.shape=circle,decor.filled=full}{k3,v1,k1}
  \fmf{dashes}{l1,k1}\fmf{dashes,label=$P$,la.s=right,la.d=20}{k3,m1}
  \fmf{dashes,label=$V^0$,la.s=right,la.d=10,tension=0.4}{m1,v1} 
  \fmf{boson,tension=0.6}{v1,a}\fmf{fermion}{r4,a,r3}
  \fmf{dashes}{m1,r1}
  \fmflabel{$D$}{l1}\fmflabel{$P$}{r1}
  \fmflabel{$l^+$}{r4}\fmflabel{$l^-$}{r3}
  \end{fmfgraph*} }
\end{fmffile}
}
    }       

\caption{Long distance weak annihilation diagrams for $D\to Pl^+l^-$ decays in the hybrid
model. 
The bremsstrahlung diagrams are gathered in Fig. (b). The  diagrams that do not correspond 
to  bremsstrahlung are
gathered in Fig. (a). The parameters,  given in the frames  by the
verteces, indicate which terms in the Lagrangian (\ref{hybrid}) and weak current
(\ref{current.heavy}) are responsible for the couplings.  
The box denotes the action of the weak nonleptonic effective Lagrangian (\ref{eff}). 
The box contains two dots each denoting a weak current in the Lagrangian (\ref{eff}).}
\label{fig36}
\end{figure}

\item
The {\bf long distance weak annihilation}  corresponds to the quark level mechanism shown in Fig.  \ref{fig2}. It is represented by the diagrams in Fig.  \ref{fig36} in the hybrid model. One weak current has the flavour of the initial $D$ meson, while  the other has the flavour of the final $P$ meson.  The vector mesons do not  enter as the intermediate states (except for vector meson $V^0$ which converts to a photon) since the parity is conserved in $D\to Pl^+l^-$ decay. The terms proportional to $\epsilon^{\mu\nu\alpha\beta}$ in the Lagrangian do not contribute for the same reason\footnote{There are only three independent kinematical variables for a $D(p)\to P\gamma^*(q,\epsilon)$ decay and the their product with the antisymmetric tensor $\epsilon^{\mu\nu\alpha\beta}$ vanishes.}. The bremsstrahlung diagrams are given in Fig.  \ref{fig36}b and are present only when $D$ and $P$ are charged. The diagrams, where the photon is emitted before the weak transition and do not correspond to bremsstrahlung, are gathered in Fig.  \ref{fig36}a.

First I consider {\bf bremsstrahlung} given by the diagrams in Fig. \ref{fig36}b. In view of the general discussion in Section 3.3.3, the bremsstrahlung amplitude for $D\to P\gamma^*$ diagrams in Fig.  \ref{fig24} can be parameterized in terms of the general electromagnetic form factors $G_D$, $G_{Dd}$, $G_{P}$ and $G_{Pd}$ as (\ref{3.125})
\begin{align}
\label{5.201}
|{\cal A}&[D(p)\to P\gamma^*(q,\epsilon)]|=\frac{G_F}{\sqrt{2}}eV_{\!C\!K\!M}V_{\!C\!K\!M}^\prime f_Df_{P}\nonumber\\
&\times\epsilon_\mu^*\bigl[\frac{G_D^{\mu\nu}(q^2)m_{P}^2-G_{P}^{\mu\nu}(q^2)m_D^2}{m_D^2-m_{P}^2}(p+p^\prime)_\nu+G_{Dd}^{\mu\nu}(q^2)p^\prime_\nu+G_{Pd}^{\mu\nu}(q^2)p_\nu\bigr]~.
\end{align}
In Section 3.3.3 I have shown that this amplitude is manifestly gauge invariant  for the case of the flat form factors, the polar form factors and their linear combination (\ref{3.123}). The bremsstrahlung diagrams in Fig.  \ref{fig36}b, as given by the hybrid model, indicate that $G_{Dd}\!=\!1$ is flat, $G_{P}$ has resonant shape and $G_{Pd}$ has a resonant and a flat part\footnote{In order to satisfy the condition $G_i(q^2\!=\!0)$  one has to take $\Gamma_{V^0}(q^2\!=\!0)=0$.}:
\begin{align}
\label{5.202}
G_{\pi^+}^{\mu\nu}(q^2)&=\frac{m_{\rho}^2~\bigl[g^{\mu\nu}-\frac{q^\mu q^\nu}{m_{\rho}^2}\bigr]}{m_{\rho}^2-q^2-i\Gamma_{\rho}m_{\rho}}~,\\
G_{\pi^+d}^{\mu\nu}(q^2)&=2G_{\pi^+}^{\mu\nu}(q^2)-g^{\mu\nu}~~,\nonumber\\
G_{K+}^{\mu\nu}(q^2)&=\frac{m_{\rho}^2~\bigl[g^{\mu\nu}-\frac{q^\mu q^\nu}{m_{\rho}^2}\bigr]}{2(m_{\rho}^2-q^2-i\Gamma_{\rho}m_{\rho})}+\frac{m_{\omega}^2~\bigl[g^{\mu\nu}-\frac{q^\mu q^\nu}{m_{\omega}^2}\bigr]}{6(m_{\omega}^2-q^2-i\Gamma_\omega m_\omega)}+\frac{m_{\phi}^2~\bigl[g^{\mu\nu}-\frac{q^\mu q^\nu}{m_{\phi}^2}\bigr]}{3(m_{\phi}^2-q^2-i\Gamma_\phi m_\phi)}~,\nonumber\\
 G_{K^+d}^{\mu\nu}(q^2)&=2G_{K^+}^{\mu\nu}(q^2)-g^{\mu\nu}~.\nonumber
\end{align}
The shape of the form factor $G_D(q^2)$ depends on the parameter $\kappa$ (\ref{hybrid}) and it is flat for $\kappa=0$ and resonant for $\kappa=1$. The parameter $\kappa$ could not be determined from the available experimental data and it was left undetermined. The contribution of the form factor $G_D$ to the bremsstrahlung amplitude (\ref{5.201}) is fortunately proportional to $m_P^2$ and  will be neglected in view of $m_P^2\ll m_D^2$. In this case, the bremsstrahlung amplitude (\ref{5.201}) with form factors given in (\ref{5.202}) vanishes exactly
\begin{equation}
{\cal A}_{Bremss.}=0~.
\end{equation}

Now I consider the weak annihilation diagrams, which do not correspond to brems\-stra\-hlung. The diagrams, where the {\bf photon is emitted before the weak transition}, are gathered in Fig.  \ref{fig36}a. The diagrams $III$ and $III^D$ are induced by the terms proportional to parameter $\beta$ in the Lagrangian (\ref{hybrid}); their amplitude is proportional to $m_P^2$ and therefore small. The diagrams $IV$ and $IV^D$ are given by the terms proportional to $\alpha_1$ and $\alpha_2$ in the weak current (\ref{current.heavy}). The significance of the diagrams $III^D$ and $IV^D$ will be discussed in connection with gauge invariance bellow. The hybrid model does not render any diagrams where the {\bf photon is emitted after the weak transition}. Such diagrams would involve the weak transition $D\to R$  followed by the strong decay $R\to V^0P$ and electromagnetic decay $V^0\to l^+l^-$. Due to the parity conservation in $D\to R$ transition, the resonance $R$ should be pseudoscalar or axial vector meson.  The hybrid model  does not contain axial vector degrees of freedom and it is unfortunately  not very powerful in analyzing this particular mechanism. Among the pseudoscalar resonances $R$, the resonances $R=P$  have already been incorporated to the bremsstrahlung contribution. The resonances $R\not=P$ do not contribute as the strong coupling $R-V^0-P$ vanishes due to the isospin and $G$ parity assignments of $R$, $P$ and $V^0=\rho,\omega,\phi$ states.  
\end{itemize}

\subsubsection{Discussion of gauge invariance}

First let me review how the gauge invariance was assured in  $B_c\to B_u^*\gamma$, $D\to V\gamma$ and $D\to Vl^+l^-$ decays. The parity conserving part of the amplitude was automatically gauge invariant. The parity violating mechanism was incorporated via the nonleptonic decay $P\to VV^0$ followed by $V^0\to \gamma^*$ ($V^0=\rho,~\omega,~\phi$). The $VV^0$ intermediate state involved three helicity amplitudes $++,~--,~00$ discussed in Section 4.2. The real photon in the final state can not have longitudinal polarization and the helicity state $00$ had to be discarded when the decay $P\to  V\gamma$ was considered. This was achieved by relating the form factors $A_1$ and $A_2$ (\ref{4.5}, \ref{4.5a}) that parameterize the nonleptonic decay $P\to VV^0$ in the factorization approximation. 

In the case of  $D\to P\gamma^*\to Pl^+l^-$ decay this idea would be implemented by considering the nonleptonic decay $D\to PV^0$ followed by the electromagnetic transition $V^0\to l^+l^-$. As the nonleptonic decay $D\to PV^0$ involves only one $PV^0$ helicity state, namely $00$, one can not achieve the gauge invariance by relating different form factors that parameterize the nonleptonic decay $P\to VV^0$ in the factorization approximation. Instead, I will take advantage of the fact that the hybrid model is manifestly gauge invariant until the $SU(3)$ flavour breaking effects are incorporated in Section 5.1.8. For this discussion, it is particularly important, that the terms  of the form ${\cal V}^D_\mu-\rho_\mu$ in (\ref{hybrid}, \ref{current.light}, \ref{current.heavy}) are invariant under the electromagnetic gauge transformation given in (\ref{5.19}). The neutral fields from the matrix $\rho$ couple to the photon via vector meson dominance (\ref{VMDexact}), while the vector current ${\cal V}^D_\mu=ie_0{\cal Q}A_\mu+\tfrac{1}{2}(\xi^\dagger \partial_\mu\xi+\xi \partial_\mu\xi^\dagger)$ (\ref{5.20}) contains the photon field $A_\mu$. The term  $({\cal V}^D_\mu-\rho_\mu)_{ss}$, for example, effectively incorporates the electromagnetic field through
\begin{equation}
\label{5.214} 
-i\tfrac{1}{3}e_0A_\mu-i\tfrac{1}{3}e_0\tfrac{m_\phi^2}{q^2-m_\phi^2}\bigl[g^{\mu\nu}-\tfrac{q^\mu q^\nu}{m_\phi^2}\bigr]A_\nu~,
\end{equation}
which is manifestly gauge invariant and equal to zero for the real photon. The  terms proportional to ${\cal V}^D_\mu-\rho_\mu$ give rise to the photon emission via the resonance and the direct photon emission, so that both contributions cancel for the case of the real photon. 
The corresponding pairs of diagrams  are shown in Figs.  \ref{fig35} and  \ref{fig36}a: diagrams $II$ and $II^D$ are given by the term proportional to $\beta$ in (\ref{hybrid}); diagrams $III$ and $III^D$ are given by the term proportional to $g$ in (\ref{hybrid});    
diagrams $I$, $I^D$, $IV$ and $IV^D$ are given by the terms proportional to $\alpha_1$ and $\alpha_2$ in the weak current (\ref{current.heavy}). The pairs of the diagrams ensure the gauge invariance of the amplitude for $D\to Pl^+l^-$. In the same way the gauge invariance for the decays $D\to Vl^+l^-$ and $D\to V\gamma$ could be ensured without imposing the relations among the form factors $A_1$ and $A_2$ (\ref{5.154}). This mechanism would not significantly change  $D\to Vl^+l^-$ rates compared to the predictions given in Section 5.4. The predictions for $D\to V\gamma$ rates would be however dramatically decreased: the parity violating part of the $D\to V\gamma$ amplitude would vanish; the parity conserving part of the long distance penguin amplitude for $D\to V\gamma$ would vanish as well\footnote{In Table \ref{tab.gamma1} the only nonzero amplitudes would be $A_{PC}^{III,IV}$ and $A_{PC}^{VI}$, which have opposite signs in all the decays. They have similar magnitude in some decays and almost cancel.}. In order to generate nonzero parity violating amplitudes to $D\to V\gamma$ decays, we made the phenomenologically motivated assumption based on the vector meson dominance mechanism in Section 5.4. The parity violating contribution was incorporated  via the nonleptonic decay $D\to VV^0$ followed by $V^0\to \gamma$. At the same time we discarded the   contributions where the photon  arises from the current ${\cal V}^D$ \cite{FPS1,FS,FPS2}. 

\subsection{Short distance contributions}

The short distance contribution is present only in the Cabibbo suppressed $D\to Pl^+l^-$ decays and is induced by the flavour changing transition $c\to ul^+l^-$. The amplitude (\ref{asd}) is the matrix element of ${\cal L}^{c\to ul^+l^-}$ (\ref{o9})
$${\cal A}^{SD}(D\to Pl^+l^-)\!=\!\langle l^+l^- P|:i{\cal L}^{c\to ul^+l^-}\!\!:|D\rangle\!=\!-i\frac{G_F}{\sqrt{2}}{e^2\over 8\pi^2}c_9^{eff}\langle P|\bar u_{\alpha}\gamma_{\mu}(1-\gamma_5)c_{\alpha}|D\rangle\langle l^+l^-|\bar l\gamma^{\mu}l|0\rangle~$$
with $c_9^{eff}=0.24{+0.01\atop -0.06}$ (\ref{c9}) in the standard model. Possible physics beyond the standard model could alter the value of $c_9^{eff}$ as well as the form of the effective  Lagrangian ${\cal L}^{c\to ul^+l^-}$. 
The necessary matrix elements $\langle P|\bar u\gamma^{\mu}(1-\gamma_5)c|D\rangle$ can be expressed in terms of the form factors $f_1$ and $f_0$ (\ref{defP}) and the form factor $f_0$ does not contribute in the decay to a lepton pair. The form factor $f_1$ in the hybrid model is given by the expression (\ref{formP}). 

\subsection{The amplitudes}

The sum of the long distance amplitude, given by the diagrams in Figs.  \ref{fig35} and  \ref{fig36}, and the short distance amplitude is
\begin{equation}
{\cal A}[D(p)\to Pl^+(p^+)l^-(p^-)]=i\frac{G_F}{\sqrt{2}}e_0^2~\bar u(p_-) \!{\not \!p} v(p_+)~A(q^2)
\end{equation}
with
\begin{align*}
A(q^2)&=A^{SD}(q^2)+A^{LD}(q^2)~,\\
A^{LD}(q^2)&=A^{LD}_{peng.}(q^2)+A^{LD}_{annih.}(q^2)~,\\
A^{SD}(q^2)&=-\frac{1}{4\pi^2}~c_9^{eff}~f_1(q^2)~,\\
A^{LD}_{peng}(q^2)&=a_2V_{cs}^*V_{us}\frac{1}{q^2}f_1(q^2)N_1(q^2)~,\\
A^{LD}_{annih.}(q^2)&=f_{Cabb}^{(i)}\frac{1}{q^2}M_1^{(i)}(q^2)f_P\biggl[f_D\frac{m_P^2}{m_D^2-m_P^2}\beta-\sqrt{m_D}\bigl(\alpha_1-\frac{m_D^2+m_P^2-q^2}{2m_D^2}\alpha_2\bigr)\biggr]\frac{\tilde g_V}{\sqrt{2}}~.
\end{align*}
The form factor $f_1(q^2)$ and the coefficient $N_1(q^2)$ are given by
\begin{align*}
f_1(q^2)&=K_{DP}^{(i)}\bigl[-\frac{f_D}{2}+gf_{D^{*0}}\frac{m_{D^{*0}}\sqrt{m_{D^{*0}}m_D}}{q^2-m_{D^{*0}}}\bigr]~,\\
N_1(q^2)&={g_{\rho}^2\over q^2-m_{\rho}^2+i\Gamma_{\rho}m_{\rho}}-{g_{\omega}^2\over 3(q^2-m_{\omega}^2+i\Gamma_{\omega}m_{\omega})}-{2g_{\phi}^2\over 3(q^2-m_{\phi}^2+i\Gamma_{\phi}m_{\phi})}\\
&+{g_{\rho}^2\over m_{\rho}^2}-{g_{\omega}^2\over 3m_{\omega}^2}-{2g_{\phi}^2\over 3m_{\phi}^2}~.
\end{align*}
The Cabibbo factors $f_{Cabb}^{(i)}$ and the coefficients $M_1(q^2)$ and $K_{DP}^{(i)}$ are given in Table \ref{tab.pll} for eight $D\to Pl^+l^-$ decays. The coefficients $M_1^{(i)}$ are given in terms of  $M^{D^0}_1$, $M^{D^+}_1$ and $M^{D_s^+}$ 
\begin{eqnarray}
\label{5.210}
M^{D^0}_1\!\!\!&=&\!\!\!{g_{\rho}\over q^2-m_{\rho}^2+i\Gamma_{\rho}m_{\rho}}+{g_{\omega}\over 3(q^2-m_{\omega}^2+i\Gamma_{\omega}m_{\omega})}+{g_{\rho}\over m_{\rho}^2}+{g_{\omega}\over 3m_{\omega}^2}~,
\nonumber\\
M^{D^+}_1\!\!\!&=&\!\!\!-{g_{\rho}\over q^2-m_{\rho}^2+i\Gamma_{\rho}m_{\rho}}+{g_{\omega}\over 3(q^2-m_{\omega}^2+i\Gamma_{\omega}m_{\omega})}-{g_{\rho}\over m_{\rho}^2}+{g_{\omega}\over 3m_{\omega}^2}~,
\nonumber\\
M^{D_s^+}_1\!\!\!&=&\!\!\!-{2g_{\phi}\over 3(q^2-m_{\phi}^2+i\Gamma_{\phi}m_{\phi})}-{2g_{\phi}\over 3m_{\phi}^2}~.
\end{eqnarray}
Note that  $N_1(0)=M_1(0)=0$ and there is no pole arising from the photon propagator at $q^2=0$. The coefficients $N_1(q^2)$ and $M_1(q^2)$ are related to the coefficients $N(q^2)$ and $M(q^2)$ (\ref{jklm}) used in $D\to Vl^+l^-$ decays via $N_1(q^2)=N(q^2)-N(0)$ and $M_1(q^2)=M(q^2)-M(0)$. The difference in two approaches arises from the direct photon emission induced by the field ${\cal V}^D$, which cancels the  resonant emission at $q^2\!=\!0$ (\ref{5.214}).

\begin{table}[h]
\begin{center}
\begin{tabular}{|c|c||c|c|c|}
\hline
$i$ & $D\to P l^+l^-$ & $f_{Cabb}^{(i)}$  & $M^{(i)}$  & $K_{DP}^{(i)}$\\
\hline \hline
$1$ & $ D^0 \to {\bar K}^{0} l^+l^-$ &$ a_2\cos^2\theta_C$ & $M_1^{D^0}$ & $0$   \\
\hline
$2$ & $ D_s^+ \to \rho^+ l^+l^-$ & $a_1\cos^2\theta_C$ & $M_1^{D_s^+}$ & $0$  \\
\hline
\hline
$3$ & $ D^0 \to \pi^{0}l^+l^-$ &$ -a_2\sin\theta_C\cos\theta_C$ & $-M_1^{D^0}/\sqrt{2}$ & $1/(\sqrt{2}f_\pi)$ \\
\hline
$4$ & $ D^0 \to \eta l^+l^-$ &$ a_2\sin\theta_C\cos\theta_C$ & $M_1^{D^0}(K^s_\eta-K^d_\eta)$ & $K^d_\eta/f$  \\
\hline
$5$ & $ D^+ \to \pi^+ l^+l^-$ &$ -a_1\sin\theta_C\cos\theta_C$ &  $M_1^{D^+}$ & $1/f_\pi$  \\
\hline
$6$ & $ D_s^+ \to K^{+ }l^+l^-$ &$ a_1\sin\theta_C\cos\theta_C$ &  $M_1^{D_s^+}$  & $1/f_K$ \\
\hline
\hline 
$7$ & $ D^+ \to K^{+} l^+l^-$ &$ -a_1\sin^2\theta_C$ & $M_1^{D^+}$ & $0$ \\
\hline
$ 8$ & $ D^0 \to K^{0} l^+l^-$ &$ -a_2\sin^2\theta _C$ &  $M_1^{D^0}$ & $0$ \\
\hline
\end{tabular}
\caption{ The Cabibbo factors $f_{Cabb}^{(i)}$, the coefficients $K_{DP}^{(i)}$ and the functions $M_1^{(i)}$ for eight $D\to Pl^+l^-$ decays.}
\label{tab.pll}
\end{center}
\end{table}

\subsection{The results}

The allowed kinematical region for the di-lepton mass in the $D\to Pl^+l^-$ decay is $m_{ll}=[2m_l,m_D\!-\!m_P]$. 
The long distance contribution has resonant shape with poles at the di-lepton masses $m_{ll}=m_{\rho^0},~m_\omega,~m_\phi$. There is no  pole at zero di-lepton mass since the decay $D\to P\gamma$ is forbidden. The short distance contribution is rather flat in terms of the di-lepton mass.   
The spectrums of $D\to Pe^+e^-$ and $D\to P\mu^+\mu^-$ decays  in terms of the di-lepton mass are practically identical. The difference in their rates is due to  the kinematical region $m_{ll}=[2m_e,2m_\mu]$ and is negligible. The rates for  $D\to Pe^+e^-$ and $D\to P\mu^+\mu^-$ decays are practically identical and I do not consider them separately.  The predicted branching ratios for eight decays are given in Table \ref{tab.pll1} together with the available experimental data \cite{PDG,pll.exp}. The long distance contributions are given in column 5. The penguin and weak annihilation parts of the long distance contribution are given in columns 3 and 4, respectively. The short distance contributions, as predicted by the standard model, are given in the second column and are smaller than the long distance contributions in all  decays. 

\begin{table}[!htb]
\begin{center}
\begin{tabular}{|c||c||c|c|c||c|c|}
\hline
$D\to P l^+l^-$ & $\boldsymbol{Br^{SD}}$  &    $Br^{LD}_{peng.}$&$Br^{LD}_{annih.}$&$\boldsymbol{Br^{LD}}$&$\boldsymbol{Br^{exp}}$&$\boldsymbol{Br^{exp}}$\\
 & $l=\mu,~e$& $l=\mu,~e$& $l=\mu,~e$& $l=\mu,~e$&$l=e$&$l=\mu$\\
\hline \hline
$ D^0 \to {\bar K}^{0} l^+l^-$ & $0$&$0$&$5.0\cdot 10^{-7}$&$5.0\cdot 10^{-7}$&$<1.1\cdot 10^{-4}$&$<2.6\cdot 10^{-4}$\\
\hline
 $ D_s^+ \to \pi^+ l^+l^-$ & $0$&$0$&$6.2\cdot 10^{-6}$&$6.2\cdot 10^{-6}$&$<2.7\cdot 10^{-4}$&$<1.4\cdot 10^{-4}$\\
\hline
\hline
$ D^0 \to \pi^{0}l^+l^-$ & $2.6\cdot 10^{-9}$&$3.9\cdot 10^{-7}$&$1.2\cdot 10^{-8}$&$3.9\cdot 10^{-7}$&$<4.5\cdot 10^{-5}$&$<1.8\cdot 10^{-4}$\\
\hline
$ D^0 \to \eta l^+l^-$ & $8.6\cdot 10^{-10}$ &$1.2\cdot 10^{-7}$&$2.1\cdot 10^{-8}$&$1.4\cdot 10^{-7}$&$<1.1\cdot 10^{-4}$&$<5.3\cdot 10^{-4}$\\
\hline
$ D^+ \to \pi^+ l^+l^-$ & $1.3\cdot 10^{-8}$&$1.9\cdot 10^{-6}$&$1.4\cdot 10^{-7}$&$1.7\cdot 10^{-6}$&$<5.2\cdot 10^{-5}$&$<1.5\cdot 10^{-5}$\\
\hline
$ D_s^+ \to K^{+ } l^+l^-$ & $3.7\cdot 10^{-9}$&$5.6\cdot 10^{-7}$&$3.2\cdot 10^{-7}$&$8.3\cdot 10^{-8}$&$<1.6\cdot 10^{-3}$&$<1.4\cdot 10^{-4}$\\
\hline
\hline 
$ D^+ \to K^{+} l^+l^-$ & $0$&$0$&$8.2\cdot 10^{-9}$&$8.2\cdot 10^{-9}$&$<2.0\cdot 10^{-4}$&$<4.4\cdot 10^{-5}$\\
\hline
$ D^0 \to K^{0} l^+l^-$ &$0$&$0$&$1.3\cdot 10^{-9}$&$1.3\cdot 10^{-9}$&&\\
\hline
\end{tabular}
\caption{The branching ratios for eight $D\to Pl^+l^-$ decays. The predictions for the short distance contributions in the standard model are given in column 2, while the predictions for the long distance contributions are given in column 5.  The experimental upper bounds are given in the last two columns \cite{PDG,pll.exp}. The penguin and the weak annihilation parts of the long distance contribution are given in the columns 3 and 4, respectively.}
\label{tab.pll1}
\end{center}
\end{table}

\begin{figure}[!htb]
\begin{center}
\includegraphics[scale=.4]{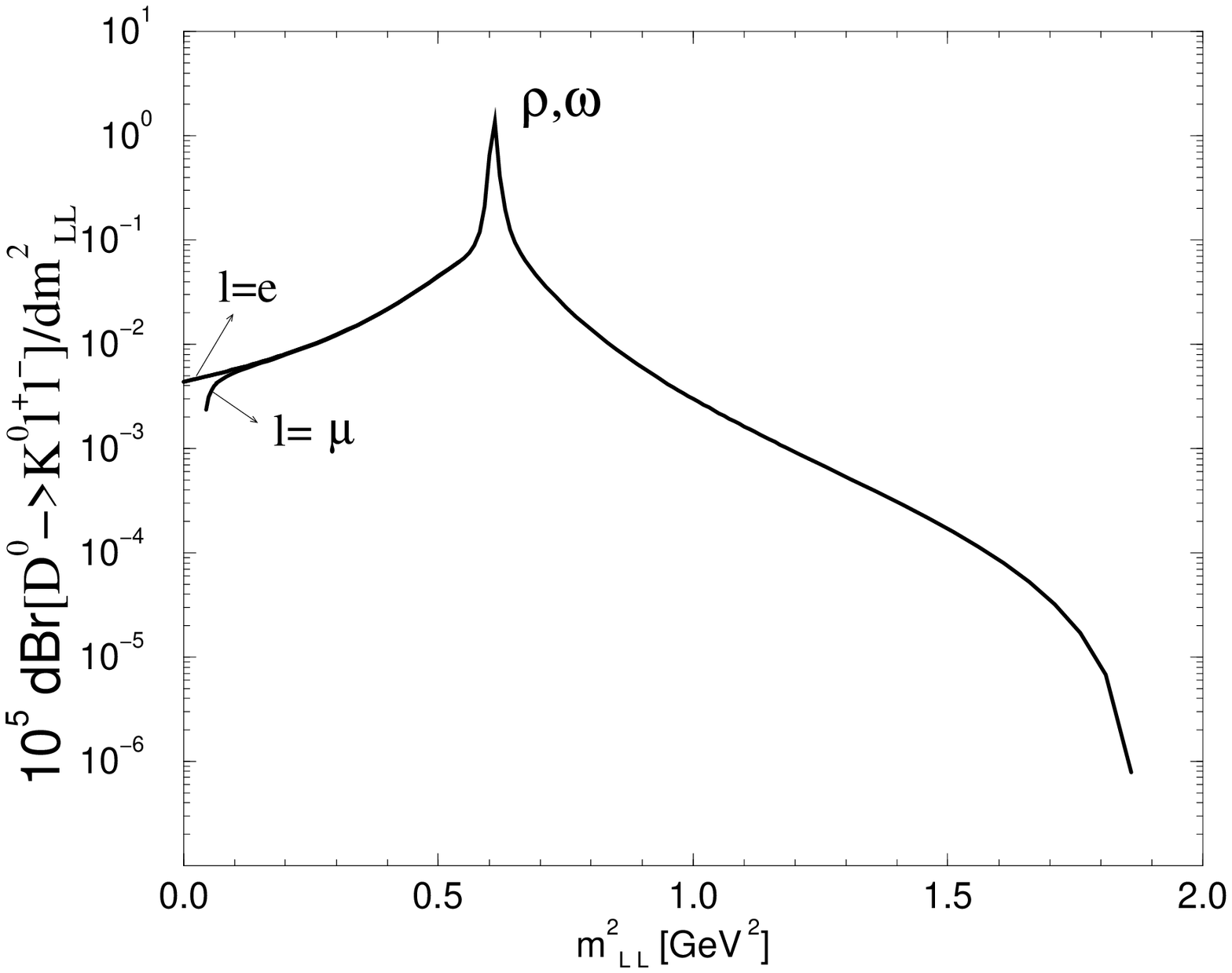} 
\quad
\includegraphics[scale=.4]{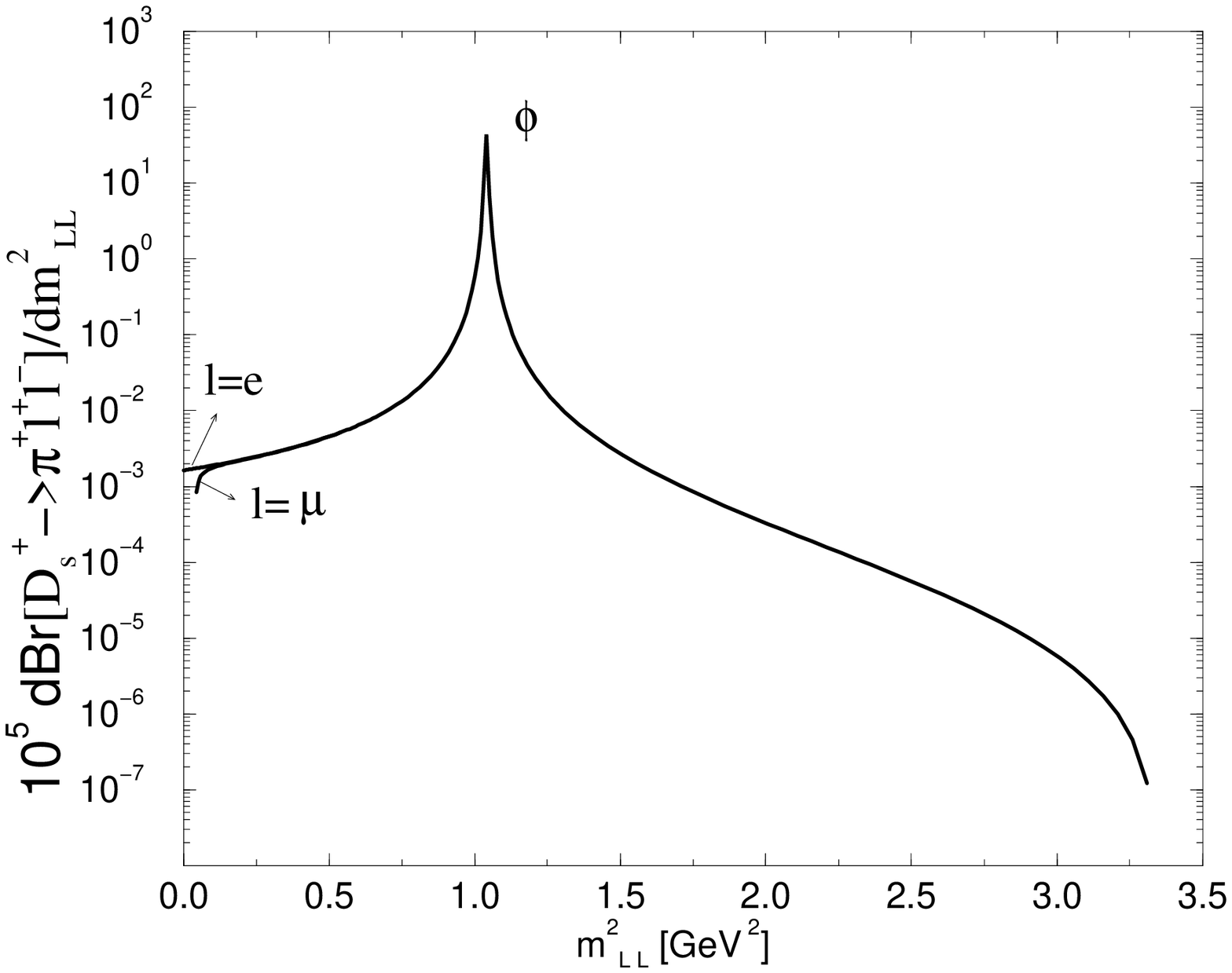} 
\\
\includegraphics[scale=.4]{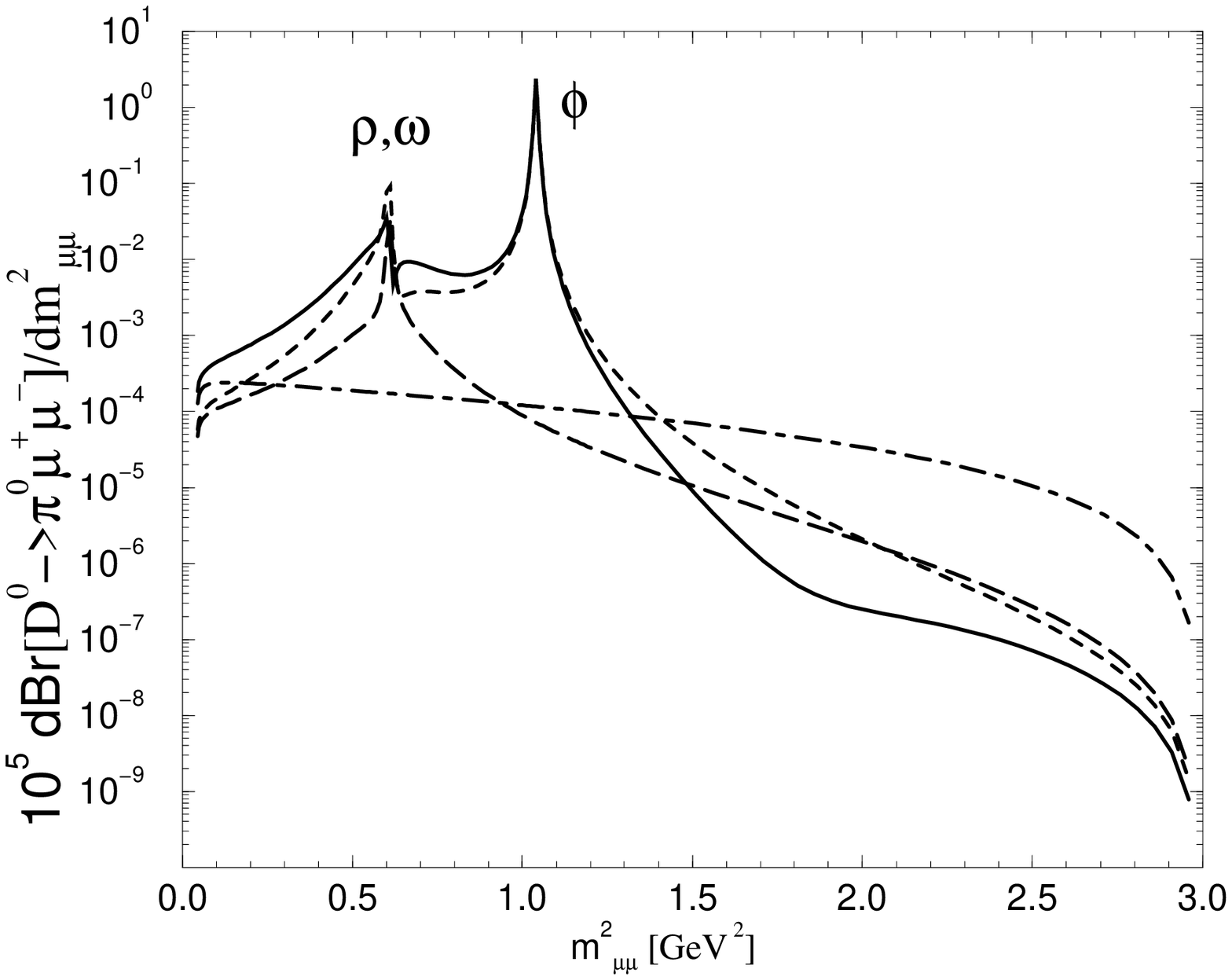} 
\quad
\includegraphics[scale=.4]{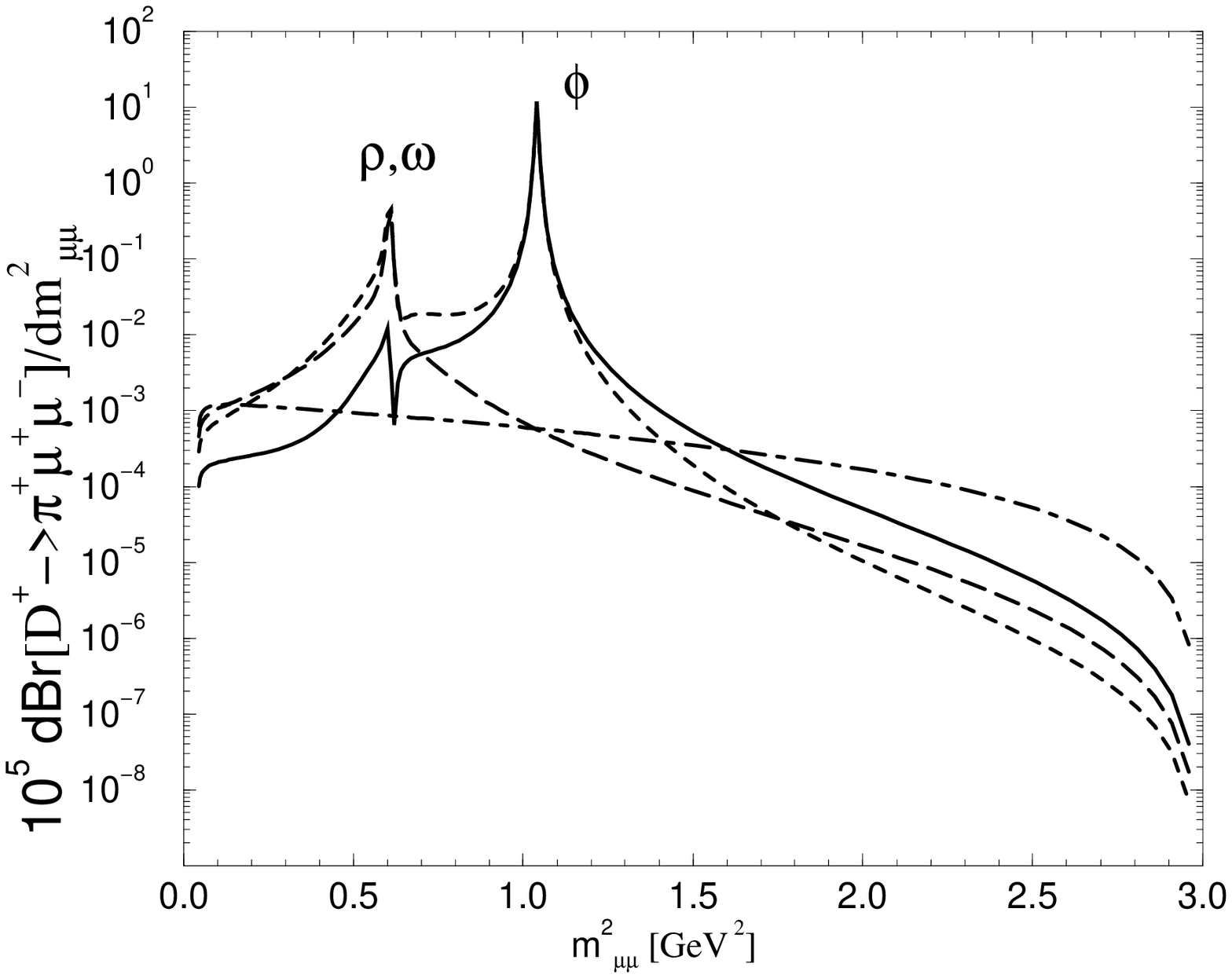} 
 \caption{The differential branching ratios $dBr/dm^2_{ll}$ as a function of the invariant di-lepton mass $m_{ll}^2$ for the  decays $D^0\to \bar K^{0}l^+l^-$, $D_s^+\to \pi^+l^+l^-$, $D^0\to \pi^0l^+l^-$ and $D^+\to \pi^+l^+l^-$. The solid lines denote the long distance contribution, while the dot-dashed lines denote the short distance contribution. In the last two figures, the dotted  and dashed lines indicate the penguin and the weak annihilation parts   of the long distance contribution, respectively.} 
\label{fig38}
\end{center}
\end{figure}

In Fig.  \ref{fig38} I plot the differential branching ratios $dBr/dm_{\mu\mu}^2$ in terms of the di-lepton mass squared $m_{\mu\mu}^2$ for the two Cabibbo allowed decays $D^0\to \bar K^0\mu^+\mu^-$ and $D_s^+\to \pi^+\mu^+\mu^-$.  I plot also two Cabibbo suppressed decays $D^0\to \pi^0\mu^+\mu^-$ and $D^+\to \pi^{+}\mu^+\mu^-$, in which the kinematical upper bound on di-lepton mass $m_{ll}^{max}\!=\!m_D\!-\!m_P$ is as high as possible. 
The full line presents the long distance contribution, while the dotted and dashed lines present the  penguin and weak annihilation parts of the long distance contribution, respectively. In the Cabibbo allowed decays I show  also the long distance contribution for the case of the electron and positron  in the final state. The short distance contribution to $D^0\to \pi^0\mu^+\mu^-$ and $D^+\to \pi^{+}\mu^+\mu^-$ decays is represented by the dot-dashed line. The long distance contribution dies out in the kinematical region above the resonances and the short distance contribution becomes dominant. The decays $D\to \pi l^+l^-$ at high di-lepton mass present the unique opportunity to probe the flavour changing neutral transition $c \to ul^+l^-$ in the future. As the pion is the lightest hadron state, this interesting kinematical region is not present in other  $D\to Xl^+l^-$ decays.

Let me examine the kinematical region of high di-lepton mass in $D\to \pi l^+l^-$ decays more closely. In this region the excited states of the vector mesons $\rho^0$, $\omega$ and $\phi$ may become important. By means of the duality their contribution is partly  incorporated in the short distance part of the amplitude. I make a very rough estimate of the additional long distance contributions due to the excited states $\rho^0(1450)$, $\omega(1420)$ and $\phi(1680)$. The knowledge of their couplings to photons, as well as to other particles, are poor at present  and I assume that they couple with the same couplings as the corresponding ground state vector mesons $\rho^0$, $\omega$ and $\phi$. This assumption probably overestimates their contribution.  At the sime time I take their measured masses $m_{V^{\prime}}$ and widths $\Gamma_{V^{\prime}}$ given in Table \ref{tab.excited}. 
\begin{table}[h]
\begin{center}
\begin{tabular}{|c||c|c|}
\hline
& $m_{V^\prime}$[GeV] & $\Gamma_{V^\prime}$[GeV] \\
\hline
$\rho(1450)$&$ 1.465$&$0.310$\\
$\omega(1420)$&$1.419$&$0.174$\\
$\phi(1680)$&$1.649$&$0.150$\\
\hline
\end{tabular}
\caption{The masses and widths of the excited vector mesons.}
\label{tab.excited}
\end{center}
\end{table}
In this case the coefficients $N_1$ and $M_1$ are replaced in the formulas above by
\begin{align*}
 N_1&\to N_1+{g_{\rho}^2\over q^2-m_{\rho^\prime}^2+i\Gamma_{\rho^\prime}m_{\rho^\prime}}-{g_{\omega}^2\over 3(q^2-m_{\omega^\prime}^2+i\Gamma_{\omega^\prime}m_{\omega^\prime})}-{2g_{\phi}^2\over 3(q^2-m_{\phi^\prime}^2+i\Gamma_{\phi^\prime}m_{\phi^\prime})}\\
&\qquad +{g_{\rho}^2\over m_{\rho^\prime}^2}-{g_{\omega}^2\over 3m_{\omega^\prime}^2}-{2g_{\phi}^2\over 3m_{\phi^\prime}^2}~,\\
M^{D^0}_1&\to M^{D^0}_1+{g_{\rho}\over q^2-m_{\rho^\prime}^2+i\Gamma_{\rho^\prime}m_{\rho^\prime}}+{g_{\omega}\over 3(q^2-m_{\omega^\prime}^2+i\Gamma_{\omega^\prime}m_{\omega^\prime})}+{g_{\rho}\over m_{\rho^\prime}^2}+{g_{\omega}\over 3m_{\omega^\prime}^2}~,
\nonumber\\
M^{D^+}_1&\to M^{D^+}_1-{g_{\rho}\over q^2-m_{\rho^\prime}^2+i\Gamma_{\rho^\prime}m_{\rho^\prime}}+{g_{\omega}\over 3(q^2-m_{\omega^\prime}^2+i\Gamma_{\omega^\prime}m_{\omega^\prime})}-{g_{\rho}\over m_{\rho^\prime}^2}+{g_{\omega}\over 3m_{\omega^\prime}^2}~,
\nonumber\\
M^{D_s^+}_1&\to M^{D_s^+}-{2g_{\phi}\over 3(q^2-m_{\phi^\prime}^2+i\Gamma_{\phi^\prime}m_{\phi^\prime})}-{2g_{\phi}\over 3m_{\phi^\prime}^2}~.
\end{align*}
The differential branching ratios for the charged and neutral $D\to \pi \mu^+\mu^-$ decays are given in Fig.  \ref{fig39}. The dashed line represents the long distance part that incorporates the ground state and the excited vector mesons. 
The solid line represent the long distance part that incorporates only the ground state vector mesons. The dashed-dotted line represent the short distance contribution in the standard model.  The excited vector resonances  $\rho^0(1450)$, $\omega(1420)$, $\phi(1680)$ are much more broad than the ground state vector resonances $\rho^0$, $\omega$, $\phi$ and their contribution to the $D\to \pi l^+l^-$ rates is much smaller. The short distance contribution is dominant at high di-lepton mass in spite of the long distance channels via the excited vector mesons. The integrated $D\to \pi l^+l^-$ rates  over the kinematical region of high di-lepton mass are shown in  Table \ref{tab.pll2}. The short distance part is given in the second column. The long distance part, which does not incorporate the excited vector mesons, is given in the third column. The long distance part, which incorporate the ground state and the excited vector mesons, is given in the last column. The short distance mechanism induced by the flavour changing neutral transition $c\to ul^+l^-$  is indeed dominant in this kinematical region.

\begin{figure}[h]
\begin{center}
\includegraphics[scale=.4]{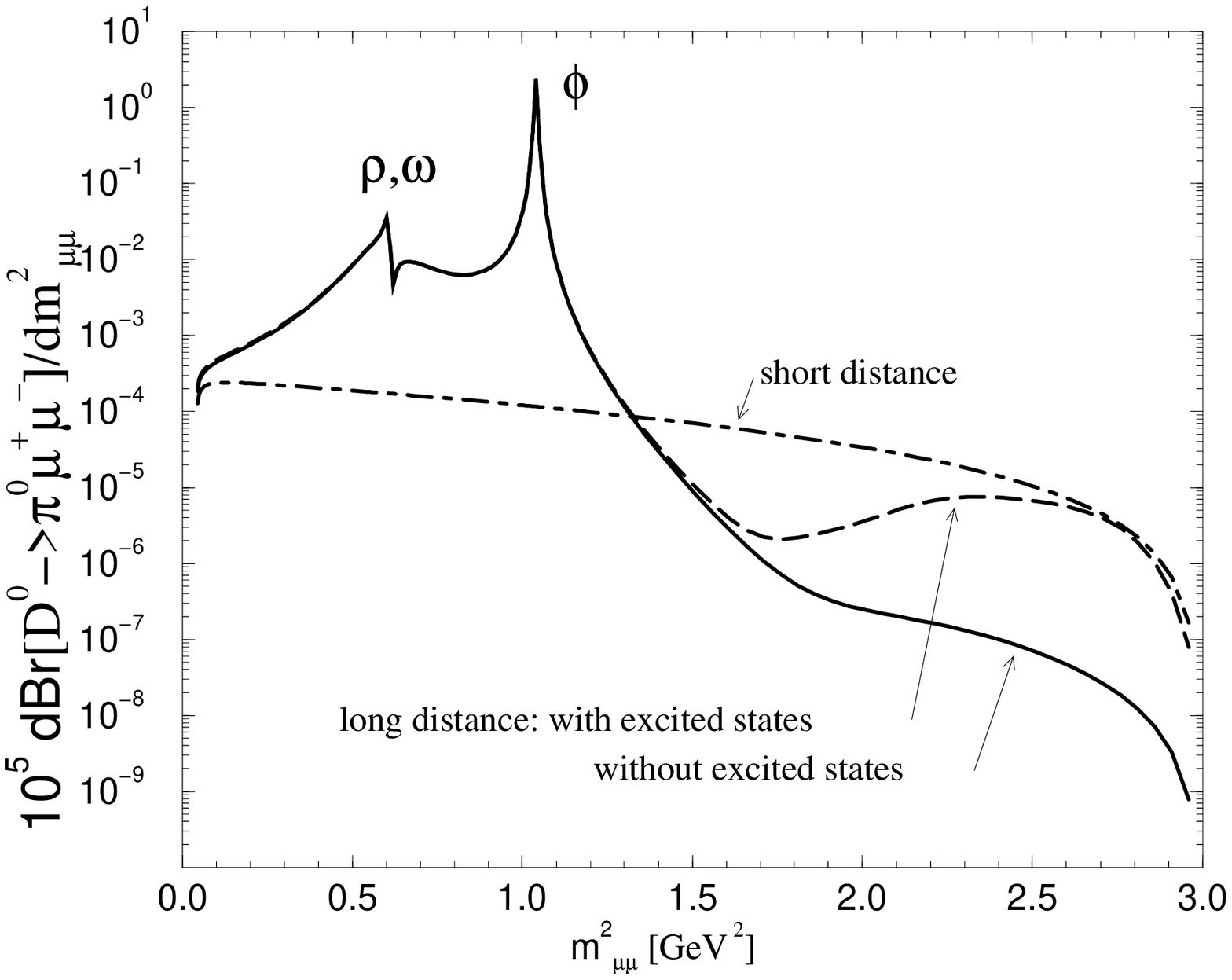} 
\quad
\includegraphics[scale=.4]{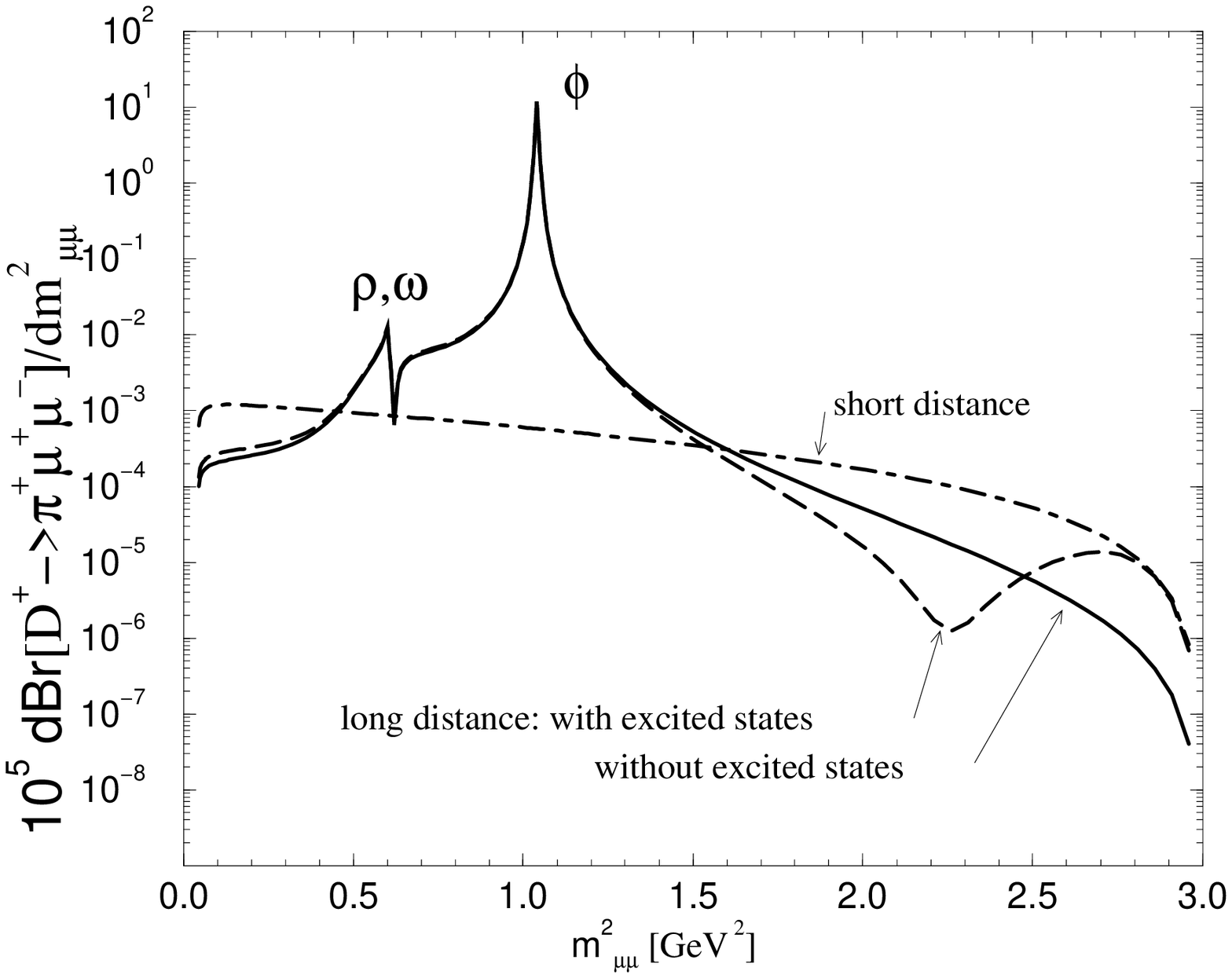} 
 \caption{The differential branching ratios $dBr/dm^2_{\mu\mu}$ as a function of the invariant di-lepton mass $m_{\mu\mu}^2$ for the  decays $D^0\to \pi^{0}\mu^+\mu^-$ and $D^+\to \pi^+\mu^+\mu^-$. The dot-dashed line denotes the short distance contribution due to the $c\to ul^+l^-$ transition as predicted by the standard model. The long distance contribution is calculated in two different ways: the dashed line indicates the long distance contribution, which takes into account the exited vector mesons $\rho^\prime$, $\omega^\prime$ and $\phi^\prime$; the solid line indicates the long distance contribution, which does not take into account the exited vector mesons. The short distance mechanism dominates over the long distance mechanism in the region $m_{\mu\mu}^2>1.6$ GeV$^2$. The spectrum for the case of electron and positron in the final state  is almost identical in the region $m_{ee}>2m_{\mu}$. } 
\label{fig39}
\end{center}
\end{figure}

\begin{table}[h]
\begin{center}
\begin{tabular}{|c||c|c|c|}
\hline
& short  & long distance & long distance \\
&   distance & without $\rho^{\prime}$, $\omega^\prime$, $\phi^\prime$ & with $\rho^\prime$, $\omega^\prime$, $\phi^\prime$\\
\hline
\hline
$ Br(D^0 \to \pi^{0}l^+l^-)[m_{ll}^2>1.5~{\rm GeV}^2]$ & $3.8\cdot 10^{-10}$&$9.9\cdot 10^{-12}$&$6.5\cdot 10^{-11}$\\
\hline
$ Br(D^+ \to \pi^+ l^+l^-)[m_{ll}^2>1.7~{\rm GeV}^2]$&$1.3\cdot 10^{-9}$&$4.3\cdot 10^{-10}$&$2.3\cdot 10^{-10}$\\
\hline
\end{tabular}
\caption{The $D\to \pi l^+l^-$ branching ratios integrated over the kinematical region of high di-lepton mass. The short distance part, as predicted by the standard model, is given in the second column. The long distance part, which does not incorporate the excited vector mesons, is given in the third column. The long distance part, which incorporates the ground state and the excited vector mesons, is given in the last column.}
\label{tab.pll2}
\end{center}
\end{table}

\vspace{0.2cm}

I conclude by stressing that the $D\to Pl^+l^-$ rates are dominated by the long distance contributions in the standard model. 
The $D^+\to \pi^+l^+l^-$ and $D_s^+\to \pi^+l^+l^-$ decays are predicted with the highest rates and may soon be observed (see Table \ref{tab.pll1}). 

The decay channels  $D^0\to \pi^0 l^+l^-$ and $D^+\to \pi^+ l^+l^-$ are of special interest. In the kinematical region of high di-lepton mass they are dominantly induced by the flavour changing neutral transition $c\to u l^+l^-$. The short distance mechanism, as predicted by the standard model,  dominates over the long distance mechanism in this kinematical region. These decays present the unique opportunity to probe the flavour changing neutral transition $c\to ul^+l^-$, which has enhanced sensitivity to the physics beyond the standard model.

\chapter{Conclusion}

I have explored the possibility of using the heavy meson decays as probes for the flavour changing neutral transitions $c\to u\gamma$, $c\to ul^+l^-$ and $c\bar u\to l^+l^-$, which are very rare in the standard model and present a good test for  physics beyond it.  I have calculated the standard model predictions for the heavy meson decay channels of interest and explored their sensitivity to several scenarios of physics beyond the standard model: the models with the extended Higgs sector, the minimal and non-minimal supersymmetric standard model, standard model with an extension of the fourth generation and left-right symmetric models. The  effects of new physics on the  decays $c\to u\gamma$, $c\to ul^+l^-$ and $c\bar u\to l^+l^-$ are found to be severely constrained by the present experimental upper bound on $\Delta m_D$. 

\vspace{0.2cm}

A hadron decay induced by the flavour changing neutral transition at short distances may also be induced by the long distance mechanisms. The long distance contribution can severely overshadow the short distance contribution of interest in a hadron decay. 

The baryon decays, discussed in Chapter 3,  are found to be dominated by the long distance effects. The only channels, which are not expected to be dominated by the long distance contribution, are   $\Xi_{cc}^{++}\to \Sigma_c^{++}\gamma$ and $\Omega_{ccc}^{++}\to \Xi_{cc}^{++}\gamma$ decays. These decays could serve for probing the  $c\to u\gamma$ transition, but they are too exotic for the experimental investigation at present.

Among the meson decays, the decay channels of the pseudoscalar heavy meson states are the most appropriate to study the rare weak channels.  

\begin{itemize}
\item
The $B_c\to B_u^*\gamma$ decay is proposed as the most suitable channel to study the $c\to u\gamma$ transition. Different contributions to this decay have been predicted using  the Isgur-Scora-Grinstein-Wise model. The long distance part of the branching ratio  is predicted at  $(7.5 {+7.7\atop -4.3})\cdot 10^{-9}$
and it is small due to the factor $V_{cb}^*V_{ub}$ in the amplitude. The uncertainty arises from the value of parameter $C_{MVD}$ (\ref{cvmd}), which is proportional to the $SU(3)$ flavour breaking parameter. The short distance part of the branching ratio, which arises via the $c\to u\gamma$ transition, is predicted at $4.7\cdot 10^{-9}$ in the standard model. The short and long distance contributions give branching ratios of comparable size $\sim 10^{-8}$, which in principle allows to use the $B_c\to B_u^*\gamma$ decay for probing the $c\to u\gamma$ transition. The total branching branching ratio is predicted at $(8.5 {+5.8 \atop -2.5})\cdot 10^{-9}$ and the experimental detection of this decay at the branching ratios well above $10^{-8}$ would clearly indicate a signal of physics beyond the standard model.  This decay is especially sensitive to the scenarios that could significantly enhance the $c\to u\gamma$ rate compared to the standard model predictions. Its branching ratio could be enhanced up to  $4\cdot 10^{-6}$ in some versions of the nonminimal supersymmetric model. Such effect would give distinctive experimental signature  when the measurements reach the corresponding sensitivity. The effects arising from the fourth generation could enhance the short distance part of the branching ratio up to $3\cdot 10^{-7}$. The effects of the extended Higgs sector and that of the left-right symmetry  are negligible compared to the standard model contributions. Different scenarios could be probed at The Large Hadron Collider, which is expected to produce $2.1\cdot 10^{8}$ $B_c$ mesons with $p_T(B_c)>20$ GeV at the integrated luminosity $100$ fb$^{-1}$.
\item
Among the charm decays, which can be induced via the transitions $c\to u\gamma$ and $c\to ul^+l^-$, the most interesting are the Cabibbo suppressed  decay channels $D\to V\gamma$, $D\to Vl^+l^-$ and $D\to Pl^+l^-$ ($V$ and $P$ denote light vector and pseudoscalar mesons, respectively). 
 I have systematically studied also all the Cabibbo allowed and the Cabibbo doubly suppressed decays of this kind, which arise only via the long distance mechanisms. 
To study the charm meson decays I  adapted the hybrid model, which combines the heavy quark effective theory and chiral perturbation theory, and  proposed a mechanism to incorporate the long distance contributions in a manifestly gauge invariant way. 

The predictions for the long distance parts of the branching ratios ($Br_{LD}$) for nine $D\to V\gamma$ decays are given in the second column of Table \ref{tab.gamma2}, while the short distance parts of the branching ratios for the Cabibbo suppressed decays are of the order of $10^{-9}$ (\ref{5.213}) in the standard model. 
The predicted long and short distance parts of branching ratios ($Br_{LD}$ and $Br_{SD}$) for $D\to Vl^+l^-$ and $D\to Pl^+l^-$ decays are given  Tables \ref{tab.vll2}, \ref{tab.vll3} and \ref{tab.pll1}. The main uncertainty in the predicted rates arises from the validity of the model and is estimated to be of the order of $50\%$. The branching ratios for the charm meson decays are found to be dominated by the long distance contributions. They are not  sensitive to the flavour changing neutral transitions $c\to u\gamma$ and $c\to ul^+l^-$ unless these transitions are significantly enhanced by some mechanism beyond the standard model. Different scenarios of the physics beyond the standard model could alter the short distance contributions, but none of the scenarios, discussed in Chapter 2, can not enhance them above the long distance ones. The non-minimal supersymmetric model  could perhaps enhance the short distance parts of the branching ratios for the Cabibbo suppressed $D\to V\gamma$ decays  up to about $10^{-6}$. 
Such enhancement would still be difficult to separate from the long distance contributions in the measured rate, given the present theoretical and experimental uncertainties.  

A window for probing the $c\to ul^+l^-$ transition  is, however, found  at the kinematical region of high di-lepton masses ( $m_{ll}\simeq[1.2~{\rm GeV},m_D-m_\pi]$ ) in $D\to \pi l^+l^-$ decays. The Fig. \ref{fig39} and Table \ref{tab.pll2} indicate that the short distance contribution, as predicted by the standard model,  dominates over the long distance contribution in this kinematical region. The $c\to ul^+l^-$ rate is found to be rather insensitive to the effects of the scenarios of new physics discussed in Chapter 2 due to strong constraints imposed by the present experimental upper bound on $\Delta m_D$. In spite of this fact, it would be interesting to confirm the standard model prediction for the $c\to ul^+l^-$ rate in this unique channel. A more detailed study of the end-point kinematical region in $D\to \pi l^+l^-$ decays has to be carried out for this purpose.

The predicted rates in Tables \ref{tab.gamma2}, \ref{tab.vll2}, \ref{tab.vll3} and \ref{tab.pll1} raise our hopes that the decays $D^0\to \bar K^{*0}\gamma$, $D_s\to \rho^+\gamma$ , $D^+\to\rho^+\gamma$, $D_s^+\to K^{*+}\gamma$, $D^0\to \bar K^{*0}l^+l^-$, $D_s^+\to \rho^+l^+l^-$, $D_s^+\to \pi^+l^+l^-$ and $D^+\to\pi^+l^+l^-$ will be observed soon. Let me point out, that there are no existing experimental upper bounds for the Cabibbo allowed channels $D_s^+\to \rho^+\gamma$ and $D_s^+\to \rho^+l^+l^-$, which are predicted at the highest rates. 
The experimental observation of the rare charm decay channels would improve our knowledge on the long distance dynamics in  the heavy meson decays and would help to disentangle the  similar long distance contributions in the decays of beauty mesons.

\end{itemize}

\vspace{0.1cm}

In addition several  interesting results  have arisen  through this work:

\begin{itemize}
\item 
The long distance penguin contribution is induced by the $SU(3)$ flavour breaking part of the action. This contribution is particularly small in the case of the decays $D\to V\gamma$ and $B_c\to B_u^*\gamma$, where a remarkable $SU(3)$ cancellation is carried from the quark level to the hadronic level (\ref{cvmd}, \ref{5.53}).  
\item
The bremsstrahlung part of the amplitude for a general decay of the form $P\to V\gamma$ is found to vanish (\ref{3.brem}). The bremsstrahlung amplitudes for  $D\to Pl^+l^-$ decays, as predicted by the hybrid model,  are found to vanish in the limit $m_P^2\ll m_D^2$. The bremsstrahlung amplitudes for  $D\to Vl^+l^-$ decays vanish in the exact $SU(3)$ flavour limit.
\item
The charm meson nonleptonic two body decays, in which the annihilation contribution is negligible and in which the final state contains a single isospin, are well understood in terms of the hybrid model. The predicted and the measured rates are summarized in Table \ref{5.10.tab} and agree well given the simplicity of the model and the present experimental uncertainties. 
\end{itemize}

I would like to conclude by prompting the experimental colleagues to look for the interesting  decay channels proposed in this work. This investigation may lead to  new insights  into the fundamental interactions of the elementary particles. At the same time, it will improve our understanding of  the strong interactions in the weak decays of the heavy mesons.

\appendix

\chapter{Evolution of effective operators
         with the renormalization scale $\boldsymbol{\mu}$}

\section{Evolution of operators $O_1$ and $O_2$}

The evolution of the coefficients $c_1$ and $c_2$ in the effective Lagrangian 
\begin{eqnarray}
\label{b.17}
{\cal L}&=& -{G_F\over \sqrt{2}}[c_1(\mu)O_1(\mu)+c_2(\mu)O_2(\mu)]\\
O_{1}&=&\bar q_3^{\alpha}\gamma_{\mu}(1-\gamma_5)q_1^{\alpha}~\bar q_4^{\beta}\gamma^{\mu}(1-\gamma_5)q_2^{\beta}\nonumber\\
O_{2}&=&\bar q_3^{\alpha}\gamma_{\mu}(1-\gamma_5)q_2^{\alpha}~\bar q_4^{\beta}\gamma^{\mu}(1-\gamma_5)q_1^{\beta}\nonumber
\end{eqnarray}
with the renormalization scale $\mu$ is studied in this section. The  matrix elements  $\gamma_{ij}$ defined by (\ref{evolution.77})
\begin{equation}
\label{b.1}
\mu {d\over d\mu} c_j(\mu)\equiv \gamma_{ij}(\mu)c_i(\mu)
\end{equation}
 for $i,j=1,2$ are explicitly calculated by neglectiong the  mixing of $O_1(\mu)$ and $O_2(\mu)$ with other operators in (\ref{operatorji}). 

\vspace{0.2cm}

Let me start by the first part of ${\cal L}$ (\ref{b.17})
\begin{equation}
\label{b.10}
{\cal L}_1=-{G_F\over \sqrt{2}}c_1O_1~.
\end{equation}
The quantities without the subscript $''B''$ (for bare) denote the renormalized quantities and the strong one-loop corrections to ${\cal L}_1$ in Figs.  \ref{fig7}b,  \ref{fig7}c and  \ref{fig7}d give infinite results. 
The bare Lagrangian at the one-loop level is obtained by calculating the necessary counterterms that cancel the divergences of the diagrams in Fig.  \ref{fig7}. In minimal substraction scheme combined with the dimensional regulatiozation in $d=4-\epsilon$ dimensions, the infinite parts of the form $1/\epsilon$ are substracted \cite{veltman}. First let me calculate the necessary counterterms for the diagrams in Figs.  \ref{fig7}c and  \ref{fig7}d.

\subsubsection{The renormalization of the diagrams in Figs.  \ref{fig7}c and  \ref{fig7}d}

The diagrams in Figs.  \ref{fig7}c and  \ref{fig7}d present the strong corrections to the currents $j_{1,\mu}=\bar q_3^{\alpha}\gamma_{\mu}(1-\gamma_5)q_1^{\alpha}$ and $j_2^{\mu}=\bar q_4^{\beta}\gamma^{\mu}(1-\gamma_5)q_2^{\beta}$, respectively. The diagrams in Fig.  \ref{fig7}c do not have any effect on the current $j_2$ and we will explicitely show that they do not affect the current $j_1$ either \footnote{This is a concequence of the Ward-Takashi identity.}.

The infinities arising from the {\bf first two diagrams} of Fig.  \ref{fig7}c are cancelled by adding the counter terms $\bar q_i(iB\!\!\not{\!\partial}-Am)q_i$ where 
$$-i\Sigma(p)-i(Am-B\!\!\not {\!p})={\rm finite}$$
and $-i\Sigma(p)$ is the amplitude of the diagram in Fig.  \ref{fig20}
$$-i\Sigma(p)=\mu^{\epsilon}(-i{g_s\over 2}\lambda^a)^2\int {d^dk\over (2\pi)^d}{-i\over k^2}\gamma^{\mu}{i(\!\!\not{\! p}+\!\!\not {\! k}+m)\over (p+k)^2-m^2}\gamma_{\mu}=-i{4\over 3} (4m-\!\!\!\not {\! p}){g_s^2\over 8\pi^2\epsilon}~+~{\rm finite}~,$$
so
\begin{equation}
\label{b.12}
A=-{16\over 3}~{g_s^2\over 8\pi^2\epsilon}\quad {\rm and}\quad B=-{4\over 3}~{g_s^2\over 8\pi^2\epsilon}~.
\end{equation}
The infinity arising from the {\bf last diagram} of Fig.  \ref{fig7}c is cancelled by adding a counterterm $-{G_F\over \sqrt{2}}c_1DO_1$ with 
$$O_1^{loop~c}+DO_1={\rm finite}$$
and $O_1^{loop~c}$ denotes the effective operator for the last loop  diagram in Fig.  \ref{fig7}c
\begin{align*}
 O_1^{loop~c}&=\mu^{\epsilon}(-i{g_s\over 2}\lambda^a)^2~\bar q_3\int\!\! {d^dk\over (2\pi)^d}{-i\over k^2}\gamma^{\nu} {i(\!\not {\!p_3}+\!\!\not {\!k}+m_3)\over (p_3+k)^2-m_3^2}\gamma_{\mu}(1-\gamma_5) {i(\!\not {\!p_1}+\!\!\not {\! k}+m_1)\over (p_1+k)^2-m_1^2}~q_1~j_2^{\mu}\\
&={4\over 3}{g_s^2\over 8\pi^2\epsilon}O_1+~{\rm finite}~,
\end{align*}
so
\begin{equation}
\label{b.13}
D=-{4\over 3}~{g_s^2\over 8\pi^2\epsilon}~.
\end{equation}

\begin{figure}[h]

\centering
\mbox{
\begin{fmffile}{f20n}
  \fmfframe(0,0)(0,0){
  \begin{fmfgraph*}(35,20)
  \fmfpen{thin}
  \fmfleft{l1}\fmfright{r1}
  \fmf{fermion,tension=1}{l1,v1}
  \fmf{fermion,tension=0.3,label=$p+k$,la.si=right}{v1,v2}
  \fmf{fermion}{v2,r1}
   \fmffreeze
   \fmf{gluon,label=$k$,la.s=left,left}{v1,v2}
   \fmflabel{$p$}{l1}\fmflabel{$p$}{r1}
  \end{fmfgraph*} }
\end{fmffile}
     }
\caption{Renormalization of the quark mass and the wave function due to the strong interactions.} 
\label{fig20}
\end{figure}
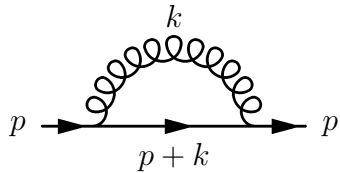

If the diagrams in Fig.  \ref{fig7}c presented the only strong correction to ${\cal L}_1$ (\ref{b.10}) at this order, then the corresponding bare Lagrangian ${\cal L}_1$ would be given by\footnote{The power of $\mu$ in the first term is obtained by equating the mass dimensions of the left and right hand side and by taking $[{\cal L}]=d$, $[q]=(d-1)/2$, $[G_F]=-2$.}
\begin{align}
\label{b.11}
{\cal L}_1&=-{G_F\over \sqrt{2}}c_1(\mu)(1+D)\mu^{\epsilon}\bar q_3^{\alpha}\gamma_{\mu}(1-\gamma_5)q_1^{\alpha}j_2^{\mu}+\sum_{i=1,3}\bar q_i[i(1+B)\!\!\not{\!\partial}-(1+A)m]q_i \\
&=-{G_F\over \sqrt{2}}c_1(\mu){1+D\over 1+B}\mu^{\epsilon}~\bar q_{B3}^{\alpha}\gamma_{\mu}(1-\gamma_5)q_{B1}^{\alpha}j_{B2}^{\mu}+\sum_{i=1,3}\bar q_{Bi}[i\!\!\not{\!\partial}-m_B]q_{Bi}~.\nonumber
\end{align}
 In the last line, the Lagrangian is rewritten in terms of the bare quantities
\begin{equation}
\label{b.11a}
q={q_B\over \sqrt{1+B}}\qquad{\rm and}\qquad m={1+B\over 1+A}m_B
\end{equation}
and since $B=D$ (\ref{b.12}, \ref{b.13})
\begin{equation}
\label{b.23}
{\cal L}_1=-{G_F\over \sqrt{2}}c_1(\mu)\mu^{\epsilon}O_{B1}\stackrel{\epsilon\to 0}{\longrightarrow}-{G_F\over \sqrt{2}}c_1 O_{B1}=-{G_F\over \sqrt{2}}c_1 O_{1}~,
\end{equation}
the renormalized Lagrangian (\ref{b.23}) has te same form as the tree level Lagrangian (\ref{b.10}). The Lagrangian ${\cal L}_1$ (\ref{b.10}) does not get renormalized by the diagrams in Fig.  \ref{fig7}c. Similar conclusion follows for the diagrams in Fig.  \ref{fig7}d.

\begin{figure}[!htb]

\centering
\mbox{
\subfigure[]
{
\begin{fmffile}{f7a}
  \fmfframe(3,3)(3,3){
  \begin{fmfgraph*}(30,20)
  \fmfpen{thin}  
  \fmfleft{l1,l2}\fmfright{r1,r2}
  \fmfrpolyn{shaded,tension=4}{k}{6}   
  \fmf{fermion}{l1,k1,r1}
  \fmf{fermion}{l2,k4,r2}
  \fmfv{decor.size=1.2thick,decor.shape=circle,decor.filled=full}{k1,k4}
  \fmflabel{$q_1(p_1)$}{l2}\fmflabel{$q_2(p_2)$}{l1}
  \fmflabel{$q_3(p_3)$}{r2}\fmflabel{$q_4(p_4)$}{r1}
  \end{fmfgraph*} }
\end{fmffile}
}
     }
\mbox{
\subfigure[]
{
\begin{fmffile}{f7b1}
  \fmfframe(6,3)(6,3){
  \begin{fmfgraph*}(25,25)
  \fmfpen{thin}  
  \fmfleft{l1,l2}\fmfright{r1,r2}
  \fmfrpolyn{shaded,tension=3}{k}{6}   
  \fmf{fermion}{l1,k1}
  \fmf{fermion,label=$p_4-k$,la.s=right,tension=0.5}{k1,m1}
  \fmf{fermion}{m1,r1}
    \fmf{fermion}{l2,k4}
    \fmf{fermion,label=$p_3+k$,la.s=left,tension=0.5}{k4,m2}
    \fmf{fermion}{m2,r2}
  \fmf{gluon,label=$k$,la.s=right,tension=0.5}{m1,m2}
  \fmfv{decor.size=1.2thick,decor.shape=circle,decor.filled=full}{k1,k4}
  \fmflabel{$p_1$}{l2}\fmflabel{$p_2$}{l1}
  \fmflabel{$p_3$}{r2}\fmflabel{$p_4$}{r1}
  \end{fmfgraph*} }
\end{fmffile}
\quad
\begin{fmffile}{f7b2n}
  \fmfframe(6,3)(6,3){
  \begin{fmfgraph*}(25,25)
  \fmfpen{thin}  
  \fmfleft{l1,l2}\fmfright{r1,r2}
  \fmfrpolyn{shaded,tension=3}{k}{6}   
  \fmf{fermion}{l1,m1}
  \fmf{fermion,label=$p_2+k$,la.s=right,tension=0.5}{m1,k1}
  \fmf{fermion}{k1,r1}
    \fmf{fermion}{l2,m2}
    \fmf{fermion,label=$p_1-k$,la.s=left,tension=0.5}{m2,k4}
    \fmf{fermion}{k4,r2}
  \fmf{gluon,label=$k$,la.s=left,tension=0.5}{m1,m2}
  \fmfv{decor.size=1.2thick,decor.shape=circle,decor.filled=full}{k1,k4}
  \fmflabel{$p_1$}{l2}\fmflabel{$p_2$}{l1}
  \fmflabel{$p_3$}{r2}\fmflabel{$p_4$}{r1}
  \end{fmfgraph*} }
\end{fmffile}
\quad
\begin{fmffile}{f7b3}
  \fmfframe(3,3)(3,3){
  \begin{fmfgraph*}(25,25)
  \fmfpen{thin}  
  \fmfleft{l1,l2}\fmfright{r1,r2}
  \fmfrpolyn{shaded,tension=3}{k}{6}   
  \fmf{fermion}{l1,m1}
  \fmf{fermion,tension=0.5}{m1,k1}
  \fmf{plain}{k1,r1}
    \fmf{fermion}{l2,k4}
    \fmf{fermion,tension=0.5}{k4,m2}
    \fmf{fermion}{m2,r2}
  \fmf{gluon,tension=0.5,right}{m1,m2}
  \fmfv{decor.size=1.2thick,decor.shape=circle,decor.filled=full}{k1,k4}
  \end{fmfgraph*} }
\end{fmffile}
\quad
\begin{fmffile}{f7b4}
  \fmfframe(3,3)(3,3){
  \begin{fmfgraph*}(25,25)
  \fmfpen{thin}  
  \fmfleft{l1,l2}\fmfright{r1,r2}
  \fmfrpolyn{shaded,tension=3}{k}{6}   
  \fmf{fermion}{l1,k1}
  \fmf{fermion,tension=0.5}{k1,m1}
  \fmf{fermion}{m1,r1}
    \fmf{fermion}{l2,m2}
    \fmf{fermion,tension=0.5}{m2,k4}
    \fmf{plain}{k4,r2}
  \fmf{gluon,tension=0.5,right}{m1,m2}
  \fmfv{decor.size=1.2thick,decor.shape=circle,decor.filled=full}{k1,k4}
  \end{fmfgraph*} }
\end{fmffile}
}
     }
\mbox{
\subfigure[]
{
\begin{fmffile}{f7c1}
  \fmfframe(3,3)(3,3){
  \begin{fmfgraph*}(30,25)
  \fmfpen{thin}  
  \fmfleft{l1,l2}\fmfright{r1,r2}
  \fmfrpolyn{shaded,tension=4}{k}{6}   
  \fmf{fermion}{l2,m1}\fmf{plain,tension=0.8}{m1,m2}\fmf{fermion}{m2,k4,r2}
  \fmf{fermion}{l1,k1,r1}
  \fmffreeze
  \fmf{gluon,tension=0.5,left}{m1,m2}
  \fmfv{decor.size=1.2thick,decor.shape=circle,decor.filled=full}{k1,k4}
  \end{fmfgraph*} }
\end{fmffile}
\quad
\begin{fmffile}{f7c2}
  \fmfframe(3,3)(3,3){
  \begin{fmfgraph*}(30,25)
  \fmfpen{thin}  
  \fmfleft{l1,l2}\fmfright{r1,r2}
  \fmfrpolyn{shaded,tension=4}{k}{6}   
  \fmf{fermion}{l2,k4,m1}\fmf{plain,tension=0.8}{m1,m2}\fmf{fermion}{m2,r2}
  \fmf{fermion}{l1,k1,r1}
  \fmffreeze
  \fmf{gluon,tension=0.5,left}{m1,m2}
  \fmfv{decor.size=1.2thick,decor.shape=circle,decor.filled=full}{k1,k4}
  \end{fmfgraph*} }
\end{fmffile}
\quad
\begin{fmffile}{f7c3}
  \fmfframe(3,3)(3,3){
  \begin{fmfgraph*}(30,25)
  \fmfpen{thin}  
  \fmfleft{l1,l2}\fmfright{r1,r2}
  \fmfrpolyn{shaded,tension=3}{k}{6}   
  \fmf{fermion}{l2,m1}\fmf{plain,tension=1}{m1,k4,m2}\fmf{fermion}{m2,r2}
  \fmf{fermion}{l1,k1,r1}
  \fmffreeze
  \fmf{gluon,tension=0.5,left}{m1,m2}
  \fmfv{decor.size=1.2thick,decor.shape=circle,decor.filled=full}{k1,k4}
  \end{fmfgraph*} }
\end{fmffile}
}
   }
\mbox{
\subfigure[]
{
\begin{fmffile}{f7d1}
  \fmfframe(3,3)(3,3){
  \begin{fmfgraph*}(30,25)
  \fmfpen{thin}  
  \fmfleft{l1,l2}\fmfright{r1,r2}
  \fmfrpolyn{shaded,tension=4}{k}{6}   
  \fmf{fermion}{l1,m1}\fmf{plain,tension=0.8}{m1,m2}\fmf{fermion}{m2,k1,r1}
  \fmf{fermion}{l2,k4,r2}
  \fmffreeze
  \fmf{gluon,tension=0.5,right}{m1,m2}
  \fmfv{decor.size=1.2thick,decor.shape=circle,decor.filled=full}{k1,k4}
  \end{fmfgraph*} }
\end{fmffile}
\quad
\begin{fmffile}{f7d2}
  \fmfframe(3,3)(3,3){
  \begin{fmfgraph*}(30,25)
  \fmfpen{thin}  
  \fmfleft{l1,l2}\fmfright{r1,r2}
  \fmfrpolyn{shaded,tension=4}{k}{6}   
  \fmf{fermion}{l1,k1,m1}\fmf{plain,tension=0.8}{m1,m2}\fmf{fermion}{m2,r1}
  \fmf{fermion}{l2,k4,r2}
  \fmffreeze
  \fmf{gluon,tension=0.5,right}{m1,m2}
  \fmfv{decor.size=1.2thick,decor.shape=circle,decor.filled=full}{k1,k4}
  \end{fmfgraph*} }
\end{fmffile}
\quad
\begin{fmffile}{f7d3}
  \fmfframe(3,3)(3,3){
  \begin{fmfgraph*}(30,25)
  \fmfpen{thin}  
  \fmfleft{l1,l2}\fmfright{r1,r2}
  \fmfrpolyn{shaded,tension=3}{k}{6}   
  \fmf{fermion}{l1,m1}\fmf{plain,tension=1}{m1,k1,m2}\fmf{fermion}{m2,r1}
  \fmf{fermion}{l2,k4,r2}
  \fmffreeze
  \fmf{gluon,tension=0.5}{m1,m2}
  \fmfv{decor.size=1.2thick,decor.shape=circle,decor.filled=full}{k1,k4}
  \end{fmfgraph*} }
\end{fmffile}
}
   }
\caption{The strong corrections to the effective operator $O_1$ 
(\ref{3.1}, \ref{b.17}) at the order of $\alpha_s$. The operator $O_1$ 
(\ref{3.1}, \ref{b.17}) is denoted by the hexagon. The two dots in the hexagon 
denote the action of the weak current given by $\gamma^\mu(1-\gamma_5)$.  }  
\label{fig7} 
\end{figure}
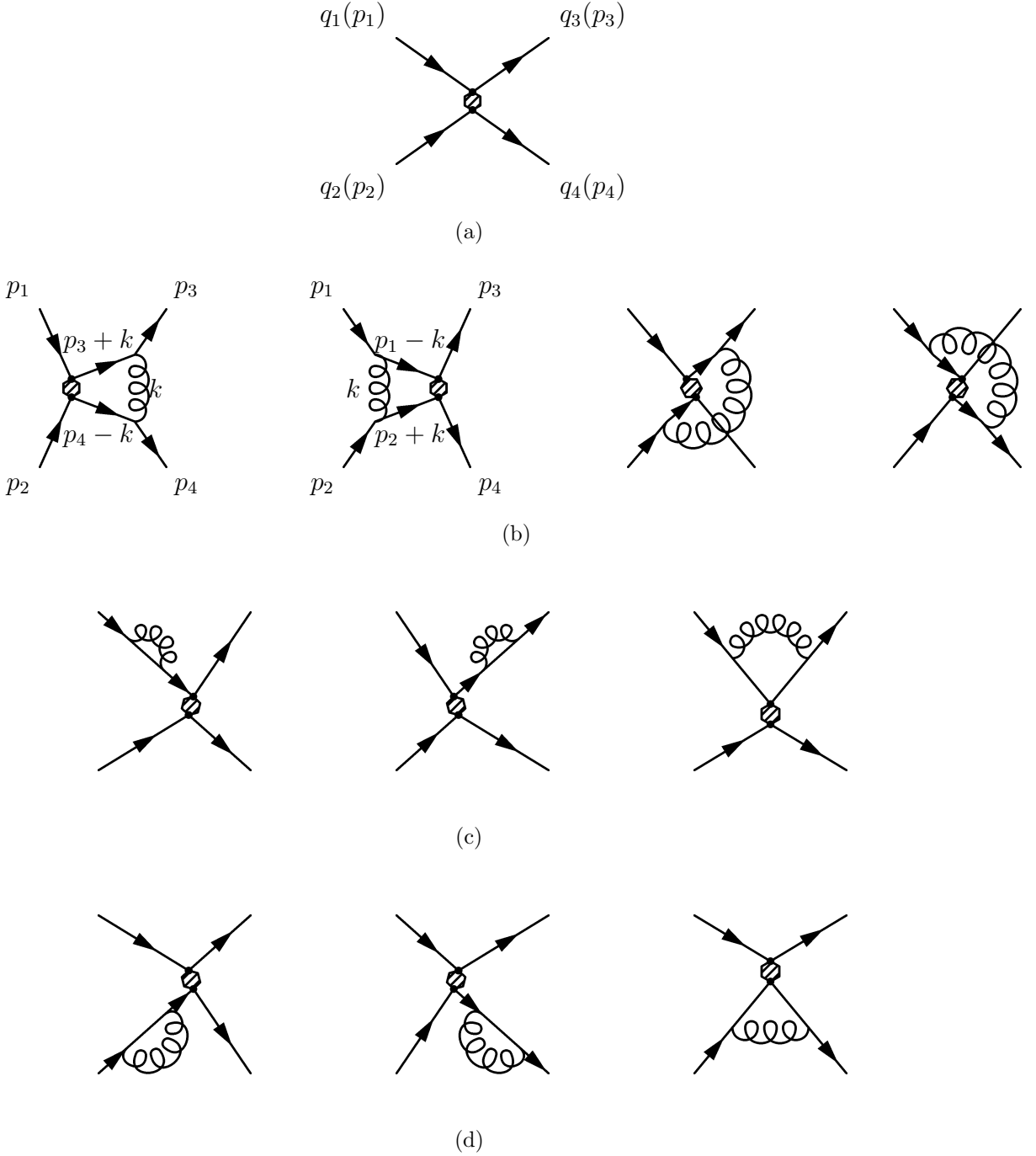

\subsubsection{The renormalization of the diagrams in Fig  \ref{fig7}b}

The divergent parts arising from the diagrams in Fig.  \ref{fig7}b are cancelled by adding the counterterm $-{G_F\over \sqrt{2}}c_1(E O_1+FO_2)$ with the condition 
\begin{equation}
\label{b.15}
O_1^{loop~b}+EO_1+FO_2={\rm finite}~.
\end{equation}
The ${O_1}^{loop ~b}$ denotes the effective operator for the loop diagrams in Fig.  \ref{fig7}b, which are induced by $O_1$
\begin{align}
\label{b.14}
O_1&^{loop~b}=(-i{g_s\over 2})^2\mu^{\epsilon}\int {d^dk\over (2\pi)^d}{-i\over k^2}\\
\times\biggl[&-\bar q_3\gamma^{\nu}\lambda^ai{\!\!\not {\! k}+\!\!\not {\! p_3}+m_3\over (k+p_3)^2-m_3^2} \gamma^{\mu}(1-\gamma_5)q_1\bar q_4\lambda_a\gamma_{\nu}i{\!\!\not {\! k}-\!\!\not {\! p_4}+m_4\over (k-p_4)^-m_4^2} \gamma_{\mu}(1-\gamma_5)q_2\nonumber\\
&-\bar q_3\gamma^{\mu}\lambda^ai{\!\!\not {\! k}-\!\!\not {\! p_1}+m_1\over (k-p_1)^2-m_1^2} \gamma^{\nu}(1-\gamma_5)q_1\bar u_4\lambda_a\gamma_{\mu}i{\!\!\not {\! k}+\!\!\not {\! p_2}+m_2\over (k+p_2)^-m_2^2} \gamma_{\nu}(1-\gamma_5)u_2\nonumber\\
& +\bar u_3\gamma^{\mu}\lambda^ai{\!\!\not {\! k}-\!\!\not {\! p_1}+m_1\over (k-p_1)^2-m_1^2} \gamma^{\nu}(1-\gamma_5)q_1\bar q_4\lambda_a\gamma_{\nu}i{\!\!\not {\! k}-\!\!\not {\! p_4}+m_4\over (k-p_4)^-m_4^2} \gamma_{\mu}(1-\gamma_5)q_2\nonumber\\
& +\bar q_3\gamma^{\nu}\lambda^ai{\!\!\not {\! k}+\!\!\not {\! p_3}+m_3\over (k+p_3)^2-m_3^2} \gamma^{\mu}(1-\gamma_5)q_1\bar q_4\lambda_a\gamma_{\mu}i{\!\!\not {\! k}+\!\!\not {\! p_2}+m_2\over (k+p_2)^-m_2^2} \gamma_{\nu}(1-\gamma_5)q_2~\biggr]\nonumber\\
&\!\!\!\!\!\!=(-i{g\over 2})^2~I^{\alpha\beta}~\bar q_3[\gamma^{\mu}\gamma_{\alpha}\gamma^{\nu}-\gamma^{\nu}\gamma_{\alpha}\gamma^{\mu}]\lambda^a(1-\gamma_5)q_1~\bar q_4[\gamma_{\nu}\gamma_{\beta}\gamma_{\mu}-\gamma_{\mu}\gamma_{\beta}\gamma_{\nu}]\lambda_a(1-\gamma_5)q_2\nonumber+{\rm finite}
\end{align}
with
\begin{align}
\label{b.11b}
I^{\alpha\beta}&=i\mu^{\epsilon}\int {d^dk\over (2\pi)^d}{(k+p_A)^{\alpha}(k+p_B)^{\beta}\over [(k+p_A)^2-m_A^2][(k+p_B)^2-m_B^2]k^2}\\
&=2i\mu^{\epsilon}\int_0^1 \!dx\!\int_0^{1-x}\! \!dy{d^dk\over (2\pi)^d}\biggl[{k^{\alpha}k^{\beta}\over (k^2-F^2)^3}+{[p_A-yp_A+(x+y)p_B]^{\alpha}[p_B-yp_A+(x+y)p_B]^{\beta}\over(k^2-F^2)^3}\biggr]\nonumber\\
&={\mu^{\epsilon}\over 2(4\pi)^{2-\epsilon/2}}\int_0^1 dx\int_0^{1-x} dy\{-g^{\alpha\beta}\Gamma(\epsilon/2)(F^2)^{-\epsilon/2}\}+{\rm finite}=-{g^{\alpha\beta}\over 4}~{1\over 8\pi^2\epsilon}+{\rm finite}~.\nonumber
\end{align}
Inserting $I^{\alpha\beta}$ to (\ref{b.14})
\begin{align*}
O_1^{loop~b}=-{1\over 4} ~{1\over 8\pi^2\epsilon}(-i{g_s\over 2})^2
&\bar q_3[\gamma^{\mu}\gamma^{\beta}\gamma^{\nu}-\gamma^{\nu}\gamma^{\beta}\gamma^{\mu}]\lambda^a(1-\gamma_5)q_1\nonumber\\
\times &\bar q_4[\gamma_{\nu}\gamma_{\beta}\gamma_{\mu}-\gamma_{\mu}\gamma_{\beta}\gamma_{\nu}]\lambda_a(1-\gamma_5)q_2+{\rm finite}
\end{align*}
 and using $\gamma^{\mu}\gamma^{\beta}\gamma^{\nu}-\gamma^{\nu}\gamma^{\beta}\gamma^{\mu}=2i\epsilon^{\mu\beta\nu\beta}\gamma_{\beta}\gamma_5$, $\lambda^a_{ij}\lambda_{kl}^a=-2/3\delta_{ij}\delta_{kl}+2\delta_{il}\delta_{jk}$ and the Fierz rearangement we get
$$O_1^{loop~b}=\mu^{\epsilon}(O_1-3O_2){g_s^2\over 8\pi^2\epsilon}+{\rm finite}~.$$
With (\ref{b.15})
$$E=-{g_s^2\over 8\pi^2\epsilon}\qquad {\rm and}\qquad F=3~{g_s^2\over 8\pi^2\epsilon}$$
 the bare Lagrangian is given by 
\begin{equation*}
\label{b.16}
{\cal L}_1=-{G_F\over \sqrt{2}}c_1(\mu)\mu^{\epsilon}[O_1+EO_1+FO_2]=-{G_F\over \sqrt{2}}c_1(\mu)\mu^{\epsilon}[O_{B1}+EO_{B1}+FO_{B2}]~.
\end{equation*}
Each of the operators $O_1$, $O_2$ was replaced by the bare operator $O_{B1}$, $O_{B2}$ since the two currents in the operators do not get renormalized, as we have seen above. 
The Lagrangian (\ref{b.16}) is the one-loop renormalized Lagrangian corresponding to the tree level Lagrangian (\ref{b.10}). 

\vspace{0.2cm}

Analogously, the one-loop renormalization of the tree level  Lagrangian ${\cal L}_2=-{G_F\over \sqrt{2}}c_2O_2$ turns out to be 
\begin{align*}
\label{b.24}
{\cal L}_2=-{G_F\over \sqrt{2}}c_1(\mu)\mu^{\epsilon}[O_{B2}+EO_{B2}+FO_{B1}]~.
\end{align*}

\vspace{0.2cm}

Finaly, the one-loop renormalized Lagrangian corresponding to the tree level Lagrangian (\ref{b.17}) is given by the sum of (\ref{b.16}) and (\ref{b.24}) 
\begin{align}
{\cal L}&=-{G_F\over \sqrt{2}}\mu^{\epsilon}\biggl([c_1(\mu)(1+E)+c_2(\mu)F]O_{B1}+
[c_2(\mu)(1+E)+c_1(\mu)F]O_{B2}\biggr)\\
&=-{G_F\over \sqrt{2}}\mu^{\epsilon}\biggl(\biggl[c_1(\mu)(1-{g_s^2\over 8\pi^2\epsilon})+3c_2(\mu){g_s^2\over 8\pi^2\epsilon}\biggr]O_{B1}+
\biggl[c_2(\mu)(1-{g_s^2\over 8\pi^2\epsilon})+3c_1(\mu){g_s^2\over 8\pi^2\epsilon}\biggr]O_{B2}\biggr)\nonumber
\end{align}
and since the bare Lagrangian and the bare operators $O_{B1}$ and $O_{B2}$ are $\mu$ independent
\begin{equation}
\label{b.18}
\mu{d\over d\mu}\biggl[c_1(\mu)(1-{g_s^2\over 8\pi^2\epsilon})+3c_2(\mu){g_s^2\over 8\pi^2\epsilon}\biggr]=0~~, \qquad \mu{d\over d\mu}\biggl[c_2(\mu)(1-{g_s^2\over 8\pi^2\epsilon})+3c_1(\mu){g_s^2\over 8\pi^2\epsilon}\biggr]=0~.
\end{equation}
The equations must be solved for an arbitrary value of the regulator $\epsilon$, which is eventualy sent to zero.  The coefficient for each term in the Laurent series $1/\epsilon^n$ of (\ref{b.18}) must vanish. In order to be able to solve the coupled equations, any quantity $a$, with a value $a(\epsilon\!\!=\!\!0)$ in four dimensions, has to be defined in arbitrary demensions $a(\epsilon)=a_0+a_1\epsilon+..$ .  The matrix  $\gamma_{ij}$,  defined by (\ref{b.1}), is calculated from the equations (\ref{b.18}) following the general procedure explained in the equations (18.6.1)-(18.6.8) of \cite{weinbergII}
$$\gamma={g_s^2\over 8\pi^2}\begin{pmatrix}-1&3\\
                                       3&-1\end{pmatrix}~.$$

\section{Anomalous dimension $\gamma_{77}$}

The one-loop strong corrections to the effective Lagrangian  (\ref{o7})
\begin{equation}
{\cal L}=-{4G_F\over \sqrt{2}}c_7O_7~~, \qquad
O_7={e\over 32\pi^2}m_1~\bar q_2\sigma_{\mu\nu}(1+\gamma_5)q_1F^{\mu\nu}
\end{equation}
are shown in Fig.  \ref{fig4}. The bare Lagrangian at the one-loop level is obtained by calculating the necessary counterterms that cancel the divergences arising from  the diagrams in Fig.  \ref{fig4}. In minimal substraction scheme combined with the dimensional regularization in $d=4-\epsilon$ dimensions, the infinite parts of the form $1/\epsilon$ are substracted. 

 \subsubsection{The renormalization of diagram in Fig.  \ref{fig4}a}

The effective operator $O_7^{loop ~a}$ for the one loop diagram in Fig.  \ref{fig4}a is
\begin{align*}
O_7^{loop ~a}&={e\over 32\pi^2}m_1(-i{g_s\over 2}\lambda^a)^2\mu^{\epsilon}\int {d^dk\over (2\pi)^d}{-i\over k^2}\bar q_2\gamma^{\delta}{i(\!\!\not {\! p_2}+\!\!\not {\! k}+m_2)\over (p_2+k)^2-m_2^2} \sigma^{\mu\nu}(1+\gamma_5){i(\!\!\not {\! p_1}+\!\!\not {\! k}+m_1)\over (p_1+k)^2-m_1^2}\gamma_{\delta}q_1F_{\mu\nu}\\
&={e\over 32\pi^2}m_1(-i{g_s\over 2}\lambda^a)^2\bar q_2\gamma^{\delta}\gamma_{\alpha} \sigma^{\mu\nu}(1+\gamma_5)\gamma_{\beta}q_1F_{\mu\nu}I^{\alpha\beta}+{\rm finite}~.
\end{align*}
The $1/\epsilon$ part of $I^{\alpha\beta}$ (\ref{b.11b}) is proportional to $g^{\alpha\beta}$ and so
$$O_7^{loop ~a}={\rm finite}$$
since $g^{\alpha\beta}\gamma_{\alpha} \sigma^{\mu\nu}(1+\gamma_5)\gamma_{\beta}=\gamma^{\beta}\sigma^{\mu\nu}\gamma_{\beta}(1-\gamma_5)=0$. The diagram in Fig.  \ref{fig4}a is finite and no counterterm is needed.

 \subsubsection{The renormalization of diagrams in Fig.  \ref{fig4}b}

The diagrams in Fig.  \ref{fig4}b contain the quark self energies loops and are renormalized by adding the counterterms $\bar q_i(iB\!\!\not{\!\partial}-Am)q_i$ where $A$ and $B$ were determined in (\ref{b.12}). The complete bare Lagrangian expressed in terms of the bare quantities via (\ref{b.11a}) is\footnote{The coefficient $\rho_7=-1/2$ is obtained by equating the mass dimensions of the left and right side $[{\cal L}]=-2-\rho_7\epsilon+1+2[q]+[F_{\mu\nu}]$ with $[{\cal L}]=d$, $[q]=(d-1)/2$ and $[F^{\mu\nu}]=d/2$. The result  turns out to be independent of $\rho_7$. The strong interactions do not renormalize the electric field and the electric charge, but the bare and the renormalized electric charges have differnet dimensions: $e_B=\mu^{\epsilon/2}e$.}
\begin{align*}
{\cal L}&=-{4G_F\over \sqrt{2}}c_7(\mu)\mu^{-\rho_7\epsilon}O_7+\sum_{i=1,2}\bar q_i[i(1+B)\!\!\not{\!\partial}-(1+A)m]q_i \\
&=-{4G_F\over \sqrt{2}}c_7(\mu)\mu^{-\rho_7\epsilon}\mu^{-\epsilon/2}{e_B\over 32\pi^2}{1+B\over 1+A}m_{B1}{\bar q_{B2}\sigma_{\mu\nu}(1+\gamma_5)q_{B1}\over 1+B}F^{\mu\nu}_B+\sum_{i=1,2}\bar q_{Bi}[i\!\!\not{\!\partial}-m_B]q_{Bi}~\\
&=-{4G_F\over \sqrt{2}}c_7(\mu)\mu^{-\rho_7\epsilon}\mu^{-\epsilon/2}{1\over 1+A}O_{B7}+\sum_{i=1,2}\bar q_{Bi}[i\!\!\not{\!\partial}-m_B]q_{Bi}
\end{align*}
and it is $\mu$ independent so
\begin{equation}
\label{b.21}
\mu{d\over d\mu}\biggl[c_7(\mu)\mu^{-\rho_7\epsilon}\mu^{-\epsilon/2}{1\over 1+A}\biggr]=0 \quad {\rm or}\quad
\mu{d\over d\mu}\biggl[c_7(\mu)\mu^{-\rho_7\epsilon}\mu^{-\epsilon/2}\biggl(1+{16\over 3}{g_s^2\over 8\pi^2\epsilon})\biggr]=0~.
\end{equation}
The only contribution to the running of the Willson coefficient $c_7$ comes from  the renormalization of the mass in the definition of $O_7$, which is responsible for the factor $1/(1+A)$ (\ref{b.12}). 
The coefficient for each term in the Laurent series $1/\epsilon^n$ of (\ref{b.21}) must vanish.  Using a general procedure given in equations (18.6.1)-(18.6.8) of \cite{weinbergII}, the coupled equations (\ref{b.21}) give
$$\gamma_{77}={16\over 3}~{g_s^2\over 8\pi^2}~.$$

\chapter{General forms of various vertices}

\subsubsection{Vertex $\boldsymbol{q_1-q_2-\gamma^*}$}

The amplitude for the vertex $q_1-q_2-\gamma^*$, where $\gamma^*$ denotes a real or a virtual photon,  has a general form
${\cal  A}[q_1\to q_2\gamma^*(q,\epsilon)]=\epsilon^{\mu}\langle q_2|J_{\mu}^{em}|q_1\rangle$
with a Lorentz decomposition
\begin{equation}
\label{c.2}
 \langle q_2|J_{\mu}^{em}(q)|q_1(p)\rangle=\bar u_2[iq^{\nu}\sigma_{\mu\nu}(A_1+A_2\gamma_5)+\gamma_{\mu}(A_3+A_4\gamma_5)+q_{\mu}(A_5+A_6\gamma_5)]u_1~.
\end{equation}
In a theory with the left-handed charged current interaction and in the limit of the massless quark $q_2$, the quark $q_2$ should necessarily be left-handed. This  implies $A_1=A_2$, $A_3=-A_4$, $A_5=A_6$ and the electromagnetic gauge condition  $\partial^{\mu}J_{\mu}^{em}=0$ implies  $-A_3/q^2=A_5/m_1$. The amplitude can then be generally written as 
\begin{equation}
\label{c.3}
{\cal  A}[q_1\to q_2\gamma^*(q,\epsilon)]=\epsilon^{\mu}\bar u_2[Aim_1q^{\nu}\sigma_{\mu\nu}(1+\gamma_5)+B(q^2\gamma_{\mu}-q_{\mu}\!\!\not {\!q})(1-\gamma_5)]u_1
\end{equation}
and for on-shell photons with $q^2=0$ and $q\epsilon=0$
\begin{equation}
\label{c.6}
{\cal  A}[q_1\to q_2\gamma(q,\epsilon)]=i A~m_1\bar u_2 \sigma_{\mu\nu}(1+\gamma_5)u_1 \epsilon^{\mu}q^{\nu}~.
\end{equation}
Tha coefficients $A$ and $B$ are functions of Lorentz invariant products of the momenta and have to be evaluated for the specific processes. The amplitude is of the second order in the external momenta and has to be evaluated at least to that order.

\subsubsection{Vertex $\boldsymbol{q_1-q_2-Z^*}$}

The  vertex $q_1-q_2-Z^*$ has a decomposition similar to (\ref{c.2}) but there is no gauge condition to be imposed. This vertex can be calculated in the zeroth order in external momenta \cite{IL} and only the second term in the general expression (\ref{c.2}) survives. In the limit  $m_2\to 0$, the vertex has the form
\begin{equation}
\label{c.4}
{\cal  A}(q_1\to q_2Z^*)=C ~\bar u_2\gamma_{\mu}(1-\gamma_5)u_1~\epsilon^{\mu}.
\end{equation}

\subsubsection{The  $\boldsymbol{q_1-q_2-l^+-l^-}$ box diagram}

The $W$ box diagram in Fig.  \ref{fig8}b couples only left-handed fermions has the amplitude of the general form 
\begin{equation}
\label{c.5}
{\cal A}^{box}=D~ \bar u_2\gamma_{\mu}(1-\gamma_5)u_1~\bar u_l\gamma^{\mu}(1-\gamma_5)u_l~.
\end{equation}

\chapter{The two Higgs doublet model}

The details of the two Higgs doublet model, presented in Section 2.2.1, are given here. The five physical Higgs bosons of this model are identified first. Then the couplings of the neutral physical Higgs bosons and the up-like quarks are derived for the general two Higgs doublet model given by (\ref{2.32}). 

Among the eight real fields contained in 
$$\Phi_1={\phi_1^+\brack {v_1+H_1+iA_2\over \sqrt{2}}}\qquad{\rm and}\qquad \Phi_2={\phi_2^+\brack {v_2+H_2+iA_2\over \sqrt{2}}}~,$$
two charged and one neutral field are would-be Goldstone bosons ($G^{\pm}$ and $G^0$) of the $SU(2)_L\times U(1)_Y\to U(1)_{em}$ breaking. They give the masses to $W^{\pm}$ and $Z$ gauge bosons. The would-be Goldstone bosons can identified by noting that they couple linearly to  $W$ or  $Z$ bosons in the $|D_{\mu}\Phi_1|^2+|D_{\mu}\Phi_2|^2$ part of the Lagrangian. In the notation of \cite{bilenky}, this part of the Lagrangian is given by 
\begin{align*}
&|D_{\mu}\Phi_1|^2+|D_{\mu}\Phi_2|^2=|(\partial_{\mu}+i{1\over 2}g\vec \tau \vec A_{\mu}+i{1\over 2}g^{\prime}B_{\mu})\Phi_1|^2+|(\partial_{\mu}+i{1\over 2}g\vec \tau \vec A_{\mu}+i{1\over 2}g^{\prime}B_{\mu})\Phi_2|^2\nonumber\\
&={ig\over \sqrt{2}}~[W^+(v_1\partial_{\mu}\phi_1^-+v_2\partial_{\mu}\phi_2^-)-W^-(v_1\partial_{\mu}\phi_1^++v_2\partial_{\mu}\phi_2^+)+iZ^{\mu}(v_1\partial_{\mu}A_1+v_2\partial_{\mu}A_2)/\cos\theta_W]~+...
\end{align*}
and the dots represent the terms that are not the products of one  Higgs and one gauge boson field.  Defining the angle $\beta$ by $\tan \beta\equiv v_2/v_1$, the identified would-be Goldstone boson fields are expressed as
$$G^{\pm}=\cos\beta\phi_1^{\pm}+\sin\beta\phi_2^{\pm}\qquad {\rm and}\qquad G^0=\cos\beta A_1+\sin\beta A_2~.$$
The charged scalar $H^{\pm}$ and the neutral pseudoscalar $A_0$ physical fields are the orthogonal combinations
\begin{equation}
\label{d.1}
H^{\pm}=\cos\beta\phi_2^{\pm}-\sin\beta\phi_1^{\pm}\qquad {\rm and}\qquad A^0=\cos\beta A_2-\sin\beta A_2~.
\end{equation}
The neutral scalar fields $H^0$ and $h^0$ with the well-defined mass are the linear combinations of $H_1$ and $H_2$
\begin{equation}
\label{d.2}
H^0=\cos\alpha H_1+\sin\alpha H_2\qquad{\rm and}\qquad h^0=\cos \alpha H_2-\sin\alpha H_1
\end{equation}
and the mixing angle $\alpha$ depends on the values of the couplings parameterizing the Higgs self interactions \cite{higgs}.

Now I turn to the couplings of the neutral Higgses to the up-like quarks in the general two Higgs doublet model given by Eq. (\ref{2.32}). 
For the purpose of $c\to u$ transitions, one is interested only in the flavour non-diagonal part of the Lagrnangian. One can choose to work in the basis, where the weak  $U^{\prime}$ and the mass  $U$ eigenstates match and  $\Gamma^u$ (\ref{2.31}) is a diagonal quark mass matrix.  Inserting (\ref{d.1}) and (\ref{d.2}) to the Yukawa Lagrangian (\ref{2.32}) we get 
\begin{align}
\label{d.3}
&{\cal L}_U={1\over \sqrt{2}}\bar U~[\lambda^u_1v_1+\lambda^u_2v_2+(\cos\alpha\lambda^u_2-\sin\alpha\lambda_1^u)h^0+(\cos\alpha\lambda^u_1+\sin\alpha\lambda_2^u)H^0\\
&\qquad\qquad\qquad+i(\sin\beta\lambda^u_1-\cos\beta\lambda^u_2)A^0\gamma_5]~U\nonumber\\
&={1\over \sqrt{2}}\bar U\lambda^u_2~[(\cos\alpha+\sin\alpha\tan\beta)h^0+(\sin\alpha-\cos\alpha\tan\beta)H^0-i(\cos\beta+\sin\beta\tan\beta)A^0\gamma_5]~U~.\nonumber
\end{align}
In the second line $\lambda_1^u$ was expressed in terms of $\lambda_2^u$ and $\Gamma^u$ via (\ref{2.31}) and the flavour diagonal part was omitted.  

The physical Higgs bosons have different couplings to $c$ and $u$ quarks given by (\ref{d.3}).  They also have different masses depending on the couplings parameterizing the Higgs self interactions \cite{higgs}. Since the values of the parameters are unknown so far, I simplify the discussion in Section 2.2.1  and set all the $c-u-H$ couplings and masses to be equal.  

\chapter{The $\boldsymbol{D^0-\bar D^0}$ mixing}

The  $D^0-\bar D^0$ mixing is not the subject of the present work and has been extensively studied by other authors \cite{dmix}. However, the experimental results on $\Delta m_D$ (the mass differentce of $D^0$ and  $\bar D^0$ mass eigenstates)  severely constrain the parameter space of new physics scenarios, especially in the sector of the flavour changing transitions among the $c$ and $u$ quarks.  The experimental upper limit  $\Delta m_D<1.6\cdot 10^{-13}$ GeV \cite{PDG} was used in Section 2.3 in order to get the upper bounds on  various parameters related to  new physics. These bounds are then used to predict the effects of the new mechanisms to other FCN transitions among $c$ and $u$ quarks. It is therefore appropriate to comment the standard model predictions for $\Delta m_D$ at this point. 

The {\bf short distance part} of $\Delta m_D$ is due to the $W$ box diagrams in Fig.  \ref{fig15}a. The amplitude  is strongly GIM suppressed by $V_{ci}^*V_{ui}V_{cj}V_{uj}^*m_i^2m_j^2/m_W^4$ and renders exceedingly small mass difference  \cite{dmix,hewett}
$$\Delta m_D^{SD}\simeq 5\cdot 10^{-18}\ {\rm GeV}~.$$

The short distance box diagrams are not the only ones that contribute to the mass difference. Since the light quarks with rather large CKM factors to the charm can propagate between the $D^0$ and $\bar D^0$, one expects relatively important {\bf long distance contributions} to the mixing \footnote{The situation is very different in $K^0-\bar K^0$ and $B^0-\bar B^0$ mixing, where the important effect comes from the heavy quark inside the box diagram loop: the charm quark in $K^0-\bar K^0$ and the top quark in $B^0-\bar B^0$. In the case of the kaon mixing, the coupling to light hadron intermediate states is still large and the long distance contributions are of the same order of magnitude as the short distance ones.}. The intermediate propagating degrees of freedom are light hadrons rather than light quarks in this case. The long distance arises from the propagation of the intermediate hadronic states to which both $D^0$ and $\bar D^0$ can decay. There will be one, two, three, etc. particle intermediate states. The important contribution comes from the intermediate two particle states $\pi^+\pi^-$, $K^+K^-$, $\pi^+K^-$ and $K^+\pi^-$ shown in Fig.  \ref{fig15}b. The two weak vertices in Fig.  \ref{fig15}b are induced by the effective nonleptonic weak Lagrangian (\ref{eff}). The long distance contributions have been calculated \cite{dmix} via the dispersive approach giving the mass differnece of the order of   
$$\Delta m_D^{LD}\simeq 10^{-16}\ {\rm GeV}~.$$
The heavy quark effective theory resluts lead to the mass differnece of the order of   
$$\Delta m_D^{LD}\simeq 10^{-17}\ {\rm GeV}.$$
 Although the standard model predictions for $\Delta m_D$ are quite uncertain, it is clear that they are far bellow the present experimental upper bound $\Delta m_D<1.6\cdot 10^{-13}$ GeV \cite{PDG}. As $\Delta m_D$ is small in the standard model, the possible effects of new physics can be relatively important. Experimentally unexplored window for $\Delta m_D$ between $10^{-16}$ and $10^{-13}$ GeV still offers  a unique opportunity to discover new effects in this sector. Different scenarios of physics beyond the standard model and its effects on $\Delta m_D$ are discussed in Section 2.2.

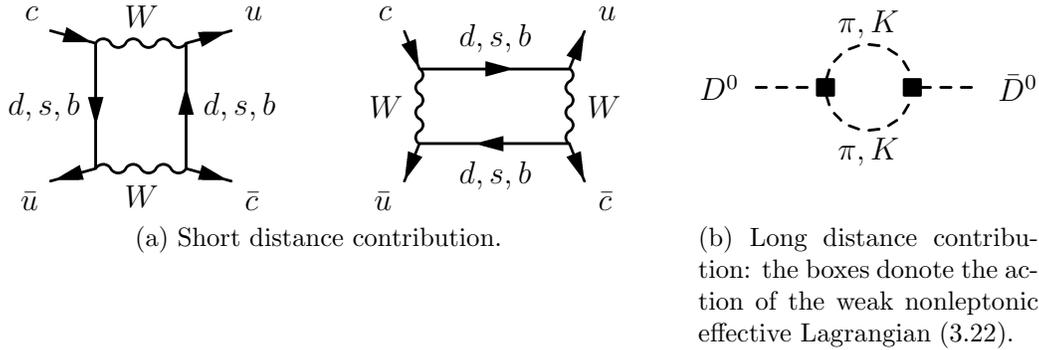
\begin{figure}[h]

\centering
\mbox{
\subfigure[Short distance contribution.]
{
\begin{fmffile}{fig15a1}
\fmfframe(3,0)(3,0){
\begin{fmfgraph*}(30,20)
\fmfpen{thin}
\fmfleft{i1,i2}
\fmflabel{$c$}{i2}
\fmflabel{$\bar u$}{i1}
\fmfright{o1,o2}
\fmflabel{$u$}{o2}
\fmflabel{$\bar c$}{o1}
\fmf{fermion}{i2,v4}
\fmf{fermion,label=$d,,s,,b$,label.side=right,tension=.1}{v4,v1}
\fmf{fermion}{v1,i1}
\fmf{fermion}{v2,o1}
\fmf{fermion,label=$d,,s,,b$,tension=.1}{v2,v3}
\fmf{fermion}{v3,o2}
\fmf{boson,label=$W$,tension=.5}{v1,v2}
\fmf{boson,label=$W$,tension=.5}{v3,v4}
\end{fmfgraph*}   }
\end{fmffile}
\quad
\begin{fmffile}{fig15ab}
\fmfframe(3,0)(3,0){
\begin{fmfgraph*}(30,20)
\fmfpen{thin}
\fmfleft{i1,i2}
\fmflabel{$c$}{i2}
\fmflabel{$\bar u$}{i1}
\fmfright{o1,o2}
\fmflabel{$u$}{o2}
\fmflabel{$\bar c$}{o1}
\fmf{fermion}{i2,v4}
\fmf{fermion,label=$d,,s,,b$,label.side=left,tension=.1}{v4,v3}
\fmf{fermion}{v1,i1}
\fmf{fermion}{v2,o1}
\fmf{fermion,label=$d,,s,,b$,la.s=left,tension=.1}{v2,v1}
\fmf{fermion}{v3,o2}
\fmf{boson,label=$W$,tension=.5,la.s=left}{v1,v4}
\fmf{boson,label=$W$,tension=.5,la.s=right}{v2,v3}
\end{fmfgraph*}   }
\end{fmffile}
}
\subfigure[Long distance contribution: the boxes donote the action of the weak nonleptonic effective Lagrangian (\ref{eff}).]
{
\begin{fmffile}{f15b}
  \fmfframe(7,0)(7,0){
  \begin{fmfgraph*}(30,25)
  \fmfpen{thin}
  \fmfleft{l1}\fmfright{r1}
  \fmf{dashes,tension=1}{l1,v1}
  \fmf{dashes,left,label=$\pi,,K$,la.s=left,la.d=20,tension=0.4}{v1,v2}
  \fmf{dashes,right,label=$\pi,,K$,la.s=right,la.d=20,tension=0.4}{v1,v2}
  \fmf{dashes}{v2,r1}
  \fmflabel{$D^0$}{l1}\fmflabel{$\bar D^0$}{r1}
  \fmfv{de.sh=square,de.si=3thick,de.f=full}{v1,v2}
  \end{fmfgraph*} }
\end{fmffile}
}
     }
\caption{The short and long distance mechanisms responsible for the $D^0-\bar D^0$ mixing.}
\label{fig15}
\end{figure}

\chapter{Transformation properties of the hadronic fields in the Heavy meson chiral Lagrangian approach}

The transformation properties of the hadronic fields in the Heavy meson chiral Lagrangian approach are gathered here.  

\subsubsection{Lorentz transformation}

Under the Lorentz transformation $x\to \Lambda x$ the fields transform as
\begin{alignat*}{2}
H_a(x)&\to D(\Lambda)H_a(\Lambda x)D(\Lambda)^{-1}~,&\qquad\qquad \bar H_a(x)&\to D(\Lambda)\bar H_a(\Lambda x)D(\Lambda)^{-1}~,\\
\xi(x)&\to \xi(\Lambda x) ~,&\qquad\qquad \xi^\dagger(x)&\to \xi^\dagger(\Lambda x)~,\\
\rho^\mu(x)&\to \Lambda^{\mu}_{~\nu}\rho^\nu(\Lambda x)~,\\
\end{alignat*}
where $D(\Lambda)$ is an element of the $4\times 4$ matrix representation of the Lorentz group.

\subsubsection{Parity}

Under parity transformation ${\cal P}(t,\vec x)=(t,-\vec x)$ the fields transform as
\begin{alignat*}{2}
H_a(x)&\to \gamma^0 H_a({\cal P}x)\gamma^0 ~,&\qquad\qquad \bar H_a(x)&\to \gamma^0 \bar H_a({\cal P}x)\gamma^0~,\\
\xi_{ab}(x)&\to \xi^\dagger_{ba}({\cal P}x)~,&\qquad \qquad \xi_{ab}^\dagger(x)&\to \xi_{ba}({\cal P}x)~,\\
\rho^\mu(x)&\to {\cal P}^\mu_{~\nu}\rho^\nu({\cal P}x)\\
\end{alignat*}
and the fields ${\cal A}^\mu$ and ${\cal V}^\mu$ defined in (\ref{5.18}) transform as 
$${\cal A}^\mu(x)\to -{\cal P}^\mu_{~\nu}{\cal A}^\nu({\cal P}x)~,\qquad\quad
{\cal V}^\mu(x)\to {\cal P}^\mu_{~\nu}{\cal V}^\nu({\cal P}x)~.$$

\subsubsection{Heavy quark spin transformation}

Under the rotation of the heavy quark spin $Q\to SQ$, where the $S$ is $4\times 4$ matrix representation of the $SU(2)$ spin transformations, the meson fields transform as
\begin{alignat*}{2}
H_a&\to S H_a~,&\qquad \qquad \bar H_a&\to \bar H_a S^{\dagger}~,\\
\xi&\to \xi~,&\qquad\qquad \xi^\dagger&\to \xi^\dagger~,\\
\rho^\mu&\to \rho^\mu~.
\end{alignat*} 

\subsubsection{Chiral transformation}

Under the global chiral $SU(3)_L\times SU(3)_R$ transformation, the mesonic fields transform as
\begin{alignat*}{2}
H_a&\to H_b U_{ba}^\dagger(x)~,&\qquad\qquad \bar H_a&\to U_{ab}(x)\bar H_b~,\\
\xi&\to g_L\xi U^\dagger(x)=U(x)\xi g_R^\dagger~,&\qquad\qquad \xi^\dagger &\to U(x)\xi^\dagger g_L^\dagger=g_R\xi^\dagger U^\dagger (x)~,\\
\rho^\mu&\to U(x)\rho^\mu U^\dagger(x)+U(x)\partial^\mu U^\dagger(x)~
\end{alignat*}
with $g_L\in SU(3)_L$ and $g_R\in SU(3)_R$. The transformation  $U(x)\in SU(3)_V$ is defined by the equation $g_L\xi U^\dagger(x)=U(x)\xi g_R^\dagger$ in the second line and depends on $g_L$, $g_R$ and the field $\xi(x)$. In the case of the transformation in $SU(3)_V$ subgroup,  $U=g_L=g_R$.

\subsubsection{Transformation $\boldsymbol{[SU(3)_L\times SU(3)_R]_{global}\times[SU(3)_V]_{local}}$}

The theory based on a direct product of the chiral group $ [SU(3)_L\times SU(3)_R]_{global}$ and the hidden group $[SU(3)_V]_{local}$ group is used to incorporate the light vector resonances  in Section 5.1.4. The fields transform under the elements of this group   as
\begin{alignat*}{2}
H_a&\to H_b h_{ba}^\dagger(x)~,&\qquad\qquad \bar H_a&\to h_{ab}(x)\bar H_b~,\\
\xi_L&\to g_L\xi_L h^\dagger(x)~,&\qquad\qquad \xi_L^\dagger &\to h(x)\xi_L^\dagger g_L^\dagger~,\\
\xi_R&\to g_R\xi_L h^\dagger(x)~,&\qquad\qquad \xi_R^\dagger &\to h(x)\xi_R^\dagger g_R^\dagger~,\\
\rho^\mu&\to h(x)\rho^\mu h^\dagger(x)+h(x)\partial^\mu h^\dagger(x)~,
\end{alignat*}
where $g_L\in SU(3)_L$, $g_R\in SU(3)_R$ and $h(x)\in [SU(3)_V]_{local}$. 

\subsubsection{Electromagnetic local gauge transformation}

Under a local electromagnetic gauge transformation the fields transform as
\begin{alignat*}{2}
\xi&\to g_0(x)\xi g_0^\dagger(x)~,&\qquad\qquad\xi^\dagger&\to g_0(x)\xi^\dagger g_0^\dagger(x)~,\nonumber\\
H_a&\to e^{ie_0{\cal Q}^\prime \alpha(x)}H_ag^{\dagger}_{0 ba}(x)~,&\qquad\qquad\bar H_a&\to g_{0ba}(x)\bar H_a e^{-ie_0{\cal Q}^\prime \alpha(x)}\nonumber~,\\
\rho_\mu&\to g_0(x)\rho_\mu g_0^\dagger(x)+g_0(x)\partial_\mu g_0^\dagger(x)~,\\
A_\mu&\to g_0(x)A_\mu g_0^\dagger(x)+g_0(x)\partial_\mu g_0^\dagger(x)~,
\end{alignat*}
 where $g_0=\exp(ie_0{\cal Q}\alpha(x))$ with ${\cal Q}=diag(2/3,-1/3,-1/3)$ and $e_0{\cal Q}^\prime$ is the charge of the heavy quark.

\chapter{The effective weak current for the heavy quark and a light antiquark}

In this appendix the most general effective current $(J_W^{heavy})_{a,\lambda}=\bar q_a\gamma_\lambda(1-\gamma_5)c$ at the order  $(\Delta k/m_H)^0$ and $E/\Lambda_\chi$ is derived following \cite{BFO1,pl4}. The current $J_W^{heavy}$  transforms according to the representation $(\bar 3_L,1_R)$ under the chiral transformation $SU(3)_L\times SU(3)_R$. It is expressed in terms of the field $H_a$ so that the heavy quark symmetry is manifest. The current $J_W^{heavy}$ should be linear in the heavy meson field $H_a$ as the current $\bar q_a\gamma^\mu(1-\gamma_5)c$ is linear in the field $c$.  At the order $(\Delta k/m_H)^0$ and $E/\Lambda_\chi$ there is no derivative on the heavy meson field $H$ and up to one deirivative on the light field $\xi$. A general current with these properties and the $V-A$ structure is given by 
\begin{equation}
\label{5.26}
(J_W^{heavy})_{a,\lambda}=Tr[\gamma_\lambda(1-\gamma_5)H(v)M\xi^\dagger]_a~,
\end{equation}
where a shorthand notation  $H=(H_1,H_2,H_3)$  is used and $H_a$ is defined in (\ref{ha}). The matrix $M$ encorporates the light fields at the order $E/\Lambda_\chi$. The current  $J_W^{heavy}$ transforms according to the representation  $(\bar 3_L,1_R)$, so the matrix $M$ should transform as $M\to U(x)MU^\dagger(x)$ under the chiral transformation. The matrix $M$ can be generaly expanded in  the $4\times 4$ space of the Dirac indeces as 
$$M=A^{\prime\prime}+B^{\prime\prime}\gamma_5+C_\mu^{\prime\prime}\gamma^\mu+D_\mu^{\prime\prime}\gamma^\mu\gamma_5+E_{\mu\nu}^{\prime\prime}\sigma^{\mu\nu}~$$
and the coefficients incorporate the light meson fields. 
Defining 
$$A^\prime=A^{\prime\prime}+B^{\prime\prime}~,\qquad B_\mu^\prime=C_\mu^{\prime\prime}+D_\mu^{\prime\prime}~,\qquad C^{\prime}_{\mu\nu}=iE_{\mu\nu}^{\prime\prime}$$
and
$$j_\alpha=Tr[\gamma_\alpha(1-\gamma_5)H(v)]~,\qquad j=Tr[(1-\gamma_5)H(v)]~,$$
 the current $(J_W^{heavy})_\lambda$ (\ref{2.26}) can be explicitly written as
$$(J_W^{heavy})_\lambda=j_\lambda(A^\prime-v^\mu B^\prime_\mu)\xi^\dagger+j^\mu(B^\prime_\mu v_\lambda+C^\prime_{\mu\lambda}-C^\prime_{\lambda\mu})\xi^\dagger-i\epsilon_\lambda^{~\alpha\mu\nu}j_\alpha(C^\prime_{\mu\nu}+B^\prime_\mu v_\nu)\xi^\dagger+j B^\prime_\lambda u^\dagger~.$$
Renaming the coefficients 
$$A=A^\prime-v^\mu B_\mu^\prime~,\qquad B_{\mu\nu}=C^\prime_{\mu\nu}+B^\prime_\mu v_\nu$$
we get 
$$ (J_W^{heavy})_\lambda=j^\mu(Ag_{\mu\lambda}+B_{\mu\lambda}-B_{\lambda\mu}+i\epsilon_{\mu\lambda\alpha\beta}B^{\alpha\beta})\xi^\dagger~.$$
The coefficients have to transform as $A\to UAU^\dagger$ and $B_{\mu\nu}\to UB_{\mu\nu}U^\dagger$ under the chiral transformation and have to be of the order of $(E/\Lambda_\chi)^0$ or  $E/\Lambda_\chi$. They can be expressed in terms of the operators 
\begin{equation}
O^{(1)}_\mu=\rho_\mu-{\cal V}_\mu~,\qquad O^{(2)}_\mu={\cal A}_\mu~,\qquad O^{(3)}_\mu=\partial_\mu+{\cal V}_\mu~,
\end{equation}
 which are of the order of $E/\Lambda_\chi$ and 
transform as $O^{(i)}_\mu\to UO^{(i)}_\mu U^\dagger$. Due to the relation $O_\mu^{(3)}\xi^\dagger=-O_\mu^{(2)}\xi^\dagger$, the operator $O^{(3)}$ does not lead to an independent term and can be omitted, so
$$A=A_0+v^\mu(A_1 O_\mu^{(1)}+A_2 O_\mu^{(2)})~,\qquad B_{\mu\lambda}=v_\mu(B_1 O_\lambda^{(1)}+B_2 O_\lambda^{(2)})~.$$
In the end we have five parameters $A_{0}$, $A_1$, $A_2$, $B_a$ and $B_2$. Expressing them in terms of a more convenient set \cite{BFO1,pl4}
$$A_0=\tfrac{1}{2}i\alpha~,\quad A_1=\alpha_1-\alpha_2~,\quad A_2=\alpha_4-\alpha_3~, \quad B_1=\alpha_1~,\quad B_2=-\alpha_3$$
the final expression for the effective weak current is
\begin{align}
(J_W^{heavy})_\lambda&=\tfrac{1}{2}i\alpha Tr[\gamma_\lambda(1-\gamma_5)H]\xi^\dagger\\
&-\alpha_1 Tr[(1-\gamma_5)H](\rho-{\cal V})_\lambda\xi^\dagger-\alpha_2Tr[\gamma_\lambda(1-\gamma_5)H]v^\alpha (\rho-{\cal V})_\alpha \xi^\dagger\nonumber\\
&+\alpha_3Tr[(1-\gamma_5)H]{\cal A}_{\lambda}\xi^\dagger+\alpha_4Tr[\gamma_\lambda(1-\gamma_5)H]v^\alpha{\cal A}_\alpha \xi^\dagger\nonumber\\
&+Tr[\gamma^{\delta}(1-\gamma_5)H]\bigl(g_{\delta\lambda}v_\alpha-g_{\delta\alpha}v_\lambda-ig_{\delta\mu}\epsilon^{\mu}_{~\lambda\alpha\beta}v^\beta\bigr)\{\alpha_1(\rho-{\cal V})_\alpha-\alpha_3{\cal A}_{\alpha}\}\xi^\dagger~.\nonumber
\end{align}

\pagestyle{empty}

\newpage

$~$

\vspace{1.5cm}

\centerline{\bf List of Publications}

\vspace{.5cm}

{\it Nonleptonic two-body charmed meson decays in an effective model for their 
semileptonic decays}, B. Bajc, S. Fajfer, R. J. Oakes and S. Prelov\v sek, 
Phys. Rev. D 56 (1997) 7207.

\vspace{0.5cm}

{\it Resonant and nonresonant contributions to the weak $D\to Vl^+l^-$ decays}, 
S. Fajfer, S. Prelov\v sek and P. Singer, 
Phys. Rev. D 58 (1998) 094038.

\vspace{0.5cm}

{\it Long distance contributions in $D\to V\gamma$ decays}, 
S. Fajfer, S. Prelov\v sek and P. Singer, 
Eur. Phys. J. C 6 (1999) 471.

\vspace{0.5cm}

{\it FCNC transitions $c\to u\gamma$ and $s\to d\gamma$ in $B_c\to B_u^*\gamma$ 
and $B_s\to B_d^*\gamma$ decays}, 
S. Fajfer, S. Prelov\v sek and P. Singer, 
Phys. Rev. D 59 (1999) 114003.

\vspace{0.5cm}

{\it The CP violating asymmetry in $B^{\pm}\to M\bar M\pi^{\pm}$ decays}, 
B. Bajc, S. Fajfer, R. J. Oakes, T. N. Pham and S. Prelov\v sek,
Phys. Lett. B 447 (1999) 313.

\vspace{0.5cm}

{\it The controversy in the $\gamma\gamma\to\rho\rho$ process: potential 
scattering or $qq\bar q\bar q$ resonance?}, 
B. Bajc, S. Prelov\v sek and M. Rosina, 
Z. Phys. A 356 (1996) 187.

\vspace{0.5cm}

{\it Can FCNC transition $c\to ul^+l^-$ be seen in $D\to Vl^+l^-$ decays?}, 
 S. Fajfer, S. Prelov\v sek and P. Singer, 
Nucl. Phys. B (Proc. Suppl.) 75B (1999) 141.

\vspace{0.5cm}

{\it Radiative decays of $D$ mesons},
S. Fajfer, S. Prelov\v sek and P. Singer,
 Nucl. Phys. B (Proc. Suppl.) 75B (1999) 138.

\vspace{0.5cm}

{\it Signal for CP violation in $B^{\pm}\to P\bar P \pi^{\pm}$ decays},
B. Bajc, S. Fajfer, R. J. Oakes, T. N. Pham and S. Prelov\v sek,
 Nucl. Phys. B (Proc. Suppl.) 75B 
(1999) 294.

\vspace{0.5cm}

{\it Semileptonic and nonleptonic charmed meson decays in an effective model},  B. Bajc, S. Fajfer, R. J. Oakes and S. Prelov\v sek,
 published in proceedings of  "Lepton-photon interactions", Hamburg  (1997).

\vspace{0.5cm}

{\it Probing $c\to u\gamma$ in $B_c\to B_u^*\gamma$ decay},  
S. Prelov\v sek, S. Fajfer and P. Singer,
published in proceedings of ``Electro-weak interactions and unified theories'', Rencontres de 
Moriond  (1999).

\vspace{0.5cm}

{\it Probing $c\to u\gamma$ in $B_c\to B_u^*\gamma$ decay},  
S. Prelov\v sek, S. Fajfer and P. Singer,
published in proceedings of ``Heavy Flavours 8'', Southampton (1999).


\newpage



\begin{thebibliography}{10}
\bibitem{SM} S. L. Glashow, Nucl. Phys. 22 (1961) 579; S. Weinberg, Phys. Rev. 
Lett. 19 (1967) 1264; A. Salam, {\it Elemetrary particle theory}, edited by N. 
Svartholm, Almqvist and Wiksell, Stockholm (1968).   
\bibitem{CKM} N. Cabibbo, Phys. Rev. Lett. 10 (1963) 531; M. Kobayashi and T. 
Maskawa, Prog. Theo. Phys. 49 (1973) 652.
\bibitem{PDG} C. Caso {\it et al.}, Review of Particle Physics, Eur. Phys. J. C 3 (1998) 1.
\bibitem{SK} Y. Fukuda {\it et al.}, The Super-Kamiokande Coll., Phys. Rev. D 81
(1998) 1562.
\bibitem{cheng.li} T. P. Cheng and L. F. Li, {\it Gauge theory of elementary particle physics},
 Oxford University Press (1984). 
\bibitem{GIM} S. Glashow, J. Iliopoulos and L. Maiani, Phys. Rev. D 2
(1970) 1285; M. K. Gaillard and B. W. Lee, Phys. Rev. D 10 (1974) 897; M. K. Gaillard, B. W. Lee and R. E. Shrock, Phys. Rev. D 13 (1976) 2674. 
\bibitem{rare.kaon} G. Ecker, A. Pich and E. de Rafael, Nucl. Phys. B 291 (1987) 692; A. I. Vainstein, V. I Zakharov, L. Okun and M. Shifman, Journal of Nuclear Physics 24 (1976) 820; L. Bergstr\"om and P. Singer, Phys. Rev. Lett. 55 (1985) 2633; Phys. Rev. D 43 (1991) 1568.
\bibitem{lichard.brem} P. Lichard, {\tt hep-ph/9904265}.
\bibitem{burdman1} G. Burdman, {\tt hep-ph/9811457}.
\bibitem{isidori} G. Isidori, {\tt hep-ph/9902235}.
\bibitem{KTeV} J. Adams {\it et al.}, KTeV Coll., {\tt hep-ex/9806007}.
\bibitem{E787} S. Adler {\it et al.}, E787 Coll., Phys. Rev. Lett. 79 (1997) 2204.
\bibitem{b.s.gamma} R. Ammar {\it et al.}, CLEO Coll., Phys. Rev. Lett. 71 
(1993) 674; M. S. Alam {\it et al.}, CLEO Coll., Phys. Rev. Lett. 74 (1995) 
2885; R. Barate, ALEPH Coll., Phys. Lett. B 429 (1998) 169.
\bibitem{Soares96} J. Soares, Phys. Rev. D 53 (1996) 241.  
\bibitem{GP} E. Golowich and S. Pakvasa, Phys. Rev. D 51 (1995) 1215.
\bibitem{DHT} N. G. Deshpande, X. G. He and J. Trampetic, Phys. Lett. B 367 
(1996) 362.

\bibitem{b.s.gamma.new} T. Skwarnicki, CLEO Coll., presented at XXIXth 
International Conference on High Energy Physics at Vancouver. 
\bibitem{blue} P. F. Harrison and H. R. Quinn, The BABAR Physics Book, SLAC, 
1998.
\bibitem{tc} T. Han and J. L. Hewett, Phys. Rev. D 60 (1999) 074015 ; D. Atwood, L. Reina and A. Soni, Phys. Rev. D 53 (1996) 1199; U. Mahanta and A. Ghosal, Phys. Rev. D 57 
(1998) 1735; V. Barger, K. Hagivara, Phys. Rev. D 37 (1988) 3320; F. Aguila, J. 
A. A. Saavedra and R. Miquel, Phys. Rev. Lett. 82 (1999) 1628; G. Eilam, talk 
given at triangular symposium, Zagreb, June 1999.
\bibitem{CS} T. P. Cheng and M. Sher, Phys. Rev. D 35 (1987) 3484.
\bibitem{BGHP} G. Burdman, E. Golowich, J. Hewett and S. Pakvasa, Phys. Rev. D52 (1995) 6383.
\bibitem{GHMW} C. Greub, T. Hurth, M. Misiak and D. Wyler, Phys. Lett. B 382 
(1996) 415.
\bibitem{KSW} A. Khodjamirian, G. Stoll and D. Wyler, Phys. Lett. B 358 (1995) 129. 
\bibitem{FPS1} S. Fajfer, S. Prelovsek and P. Singer, Eur. Phys. J. C 6 (1999) 471.
\bibitem{FS} S. Fajfer and P. Singer, Phys. Rev. D 56 (1997) 4302.
\bibitem{genova1} S. Fajfer, S. Prelovsek and P. Singer, Nucl. Phys. B (Proc. 
Suppl.) 75B (1999) 138.
\bibitem{CLEO1} D. Asner {\it el al.}, CLEO Coll., Phys. Rev. D 58 (1998) 092001.
\bibitem{FPS3} S. Fajfer, S. Prelovsek and P. Singer, Phys, Rev. D 59 (1999) 
114003.  
\bibitem{moriond} S. Fajfer, S. Prelovsek and P. Singer, to be published in 
proceedings of Recontres de Moriond 1999, {\it Electroweak interactions and 
unified theories}; {\tt hep-ph/9905304}.
\bibitem{southampton} S. Fajfer, S. Prelovsek and P. Singer, to be published in proceedings of Heavy Flavours 8, Southampton, 1999; {\tt hep-ph/9911389}.
\bibitem{cracow}  P. Singer, to be published in 
proceedings of XXXIX Cracow school of Theoretical Physics, Zakopane,  1999; {\tt hep-ph/9911215}.
\bibitem{ISGW1} N. Isgur, D. Scora, B. Grinstein and M. B. Wise, Phys. Rev. D 39 (1989) 799. 
\bibitem{AS} T. M. Aliev and M. Savci, {\tt hep-ph/9908203}.
\bibitem{FPS2} S. Fajfer, S. Prelovsek and P. Singer, Phys. Rev. D 58
(1998) 094038.
\bibitem{genova2} S. Prelovsek, S. Fajfer and P. Singer, Nucl. Phys. B (Proc. 
Suppl.) 75B (1999) 141.  
\bibitem{lattice} I. Montvay and G. Munster, {\it Quantum fields on lattice}, 
Cambridge University Press, Cambridge (1994).
\bibitem{sum.rules} M. Shifman, A. Vanstein and V. Zakharov, Nucl. Phys. B 147 (1979) 385, 488, 519.
\bibitem{GL} J. Gasser and H. Leutwyler, Ann. Phys. (N. Y.) 158 (1984) 142.
\bibitem{HQET1} N. Isgur, M. B. Wise, Phys. Lett. B 232 (1989) 113; 237 (1990) 527.
\bibitem{neubert} M. Neubert, Phys. Rep. 245 (1994) 259.
\bibitem{heavy.chiral} M. B. Wise, Phys. Rev. D 45 (1992) R2188; G. Burdman and J. F. Donoghue, Phys. Lett. B 280 (1992) 287.
\bibitem{BFO1} B. Bajc, S. Fajfer and R. J. Oakes, Phys, Rev. D 53 (1996) 4957.
\bibitem{BFOP} B. Bajc, S. Fajer, R. J. Oakes and S. Prelovsek, Phys, Rev. D 56 (1997) 7207.
\bibitem{dmix} L. Wolfenstein, Phys. Lett. B 164 (1985) 170; J. F. Donoghue, E. Golowich, B. R. Holstein and J. Trampetic, Phys. Rev. D 33 (1986) 179; H. Georgi, Phys. Lett. B 297 (1992) 353; T. Ohl, G. Ricciardi and E. H. Simmons, Nucl. Phys. B 403 (1993) 605; E. Golowich and A. A. Petrov, Phys. Lett. B 427 (1998) 172.
\bibitem{IL}  T. Inami and C. S. Lim, Prog. Theor. Phys. 65 (1981) 297.
\bibitem{KP} Q. H. Kim and X. Y. Pham, Phys. Rev. D 61 (2000) 013008.
\bibitem{SVZ} M. A. Shifman, A. I. Vainstein and V. I. Zakharov, Phys. Rev. D 18 (1978) 2583. 
\bibitem{willson} K. G. Willson, Phys. Rev. 179 (1969) 1499.
\bibitem{buras} A. J. Buras and M. Linder, Heavy Flavours II: Chapter 2, Advanced Series on Directions in High Energy Physics, Vol. 15, World Scietific Publishin Co. (1998).
\bibitem{grinstein} B. Grinstein, R. Springer and M. B. Wise, Nucl. Phys. B 339 (1990) 269.
\bibitem{GHW} C. Greub, T. Hurth and D. Wyler, Phys. Rev. D 54 (1996) 3350.
\bibitem{pakvasa} S. Pakvasa, {\tt hep-ph/9705397}.
\bibitem{GW} S. Glashow and S. Weinberg, Phys. Rev. D 15 (1977) 1958.
\bibitem{CNS} J. L. D. Cruz, J. J. G. Nava and G. L. Castro, Phys. Rev. D 51 (1995) 5263.
\bibitem{burdman} G. Burdman, FERMILAB-Conf-95/281-T; L. Hsll and S. Weinberg, Phys. Rev. D 48 (1993) 979.
\bibitem{CMM} G. L. Castro, R. Martinez and J. H. Munoz, Phys. Rev. D 58 (1998)  033003.
\bibitem{SUSY}  J. P. Deredinger, {\it Globaly supersymmetric theories in four and two dimensions}, ETH-TH/90-21; S. Martin, {\tt hep-ph/9709356}; J. Wess and J. Bagger, {\it Supersymmetry and Supergravity}, Princeton University Press  (1983); H. P. Nilles, Phys. Rep. 110 (1984) 1; M. F. Sohnius, Phys. Rep. 128 (1985) 39.
\bibitem{CM} S. Coleman and J. Mandula, Phys. Rev. 159 (1967) 1251; R. Haag, J. Lopuszanski and M. Sohnius, Nucl. Phys. B 88 (1975) 257.
\bibitem{BGM} I. Bigi, F. Gabbiani and A. Masiero, Z. Phys. C 48 (1990) 633.
\bibitem{duncan} M. J. Duncan, Nucl. Phys. B 221 (1983) 285.
\bibitem{BHLP} K. S. Babu, X.-G. He, X.-Q. Li and S. Pakvasa, Phys. Lett. B 205 (1988) 540.
\bibitem{hewett} J. L. Hewett, {\tt hep-ph/9409379} and {\tt hep-ph/9505246}.
\bibitem{eeg} J. O. Eeg, Z. Phys. C 46 (1990) 665.
\bibitem{singer.baryon} P. Singer, Nucl. Phys. B (Proc. Suppl.) 50 (1996) 202. 
\bibitem{VMD} Y. Nambu, Phys. Rev. 106 (1957) 1366; W. R. Frazer and J. R. Fulco, Phys. Rev. Lett. 2 (1959) 2; J. J. Sakurai, Ann. Phys. (N. Y.) 11 (1960) 1; M. Gell-Mann and F. Zachraisen, Phys. Rev. 124 (1961) 953. 
\bibitem{lichard} P. Lichard, Phys. Rev. D 55 (1997) 5385.
\bibitem{buras1} A. J. Buras, Nucl. Phys. B 434 (1995) 606.
\bibitem{buras:neubert} A. J. Buras and M. Linder, Heavy Flavours II: Chapter 4, Advanced Series on Directions in High Energy Physics, Vol. 15, World Scietific Publishin Co. (1998).
\bibitem{BSW} M. Bauer, B. Stech and M. Wirbel, Z. Phys. C 34 (1987) 103. 
\bibitem{neubert2} M. Neubert, Nucl. Phys. Proc. Suppl. 64 (1998) 474, {\tt hep-ph/9707368}.
\bibitem{aiphen} references [11]-[16] of \cite{buras1}; H.-Y. Cheng, K.-C. Yang, Phys. Rev. D 59 (1999) 092994.  
\bibitem{NRSX} M. Neubert, V. Rieckert, B. Stech and Q. P. Xu, in: Heavy Flavours, eds. A. J. Buras and M. Linder (World Scientific, Singapore, 1992) p. 286.
\bibitem{Soares} J. M. Soares, Phys. Rev. D  54 (1996) 6837.
\bibitem{ISGW} N. Isgur, D. Scora, B. Grinstein and M. B. Wise, Phys. Rev. D 
 39 (1989) 799.
\bibitem{IW1} N. Isgur and M. B. Wise, Phys. Rev. D 42 (1990) 2388.
\bibitem{ISGW2} D. Scora and N. Isgur, Phys. Rev. D  52 (1995) 2783.
\bibitem{itzykson} C. Itzykson and J. B. Zuber, {\it Quantum field theory}, 
Mc-Graw-Hill, New York (1985).
\bibitem{CDF} F. Abe {\it et al.}, CDF Coll., Phys. Rev. Lett. 81 (1998)  2432; Phys. Rev. D 58 (1998) 112004.
\bibitem{AIP} T. M. Aliev, E. Iltan and N. K. Pak, Phys. Lett. B 329 (1994) 123.
\bibitem{AS1} T. M. Aliev and M. Savci, J. Phys. G 24 (1998) 2223. 
\bibitem{casalbuoni} R. Casalbuoni, A. Deandrea, N. Di Bartolomeo, R. Gatto, F. Feruglio and G. Nardulli, Phys. Rep. 281 (1997) 145.
\bibitem{cheung} K. Cheung, Phys. Rev. Lett 71 (1993) 3413.

\bibitem{hidden} M. Bando, T. Kugo, S. Uehara, K. Yamawaki and T. Yanagida, Phys. Rev. Lett 54 (1985) 1215; M. Bando, T. Kugo and K. Yamawaki, Nucl. Phys. B259 (1985) 493; Phys. Rep. 164 (1988) 217; Prog. Theor. Phys. 73 (19985) 1541.
\bibitem{HQET2} H. Georgy, Phys. Lett. B 240 (1990) 447.
\bibitem{SS} J. Schechter and A. Subbaraman, Phys. Rev. D 48 (1993) 332.

\bibitem{CWZ} S. Coleman, J. Wess and B. Zumino, Phys. Rev.  177 (1969) 2239.
\bibitem{chiral} J. Gasser and H. Leutwyler, Ann. Phys. 158 (1984) 142; Nucl. Phys. 250 (1985) 465.
\bibitem{eta} Phys. Rev. Lett. 64 (1990) 172; T. Hatsuda and T. Kunihiro, Phys. Rep. 247 (1994) 221.
\bibitem{heavy.chiral2} T.-M. Yan, H.-Y. Cheng, C.-Y. Cheung, G.-L. Lin, Y. C. Lin and H.-L. Yu, Phys. Rev. D 46 (1992) 1148.
\bibitem{heavy.chiral3} G. Burdman and J. F. Donoghue, Phys. Rev. Lett. 68 (1992) 2887.  
\bibitem{hidden1} R. Casalbuoni, A. Deandrea, N. D. Bartolomeo, R. Gatto, F. Feruglio and G. Nardulli, Phys. Lett. B 292 (1992) 371.
\bibitem{wise1} M. B. Wise, Lectures given at CCAST Symposium on particle physics at Fermi Scale, {\tt hep-ph/9306277}.
\bibitem{VVP} A. Bramon, A. Grau and G. Pancheri, Phys. Lett. B 344  (1995) 240; M. Hashimoto, Phys. Rev. D 54 (1996) 5611. 
\bibitem{CFN} P. Colangelo, F. De Fazio and G. Nardulli, Phys. Lett. B 316 (1993) 555.
\bibitem{BFO2} B. Bajc, S. Fajfer and R. J. Oakes, Phys. Rev. D 51 (1995) 2230.
\bibitem{WZ} J. Wess and B. Zumino, Phys. Lett. B 37 (1971) 95.
\bibitem{witten} E. Witten, Nucl. Phys. B 223 (1983) 422.
\bibitem{PDG96} R. M. Barnett {\it et al.}, Review of Particle Physics,  Phys. Rev. D 54 (1996) 1.
\bibitem{pl4} B. Bajc, S. Fajfer, R. J. Oakes and T. N. Pham, Phys. Rev. D 58 (1998) 054009.
\bibitem{VRI} M. Luke and A. Manohar, Phys. Lett. B 286 (1992) 348.
\bibitem{grinstein1} C. G. Boyd and B. Grinstein, Nucl. Phys. B 442 (1995) 205.
\bibitem{casalbuoni1}  R. Casalbuoni, A. Deandrea, N. D. Bartolomeo and R. Gatto, Phys. Lett. B 299 (1993) 139. 
\bibitem{burchat} J. D. Richman and P. R. Burchat, Rev. Mod. Phys. 67 (1995) 893.
\bibitem{f.lattice} C. Bernard {\it et al.}, MILC Coll., {\tt hep-lat/9909121}; Abada {\it et al.}, {\tt hep-lat/9910021}.
\bibitem{stewart} I. W. Stewart, Nucl. Phys. B 529 (1998) 62.
\bibitem{nonleptonic} E. Braaten, R. J. Oakes and Sze-Man Tse, Int. J. Mod. Phys. A 5 (1990) 2737; A. N. Kamal and Q. P. Xu, Phys. Rev. D  49 (1994)  1526; A. N. Kamal, Q.P. Xu and A. Czarnecki, Phys. Rev.  D 49 (1994) 1330; A. N. Kamal and T. N. Pham, Phys. Rev.  D   50 (1994) 6849; R. C. Verma, A. N. Kamal and M. P. Khanna, Z. Phys. C  65 (1995)  
 255;  A.N. Kamal and A. B. Santra, Z. Phys. C   71 (1996)  101;  A. N. Kamal, A. B. Santra, T. Uppal, R.C. Verma, 
Phys. Rev. D   53 (1996) 2506;  M. Gourdin, A. N. Kamal, Y. Y. Keum, X. Y. Pham,  
Phys. Let. B  333 (1994) 507;  M. Gourdin, A. N. Kamal, Y. Y. Keum, X. Y. Pham,  
Phys. Let. B   339 (1994) 173; F. Buccella, M. Lusignoli, G. Miele, A. Pugliese, 
Z. Phys. C  55 (1992) 243;  F. Buccella, M. Lusignoli, G. Mangano, G. Miele, A. Pugliese, 
and P. Santorelli, Phys. Lett. B   302 (1993)  319;  F. Buccella, M. Lusignoli, G. Miele, A. Pugliese, 
and P. Santorelli, Phys. Rev. D   3478 (1995)  319; F. Close and H. Lipkin, Phys. Lett. B 405 (1997) 157; M. Wirbel, B. Stech and M. Bauer,
Z. Phys. C  29 (1985) 637; S. Fajfer and J. Zupan, Int. J. Mod. Phys. A 14 (1999) 4161; F. Buccella, M. Lusignoli, A. Pugliese, 
Phys. Lett. B    379 (1996) 249;  Y. Hinchliffe and T. Keading, Phys. Rev. D 54  (1996) 914.
\bibitem{kamal} El hassan El aaoud and A. N. Kamal, Phys. Rev. D 59 (1999) 114013. 
\bibitem{radiative.exp1} D. M. Asner {\it et al.}, CLEO Coll., Phys. Rev. D 58 (1998) 092001; M. Selen, Bull. Am. Phys. Soc. 39 (1994) 1147.
\bibitem{radiative.exp2} S. Ratti: private communication on experiment E832-FOCUS.
\bibitem{vll.exp} A. Freyberger {\it et al.}, CLEO Coll., Phys. Rev. Lett. 76 (1996) 3065, ibid. 77 (1996) 2147; K. Kodama {\it et al.}, E653 Coll, Phys. Lett. B 345 (1995) 85; P. Hass {\it et al.}, CLEO Coll.,  Phys. Rev. Lett. 60 (1988) 1614.
\bibitem{schwartz} A. J. Schwartz, Mod. Phys. Lett. A 8 (1993) 967.
\bibitem{gamma.quark} H.-Y. Cheng {\it et al.}, Phys. Rev. D 51 (1995) 1199; P. Asthana and A. N. Kamal, Phys. Rev. D 43 (1991) 278, R. F. Lebed, {\tt hep-ph/9908414}.
\bibitem{BFO3} B. Bajc, S. Fajfer and R. J. Oakes, Phys. Rev. D 54 (1996) 5883.
\bibitem{pll.exp} E. M. Aitala {\it et al.}, E791 Coll., Phys. Lett. B 462 (1999) 401.
\bibitem{singer} P. Singer and D.-X. Zhang, Phys. Rev. D 55 (1997) R1127.
\bibitem{veltman} M. Veltman, {\it Diagrammatica}, Cambridge University Press, Cambridge, (1994). 
\bibitem{weinbergII} S. Weinberg, {\it The Quamtum Theory of fields}, Vol. II, Cambridge University Press (1996).
\bibitem{bilenky} S. M. Bilenky, {\it Introduction to Feynman Diagrams and Electroweak Interaction Physics}, Editions Frontiers (1994).
\bibitem{higgs} J. F. Gunion, H. E. Haber, G. L. Kane and S. Dawson, {\it The Higgs Hunter's Guide}, Addison-Wesley, New York (1990).



\end{thebibliography}
\end{document}